\documentclass[11pt,english,a4paper]{article}
\pdfoutput=1
\usepackage{amsfonts} 
\usepackage{amsmath}
\usepackage{amsthm}
\usepackage{graphicx}
\usepackage{amssymb}
\usepackage{latexsym,wasysym}
\usepackage{mathrsfs}
\usepackage{slashed}
\usepackage{stmaryrd}
\usepackage{cite}

\usepackage{color}
\definecolor{lightblue}{rgb}{0,0.25,0.75}
\definecolor{dullblue}{rgb}{0.15,0.25,0.45}
\definecolor{darkblue}{rgb}{0,0,0.55}
\definecolor{bluegreen}{rgb}{0,0.3,0.35}
\definecolor{darkgreen}{rgb}{0,0.35,0}
\definecolor{darkred}{rgb}{0.5,0,0}

\usepackage[linktocpage=true]{hyperref}
\hypersetup{
colorlinks=true,
citecolor=darkblue,
linkcolor=black,
urlcolor=darkblue,
pdfauthor={},
pdftitle={},
breaklinks=trueBorsatz:2012:all-loop
}

\usepackage{titlesec}
\titlespacing\subsection{0pt}{10pt plus 4pt minus 2pt}{8pt plus 2pt minus 2pt}
\titlespacing\subsubsection{0pt}{10pt plus 4pt minus 2pt}{8pt plus 2pt minus 2pt}
\titlespacing\paragraph{0pt}{6pt plus 4pt minus 2pt}{10pt plus 2pt minus 2pt}

\newcommand{\nc}{\newcommand}

\nc{\p}{\partial}

\nc{\ot}{\otimes}
\nc{\op}{\oplus}
\nc{\ol}{\overline}
\nc{\un}{\underline}

\nc{\SL}{{\scriptstyle \ms{L}}}
\nc{\SR}{{\scriptstyle \ms{R}}}
\nc{\sL}{{\scriptscriptstyle \ms{L}}}
\nc{\sR}{{\scriptscriptstyle \ms{R}}}
\nc{\SD}{{\scriptstyle \ms{D}}}
\nc{\sD}{{\scriptscriptstyle \ms{D}}}
\nc{\SN}{{\scriptstyle \ms{N}}}
\nc{\sN}{{\scriptscriptstyle \ms{N}}}
\nc{\ST}{{\scriptstyle \ms{T}}}
\nc{\sT}{{\scriptscriptstyle \ms{T}}}

\nc{\SB}{{\scriptstyle \ms{B}}}
\nc{\sB}{{\scriptscriptstyle \ms{B}}}
\nc{\SA}{{\scriptstyle \ms{A}}}
\nc{\sA}{{\scriptscriptstyle \ms{A}}}
\nc{\SC}{{\scriptstyle \ms{C}}}
\nc{\sC}{{\scriptscriptstyle \ms{C}}}

\nc{\Sa}{{\scriptstyle \ms{a}}}
\nc{\sa}{{\scriptscriptstyle \ms{a}}}

\newcommand{\tI}{{\scriptscriptstyle{I}}}

\newcommand{\tK}{{\scriptscriptstyle{K}}}

\nc{\mc}{\mathcal}
\nc{\ms}{\mathsf}
\nc{\mf}{\mathfrak}
\nc{\mb}{\mathbf}
\nc{\bb}{\mathbb}
\nc{\mr}{\mathrm}

\nc{\al}{\alpha}
\nc{\bet}{\beta}
\nc{\eps}{\epsilon}
\nc{\del}{\delta}
\nc{\ga}{\gamma}
\nc{\Ga}{\Gamma}
\nc{\ka}{\kappa}
\nc{\la}{\lambda}
\nc{\om}{\omega}
\nc{\si}{\sigma}
\nc{\Si}{\Sigma}
\nc{\Ups}{\upsilon}
\nc{\vphi}{\varphi}

\nc{\id}{\mathrm{id}}
\nc{\gr}{\mathrm{gr}}

\nc{\Ug}{U\mathfrak{g}}
\nc{\Ub}{U\mathfrak{b}}

\nc{\Hk}{\mathsf{H}}
\nc{\ombH}{\overline{\mathbf{H}}}

\nc{\ud}{\underline}
\nc{\tl}{\tilde}
\nc{\wt}{\widetilde}
\nc{\wh}{\widehat}

\nc{\End}{\mathrm{End}}
\nc{\Ext}{\mathrm{Ext}}
\nc{\Hom}{\mathrm{Hom}}
\nc{\Ima}{\mathrm{Image}}
\nc{\Ind}{\mathrm{Ind}}
\nc{\Ker}{\mathrm{Ker}}
\nc{\RHom}{\mathrm{RHom}}
\nc{\Sym}{\mathrm{Sym}}

\nc{\C}{\mathbf{C}}
\nc{\N}{\mathbf{N}}
\nc{\Z}{\mathbf{Z}}

\nc{\lan}{\langle}
\nc{\ran}{\rangle}

\nc{\mfg}{\mf{g}}
\nc{\mfa}{\mf{a}}

\nc{\eqa}[1]{\begin{align}#1\end{align}}
\nc{\eqn}[1]{\begin{align*}#1\end{align*}}
\nc{\eq}[1]{\begin{equation}#1\end{equation}}
\nc{\spl}[1]{\begin{equation}\begin{aligned}#1\end{aligned}\end{equation}}
\nc\el{\nonumber\\}
\nc\nn{\nonumber}

\nc{\qu}{\quad}
\nc{\qq}{\qquad}

\nc{\ket}[1]{\,|#1\rangle}

\nc{\ident}[1]{\vspace{0.1cm}\noindent{\bf #1}}

\usepackage{color}
\usepackage[usenames,dvipsnames]{xcolor}
\nc{\red}{\color{red}}
\nc{\blu}{\color{blue}}

\nc{\ttm}[4]{\left( \begin{array}{cc} #1 & #2 \\ #3 & #4 \end{array} \right)}

\usepackage{titletoc}
\usepackage{tocloft}

\begin{document}

\pagestyle{empty}

\begin{flushright} 
\small{DMUS-MP-15/07}
\end{flushright}


\vspace{0.2cm}

\begin{center}

{\bf \Large {Integrable open spin-chains in AdS$_{3}$/CFT$_{2}$ correspondences}}

\vspace{0.75cm}

{\large Andrea Prinsloo, Vidas Regelskis and Alessandro Torrielli} 



\vspace{0.3cm}

\small{ \it Department of Mathematics, \\ University of Surrey, \\ Guildford, GU2 7XH, United Kingdom} 

\vspace{0.3cm}
\small{ \texttt{a.prinsloo, v.regelskis, a.torrielli  @surrey.ac.uk} }

\vspace{1.5cm}

\end{center}

\begin{abstract}

We study integrable open boundary conditions for $\mf{d}(2,1;\alpha)^{2}$ and $\mf{psu}(1,1|2)^{2}$ spin-chains.  Magnon excitations of these open spin-chains are mapped to massive excitations of type IIB open superstrings ending on D-branes in the $AdS_{3}\times S^{3} \times S^{3} \times S^{1}$ and $AdS_{3}\times S^{3} \times T^{4}$ supergravity geometries with pure R-R flux.  We derive reflection matrix solutions of the boundary Yang-Baxter equation which intertwine representations of a variety of boundary coideal subalgebras of the bulk Hopf superalgebra.  
Many of these integrable boundaries are matched to D1- and D5-brane maximal giant gravitons.

\end{abstract}

\newpage

\pagestyle{plain}

\setcounter{page}{0}
\pagenumbering{arabic}

\tableofcontents

\bigskip

\numberwithin{equation}{section}

\numberwithin{table}{section}


\part{Introduction}  \label{part1}

Integrability has been a remarkable discovery in AdS/CFT, leading to the matching of an infinite tower of conserved quantities on the gauge and gravity sides of these dualities \cite{Frolov:2009,Review}.  In the canonical AdS$_{5}$/CFT$_{4}$ \cite{Maldacena:1997}, the infinite-dimensional superalgebra underlying integrability is a Yangian symmetry \cite{Drinfeld, Mackay04} with level-0 Lie superalgebra a central extension of $\mf{su}(2|2)$, denoted $\mf{su}(2|2)_{\ms{c}}$. 
The universal enveloping algebra of  $\mf{su}(2|2)_{\ms{c}}$ is endowed with the structure of a Hopf algebra.
Massive excitations of the worldsheet of a closed IIB superstring on $AdS_{5}\times S^{5}$ map to $\mf{su}(2|2)_{\ms{c}}$-symmetric magnon excitations of a closed spin-chain built from states in a representation of $\mf{psu}(2,2|4)$, which is the superisometry algebra of $AdS_{5}\times S^{5}$. The integrable $S$-matrix describing two-magnon scattering is identified with the $R$-matrix of the underlying superalgebra \cite{Beisert:2008}.  Three-magnon scattering factorizes into a succession of two-magnon scattering processes -- a statement of integrability encoded in the Yang-Baxter equation.   Massive excitations of an open IIB superstring ending on a D-brane in $AdS_{5}\times S^{5}$ map to similar magnon excitations of a $\mf{psu}(2,2|4)$ open spin-chain with a distinguished boundary site \cite{HM:2006}.  The symmetry of the boundary is determined by the superisometries preserved by the D-brane which are contained in the bulk magnon symmetry algebra $\mf{su}(2|2)_{\ms{c}}$.  This boundary Lie algebra may be extended to a coideal subalgebra of the bulk Hopf superalgebra.  The scattering of a single magnon off an integrable boundary is described by a boundary $S$-matrix -- the matrix part of which is the reflection $K$-matrix \cite{HM:2007,CY:2008,Murgan:2008,CRY:2011}.  Two-magnon reflections factorize into a succession of single-magnon reflections and bulk two-magnon scattering processes -- with this boundary integrability encoded in the boundary Yang-Baxter equation, also called the reflection equation \cite{Sk:1988}.  The $R$-matrix intertwines representations of the bulk Hopf superalgebra, while the $K$-matrix intertwines representations of a boundary coideal subalgebra of this bulk superalgebra.

Extensive studies, which were initiated in \cite{Babichenko:2010}, have been made of integrability in AdS$_{3}$/CFT$_{2}$.  The dual field theories have been the subject of recent interest \cite{Strominger:2004,Pakman:2009,SSS-fieldtheory:2014,Tong:2014}, but much still remains to be understood.  On the string theory side, the IIB supergravity backgrounds $AdS_{3}\times S^{3}\times M^{4}$, with $M^{4}=S^{3\prime} \times S^{1}$ or $T^{4}$, are known \cite{Cowdall:1998,Gauntlett:1998}.  Both backgrounds are half-BPS, preserving eight left- and eight right-moving supersymmetries, and have a combination of NS-NS or R-R 3-form flux -- we focus here exclusively on the case of pure R-R flux. The $AdS_{3}\times S^{3}\times S^{3\prime} \times S^{1}$ supergravity geometry has $AdS_{3}$ radius $L$, and $S^{3}$ and $S^{3\prime}$ radii $R$ and $R^{\prime}$, which must satisfy \cite{Cowdall:1998}
\[
\frac{1}{R^{2}} + \frac{1}{{R^{\prime}}^{2}} = \frac{1}{L^{2}},
\]
implying $R = L \sec{\beta}$ and  $R^{\prime} = L \csc{\beta}$.  Here $\alpha \equiv \cos^{2}{\beta}$ is a parameter related to the relative size of the 3-spheres. 
The bosonic isometry algebra is
\[ \mf{so}(2,2) \oplus \mf{so}(4) \oplus \mf{so}(4)^{\prime} \oplus \mf{u}(1) \sim 
\left[ \mf{su}(1,1) \oplus \mf{su}(2)\oplus \mf{su}(2)^{\prime} \right]_{\sL} \oplus \left[ \mf{su}(1,1) \oplus  \mf{su}(2)\oplus \mf{su}(2)^{\prime} \right]_{\sR} \oplus \mf{u}(1), \]
which splits into two copies  (left and right) of the bosonic subalgebra $\mf{su}(1,1) \oplus \mf{su}(2) \oplus \mf{su}(2)^{\prime}$
of the Lie superalgebra $\mf{d}(2,1;\al)$. The superisometry algebra of $AdS_{3}\times S^{3}\times S^{3\prime} \times S^{1}$ is $\mf{d}(2,1;\al)^{2}  \oplus \mf{u}(1)$.  The size of the 3-sphere $S^{3\prime}$ becomes infinite in the $\alpha \to 1$ limit and a compactification\footnote{Here we ignore complications which arise from neglecting winding modes.} of the  resulting $\mathbb{R}^{3}$ gives the IIB supergravity background $AdS_{3}\times S^{3}\times T^{4}$.  The radii of $AdS_{3}$ and $S^{3}$, denoted $L$, are now the same.  The bosonic isometry algebra
\[ \mf{so}(2,2) \oplus \mf{so}(4) \oplus \mf{u}(1)^{4}  \sim 
\left[ \mf{su}(1,1) \oplus  \mf{su}(2) \right]_{\sL} \oplus \left[ \mf{su}(1,1) \oplus  \mf{su}(2) \right]_{\sR}
 \oplus \mf{u}(1)^{4}  \]
contains two copies of the bosonic algebra $ \mf{su}(1,1) \oplus \mf{su}(2)$ of the Lie superalgebra $\mf{psu}(1,1|2)$. The superisometry algebra of $AdS_{3}\times S^{3}\times T^{4}$ is $\mf{psu}(1,1|2)^{2} \oplus \mf{u}(1)^{4}$. 

Massive excitations of the worldsheet of a closed IIB superstring on $AdS_{3}\times S^{3}\times S^{3\prime} \times S^{1}$ map to 
$\mf{su}(1|1)^{2}_{\ms{c}}$-symmetric magnon excitations of a closed, alternating spin-chain built from states in two representations of 
$\mf{d}(2,1;\al)^{2}$ at odd and even sites.  There is symmetry enhancement in the $\alpha \to 1$ limit.  Massive worldsheet excitations of a closed IIB superstring on $AdS_{3}\times S^{3}\times T^{4}$ map to $[\mf{psu}(1|1)^{2} \op \mf{u}(1)]^{2}_{\ms{c}}$-symmetric magnon excitations of a closed, homogeneous spin-chain built from states in a representation of $\mf{psu}(1,1|2)^{2}$.  The universal enveloping algebras of $\mf{su}(1|1)^{2}_{\ms{c}}$ and $[\mf{psu}(1|1)^{2} \op \mf{u}(1)]^{2}_{\ms{c}}$ can be endowed with Hopf algebra structures. Integrable $S$-matrices describing two-magnon scattering were derived in \cite{BSS13,BOSST13} (see \cite{Sfondrini:2014} for a recent review and \cite{David:2008yk,David:2010yg,AB:2013} for early work). 
The $R$-matrix of the $\mf{psu}(1,1|2)^{2}$ spin-chain is essentially two copies of the $R$-matrix of the $\mf{d}(2,1;\al)^{2}$ spin-chain.
A variety of results for the string sigma model were obtained in \cite{Rughoonauth:2012qd, Sundin:2012gc, Abbott:2012dd, Beccaria:2012kb, Beccaria:2012pm, Sundin:2013ypa, Bianchi:2013nra, Engelund:2013fja, Abbott:2013ixa, Sundin:2013uca,  Bianchi:2014rfa, Wulff:2015mwa}, and a proposal for the dressing phases was put forward in \cite{Borsato:2013hoa} (see also \cite{Abbott:2014pia}). The scattering of massless excitations of the superstring worldsheet was considered in \cite{SST13,Lloyd:2013wza,Borsato:2014,BSS14,Abbott:2014rca}.  

Integrability manifests here in the form of infinite-dimensional Yangian symmetries.  Hopf algebra structures in AdS$_5$/CFT$_4$ were described in \cite{GH06,PST06}, and the full $\mf{su}(2|2)_\ms{c}$ Yangian symmetry was introduced in \cite{Be:2006} and further explored in \cite{ST:2009,BL:2014}. Yangian symmetries in AdS$_{3}$/CFT$_{2}$ were explored in \cite{BOSST13,PTW14}, but only recently fully described in \cite{Re}. Since the representation theory of $\mf{su}(1|1)^2_\ms{c}$ is relatively simple, with massive excitations being vectors in a 2-dimensional atypical (short) representation and the $R$-matrix intertwining two 4-dimensional typical (long) representations, the Yangian is not needed to obtain the $R$-matrix (see \cite{ALT:2009} for the situation in AdS$_5$/CFT$_4$). However, it is necessary to know the full Yangian symmetry to construct the Algebraic Bethe Ansatz equations. The Bethe equations for AdS$_3$/CFT$_2$ were proposed in \cite{Sax:2011} and later derived using the coordinate method in   \cite{Borsatz:2012:all-loop}.
Boundary Yangian symmetry for AdS$_5$/CFT$_4$ has undergone an extensive study in \cite{DS:2009,AN:2010,MR:2010,MR:2012,Pa:2011,MR:2011} and the boundary Bethe equations were constructed in \cite{CRY:2011,Ne:2009,Ga:2009,CY:2010:Bethe,Nepomechie-et-al:2015}. The natural next step in the exploration of integrability and the spectral problem for open superstrings in AdS$_{3}$/CFT$_{2}$ is to find the boundary Yangian symmetries and hence derive the Bethe equations.

A comprehensive study of {\it open spin-chains with integrable boundaries} in AdS$_{3}$/CFT$_{2}$ is presented in this paper. The $\mf{d}(2,1;\al)^{2}$ and $\mf{psu}(1,1|2)^{2}$ open spin-chains map to open IIB superstrings ending on D-branes in $AdS_3 \times S^3 \times S^{3\prime} \times S^1$ and $AdS_{3}\times S^{3}\times T^{4}$. In particular, half- and quarter-BPS maximal D1- and D5-brane giant gravitons \cite{Raju-et-al:2008,Janssen:2005,Prinsloo:2014} provide a variety of integrable boundaries.  We derive reflection matrices which describe single-magnon scattering off singlet and vector boundary states. As for the bulk $R$-matrix, the $K$-matrices of the $\mf{psu}(1,1|2)^{2}$ spin-chain can be built from two $K$-matrices of the $\mf{d}(2,1;\al)^{2}$ spin-chain. In the $\mf{d}(2,1;\al)^{2}$ case, the reflection matrices intertwine representations of totally supersymmetric, half-supersymmetric and non-supersymmetric boundary Lie algebras (symmetries of the D-branes), which can be extended to coideal subalgebras of the bulk Hopf superalgebra.  
Several of our reflection matrices coincide with certain $\mf{su}(1|1)$ subsectors of the reflection matrices of $\mf{psu}(2,2|4)$ open spin-chains. These map to open IIB superstrings ending on D3-brane $Y=0$ and $Z=0$ giant gravitons \cite{HM:2007} and D7-branes \cite{CY:2008} in $AdS_5 \times S^5$.  
We uncover novel hidden boundary symmetries of a chiral reflection matrix with a non-supersymmetric boundary Lie algebra -- these have no known analogue in  AdS$_{5}$/CFT$_{4}$. We also derive an achiral reflection matrix for a non-supersymmetric boundary Lie algebra generated by the magnon Hamiltonian.

The structure of this paper is as follows:  D1- and D5-brane maximal giant gravitons and their symmetries are described in Part \ref{part2}. Chapters \ref{part2-sec1} and \ref{part2-sec2} therein focus on maximal giant gravitons in $AdS_3 \times S^3 \times {S^{\prime}}^{3} \times S^1$ and $AdS_{3}\times S^{3}\times T^{4}$, respectively.  Part \ref{part3} describes the $\mf{d}(2,1;\al)^{2}$ closed and open spin-chains.  Chapter \ref{part3-sec1} contains a review of the $\mf{d}(2,1;\al)^{2}$ closed spin-chain and its $R$-matrices. Here we choose a different frame for the $\mf{U}$-deformation of the bulk Hopf superalgebra from the one used in \cite{BSS13}; this frame is more convenient for the boundary scattering theory. Chapter \ref{part3-sec2} presents our novel results for the $\mf{d}(2,1;\al)^{2}$ open spin-chain. We derive various $K$-matrices and describe the associated boundary coideal subalgebras.  The $\mf{psu}(1,1|2)^{2}$ closed and open spin-chains are the subject of  Part \ref{part4}. Chapter \ref{part4-sec1} contains a brief review of the $\mf{psu}(1,1|2)^{2}$ closed spin-chain and its $R$-matrices.  Our new results concerning $\mf{psu}(1,1|2)^{2}$ open spin-chains, $K$-matrices and boundary coideal subalgebras are presented in Chapter \ref{part4-sec2}.  Concluding remarks are contained in Part \ref{part5}.  Appendix \ref{app:A} states our spinor conventions.  Appendix \ref{app:B} describes the relevant representation theory of $\mf{d}(2,1;\al)$ and $\mf{psu}(1,1|2)$. Appendix \ref{app:C} shows various useful expressions relating to the $SO(2,2)$ and $SO(4)$ bosonic isometry groups of the supergravity backgrounds.


\part{Maximal giant gravitons}  \label{part2}


\section{Maximal giant gravitons on $AdS_{3} \times S^{3} \times S^{3\prime} \times S^{1}$}  \label{part2-sec1}

We start by giving the details of the type IIB supergravity background $AdS_{3} \times S^{3} \times S^{3\prime} \times S^{1}$ with pure R-R 3-form flux, and
describe D1- and D5-brane maximal giant gravitons based on \cite{Prinsloo:2014}.

\subsection{$AdS_{3} \times S^{3} \times S^{3\prime} \times S^{1}$ with pure R-R flux}

\paragraph{IIB supergravity solution.}

The $AdS_{3} \times S^{3} \times S^{3\prime} \times S^{1}$ background has the metric
\begin{eqnarray} \label{metric}
\nonumber && \hspace{-0.65cm} ds^{2} = 
L^{2} \left( - \cosh^{2}{\rho} \hspace{0.1cm} dt^{2} + d\rho^{2} + \sinh^{2}{\rho} \hspace{0.1cm} d\varphi^{2} \right)
\, + \, R^{2}  \left( d\theta^{2} + \cos^{2}{\theta} \, d\chi^{2} + \sin^{2}{\theta} \, d\phi^{2} \right) \hspace{0.5cm} \\
&& \hspace{-0.725cm} \hspace{0.675cm} \, + \, \hspace{0.05cm} {R^{\prime}}^{2} \left( d{\theta^{\prime}}^{2} + \cos^{2}{\theta^{\prime}} \, d{\chi^{\prime}}^{2} + \sin^{2}{\theta^{\prime}} \, d{\phi^{\prime}}^{2} \right) \, + \, \ell^{2} \hspace{0.1cm} d\xi^{2},
\end{eqnarray}
with $R = L \sec{\beta}$ and  $R^{\prime} = L \csc{\beta}$, where $\alpha \equiv \cos^{2}{\beta}$.  This geometry is symmetric under $\alpha \to 1- \alpha$  and an interchange of  $S^{3}$ and ${S^{3}}^{\prime}$. The 3-form field strength $F_{(3)} = dC_{(2)}$ is given by
\begin{eqnarray}  \label{R3}
&& \nonumber \hspace{-0.5cm} F_{(3)} = 2 L^{2} \hspace{0.15cm} dt \wedge (\sinh{\rho} \cosh{\rho} \hspace{0.1cm} d\rho) \wedge d\varphi \\
&& \hspace{-0.6cm} \hspace{0.88cm} + \hspace{0.075cm} 
2 R^{2} \, (\sin{\theta} \cos{\theta} \, d\theta) \wedge d\chi \wedge d\phi
+ 2 {R^{\prime}}^{2} \, (\sin{\theta^{\prime}} \cos{\theta^{\prime}} \, d\theta^{\prime}) \wedge d\chi^{\prime} \wedge d\phi^{\prime} \hspace{1.0cm}
\end{eqnarray}
in the case of pure R-R flux.
The Hodge dual 7-form field strength $F_{(7)} = dC_{(6)} = \ast \, F_{(3)}$ is
\begin{eqnarray} \label{R7}
\nonumber && \hspace{-0.65cm} F_{(7)} = - \,\frac{2 R^{3} {R^{\prime}}^{3} \hspace{0.025cm} \ell}{L} \hspace{0.1cm} (\sin{\theta} \cos{\theta} \hspace{0.1cm} d\theta) \wedge d\chi \wedge d\phi \wedge (\cos{\theta^{\prime}} \sin{\theta^{\prime}} \hspace{0.1cm}  d\theta^{\prime}) \wedge d\chi^{\prime} \wedge d\phi^{\prime} \wedge d\xi \\
\nonumber && \hspace{-0.65cm} \hspace{1.16cm} - \, \frac{2L^{3} {R^{\prime}}^{3} \hspace{0.025cm} \ell }{R} \hspace{0.15cm} 
dt \wedge  (\sinh{\rho} \, \cosh{\rho} \hspace{0.1cm} d\rho) \wedge d\varphi \wedge  (\sin{\theta^{\prime}} \cos{\theta^{\prime}} \hspace{0.1cm} d\theta^{\prime}) \wedge d\chi^{\prime} \wedge d\phi^{\prime} \wedge d\xi   \\
&& \hspace{-0.65cm} \hspace{1.16cm} + \, \frac{2L^{3} R^{3} \hspace{0.025cm} \ell}{R^{\prime}} \hspace{0.15cm} 
dt \wedge (\sinh{\rho} \, \cosh{\rho} \hspace{0.1cm}  d\rho) \wedge d\varphi \wedge  (\sin{\theta} \cos{\theta}  \hspace{0.1cm} d\theta) \wedge d\chi \wedge d\phi \wedge d\xi.
\end{eqnarray}
The 3-form and 5-form fluxes couple to D1- and D5-branes.  Dynamically stable giant gravitons with angular momentum on both 3-spheres were shown to exist in \cite{Prinsloo:2014}. D1- and D5-brane maximal giant gravitons provide integrable boundary conditions for open IIB superstrings on $AdS_{3}\times S^{3} \times S^{3\prime} \times S^{1}$.

\paragraph{Supersymmetry.}

The supersymmetry variations of the dilatino and gravitino are\footnote{The vielbeins $\hat{E}^{A} = E^{A}_{M} \, dx^{M}$ are given by
\begin{eqnarray}
\nonumber && \hspace{-0.65cm} \hat{E}^{0} = L \cosh{\rho} \, dt, \hspace{0.5cm} \hat{E}^{1} = L \, d\rho, \hspace{0.5cm} \hat{E}^{2} = L \sinh{\rho} \, d\varphi, \hspace{0.5cm} 
\hat{E}^{3} = R \, d\theta, \hspace{0.5cm} \hat{E}^{4} = R \, \cos{\theta} \, d\chi,  \hspace{0.5cm}
\hat{E}^{5} = R \, \sin{\theta} \, d\phi, \\
&& \hspace{-0.65cm} \nonumber \hat{E}^{6} = R^{\prime} \, d\theta^{\prime}, \hspace{0.5cm} \hat{E}^{7} = R^{\prime} \, \cos{\theta^{\prime}} \, d\chi^{\prime}, \hspace{0.5cm}
\hat{E}^{8} = R^{\prime} \, \sin{\theta^{\prime}} \, d\phi^{\prime}, \hspace{0.5cm} \hat{E}^{9} = \ell \,\, d\xi.
\end{eqnarray}
The supercovariant derivatives $\nabla_{M} = \p_{M} + \Omega^{AB}_{M}\Gamma_{AB}$, with $\hat{\Omega}^{AB} = \Omega^{AB}_{M} \, dx^{M}$ satisfying
$d\hat{E}^{A} + \hat{\Omega}^{A}_{B} \wedge \hat{E}^{B}$, are
\begin{eqnarray}
\nonumber && \hspace{-0.65cm}  \nabla_{t}  = \p_{t} + \tfrac{1}{2} \sinh{\rho} \,\, \Gamma_{01}, \hspace{0.5cm} \nabla_{\rho} =  \p_{\rho}, \hspace{0.5cm}
\nabla_{\varphi} = \p_{\varphi} - \tfrac{1}{2} \cosh{\rho} \,\, \Gamma_{12}, \hspace{0.5cm} \nabla_{\theta}  = \p_{\theta}, \hspace{0.5cm} 
\nabla_{\chi} = \p_{\chi} + \tfrac{1}{2} \sin{\theta} \,\, \Gamma_{34},   \\
&& \hspace{-0.65cm} \nonumber \nabla_{\phi} = \p_{\phi} - \tfrac{1}{2} \cos{\theta} \,\, \Gamma_{35}, \hspace{0.5cm} \nabla_{\theta^{\prime}} = \p_{\theta^{\prime}}, 
 \hspace{0.5cm}  \nabla_{\chi^{\prime}} = \p_{\chi^{\prime}} + \tfrac{1}{2} \sin{\theta^{\prime}} \,\, \Gamma_{67}, \hspace{0.5cm} 
\nabla_{\phi^{\prime}} = \p_{\phi^{\prime}} - \tfrac{1}{2} \cos{\theta^{\prime}} \,\, \Gamma_{68},  \hspace{0.5cm} \nabla_{\xi}  = \p_{\xi}.
\end{eqnarray}
}
\begin{equation} \label{susy-variations}
\delta \lambda = \frac{i}{4} \, \slashed{F}_{(3)} \, (B\varepsilon)^{\ast}, \hspace{1.0cm}
\delta \Psi_{M} = \nabla_{M} \, \varepsilon - \frac{i}{8}\,  \slashed{F}_{(3)} \, \Gamma_{M} \, (B \varepsilon)^{\ast},
\end{equation}
parameterised by $\varepsilon = \varepsilon^{1} + i \, \varepsilon^{2}$, with the 32-component Weyl-Majorana spinors $\varepsilon^{\tI}$ satisfying 
$\Gamma \, \varepsilon^{\tI} = \varepsilon^{\tI}$ and $(B\varepsilon^{\tI})^{\ast}=\varepsilon^{\tI}$, for $I\in\{1,2\}$.  The charge conjugation matrix is $C=B \, \Gamma^{0}$. Also\footnote{Here we use the notation
$\slashed{F}_{(3)} \equiv  \frac{1}{3!} \, F_{(3) \hspace{0.05cm} NRS} \, E^{N}_{A} \, E^{R}_{B} \, E^{S}_{C} \,\, \Gamma_{ABC}$.
}
\begin{equation}
\slashed{F}_{(3)} = \frac{2}{L} \,\, \Gamma^{012} \left(\mathbb{I} + \sqrt{\alpha} \hspace{0.15cm} \Gamma^{012}\, \Gamma^{345} + \sqrt{1-\alpha} \hspace{0.15cm} \Gamma^{012} \, \Gamma^{678} \right) \, = \,  \frac{4}{L} \,\, \Gamma^{012} \, K^{+}(\alpha), \hspace{0.75cm}
\end{equation}
with our gamma matrix conventions shown in Appendix \ref{app:A}.  Here
\begin{equation}
K^{\pm}(\alpha) \equiv \frac{1}{2} \left[ \mathbb{I} \, \pm \, \left( \sqrt{\alpha} \hspace{0.15cm} \Gamma^{012} \, \Gamma^{345} + \sqrt{1-\alpha} \hspace{0.15cm} \Gamma^{012} \, \Gamma^{678} \right) \right].
\end{equation}

The \emph{gravitino Killing-spinor equation} $\delta \Psi_{M} = 0$ implies a solution of the form  \cite{Prinsloo:2014}
\begin{equation} \label{killing-spinor-S3xS3}
\varepsilon \big(x^{M}\big) \, = \, \mathcal{M}^{+} \big(x^{M}\big) \hspace{0.1cm} \varepsilon^{+}  + \, \mathcal{M}^{-}\big(x^{M}\big) \hspace{0.1cm} \varepsilon^{-} 
\, = \, \mathcal{M}^{+} \big(x^{M}\big) \hspace{0.1cm} (1+i) \hspace{0.1cm} \varepsilon^{\sL}  + \, \mathcal{M}^{-}\big(x^{M}\big) \hspace{0.1cm} (1+i) \hspace{0.1cm} \varepsilon^{\sR},
\end{equation}
decomposed into left- and right-movers, with
\begin{equation} \hspace{-0.25cm}
 \mathcal{M}^{\pm} \big(x^{M}\big) \, = \, e^{\pm \frac{1}{2} \hspace{0.025cm} \rho \, \Gamma_{02}} \, e^{ \frac{1}{2} \hspace{0.025cm} (\varphi \hspace{0.05cm} \pm \hspace{0.05cm} t) \, \Gamma_{12}}
\, e^{ \pm \frac{1}{2} \hspace{0.025cm} \theta   \hspace{0.025cm} \Gamma_{45}}  \, e^{ \frac{1}{2}  \hspace{0.025cm} ( \phi \hspace{0.025cm} \mp \hspace{0.05cm} \chi )  \hspace{0.05cm} \Gamma_{35}}
\, e^{ \pm \frac{1}{2} \hspace{0.025cm} \theta^{\prime}   \hspace{0.025cm}  \Gamma_{78}} \, e^{ \frac{1}{2} \hspace{0.025cm} ( \phi^{\prime} \hspace{0.025cm} \mp \hspace{0.05cm} \chi^{\prime} )  \hspace{0.05cm} \Gamma_{68}}.
\end{equation}
Here  $i \, (B\varepsilon^{\pm})^{\ast} = \pm \, \varepsilon^{\pm}$ and 
$\Gamma \, \varepsilon^{\pm} = \varepsilon^{\pm}$, with 
$ \varepsilon^{+} = \left(1+i\right) \, \varepsilon^{\sL}$ and $\varepsilon^{-} = \left(1+i\right) \, \varepsilon^{\sR}$. 
The left- and right-movers satisfy $(B\varepsilon^{\sL})^{\ast} = \varepsilon^{\sL}$ and $(B\varepsilon^{\sR})^{\ast} = -\varepsilon^{\sR}$, with 
$\Gamma \, \varepsilon^{\ms{a}} = \varepsilon^{\ms{a}}$, for $\ms{a} \in \{\SL,\SR\}$. 
The Weyl-Majorana spinors are written in terms of these left- and right-movers as
$\varepsilon^{\tI} = \varepsilon^{\sL} - i \, (-1)^{\tI} \,\varepsilon^{\sR}$.

The \emph{dilatino Killing-spinor equation} $\delta \lambda =0$ now further implies
$K^{+}(\alpha) \hspace{0.1cm} \varepsilon^{\ms{a}} = 0$, which halves the number of left- and right-moving degrees of freedom from 16 to 8, yielding a half-BPS geometry.

The spinors $\varepsilon^{\sL}$ and $\varepsilon^{\sR}$ can be decomposed into eigenstates $\varepsilon^{\sL \, b\beta\dot{\beta}}$ and 
$\varepsilon^{\sR \, b\beta\dot{\beta}}$, which have eigenvalues $(b, \beta, \dot{\beta}) = (\pm,\pm,\pm)$ of the Dirac bilinears given in (\ref{bilinears}):
\begin{equation}
i \hspace{0.1cm} \Gamma_{12} \hspace{0.1cm} \varepsilon^{\ms{a} \, b\beta\dot{\beta}} = b \hspace{0.1cm} \varepsilon^{\ms{a} \, b\beta\dot{\beta}}, \hspace{1.0cm}
i \hspace{0.1cm}  \Gamma_{35} \hspace{0.1cm} \varepsilon^{\ms{a} \, b\beta\dot{\beta}} = - \, \beta \hspace{0.1cm} \varepsilon^{\ms{a} \, b\beta\dot{\beta}}, \hspace{1.0cm}
i \hspace{0.1cm}  \Gamma_{68} \hspace{0.1cm} \varepsilon^{\ms{a} \,  b\beta\dot{\beta}} = - \, \dot{\beta} \hspace{0.1cm} \varepsilon^{\ms{a} \,  b\beta\dot{\beta}}.
\end{equation}  
The IIB supergravity background $AdS_{3}\times S^{3} \times S^{3\prime} \times S^{1}$ is thus invariant under eight left and eight right-moving supersymmetry transformations, parameterised by $\varepsilon^{\ms{a} \, b\beta\dot{\beta}}$, which satisfy
\begin{equation}
K^{+}(\alpha) \,\, \varepsilon^{\ms{a} \, b\beta\dot{\beta}} = 0 \hspace{0.5cm} \text{and hence} 
\hspace{0.5cm} K^{-}(\alpha) \,\,\varepsilon^{\ms{a} \, b\beta\dot{\beta}} = \varepsilon^{\ms{a} \,  b\beta\dot{\beta}}.
\end{equation}
This gives rise to the kappa symmetry condition $K^{+}(\alpha) \,\, \Theta^{\ms{a} \, b\beta\dot{\beta}}(\tau,\sigma) = 0$ of \cite{Babichenko:2010} when the target space superfields are pulled back to the superstring worldsheet to give $X^{M}(\tau,\sigma)$ and $\Theta^{\ms{a} \, b\beta\dot{\beta}}(\tau,\sigma)$. 
Here the spinors $\varepsilon^{\ms{a} \,  b\beta\dot{\beta}}$ parameterize translations in these spinor directions in superspace, which are generated by the supercharges $\mf{Q}_{\ms{a} \,  b\beta\dot{\beta}}$.  The full superisometry algebra is $\mf{d}(2,1;\alpha)_{\sL} \oplus \mf{d}(2,1;\alpha)_{\sR} \oplus \mf{u}(1)$, with the details of the exceptional Lie superalgebra $\mf{d}(2,1;\alpha)$ given in Appendix \ref{app:B}.  It was noted in \cite{Babichenko:2010} that this is the correct kappa symmetry gauge choice to allow for comparison between the Green-Schwarz action and their $\mathbb{Z}_{4}$-graded 
$(\mf{d}(2,1;\alpha)_{\sL} \oplus \mf{d}(2,1;\alpha)_{\sR}) \hspace{0.05cm} / \hspace{0.05cm} (\mf{su}(1,1) \oplus \mf{su}(2) \oplus \mf{su}(2))$ 
integrable coset model which describes a closed IIB superstring on $AdS_{3}\times S^{3} \times S^{3\prime} \times S^{1}$.

\subsection{Maximal giant gravitons and boundary algebras}

Massive excitations of a closed IIB superstring on $AdS_{3} \times S^{3} \times S^{3\prime} \times S^{1}$ map to magnon excitations of a $\mf{d}(2,1;\al)^{2}$ closed spin-chain.  Chapter \ref{part3-sec1} discusses how choosing a vacuum $\mathcal{Z}$ breaks the $\mf{d}(2,1;\al)^{2}$ symmetry to $\mf{su}(1|1)^{2}$, centrally extended to $\mf{su}(1|1)^{2}_{\ms{c}}$. An open IIB superstring ending on a D-brane maps to a $\mf{d}(2,1;\al)^{2}$ open spin-chain with boundaries. The boundary Lie algebra is a subalgebra of $\mf{su}(1|1)^{2}_{\ms{c}}$ determined by the D-brane symmetries which survive the choice of spin-chain vacuum. 
This can be extended to a coideal subalgebra of the Hopf superalgebra, as explained in Chapter \ref{part3-sec2}.
We classify boundary Lie algebras for D5- and D1-brane maximal giant gravitons in $AdS_{3} \times S^{3} \times S^{3\prime} \times S^{1}$.

\subsubsection{Maximal D5-brane giant gravitons}

\begin{figure}[htb!]
\begin{center}
\includegraphics[scale=0.35]{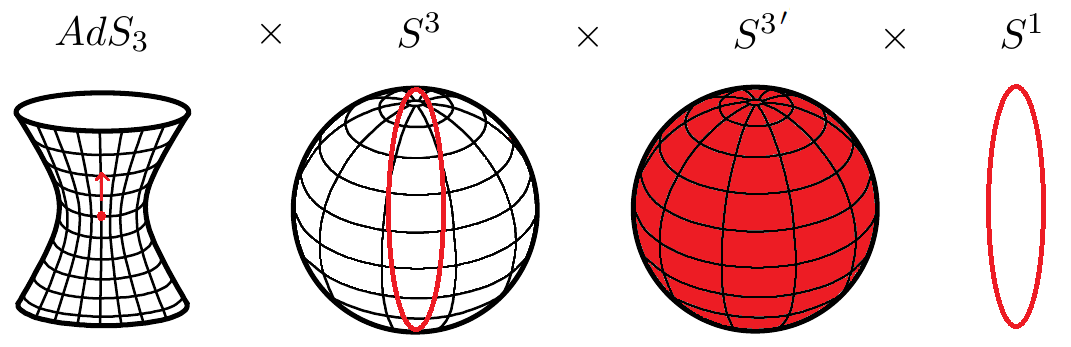}
\vspace{-0.2cm}
\caption{Half of the D5-brane maximal giant graviton wrapping $S^{1} \times {S^{3}}^{\prime} \times S^{1}$ in $AdS_{3} \times S^{3} \times {S^{3}}^{\prime} \times S^{1}$.}  \label{figure-D5}
\end{center}
\end{figure}

The D5-brane giant graviton of \cite{Prinsloo:2014} factorizes at maximum size into two D5-branes, which wrap $S^{1} \times {S^{3}}^{\prime} \times S^{1}$ and
$S^{3} \times {S^{1}}^{\prime} \times S^{1}$, and move only along the time direction in $AdS_{3}$.  Each half-BPS D5-brane preserves four left- and four right-moving supersymmetries on its worldvolume.
We focus on the first half of the maximal giant (see Figure \ref{figure-D5}).  This breaks the bosonic isometry algebra to
\[
\left(\mf{u}(1) \oplus \mf{u}(1) \right) \, \oplus \, \left(\mf{u}(1) \oplus \mf{u}(1)\right) \, \oplus \, \mf{so}(4)^{\prime} \, \oplus \, \mf{u}(1) \,\, \subset \,\,
\mf{so}(2,2) \, \oplus \, \mf{so}(4) \, \oplus \, \mf{so}(4)^{\prime} \, \oplus \, \mf{u}(1). \hspace{0.25cm}
\]  
\vspace{-0.7cm}
\vspace{-0.5cm}

\paragraph{$Z=0$ giant.} 

Let us take the $\C^{2}$ and ${\C^{\prime}}^{2}$ embedding coordinates of the 3-spheres to be
\begin{eqnarray}  \label{C2-coordinates}
\nonumber && \hspace{-0.5cm} (Z, Y) \hspace{0.21cm} = (x_{1} + ix_{2}, \, x_{3} + ix_{4}) = (R \, \cos{\theta} \, e^{i\chi}, \, R \, \sin{\theta} \, e^{i\phi}), \\ \hspace{1.0cm} 
&& \hspace{-0.5cm} (Z^{\prime}, Y^{\prime}) =  (x_{1}^{\prime} + ix_{2}^{\prime}, \, x_{3}^{\prime} + ix_{4}^{\prime}) = (R^{\prime} \cos{\theta^{\prime}} \, e^{i\chi^{\prime}} , \, R^{\prime} \sin{\theta^{\prime}} \, e^{i\phi^{\prime}}), 
\end{eqnarray}
as in \cite{Prinsloo:2014}. The maximal D5-brane giant graviton consists of $Z=0$ and $Z^{\prime}=0$ halves. Let us focus on the $Z=0$ giant which wraps the $\phi$ great circle in $S^{3}$ and the ${S^{3\prime}} \times S^{1}$ space. The worldvolume metric, obtained by setting $\rho = 0$ and $\theta = \tfrac{\pi}{2}$, is
\[ ds^{2} = -L^{2} \, dt^{2} + R^{2} \, d\phi^{2} + {R^{\prime}}^{2} \left( d{\theta^{\prime}}^{2} + \cos^{2}{\theta^{\prime}} \, d{\chi^{\prime}}^{2} + \sin^{2}{\theta^{\prime}} \, d{\phi^{\prime}}^{2} \right) \, + \, \ell^{2} \hspace{0.1cm} d\xi^{2}. \]
The bosonic symmetries of this D5-brane include time translations and rotations in $AdS_{3}$, rotations in $S^{3}$ in the $x_{1}x_{2}$ and $x_{3}x_{4}$-planes, and all rotations in ${S^{3}}^{\prime}$, generated by
\[  \mf{J}_{\ms{a} \, 0} \in \mf{u}(1)_{\ms{a}} \subset \mf{su}(1,1)_{\ms{a}}, \hspace{0.5cm} 
 \mf{L}_{\ms{a} \, 5} \in \mf{u}(1)_{\ms{a}} \subset \mf{su}(2)_{\ms{a}}, \hspace{0.5cm} 
\{ \mf{R}_{\ms{a} \, 8}, \, \mf{R}_{\ms{a} \, \pm}\} \in \mf{su}(2)_{\ms{a}}^{\prime}, \hspace{0.5cm} \text{for} \hspace{0.35cm} \ms{a} \in \{\SL,\SR\},
\]
with the splitting of $\mf{so}(2,2)$, $\mf{so}(4)$ and $\mf{so}(4)^{\prime}$ into left and right algebras shown in Appendix \ref{app:C}. The Cartan elements of the $\mf{su}(1|1)^{2}$ superalgebra, denoted
$\mf{H}_{\ms{a}} = -\mf{J}_{\ms{a} \, 0}-\alpha  \, \mf{L}_{\ms{a} \, 5}-(1-\alpha) \, \mf{R}_{\ms{a} \, 8}$, 
are thus included in the generators of bosonic symmetries of the $Z=0$ giant.

Kappa symmetry on the worldvolume of the D5-brane requires
\begin{equation} \Gamma_{056789} \hspace{0.15cm} \varepsilon =  - \, i \, (B \hspace{0.025cm} \varepsilon )^{\ast},  
\end{equation}
with the pullback of the Killing spinor (\ref{killing-spinor-S3xS3}) given by
\begin{equation}
\varepsilon = (1+i) \, \mathcal{M}^{+} (t,\phi,\theta^{\prime},\chi^{\prime},\phi^{\prime},\xi) \hspace{0.15cm} \varepsilon^{\sL} 
\, + \, (1+i) \, \mathcal{M}^{-}  (t,\phi,\theta^{\prime},\chi^{\prime},\phi^{\prime},\xi) \hspace{0.15cm} \varepsilon^{\sR}.
\end{equation}
Here $K^{+}(\alpha) \hspace{0.1cm} \varepsilon^{\ms{a}} = 0$, for $\ms{a} \in \{ \SL, \SR \}$.  This kappa symmetry condition reduces to
\begin{equation} 
\Gamma_{12} \hspace{0.1cm} \varepsilon^{\ms{a}} \hspace{0.05cm} = \hspace{0.05cm} \Gamma_{35} \hspace{0.1cm} \varepsilon^{\ms{a}}, 
\end{equation}
and hence $\varepsilon^{\ms{a} \, b\beta\dot{\beta}}$ has labels satisfying $-b = \beta$, with $\dot{\beta}$ free.  The 4+4 supersymmetries compatible with kappa symmetry on the worldvolume of the $Z=0$ giant are generated by
\[ 
\{ \mf{Q}_{\ms{a} \, -++} \equiv \mf{Q}_{\ms{a}}, \hspace{0.25cm} \mf{Q}_{\ms{a} \, -+-}, \hspace{0.25cm} 
\mf{Q}_{\ms{a} \, +--} \equiv \mf{S}_{\ms{a}}, \hspace{0.25cm} \mf{Q}_{\ms{a} \, +-+} \}.
\]
The symmetries of this D5-brane include the $\mf{su}(1|1)^{2}$ superalgebra generated by $\{\mf{Q}_{\ms{a}},\mf{S}_{\ms{a}}, \mf{H}_{\ms{a}} \}$ which acts on magnon excitations of the $\mf{d}(2,1;\al)^{2}$ spin-chain.  The boundary Lie algebra is the centrally extended $\mf{su}(1|1)^{2}_{\ms{c}}$. We say the $Z=0$ giant is completely aligned with the spin-chain vacuum $\mathcal{Z}$.

\paragraph{$Y=0$ giant.}  An $SO(4)$ transformation taking $Z$ to $Y$ corresponds to simultaneous rotations by $\theta_{13} = \theta_{24} = \tfrac{\pi}{2}$ in the $x_{1}x_{3}$ and $x_{2}x_{4}$ planes,  achieved by $U_{\sL} =  i \sigma_{1}$ and $U_{\sR} = \mathbb{I}$ (see Appendix \ref{app:C}).
The 4+4 supersymmetries on the worldvolume of the $Y=0$ giant are hence
\[
\{ \mf{Q}_{\sL \, --+}, \hspace{0.3cm}  \mf{Q}_{\sL \, ---}, \hspace{0.3cm} 
 \mf{Q}_{\sL \, ++-}, \hspace{0.3cm}  \mf{Q}_{\sL \, +++}, \hspace{0.3cm}
\mf{Q}_{\sR \, -++} \equiv \mf{Q}_{\sR}, \hspace{0.3cm} \mf{Q}_{\sR \, -+-}, \hspace{0.3cm} 
\mf{Q}_{\sR \, +--} \equiv \mf{S}_{\sR}, \hspace{0.3cm} \mf{Q}_{\sR \, +-+} \}.
\]
Thus $\{ \mf{Q}_{\sR}, \mf{S}_{\sR}, \mf{H}_{\sR}, \mf{H}_{\sL} \}$ are both worldvolume symmetries and in the $\mf{su}(1|1)^{2}_{\ms c}$ superalgebra of magnon excitations, and generate the right half-supersymmetric boundary Lie algebra $\mf{u}(1)_{\sL} \oplus \ms{su}(1|1)_{\sR}$.

\paragraph{$\bar{Y}=0$ giant.}  An $SO(4)$ transformation which takes $Z$ to $\bar{Y}$ is obtained by setting $\theta_{13}=-\theta_{24} = \tfrac{\pi}{2}$, corresponding to $U_{\sL} = \mathbb{I}$ and $U_{\sR} = -i \sigma_{1}$.  The 4+4 supersymmetries of the $\bar{Y}=0$ giant are
\[
\{ \mf{Q}_{\sL \, -++} \equiv \mf{Q}_{\sL}, \hspace{0.3cm} \mf{Q}_{\sL \, -+-}, \hspace{0.3cm} 
\mf{Q}_{\sL \, +--} \equiv \mf{S}_{\sL}, \hspace{0.3cm} \mf{Q}_{\sL \, +-+}, \hspace{0.3cm}
\mf{Q}_{\sR \, --+}, \hspace{0.3cm} \mf{Q}_{\sR \, ---}, \hspace{0.3cm} 
\mf{Q}_{\sR \, ++-}, \hspace{0.3cm} \mf{Q}_{\sR \, +++} \}.
\]  
Here $\{ \mf{Q}_{\sL}, \mf{S}_{\sL}, \mf{H}_{\sL}, \mf{H}_{\sR} \}$ generate the left half-supersymmetric boundary Lie algebra $\ms{su}(1|1)_{\sL} \oplus \mf{u}(1)_{\sR}$.

\paragraph{$\bar{Z}=0$ giant.} An $SO(4)$ rotation taking $Z$ to $\bar{Z}$ is obtained by setting $\theta_{24}=\pi$, which corresponds to $U_{\sL} = i \sigma_{1}$ and $U_{\sR} = i \sigma_{1}$. The 4+4 supersymmetries of the $\bar{Z}=0$ giant are thus
\[
\{ \mf{Q}_{\sL \, --+}, \hspace{0.3cm} \mf{Q}_{\sL \, ---}, \hspace{0.3cm} 
\mf{Q}_{\sL \, ++-}, \hspace{0.3cm} \mf{Q}_{\sL \, +++}, \hspace{0.3cm}
\mf{Q}_{\sR \, --+}, \hspace{0.3cm} \mf{Q}_{\sR \, ---}, \hspace{0.3cm} 
\mf{Q}_{\sR \, ++-}, \hspace{0.3cm} \mf{Q}_{\sR \, +++} \},
\]
none of which are in the $\mf{su}(1|1)^{2}_{\ms c}$ superalgebra, although the Cartan elements $\mf{H}_{\ms{a}}$ remain worldvolume symmetries. The boundary Lie algebra $\mf{u}(1)_{\sL} \oplus \mf{u}(1)_{\sR}$ is therefore non-supersymmetric. 

\smallskip

We also consider the $Z^{\prime}=0$ giant which is completely aligned with the spin-chain vacuum $\mathcal{Z}$ and has totally supersymmetric boundary algebra $\mf{su}(1|1)^{2}_{\ms{c}}$. Then $SO(4)^{\prime}$ transformations yield the $Y^{\prime}=0$, $\bar{Y}^{\prime}=0$ and $\bar{Z}^{\prime}=0$ giants, which give right and left half-supersymmetric, and non-supersymmetric boundary algebras,  $\mf{u}(1)_{\sL} \oplus \ms{su}(1|1)_{\sR}$, $\ms{su}(1|1)_{\sL} \oplus \mf{u}(1)_{\sR}$ and $\mf{u}(1)_{\sL} \oplus \mf{u}(1)_{\sR}$ (see Table \ref{table-D5}).

\vspace{0.1cm}
\begin{table} [htb!]
\begin{center}
\begin{tabular}{|cccc|}
\hline
\small{\text{\bf D5 giant}} & \small{\text{\bf bosonic generators}} & \small{\text{\bf supersymmetry generators}} & \small{\bf boundary algebra} \\
\hline
\small{$Z=0$} & \small{$\{ \mf{J}_{\ms{a} \, 0}, \, \mf{L}_{\ms{a} \, 5}, \, \mf{R}_{\ms{a} \, \pm}, \, \mf{R}_{\ms{a} \, 8} \}$} & 
\small{$\{ \mf{Q}_{\ms{a} - + \pm}, \, \mf{Q}_{\ms{a} + - \pm}\}$} & \small{$\mf{su}(1|1)^{2}_{\ms{c}}$} \\
\hline
\small{$Y=0$} & \small{$\{ \mf{J}_{\ms{a} \, 0}, \, \mf{L}_{\ms{a} \, 5}, \, \mf{R}_{\ms{a} \, \pm}, \, \mf{R}_{\ms{a} \, 8} \}$} & 
\small{$\{ \mf{Q}_{\sL - - \pm}, \, \mf{Q}_{\sL + + \pm}, \, \mf{Q}_{\sR - + \pm}, \, \mf{Q}_{\sR + - \pm} \}$} & 
\small{$\mf{u}(1)_{\sL} \oplus \mf{su}(1|1)_{\sR}$} \\
\small{$\bar{Y}=0$} & \small{$\{ \mf{J}_{\ms{a} \, 0}, \, \mf{L}_{\ms{a} \, 5}, \, \mf{R}_{\ms{a} \, \pm}, \, \mf{R}_{\ms{a} \, 8} \}$} & 
\small{$\{ \mf{Q}_{\sL - + \pm}, \, \mf{Q}_{\sL + - \pm}, \, \mf{Q}_{\sR - - \pm}, \, \mf{Q}_{\sR + + \pm} \}$} & 
\small{$\mf{su}(1|1)_{\sL} \oplus \mf{u}(1)_{\sR}$} \\
\small{$\bar{Z}=0$} & \small{$\{ \mf{J}_{\ms{a} \, 0}, \, \mf{L}_{\ms{a} \, 5}, \, \mf{R}_{\ms{a} \, \pm}, \, \mf{R}_{\ms{a} \, 8} \}$} & 
\small{$\{ \mf{Q}_{\sL - - \pm}, \, \mf{Q}_{\sL + + \pm}, \, \mf{Q}_{\sR - - \pm}, \, \mf{Q}_{\sR + + \pm} \}$} & 
\small{$\mf{u}(1)_{\sL} \oplus \mf{u}(1)_{\sR}$} \\
\hline
\small{$Z^{\prime}=0$} & \small{$\{ \mf{J}_{\ms{a} \, 0}, \, \mf{L}_{\ms{a} \, \pm}, \, \mf{L}_{\ms{a} \, 5},  \, \mf{R}_{\ms{a} \, 8} \}$} & 
\small{$\{ \mf{Q}_{\ms{a} - \pm +}, \, \mf{Q}_{\ms{a} + \pm -}\}$} & \small{$\mf{su}(1|1)^{2}_{\ms{c}}$} \\
\hline
\small{$Y^{\prime}=0$} & \small{$\{ \mf{J}_{\ms{a} \, 0}, \, \mf{L}_{\ms{a} \, \pm}, \, \mf{L}_{\ms{a} \, 5}, \, \mf{R}_{\ms{a} \, 8} \}$} & 
\small{$\{ \mf{Q}_{\sL - \pm -}, \, \mf{Q}_{\sL + \pm +}, \, \mf{Q}_{\sR - \pm +}, \, \mf{Q}_{\sR + \pm -} \}$} & \small{$ \mf{u}(1)_{\sL} \oplus \mf{su}(1|1)_{\sR}$} \\
\small{$\bar{Y}^{\prime}=0$} & \small{$\{ \mf{J}_{\ms{a} \, 0}, \, \mf{L}_{\ms{a} \, \pm}, \, \mf{L}_{\ms{a} \, 5}, \, \mf{R}_{\ms{a} \, 8} \}$} & 
\small{$\{\mf{Q}_{\sL - \pm +}, \, \mf{Q}_{\sL + \pm -}, \, \mf{Q}_{\sR - \pm -}, \, \mf{Q}_{\sR + \pm +} \}$} & \small{$\mf{su}(1|1)_{\sL}\oplus \mf{u}(1)_{\sR}$} \\
\small{$\bar{Z}^{\prime}=0$} & \small{$\{ \mf{J}_{\ms{a} \, 0}, \, \mf{L}_{\ms{a} \, \pm}, \, \mf{L}_{\ms{a} \, 5}, \, \mf{R}_{\ms{a} \, 8} \}$} & 
\small{$\{ \mf{Q}_{\sL - \pm -}, \, \mf{Q}_{\sL + \pm +}, \, \mf{Q}_{\sR - \pm -}, \, \mf{Q}_{\sR + \pm +} \}$} & \small{$\mf{u}(1)_{\sL} \oplus \mf{u}(1)_{\sR}$} \\
\hline
\end{tabular}
\caption{Boundary Lie algebras for D5-brane maximal giant gravitons on $AdS_{3} \times S^{3} \times S^{3\prime} \times S^{1}$.} \label{table-D5}
\end{center}
\end{table}

\subsubsection{Maximal D1-brane giant gravitons}

\begin{figure}[htb!]
\begin{center}
\includegraphics[scale=0.35]{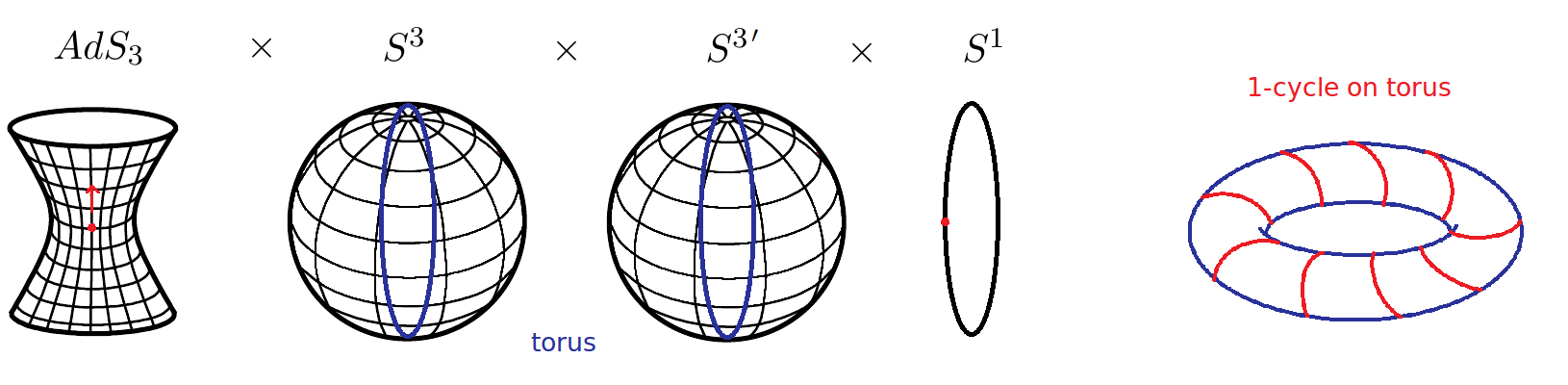} 
\vspace{-0.5cm}
\caption{The D1-brane maximal giant graviton wraps a 1-cycle on a $S^{1} \times S^{1\prime}$ in $AdS_{3} \times S^{3} \times S^{3\prime} \times S^{1}$.} \label{figure-D1}
\end{center}
\end{figure}

The maximal D1-brane giant graviton of \cite{Prinsloo:2014} wraps a 1-cycle wound around a torus $S^{1}\times S^{1\prime}$ made up of two great circles of radii $R$ and $R^{\prime}$ on the 3-spheres. This quarter-BPS D1-brane preserves two left- and two right-moving supersymmetries on its worldvolume.
As shown in Figure \ref{figure-D1}, the bosonic isometry algebra is broken by our choice of torus to
\[
\left(\mf{u}(1) \oplus \mf{u}(1) \right) \, \oplus \, \left(\mf{u}(1) \oplus \mf{u}(1)\right) \, \oplus \, (\mf{u}(1)^{\prime} \oplus \mf{u}(1)^{\prime}) \,\, \subset \,\,
\mf{so}(2,2) \, \oplus \, \mf{so}(4) \, \oplus \, \mf{so}(4)^{\prime} \, \oplus \, \mf{u}(1) \hspace{0.25cm}
\]  
and further by the D1-brane to
\[
\left(\mf{u}(1) \oplus \mf{u}(1) \right) \, \oplus \, \left(\mf{u}(1)_{\sigma} \oplus \mf{u}(1) \oplus \mf{u}(1)^{\prime}\right).
\]
 
\vspace{-0.2cm}
\vspace{-0.4cm}

\paragraph{$Z=Z^{\prime}=0$ giant.}
Let us again make use of the coordinates (\ref{C2-coordinates}).  The maximal D1-brane giant graviton is specified in \cite{Prinsloo:2014} by $\rho=0$ and  $Z=Z^{\prime}=0$,
which gives $\theta = \theta^{\prime} = \tfrac{\pi}{2}$. The worldvolume is parameterized by $(t,\sigma)$, where we define $\phi = \alpha \, \sigma$ and  $\phi^{\prime} = (1-\alpha) \, \sigma$. The D1-brane  wraps a 1-cycle on the torus $(\phi,\phi^{\prime})$. The worldvolume metric is
\[ ds^{2} = -L^{2} \, dt^{2} + L^{2} \, d\sigma^{2}. \]
The bosonic symmetries of the torus are generated by all the Cartan elements $\{  \mf{J}_{\ms{a} \, 0}, \, \mf{L}_{\ms{a} \, 5}, \, \mf{R}_{\ms{a} \, 8} \}$.
The D1-brane itself wraps a 1-cycle on this torus.  Its symmetries are time translations and rotations in  $AdS_{3}$, rotations in $S^{3}$ and $S^{3\prime}$ in the $x_{1}x_{2}$ and $x_{1}^{\prime}x_{2}^{\prime}$-planes transverse to the torus,  and translations along the 1-cycle in the worldvolume direction $\sigma$. These bosonic symmetries of the
$Z=Z^{\prime}=0$ giant are generated by 
\begin{eqnarray*} 
&& \mf{J}_{\ms{a} \, 0} \in \mf{u}(1)_{\ms{a}} \subset \mf{su}(1,1)_{\ms{a}}, \hspace{0.5cm}
\mf{L}_{\sL \,5} - \mf{L}_{\sR \,5} \in \mf{u}(1), \hspace{0.5cm} 
\mf{R}_{\sL \,8} - \mf{R}_{\sR \,8} \in \mf{u}(1)^{\prime}, \\
&& -\alpha \, ( \mf{L}_{\sL \,5} + \mf{L}_{\sR \,5}) -  (1-\alpha) \, ( \mf{R}_{\sL \, 8} +  \mf{R}_{\sR \,8} ) \in \mf{u}(1)_{\sigma}
\end{eqnarray*}
or, equivalently, by the generators
\[ \{ \mf{J}_{\ms{a} \, 0}, \hspace{0.2cm} \mf{H}_{\ms{a}}, \hspace{0.15cm}  \mf{R}_{\sL \,8} - \mf{R}_{\sR \,8}\}, 
 \hspace{0.5cm} \text{for} \hspace{0.35cm} \ms{a} \in \{\SL,\SR\}. \] 

Kappa symmetry on the worldvolume of the D1-brane requires
\begin{equation}
\Gamma_{0}\left(\sqrt{\alpha} \hspace{0.15cm} \Gamma_{5} + \sqrt{1-\alpha} \hspace{0.15cm} \Gamma_{8}\right) \varepsilon  =  - \, i \, (B  \hspace{0.025cm} \varepsilon )^{\ast},
\end{equation}
with the pullback of the Killing spinor (\ref{killing-spinor-S3xS3}) given by
\begin{equation}
\varepsilon = \mathcal{M}^{+}(t,\sigma) \,\, (1+i) \,\, \varepsilon^{\sL} \, + \, \mathcal{M}^{-}(t,\sigma) \,\, (1+i) \,\, \varepsilon^{\sR}.
\end{equation}
Here $K^{+}(\alpha) \,\, \varepsilon^{\ms{a}} = 0$, for $\ms{a} \in \{\SL, \SR\}$.  This kappa symmetry condition reduces to
\begin{equation}
\Gamma_{12} \hspace{0.1cm} \varepsilon^{\ms{a}} \hspace{0.05cm} = \hspace{0.05cm} 
\Gamma_{35} \hspace{0.1cm} \varepsilon^{\ms{a}} \hspace{0.05cm} = \hspace{0.05cm} 
\Gamma_{68} \hspace{0.1cm} \varepsilon^{\ms{a}}.
\end{equation}
The left and right-moving supersymmetries preserved on the D1-brane worldvolume are parameterized by $\varepsilon^{\ms{a} \, b\beta\dot{\beta}}$, with labels satisfying 
$-b = \beta = \dot{\beta}$.  The 2+2 supersymmetries compatible with kappa symmetry on the worldvolume of the $Z=Z^{\prime}=0$ maximal D1 giant are thus generated by 
\[
\{ \mf{Q}_{\ms{a} \, -++} \equiv \mf{Q}_{\ms{a}}, \hspace{0.3cm} \mf{Q}_{\ms{a} \, +--} \equiv \mf{S}_{\ms{a}}\}.
\]
The generators of the symmetries of this D1-brane include all the generators $\{\mf{H}_{\ms{a}}, \mf{Q}_{\ms{a}}, \mf{S}_{\ms{a}}\}$ of the $\mf{su}(1|1)^{2}$ superalgebra of magnon excitations of the $\mf{d}(2,1;\alpha)^{2}$ spin-chain, which is centrally extended to $\mf{su}(1|1)^{2}_{\ms c}$.
The $Z=Z^{\prime}=0$ giant is thus totally aligned with the spin-chain vacuum $\mathcal{Z}$ and the boundary Lie algebra is the full superalgebra 
$\mf{su}(1|1)^{2}_{\ms c}$.

Various giant gravitons and boundary Lie algebras may again be obtained by $SO(4)$ or (and) $SO(4)$ transformations on one (both) 3-spheres.  The results are summarized in Table \ref{table-D1}.
Here $\mf{u}(1)_{\pm}$ are generated by one Cartan element $\mf{H}_{\sL} \pm \mf{H}_{\sR}$.  In Chapter \ref{part3-sec2}, we will show that totally supersymmetric, right and left half-supersymmetric and non-supersymmetric boundary algebras, $\mf{su}(1|1)^{2}_{\ms c}$, $\mf{u}(1)_{\sL} \oplus \mf{su}(1|1)_{\sR}$, $ \mf{su}(1|1)_{\sL} \oplus \mf{u}(1)_{\sR}$, $\mf{u}(1)_{\sL} \oplus \mf{u}(1)_{\sR}$ and $\mf{u}(1)_{+}$,
provide integrable open boundary conditions for the $\mf{d}(2,1;\alpha)^{2}$ spin-chain; that is, we find reflection matrices which intertwine representations of these boundary Lie algebras and satisfy the boundary Yang-Baxter (reflection) equation.  However, the boundary Lie algebras $\mf{su}(1|1)_{\sL}$, $\mf{su}(1|1)_{\sR}$, $\mf{u}(1)_{\sL}$, $\mf{u}(1)_{\sR}$ and $\mf{u}(1)_{-}$  will not give rise to reflection matrices without enhanced boundary symmetry.

\begin{table} [htb!]
\hspace{-1.0cm}
\begin{tabular}{|cccc|} 
\hline
\small{\text{\bf D1 giant}} & \small{\text{\bf bosonic generators}} & \small{\text{\bf supersymmetry generators}} & \small{\text{\bf boundary algebra}} \\
\hline
\small{$Z=Z^{\prime}=0$} & \small{$\{ \mf{J}_{\ms{a} \, 0}, \, \mf{H}_{\ms{a}}, \, \mf{L}_{\sL \, 5} - \mf{L}_{\sR \, 5} \}$} & 
\small{$\{ \mf{Q}_{\ms{a} - + +}, \, \mf{Q}_{\ms{a} + - -}\}$} & \small{$ \mf{su}(1|1)^{2}_{\ms{c}}$} \\
\hline
\small{$Y=Z^{\prime}=0$} & \small{$\{ \mf{J}_{\ms{a} \, 0}, \, \mf{H}_{\sL} + 2 \alpha \hspace{0.025cm} \mf{L}_{\sL \, 5}, \, \mf{H}_{\sR}, \, 
\mf{L}_{\sL \, 5} + \mf{L}_{\sR \, 5} \}$} & 
\small{$\{ \mf{Q}_{\sL - - +}, \, \mf{Q}_{\sL + + -}, \, \mf{Q}_{\sR - + +}, \, \mf{Q}_{\sR + - -}\}$} & \small{$ \mf{su}(1|1)_{\sR}$} \\
\small{$Z=Y^{\prime}=0$} & \small{$\{ \mf{J}_{\ms{a} \, 0}, \, \mf{H}_{\sL} + 2 \alpha \hspace{0.025cm} \mf{L}_{\sL \, 5}, \, \mf{H}_{\sR}, \, 
\mf{L}_{\sL \, 5} - \mf{L}_{\sR \, 5} \}$} & 
\small{$\{ \mf{Q}_{\sL - + -}, \, \mf{Q}_{\sL + - +}, \, \mf{Q}_{\sR - + +}, \, \mf{Q}_{\sR + - -}\}$} & \small{$\mf{su}(1|1)_{\sR}$} \\
\small{$\bar{Y}=Z^{\prime}=0$} & \small{$\{ \mf{J}_{\ms{a} \, 0}, \, \mf{H}_{\sL}, \, \mf{H}_{\sR} + 2 \alpha \hspace{0.05cm} \mf{L}_{\sR \, 5}, \, 
\mf{L}_{\sL \, 5} + \mf{L}_{\sR \, 5} \}$} & 
\small{$\{\mf{Q}_{\sL - + +}, \, \mf{Q}_{\sL + - -} , \, \mf{Q}_{\sR - - +}, \, \mf{Q}_{\sR + + -} \}$} & \small{$ \mf{su}(1|1)_{\sL}$} \\
\small{$Z=\bar{Y}^{\prime}=0$} & \small{$\{ \mf{J}_{\ms{a} \, 0}, \, \mf{H}_{\sL}, \, \mf{H}_{\sR} + 2 \alpha \hspace{0.05cm} \mf{L}_{\sR \, 5}, \, 
\mf{L}_{\sL \, 5} - \mf{L}_{\sR \, 5} \}$} & 
\small{$\{ \mf{Q}_{\sL - + +}, \, \mf{Q}_{\sL + - -}, \, \mf{Q}_{\sR - + -}, \, \mf{Q}_{\sR + - +}\}$} & \small{$ \mf{su}(1|1)_{\sL}$} \\
\small{$\bar{Z}=Z^{\prime}=0$} & \small{$\{ \mf{J}_{\ms{a} \, 0}, \, \mf{H}_{\sL} - \mf{H}_{\sR}, \, \mf{H}_{\sR} + 2 \alpha \hspace{0.05cm} \mf{L}_{\sR \, 5}, \, \mf{L}_{\sL \, 5} - \mf{L}_{\sR \, 5} \}$} & 
\small{$\{ \mf{Q}_{\ms{a} - - +}, \, \mf{Q}_{\ms{a} + + -} \}$} & \small{$\mf{u}(1)_{-}$} \\
\small{$Z=\bar{Z}^{\prime}=0$} & \small{$\{ \mf{J}_{\ms{a} \, 0}, \, \mf{H}_{\sL} - \mf{H}_{\sR}, \, \mf{H}_{\sR} + 2 \alpha \hspace{0.05cm} \mf{L}_{\sR \, 5}, \, \mf{L}_{\sL \, 5} - \mf{L}_{\sR \, 5} \}$} & \small{$\{ \mf{Q}_{\ms{a} - + -}, \, \mf{Q}_{\ms{a} + - +} \}$} & \small{$\mf{u}(1)_{-}$} \\
\hline
\small{$Y=Y^{\prime}=0$} & \small{$\{ \mf{J}_{\ms{a} \, 0}, \, \mf{H}_{\ms{a}}, \, \mf{L}_{\sL \, 5} + \mf{L}_{\sR \, 5} \}$} & 
\small{$\{ \mf{Q}_{\sL - - -}, \, \mf{Q}_{\sL + + +}, \, \mf{Q}_{\sR - + +}, \, \mf{Q}_{\sR + - -} \}$} & \small{$\mf{u}(1)_{\sL} \oplus \mf{su}(1|1)_{\sR}$} \\
\small{$\bar{Y}=\bar{Y}^{\prime}=0$} & \small{$\{ \mf{J}_{\ms{a} \, 0}, \, \mf{H}_{\ms{a}}, \, \mf{L}_{\sL \, 5} + \mf{L}_{\sR \, 5} \}$} & 
\small{$\{ \mf{Q}_{\sL - + +}, \, \mf{Q}_{\sL + - -}, \, \mf{Q}_{\sR - - -}, \, \mf{Q}_{\sR + + +} \}$} & \small{$ \mf{su}(1|1)_{\sL} \oplus \mf{u}(1)_{\sR}$} \\
\small{$\bar{Z}=\bar{Z}^{\prime}=0$} & \small{$\{ \mf{J}_{\ms{a} \, 0}, \, \mf{H}_{\ms{a}}, \, \mf{L}_{\sL \, 5} - \mf{L}_{\sR \, 5} \}$} & 
\small{$\{ \mf{Q}_{\ms{a} - - -}, \, \mf{Q}_{\ms{a} + + +} \}$} & \small{$\mf{u}(1)_{\sL} \oplus \mf{u}(1)_{\sR}$} \\
\small{$\bar{Y}=Y^{\prime}=0$} & \small{$\{ \mf{J}_{\ms{a} \, 0}, \, \mf{H}_{\sL} + \mf{H}_{\sR}, \, \mf{H}_{\sR} + 2 \alpha \hspace{0.05cm} \mf{L}_{\sR \, 5}, \, \mf{L}_{\sL \, 5} + \mf{L}_{\sR \, 5} \}$} & 
\small{$\{ \mf{Q}_{\sL - + -}, \, \mf{Q}_{\sL + - +}, \, \mf{Q}_{\sR - - +}, \, \mf{Q}_{\sR + + -} \}$} & \small{$\mf{u}(1)_{+}$} \\
\small{$Y=\bar{Y}^{\prime}=0$} & \small{$\{ \mf{J}_{\ms{a} \, 0}, \,  \mf{H}_{\sL} + \mf{H}_{\sR}, \, \mf{H}_{\sR} + 2 \alpha \hspace{0.05cm} \mf{L}_{\sR \, 5},  \, \mf{L}_{\sL \, 5} + \mf{L}_{\sR \, 5} \}$} & 
\small{$\{ \mf{Q}_{\sL - - +}, \, \mf{Q}_{\sL + + -}, \, \mf{Q}_{\sR - + -}, \, \mf{Q}_{\sR + - +} \}$} & \small{$\mf{u}(1)_{+}$} \\
\small{$Y=\bar{Z}^{\prime}=0$} & \small{$\{ \mf{J}_{\ms{a} \, 0}, \, \mf{H}_{\sL}, \, \mf{H}_{\sR} + 2 \alpha \hspace{0.025cm} \mf{L}_{\sR \, 5}, \, 
\mf{L}_{\sL \, 5} + \mf{L}_{\sR \, 5} \}$} & 
\small{$\{ \mf{Q}_{\sL - - -}, \, \mf{Q}_{\sL + + +}, \, \mf{Q}_{\sR - + -}, \, \mf{Q}_{\sR + - +} \}$} & \small{$\mf{u}(1)_{\sL}$} \\
\small{$\bar{Z}=Y^{\prime}=0$} & \small{$\{ \mf{J}_{\ms{a} \, 0}, \, \mf{H}_{\sL}, \, \mf{H}_{\sR} + 2 \alpha \hspace{0.025cm} \mf{L}_{\sR \, 5}, \, 
\mf{L}_{\sL \, 5} - \mf{L}_{\sR \, 5} \}$} & 
\small{$\{ \mf{Q}_{\sL - - -}, \, \mf{Q}_{\sL + + +}, \, \mf{Q}_{\sR - - +}, \, \mf{Q}_{\sR + + -} \}$} & \small{$\mf{u}(1)_{\sL}$} \\
\small{$\bar{Y}=\bar{Z}^{\prime}=0$} & \small{$\{ \mf{J}_{\ms{a} \, 0}, \, \mf{H}_{\sL} + 2 \alpha \hspace{0.025cm} \mf{L}_{\sL \, 5}, \, \mf{H}_{\sR}, \, \mf{L}_{\sL \, 5} + \mf{L}_{\sR \, 5} \}$} & 
\small{$\{ \mf{Q}_{\sL - + -}, \, \mf{Q}_{\sL + - +}, \, \mf{Q}_{\sR - - -}, \, \mf{Q}_{\sR + + +} \}$} & \small{$\mf{u}(1)_{\sR}$} \\
\small{$\bar{Z}=\bar{Y}^{\prime}=0$} & \small{$\{ \mf{J}_{\ms{a} \, 0}, \, \mf{H}_{\sL} + 2 \alpha \hspace{0.025cm} \mf{L}_{\sL \, 5}, \, \mf{H}_{\sR} , \, \mf{L}_{\sL \, 5} - \mf{L}_{\sR \, 5} \}$} & 
\small{$\{ \mf{Q}_{\sL - - +}, \, \mf{Q}_{\sL + + -}, \, \mf{Q}_{\sR - - -}, \, \mf{Q}_{\sR + + +} \}$} & \small{$\mf{u}(1)_{\sR}$} \\
\hline
\end{tabular}
\caption{Boundary Lie algebras for D1-brane maximal giant gravitons on $AdS_{3} \times S^{3} \times S^{3\prime} \times S^{1}$.}   \label{table-D1}
\end{table}

\vspace{-0.1cm}
\section{Maximal giant gravitons on $AdS_{3} \times S^{3} \times T^{4}$}    \label{part2-sec2}

We now give the details of the type IIB supergravity background $AdS_{3} \times S^{3} \times T^{4}$ with pure R-R 3-form flux, and
discuss D1- and D5-brane maximal giant gravitons based on \cite{Raju-et-al:2008,Janssen:2005}.

\vspace{-0.1cm}
\subsection{$AdS_{3} \times S^{3} \times T^{4}$ with pure R-R flux}
 
\paragraph{IIB supergravity solution.}

The metric of the $AdS_{3} \times S^{3} \times T^{4}$ background is
\begin{equation}  
ds^{2} = 
L^{2} \left( - \cosh^{2}{\rho} \hspace{0.1cm} dt^{2} + d\rho^{2} + \sinh^{2}{\rho} \hspace{0.1cm} d\varphi^{2} \right)
\, + \, L^{2}  \left( d\theta^{2} + \cos^{2}{\theta} \, d\chi^{2} + \sin^{2}{\theta} \, d\phi^{2} \right) + \ell_{i}^{2} \hspace{0.1cm} d\xi_{i}^{2}.
\hspace{0.2cm}
\end{equation}
The 3-form field strength $F_{(3)} = dC_{(2)}$ is given by
\begin{equation}
F_{(3)} = 2 L^{2} \hspace{0.15cm} dt \wedge (\sinh{\rho} \cosh{\rho} \hspace{0.1cm} d\rho) \wedge d\varphi  + \hspace{0.075cm} 
2 L^{2} \, (\sin{\theta} \cos{\theta} \, d\theta) \wedge d\chi \wedge d\phi \hspace{1.0cm}
\end{equation}
in the case of pure R-R flux.
The Hodge dual 7-form field strength $F_{(7)} = dC_{(6)} = \ast \, F_{(3)}$ is
\begin{eqnarray}
\nonumber && \hspace{-0.65cm} F_{(7)} = - \,2 L^{2} \hspace{0.025cm} \ell_{1} \ell_{2} \ell_{3} \ell_{4} \hspace{0.15cm} (\sin{\theta} \cos{\theta} \hspace{0.1cm} d\theta) \wedge d\chi \wedge d\phi \wedge d\xi_{1} \wedge d\xi_{2} \wedge d\xi_{3} \wedge d\xi_{4} \\
&& \hspace{-0.65cm} \hspace{1.16cm} - \,2 L^{2} \hspace{0.025cm} \ell_{1} \ell_{2} \ell_{3} \ell_{4} \hspace{0.2cm}
dt \wedge  (\sinh{\rho} \, \cosh{\rho} \hspace{0.1cm} d\rho) \wedge d\varphi \wedge d\xi_{1} \wedge d\xi_{2} \wedge d\xi_{3} \wedge d\xi_{4}.
\end{eqnarray}
These 3-form and 5-form fluxes couple to the D5- and D1-brane giant gravitons of \cite{Raju-et-al:2008,Janssen:2005} which have angular momentum on the 3-sphere.  Maximal D1- and D5-brane giant gravitons provide integrable boundaries for open IIB superstrings on $AdS_{3}\times S^{3}\times T^{4}$.

\paragraph{Supersymmetry.} 

The supersymmetry variations  of the dilatino and gravitino (\ref{susy-variations}) are now written\footnote{The vielbeins $\hat{E}^{A} = E^{A}_{M} \, dx^{M}$ are 
\begin{eqnarray}
\nonumber && \hspace{-0.65cm} \hat{E}^{0} = L \cosh{\rho} \, dt, \hspace{0.5cm} \hat{E}^{1} = L \, d\rho, \hspace{0.5cm} \hat{E}^{2} = L \sinh{\rho} \, d\varphi, \hspace{0.5cm}
 \hat{E}^{3} = L \, d\theta, \hspace{0.5cm} \hat{E}^{4} = L \, \cos{\theta} \, d\chi,  \hspace{0.5cm} \hat{E}^{5} = L \, \sin{\theta} \, d\phi, \\
&& \hspace{-0.65cm} \nonumber \hat{E}^{6} = \ell_{1} \,\, d\xi_{1}, \hspace{0.5cm} \hat{E}^{7} = \ell_{2} \,\, d\xi_{2}, \hspace{0.5cm}
\hat{E}^{8} = \ell_{3} \,\, d\xi_{3}, \hspace{0.5cm}  \hat{E}^{9} = \ell_{4} \,\, d\xi_{4}  
\end{eqnarray}
and the supercovariant derivatives are given by
\begin{eqnarray}
\nonumber && \hspace{-0.65cm} \nabla_{t} \hspace{0.1cm} = \hspace{0.025cm} \p_{t} + \tfrac{1}{2} \sinh{\rho} \,\, \Gamma_{01}, \hspace{0.5cm} \nabla_{\rho} \hspace{0.15cm} = \hspace{0.02cm} \p_{\rho}, \hspace{0.5cm}
\nabla_{\varphi} \hspace{0.07cm} = \p_{\varphi} - \tfrac{1}{2} \cosh{\rho} \,\, \Gamma_{12}, \hspace{0.5cm}
\nabla_{\theta} \hspace{0.075cm} = \p_{\theta}, \hspace{0.5cm} 
\nabla_{\chi} \hspace{0.1cm} = \p_{\chi} + \tfrac{1}{2} \sin{\theta} \,\, \Gamma_{34}, \\
\nonumber && \hspace{-0.65cm}  \nabla_{\phi} \hspace{0.1cm} = \p_{\phi} - \tfrac{1}{2} \cos{\theta} \,\, \Gamma_{35}, \hspace{0.5cm} \nabla_{\xi_{i}} \hspace{0.15cm} = \p_{\xi_{i}}.
\end{eqnarray}
} in terms of
\begin{equation}
\slashed{F}_{(3)} 
= \frac{2}{L} \,\, \Gamma^{012} \left(\mathbb{I} + \Gamma^{012}\, \Gamma^{345} \right) \, = \,  \frac{4}{L} \,\, \Gamma^{012} \, K^{+}, \hspace{0.6cm} \text{where} \hspace{0.35cm} K^{\pm} \equiv \frac{1}{2} \left( \mathbb{I} \, \pm \, \Gamma^{012} \, \Gamma^{345} \right).
\end{equation}
The \emph{gravitino Killing-spinor equation} $\delta \Psi_{M} = 0$ implies a solution of the form
\begin{equation} \label{killing-spinor-T4}
\varepsilon\big(x^{M}\big) = \mathcal{M}^{+} \hspace{-0.05cm} \big(x^{M}\big)  \left(1+i\right) \varepsilon^{\sL}  +  \mathcal{M}^{-}\hspace{-0.05cm}\big(x^{M}\big)  \left(1+i\right) \varepsilon^{\sR},
\end{equation}
decomposed into left and right-movers, with
\begin{equation} \hspace{-0.25cm}
 \mathcal{M}^{\pm}\hspace{-0.05cm}\big(x^{M}\big) = e^{\pm \frac{1}{2} \hspace{0.025cm} \rho \, \Gamma_{02}} \, e^{ \frac{1}{2} \hspace{0.025cm} (\varphi \hspace{0.05cm} \pm \hspace{0.05cm} t) \, \Gamma_{12}}
\, e^{ \pm \frac{1}{2} \hspace{0.025cm} \theta   \hspace{0.025cm} \Gamma_{45}}  \, e^{ \frac{1}{2}  \hspace{0.025cm} ( \phi \hspace{0.025cm} \mp \hspace{0.05cm} \chi )  \hspace{0.05cm} \Gamma_{35}}.
\end{equation}
The \emph{dilatino Killing-spinor equation} $\delta \lambda =0$ further implies $K^{+} \hspace{0.05cm} \varepsilon^{\ms{a}} = 0$ which halves the number of left and right-moving degrees of freedom.  The spinors $\varepsilon^{\sL}$ and $\varepsilon^{\sR}$ can be decomposed into eigenstates $\varepsilon^{\sL \,  b\beta\dot{\beta}}$ and $\varepsilon^{\sR \,  b\beta\dot{\beta}}$. This Killing spinor (\ref{killing-spinor-T4}) can be seen as the $\alpha \to 1$ limit of (\ref{killing-spinor-S3xS3}).

The IIB supergravity background $AdS_{3}\times S^{3} \times T^{4}$ is thus invariant under eight left- and eight right-moving supersymmetry transformations, parameterised by $\varepsilon^{\ms{a} \, b\beta\dot{\beta}}$, which satisfy
\begin{equation}
K^{+} \,\, \varepsilon^{\ms{a} \,  b\beta\dot{\beta}} = 0 \hspace{0.5cm} \text{and hence} 
\hspace{0.5cm} K^{-} \,\,\varepsilon^{\ms{a} \,  b\beta\dot{\beta}} = \varepsilon^{\ms{a} \,  b\beta\dot{\beta}}.
\end{equation}
These supersymmetry transformations are generated by the supercharges $\mf{Q}_{\ms{a} \,  b\beta\dot{\beta}}$.  The superisometry algebra is 
$\mf{psu}(1,1|2)_{\sL} \oplus \mf{psu}(1,1|2)_{\sR} \oplus \mf{u}(1)^{4}$ with the details of the Lie superalgebra $\mf{psu}(1,1|2)$ given in Appendix \ref{app:C}.  
 
\subsection{Maximal giant gravitons and boundary algebras}
 
A closed IIB superstring on $AdS_{3} \times S^{3} \times T^{4}$ maps to a closed $\mf{psu}(1,1|2)^{2}$ spin-chain, as described in Chapter \ref{part4-sec1}, with magnon excitations transforming under a centrally extended $[\mf{psu}(1|1)^{2} \oplus \mf{u}(1)]^{2}_{\ms c}$ superalgebra specified by our choice of vacuum $\mathcal{Z}$.  An open IIB superstring ending on a D-brane maps to a $\mf{psu}(1,1|2)^{2}$ open spin-chain with the boundary Lie algebra determined by the D-brane symmetries which survive the choice of vacuum.
We consider D1- and D5-brane maximal giant gravitons which yield boundary Lie algebras, extended to the coideal subalgebras in Chapter \ref{part4-sec2}

\subsubsection{Maximal D1- and D5-brane giant gravitons}

\begin{figure}[htb!]
\begin{center}
\includegraphics[scale=0.35]{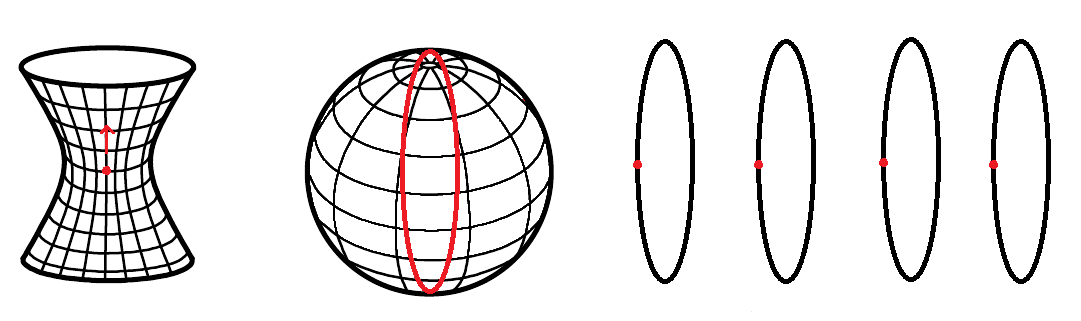} 
\vspace{-0.2cm}
\includegraphics[scale=0.35]{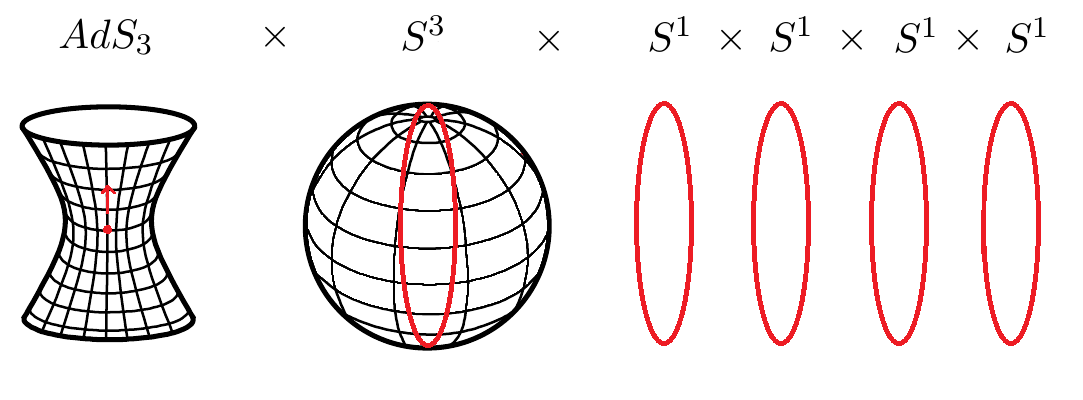} 
\vspace{-0.2cm}
\caption{The D1- and D5-brane maximal giant gravitons wrapping $S^{1}$ and $S^{1} \times T^{4}$ in $AdS_{3} \times S^{3} \times T^{4}$.} \label{figure-T4-D1D5}
\end{center}
\end{figure}

Both the maximal giant gravitons wrap a great circle in $S^{3}$, with the D1-brane point-like in the $T^{4}$ and the D5-brane wrapping the entire $T^{4}$ space \cite{Raju-et-al:2008,Janssen:2005}. 
We consider these D1 and D5-brane giants simultaneously. As shown in Figure \ref{figure-T4-D1D5}, the bosonic isometry algebra breaks to
\begin{eqnarray*}
&& \left(\mf{u}(1) \oplus \mf{u}(1) \right) \, \oplus \, \left(\mf{u}(1) \oplus \mf{u}(1)\right) \,\, \subset \,\,
\mf{so}(2,2) \, \oplus \, \mf{so}(4) \, \oplus \, \mf{u}(1)^{4} \hspace{0.5cm} \text{and} \hspace{0.5cm} \\
&& \left(\mf{u}(1) \oplus \mf{u}(1) \right) \, \oplus \, \left(\mf{u}(1) \oplus \mf{u}(1)\right) \, \oplus \, \mf{u}(1)^{4}  \,\, \subset \,\,
\mf{so}(2,2) \, \oplus \, \mf{so}(4) \, \oplus \, \mf{u}(1)^{4}. \hspace{0.25cm}
\end{eqnarray*}
\vspace{-0.8cm}
\vspace{-0.65cm}

\paragraph{$Z=0$ giants.}  

Let us take the $\C^{2}$ embedding coordinates of the 3-sphere to be
\begin{equation} 
(Z,Y) = (x_{1}+ix_{2}, \, x_{3} + i x_{4}) = (L \, \cos{\theta} \, e^{i\chi}, \, L \, \sin{\theta} \, e^{i\phi}). 
\end{equation}
The $Z=0$ giant wraps the $\phi$-circle in $S^{3}$, and is obtained by setting $\rho=0$ and $\theta = \tfrac{\pi}{2}$. The D1 and D5-brane worldvolume coordinates  are $(t,\phi)$ and $(t,\phi,\xi_{i})$. The worldvolume metrics are
\[ ds^{2} = -L^{2} \, dt^{2} + L^{2} \, d\phi^{2} \hspace{0.75cm} \text{and} \hspace{0.75cm}
ds^{2} = -L^{2} \, dt^{2} + L^{2} \, d\phi^{2} + \ell_{i}^{2} \hspace{0.1cm} d\xi_{i}^{2}. \]
The bosonic symmetries of these D-branes include time translations and rotations in $AdS_{3}$, and rotations in $S^{3}$ in the $x_{1}x_{2}$ and $x_{3}x_{4}$-planes, generated by
\[  \mf{J}_{\ms{a} \, 0} \in \mf{u}(1)_{\ms{a}} \subset \mf{su}(1,1)_{\ms{a}}, \hspace{0.5cm} 
 \mf{L}_{\ms{a} \, 5} \in \mf{u}(1)_{\ms{a}} \subset \mf{su}(2)_{\ms{a}},
 \hspace{0.5cm} \text{for} \hspace{0.35cm} \ms{a} \in \{\SL,\SR\},
\]
with the splitting of $\mf{so}(2,2)$ and $\mf{so}(4)$ in Appendix \ref{app:C}. The Cartan elements of $(\mf{psu}(1,1)^{2} \oplus \mf{u}(1))^{2}$, denoted
$\mf{H}_{\ms{a}} = -\mf{J}_{\ms{a} \, 0} -  \mf{L}_{\ms{a} \, 5}$,  
are thus  generators of bosonic symmetries of the $Z=0$ giants.

Kappa symmetry on the worldvolume of the D1- and D5-brane requires
\begin{equation}
\Gamma_{05} \, \varepsilon = - i \, (B \hspace{0.025cm} \varepsilon )^{\ast} \hspace{1.0cm} \text{and} \hspace{1.0cm} 
\Gamma_{056789} \, \varepsilon = -  i \, (B \hspace{0.025cm} \varepsilon )^{\ast},
\end{equation}
respectively, with the pullback of the Killing spinor (\ref{killing-spinor-T4}) to the worldvolume given by
\begin{equation}
\varepsilon = \mathcal{M}^{+} (t,\phi) \,\, (1+i) \,\, \varepsilon^{\sL} 
\, + \, \mathcal{M}^{-}  (t,\phi) \,\, (1+i) \,\, \varepsilon^{\sR}
\end{equation}
in both cases. Here $K^{+} \hspace{0.1cm} \varepsilon^{\ms{a}} = 0$, for $\ms{a} \in \{\SL,\SR \}$.  This kappa symmetry condition reduces to
\begin{equation}
\Gamma_{12} \hspace{0.1cm} \varepsilon^{\ms{a}} \hspace{0.05cm} = \hspace{0.05cm} \Gamma_{35} \hspace{0.1cm} \varepsilon^{\ms{a}}.
\end{equation}
Hence $\varepsilon^{\ms{a} \,  b\beta\dot{\beta}}$ satisfies  $-b = \beta$, with $\dot{\beta}$ a free label. 
The 4+4 supersymmetries compatible with kappa symmetry on the worldvolume of the D1- and D5-brane $Z=0$ giants are generated by
\[
\{ \mf{Q}_{\ms{a} \, -++} \equiv \mf{Q}_{\ms{a} \hspace{0.025cm} 1}, \hspace{0.3cm} \mf{Q}_{\ms{a} \, -+-} \equiv \mf{Q}_{\ms{a} \hspace{0.025cm} 2}, \hspace{0.3cm} 
\mf{Q}_{\ms{a} \, +--} \equiv \mf{S}_{\ms{a} \hspace{0.025cm} 1}, \hspace{0.3cm} \mf{Q}_{\ms{a} \, +-+} \equiv \mf{S}_{\ms{a} \hspace{0.025cm} 2}\}.
\]
The generators of the symmetries of these D1- and D5-branes include all the generators  $\{\mf{H}_{\ms{a}}, \, \mf{Q}_{\ms{a} \hspace{0.025cm} i}, \, \mf{S}_{\ms{a} \hspace{0.025cm} i}\}$ of the  $(\mf{psu}(1|1)^{2}\oplus\mf{u}(1))^{2}$ superalgebra of magnon excitations of the $\mf{psu}(1,1|2)^{2}$ spin-chain, centrally extended to $(\mf{psu}(1|1)^{2}\oplus\mf{u}(1))^{2}_{\ms c}$.  These $Z=0$ giants are therefore aligned with the $\mathcal{Z}$ vacuum of the spin-chain.  We expect the boundary Lie algebra to be the full superalgebra $[\mf{psu}(1|1)^{2}\oplus \mf{u}(1)]^{2}_{\ms c}$.

\paragraph{$Y=0$ giants.} 

An $SO(4)$ rotation with  $\theta_{13} = \theta_{24} = \tfrac{\pi}{2}$ takes $Z$ to $Y$, corresponding to $U_{\sL} =  i \sigma_{1}$ and $U_{\sR} = \mathbb{I}$.
The 4+4 supersymmetries on the worldvolume of the $Y=0$ giants are generated by
\begin{eqnarray}
\nonumber && \hspace{-0.65cm} \{ \mf{Q}_{\ms{\sL} \, --+}, \hspace{0.3cm} \mf{Q}_{\ms{\sL} \, ---},  \hspace{0.3cm} 
 \mf{Q}_{\ms{\sL} \, ++-}, \hspace{0.3cm} \mf{Q}_{\ms{\sL} \, +++}, \\
\nonumber && \hspace{-0.65cm}\hspace{0.18cm} \mf{Q}_{\ms{\sR} \, -++} \equiv \mf{Q}_{\ms{\sR} \hspace{0.025cm} 1}, \hspace{0.3cm} 
\mf{Q}_{\ms{\sR} \, -+-} \equiv \mf{Q}_{\ms{\sR} \hspace{0.025cm} 2}, \hspace{0.3cm} 
\mf{Q}_{\ms{\sR} \, +--} \equiv \mf{S}_{\ms{\sR} \hspace{0.025cm} 1}, \hspace{0.3cm} 
\mf{Q}_{\ms{\sR} \, +-+} \equiv \mf{S}_{\ms{\sR} \hspace{0.025cm} 2} \}.
\end{eqnarray}
The boundary Lie algebra $\mf{u}(1)_{\sL} \oplus [\mf{psu}(1|1)^{2} \oplus \mf{u}(1)]_{\sR}$ is right half-supersymmetric.

\paragraph{$\bar{Y}=0$ giants.}  An $SO(4)$ transformation with  $\theta_{13}=-\theta_{24} = \tfrac{\pi}{2}$ takes $Z$ to $\bar{Y}$, and corresponds to $U_{\sL} = \mathbb{I}$ and $U_{\sR} = -i \sigma_{1}$.  The 4+4 supersymmetries of the $\bar{Y}=0$ giants are generated by
\begin{eqnarray} 
&& \nonumber \hspace{-0.65cm}
\{ \mf{Q}_{\ms{\sL} \, -++} \equiv \mf{Q}_{\ms{\sL} \hspace{0.025cm} 1}, \hspace{0.3cm} 
\mf{Q}_{\ms{\sL} \, -+-} \equiv \mf{Q}_{\ms{\sL} \hspace{0.025cm} 2}, \hspace{0.3cm} 
\mf{Q}_{\ms{\sL} \, +--} \equiv \mf{S}_{\ms{\sL} \hspace{0.025cm} 1}, \hspace{0.3cm} 
\mf{Q}_{\ms{\sL} \, +-+} \equiv \mf{S}_{\ms{\sL} \hspace{0.025cm} 2}, \\
&& \nonumber \hspace{-0.65cm} \hspace{0.18cm}  \mf{Q}_{\ms{\sR} \, --+}, \hspace{0.3cm} \mf{Q}_{\ms{\sR} \, ---}, \hspace{0.3cm} 
\mf{Q}_{\ms{\sR} \, ++-}, \hspace{0.3cm}  \mf{Q}_{\ms{\sR} \, +++} \}.
\end{eqnarray}
The boundary Lie algebra $[\mf{psu}(1|1)^{2} \oplus \mf{u}(1)]_{\sL} \oplus \mf{u}(1)_{\sR}$ is left half-supersymmetric.

\paragraph{$\bar{Z}=0$ giants.} An $SO(4)$ rotation with $\theta_{24}=\pi$ takes $Z$ to $\bar{Z}$, and corresponds to $U_{\sL} = i \sigma_{1}$ and $U_{\sR} = i \sigma_{1}$. The 4+4 supersymmetries of the $\bar{Z}=0$ giants are generated by
\[
\{ \mf{Q}_{\sL \, --+}, \hspace{0.3cm} \mf{Q}_{\sL \, ---}, \hspace{0.3cm} 
\mf{Q}_{\sL \, ++-}, \hspace{0.3cm}  \mf{Q}_{\sL \, +++}, \hspace{0.3cm}
\mf{Q}_{\sR \, --+}, \hspace{0.3cm} \mf{Q}_{\sR \, ---}, \hspace{0.3cm} 
\mf{Q}_{\sR \, ++-}, \hspace{0.3cm} \mf{Q}_{\sR \, +++}\}.
\]
The boundary Lie algebra $\mf{u}(1)_{\sL} \oplus \mf{u}(1)_{\sR}$ is non-supersymmetric.

\smallskip

The above results for both D1- and D5-brane maximal giant gravitons are summarized in Table \ref{table-D1D5}.
The totally supersymmetric, right and left half-supersymmetric and non-supersymmetric boundary Lie algebras,
$(\mf{psu}(1|1)^{2} \oplus \mf{u}(1))^{2}_{\ms c}$, $\mf{u}(1)_{\sL} \oplus \mf{psu}(1|1)_{\sR}^{2} \oplus \mf{u}(1)_{\sR}$,
$\mf{psu}(1|1)_{\sL}^{2} \oplus \mf{u}(1)_{\sL} \oplus \mf{u}(1)_{\sR}$ and $\mf{u}(1)_{\sL} \oplus \mf{u}(1)_{\sR}$, are consistent with reflection matrices that are solutions of the boundary Yang-Baxter (reflection) equation for the $\mf{psu}(1,1|2)^{2}$ open spin-chain, as described in Chapter \ref{part4-sec2}. 

\vspace{0.1cm}
\begin{table} [htb!]
\begin{tabular}{|cccc|}
\hline
\small{\text{\bf D1/D5 giant}} & \small{\text{\bf bosonic  generators}} & \small{\text{\bf supersymmetry generators}} & \small{\text{\bf boundary algebra}} \\
\hline
\small{$Z=0$} & \small{$\{ \mf{J}_{\ms{a} \, 0}, \, \mf{L}_{\ms{a} \, 5} \}$} & 
\small{$\{ \mf{Q}_{\ms{a} - + \pm}, \, \mf{Q}_{\ms{a} + - \pm}\}$} & \small{$[\mf{psu}(1|1)^{2} \oplus \mf{u}(1)]^{2}_{\ms c}$} \\
\hline
\small{$Y=0$} & \small{$\{ \mf{J}_{\ms{a} \, 0}, \, \mf{L}_{\ms{a} \, 5} \}$} & 
\small{$\{ \mf{Q}_{\sL - - \pm}, \, \mf{Q}_{\sL + + \pm}, \, \mf{Q}_{\sR - + \pm}, \, \mf{Q}_{\sR + - \pm} \}$} & 
\small{$\mf{u}(1)_{\sL} \oplus [\mf{psu}(1|1)^{2} \oplus \mf{u}(1)]_{\sR}$} \\
\small{$\bar{Y}=0$} & \small{$\{ \mf{J}_{\ms{a} \, 0}, \, \mf{L}_{\ms{a} \, 5} \}$} & 
\small{$\{ \mf{Q}_{\sL - + \pm}, \, \mf{Q}_{\sL + - \pm}, \, \mf{Q}_{\sR - - \pm}, \, \mf{Q}_{\sR + + \pm} \}$} & 
\small{$[\mf{psu}(1|1)^{2} \oplus \mf{u}(1)]_{\sL} \oplus \mf{u}(1)_{\sR}$} \\
\small{$\bar{Z}=0$} & \small{$\{ \mf{J}_{\ms{a} \, 0}, \, \mf{L}_{\ms{a} \, 5} \}$} & 
\small{$\{ \mf{Q}_{\sL - - \pm}, \, \mf{Q}_{\sL + + \pm}, \, \mf{Q}_{\sR - - \pm}, \, \mf{Q}_{\sR + + \pm} \}$} & 
\small{$\mf{u}(1)_{\sL} \oplus \mf{u}(1)_{\sR}$} \\
\hline
\end{tabular}
\caption{Boundary Lie algebras for D1- and D5-brane maximal giant gravitons on $AdS_{3} \times S^{3} \times T^{4}$.} \label{table-D1D5}
\end{table}
\vspace{0.1cm}


\part{$\mf{d}(2,1;\alpha)^2$ spin-chains in $AdS_{3}\times S^{3} \times S^{3\prime} \times S^{1}$}  \label{part3}


\section{Integrable closed $\mf{d}(2,1;\alpha)^2$ spin-chain and scattering matrices}  \label{part3-sec1}

The bosonic isometry group of the $AdS_3 \times S^3 \times {S^3}' \times S^1$ supergravity background is
\[
SO(2,2) \times SO(4) \times SO(4)' \times U(1),
\] 
whose Lie algebra splits into left- and right-movers
\[
\mf{so}(2,2) \sim \mf{su}(1,1)_{\sL} \op \mf{su}(1,1)_{\sR}, \hspace{0.5cm} \mf{so}(4) \sim \mf{su}(2)_{\sL} \op \mf{su}(2)_{\sR}, \hspace{0.5cm}
\mf{so}(4)^{\prime} \sim \mf{su}(2)^{\prime}_{\sL} \op \mf{su}(2)^{\prime}_{\sR}.
\]
According to this splitting, the bosonic isometries can be rearranged into 
\[
[\mf{su}(1,1) \op \mf{su}(2) \op \mf{su}(2)']_{\sL} \oplus [\mf{su}(1,1) \op \mf{su}(2) \op \mf{su}(2)']_{\sR} \oplus \mf{u}(1),
\]
which constitutes the bosonic part of the full superisometry algebra 
\[
\mf{d}(2,1;\alpha)_{\sL} \oplus \mf{d}(2,1;\alpha)_{\sR} \oplus \mf{u}(1).
\]
Massive excitations of the worldsheet of a closed IIB superstring propagating on $AdS_3 \times S^3 \times S^{3\prime} \times S^1$ can be identified with the magnon excitations of an alternating double-row $\mf{d}(2,1;\al)^{2}$ closed spin-chain which transform under a centrally extended $\mf{su}(1|1)^{2}_{\ms{c}}$ algebra \cite{Sax:2011}.
The left- and right-moving excitations\footnote{These left- and right-movers are not related to the actual left- and right-moving (clockwise and counter-clockwise) modes of a closed string, but are rather string excitations charged under generators of different copies of $\mf{d}(2,1;\al)$.} decouple in the weak coupling limit.
This chapter contains a review  based on \cite{Sax:2011,BSS13,Sfondrini:2014} of this integrable $\mf{d}(2,1;\al)^{2}$ closed spin-chain and the $S$-matrix describing two-magnon scattering.

\subsection{$\mf{d}(2,1;\al)^2$ spin-chain with $\mf{su}(1|1)^{2}$ excitations}

\subsubsection{Single-row $\mf{d}(2,1;\al)$ closed spin-chain with $\mf{su}(1|1)$ excitations} \label{part3-sec1-1-1}

\paragraph{Symmetry generators.}  The $\mf{d}(2,1;\al)$ superalgebra shown in Appendix \ref{app:B} has bosonic generators
\[ \mf{J}_{0},\mf{J}_{b} \in \mf{su}(1,1), \hspace{0.75cm} \mf{L}_{\beta},\mf{L}_{5} \in \mf{su}(2), \hspace{0.75cm} \mf{R}_{\dot{\beta}},\mf{R}_{8} \in \mf{su}(2)^{\prime} \]
of $\mf{su}(1,1) \op \mf{su}(2) \op \mf{su}(2)'$, and fermionic generators $\mf{Q}_{b\hspace{0.035cm}\beta\dot{\beta}}$ labeled by $b, \beta, \dot{\beta} = {\pm}$ indices.

\paragraph{Sites.}

Two neighbouring sites, called odd and even, in the alternating single-row $\mf{d}(2,1;\al)$ spin-chain are the modules 
\[M^{(\alpha)} \equiv M(-\tfrac{\alpha}{2},\tfrac{1}{2},0) = \text{span}_{\C}\{ \ket{\phi^{(n)}_{\beta}},  \ket{\psi^{(n)}_{\dot{\beta}}} \}, \hspace{0.35cm} M^{\prime \, (1-\alpha)} \equiv M(-\tfrac{1-\alpha}{2},0,\tfrac{1}{2}) = \text{span}_{\C}\{ \ket{\phi^{(n)}_{\dot{\beta}}},  \ket{\psi^{(n)}_{\beta}} \}.
\] 
States at these sites are vectors transforming under half-BPS representations of $\mf{d}(2,1;\al)$ described in Appendix \ref{app:B}.  The odd and even sites together form the module
$M=M^{(\alpha)} \otimes M^{\prime \, (1-\alpha)}$, which carries a quarter-BPS representation of $\mf{d}(2,1;\al)$.  The vacuum state is identified with the highest weight vector
\begin{equation} 
\ket{\mc{Z}} = \ket{\phi_{+}^{(0)} \hspace{0.05cm} \phi_{+}^{\prime \, (0)}},
\end{equation} 
and the four fundamental excitations $\ket{\varphi^{r}} = \{ \ket{\phi},\ket{\psi},  \ket{\phi^{\prime}},\ket{\psi^{\prime}} \}$, with $r=1\ldots4$, are defined by
\begin{equation} 
\ket{\phi} = \ket{\phi_{-}^{(0)} \hspace{0.05cm} \phi_{+}^{\prime \, (0)}}, \hspace{0.6cm} 
\ket{\psi} = \ket{\psi_{+}^{(0)} \hspace{0.05cm} \phi_{+}^{\prime \, (0)}}, \hspace{0.6cm}
\ket{\phi^{\prime}} = \ket{\phi_{+}^{(0)} \hspace{0.05cm} \phi_{-}^{\prime \, (0)}}, \hspace{0.6cm} 
\ket{\psi^{\prime}} = \ket{\phi_{+}^{(0)} \hspace{0.05cm} \psi_{+}^{\prime \, (0)}}. 
\end{equation}
Here $\{ \ket{\phi}, \ket{\psi}\}$ and $\{ \ket{\phi^{\prime}}, \ket{\psi^{\prime}}\}$ span modules of a closed $\mf{su}(1|1)$ subalgebra%
\footnote{There are a number of other closed subsectors (see Section 6 in \cite{SST13}).}
with fermionic generators $\mf{Q} \equiv \mf{Q}_{-++}$ and $\mf{S} \equiv \mf{Q}_{+--}$, and bosonic Cartan element 
$\mf{H} = \{\mf{Q},\mf{S}\} = -\mf{J}_{0}-\alpha \mf{L}_{5} - (1-\alpha) \mf{R}_{8}$ the magnon Hamiltonian. These unprimed and primed modules are associated with energies $\alpha$ and $1-\alpha$. We note this $\mf{su}(1|1)$ subalgebra can be extended to a $\mf{u}(1|1) = \mf{u}(1) \ltimes \mf{su}(1|1)$ algebra 
\[ \{\mf{Q}, \mf{S} \} = \mf{H}, \hspace{0.85cm} 
[\mf{X},\mf{Q}] = -\tfrac{1}{2} \hspace{0.05cm} \mf{Q}, \hspace{0.85cm} 
[\mf{X},\mf{S}] = \tfrac{1}{2} \hspace{0.05cm} \mf{S} 
 \]
by the inclusion of $\mf{X} = - \hspace{0.05cm} \tfrac{1}{2} \hspace{0.05cm} \mf{L}_{5} - \tfrac{1}{2} \hspace{0.05cm} \mf{R}_{8}$,
which does not annihilate the ground state.

\paragraph{Spin-chain.}

The alternating single-row $\mf{d}(2,1;\al)$ spin-chain with $2J$ sites can be identified with the module 
$M^{\otimes J} = (M^{(\alpha)} \otimes M^{\prime \, (1-\alpha)})^{\otimes J}$. The spin-chain vacuum and fundamental excitations are
\begin{equation} 
\ket{0}  = \ket{\mc{Z}^{J}}, \hspace{0.85cm} \ket{\varphi^{r}_{(n)}}  = \ket{\mc{Z}^{n-1} \varphi^{r} \mc{Z}^{J-n}}.
\end{equation}
We construct vectors in momentum space using the standard approach to obtain low-lying single-magnon excitations
\begin{equation} 
\ket{\varphi^{r}_{p}} = \sum_{n=1}^{J} e^{ipn} \ket{\varphi^{r}_{(n)}}. 
\end{equation}
Here $\{ \ket{\phi_{p}}, \ket{\psi_{p}}\}$ and $\{ \ket{\phi^{\prime}_{p}}, \ket{\psi^{\prime}_{p}}\}$ are modules of $\mf{su}(1|1)$ with energies $\alpha$ and $1-\alpha$. The action of the fermionic generators on the unprimed single-magnon excitations is given by
\begin{equation}
\mf{Q} \ket{\phi_{p}} = \sqrt{\alpha} \, \ket{\psi_{p}}, \hspace{0.6cm} \mf{S} \ket{\phi_{p}} = 0, \hspace{0.6cm} 
\mf{Q} \ket{\psi_{p}} = 0, \hspace{0.6cm}
\mf{S} \ket{\psi_{p}} = \sqrt{\alpha} \, \ket{\phi_{p}}, 
\end{equation}
where also
\[ \mf{X}\ket{0} = -\tfrac{J}{2}\ket{0}, \hspace{0.85cm} \mf{X} \ket{\phi_{p}} = (-\tfrac{J}{2} + \tfrac{1}{2}) \ket{\phi_{p}}, \hspace{0.85cm} 
\mf{X} \ket{\psi_{p}} = -\tfrac{J}{2} \ket{\psi_{p}},  \]
and similarly for the primed single-magnon excitations with $\alpha \to 1-\alpha$.
Multi-magnon excitations are obtained using the generalized standard approach 
\[
\ket{\varphi^{r_1}_{p_1}\varphi^{r_2}_{p_2}\cdots \hspace{0.05cm} \varphi^{r_k}_{p_k}} =  \sum_{1\leq n_1<n_2<\ldots<n_k\leq J}\!\! e^{i(p_1x_1 +p_2 x_2 + \ldots + p_k x_k)}  \ket{\mc{Z}^{n_1-1}\varphi^{r_1}_{(n_1)}\mc{Z}^{n_2-n_1-1}\varphi^{r_2}_{(n_2)}\cdots \hspace{0.05cm} \varphi^{r_k}_{(n_k)}\mc{Z}^{J-n_k}} 
\]
for $2 \leq k \ll J$.  The individual excitations $\varphi^{r_i}_{(n_i)}$ are known as impurities or fields in the spin-chain, and are assumed to be well-separated.

\subsubsection{Double-row $\mf{d}(2,1;\al)^{2}$ closed spin-chain with $\mf{su}(1|1)^{2}$ excitations} \label{part3-sec3-1-2}

The alternating double-row $\mf{d}(2,1;\al)^{2}$ spin-chain is made up of left- and right-moving $\mf{d}(2,1;\al)_{\sL}$ and $\mf{d}(2,1;\al)_{\sR}$ spin-chains which decouple in the weak coupling limit.

\paragraph{Sites.}

Odd and even sites of the left and right-moving spin-chains together form the module 
$M_{\sL} \otimes M_{\sR} = M^{(\alpha)}_{\sL} \otimes M_{\sL}^{\prime \, (1-\alpha)} \otimes M^{(\alpha)}_{\sR} \otimes M_{\sR}^{\prime \, (1-\alpha)}$.
The ground state and fundamental excitations are
\begin{equation} 
\ket{\mc{Z}} = \left|{\small\begin{pmatrix} \mc{Z}_{\sL} \\ \mc{Z}_{\sR} \end{pmatrix}} \right\rangle, 
\hspace{0.85cm}
\ket{\varphi^{r}} = \left|{\small\begin{pmatrix} \varphi^{r}_{\sL} \\ \mc{Z}_{\sR} \end{pmatrix}} \right\rangle, \hspace{0.85cm} 
\ket{\bar{\varphi}^{r}} = \left|{\small\begin{pmatrix} \mc{Z}_{\sL} \\ \varphi^{r}_{\sR} \end{pmatrix}} \right\rangle,
\end{equation}
which transform under the $\mf{u}(1|1)^{2}$ algebra 
\begin{equation} 
\{ \mf{Q}_{\ms{a}}, \mf{S}_{\ms{b}} \} = \mf{H}_{\ms{a}} \, \delta_{\ms{ab}}, \hspace{0.85cm} 
[ \mf{X}_{\ms{a}}, \mf{Q}_{\ms{b}} ] = - \hspace{0.05cm} \tfrac{1}{2} \hspace{0.05cm} \mf{Q}_{\ms{a}} \, \delta_{\ms{ab}}, \hspace{0.85cm}
[ \mf{X}_{\ms{a}}, \mf{S}_{\ms{b}} ] = \tfrac{1}{2} \hspace{0.05cm} \mf{S}_{\ms{a}} \, \delta_{\ms{ab}},
\end{equation}
with $\ms{a},\ms{b} \in \{\SL, \SR\}$. Notice that $\mf{X} = \mf{X}_{\sL} - \mf{X}_{\sR}$ now does annihilate the ground state, although $\mf{X}_{\sL}$ and $\mf{X}_{\sR}$
individually do not.  We define $\mf{H} = \mf{H}_{\sL} + \mf{H}_{\sR}$ and $\mf{M} = \mf{H}_{\sL} - \mf{H}_{\sR}$, with $\mf{H}$ the magnon Hamiltonian.
We do not need to consider excitations for which the left and right-moving excitations $\varphi_{\sL}^{r}$ and $\varphi_{\sR}^{r}$ coincide, since we focus on well-separated excitations in the $J \to \infty$ limit.

\paragraph{Spin-chain.}

The alternating double-row $\mf{d}(2,1;\alpha)^{2}$ spin-chain can be identified with the module $(M_{\sL} \otimes M_{\sR} )^{\otimes J}$.  The ground state is
\begin{equation}
\ket{0} = \ket{\mc{Z}^{J}} = \left| {\small\begin{pmatrix}\mc{Z}_{\sL} \\ \mc{Z}_{\sR} \end{pmatrix}^{\!J}} \right\rangle, \hspace{0.75cm}
\end{equation}
and left- and right-moving fundamental excitations are
\begin{eqnarray}
\nonumber && \ket{\varphi^r_{(n)}} = \ket{\mc{Z}^{n-1} \varphi^{r} \mc{Z}^{J-n}} = \left|  {\small\begin{pmatrix} \mc{Z}_{\sL} \\ \mc{Z}_{\sR} \end{pmatrix}^{\!n-1}\! \begin{pmatrix}\varphi_{\sL}^r \\ \mc{Z}_{\sR} \end{pmatrix}\! \, \begin{pmatrix}\mc{Z}_{\sL} \\ \mc{Z}_{\sR} \end{pmatrix}^{\!J-n}}\right\rangle, \\
 && \ket{\bar\varphi^{r}_{(n)}} = \ket{\mc{Z}^{n-1} \bar{\varphi}^{r} \mc{Z}^{J-n}} = \left|  {\small\begin{pmatrix} \mc{Z}_{\sL} \\ \mc{Z}_{\sR} \end{pmatrix}^{\!n-1}\! \begin{pmatrix}\mc{Z}_{\sL} \\ 
\varphi_{\sR}^{r}\end{pmatrix}\! \, \begin{pmatrix} \mc{Z}_{\sL} \\ \mc{Z}_{\sR} \end{pmatrix}^{\!J-n}}\right\rangle,
\end{eqnarray}
with low-lying left- and right-moving single-magnon excitations
\begin{equation}
\ket{\varphi^r_p} = \sum_{n=1}^J e^{ipn}\, \ket{\varphi^r_{(n)}}, \hspace{1.0cm} 
\ket{\bar{\varphi}^r_p} = \sum_{n=1}^J e^{ipn}\, \ket{\bar{\varphi}^r_{(n)}}.
\end{equation}
The unprimed and primed left- and right-moving magnon excitations $\{ \ket{\phi_{p}}, \ket{\psi_{p}} \}$ and $\{ \ket{\bar{\phi}_{p}}, \ket{\bar{\psi}_{p}} \}$, and $\{ \ket{\phi^{\prime}_{p}}, \ket{\psi^{\prime}_{p}} \}$ and $\{ \ket{\bar{\phi}^{\prime}_{p}}, \ket{\bar{\psi}^{\prime}_{p}} \}$ have energies $\alpha$ and $1-\alpha$ of the magnon Hamiltonian $\mf{H}$.  The left/right-movers have mass eigenvalues $\pm \alpha$ and $\pm (1-\alpha)$ of $\mf{M}$.
The non-trivial action of the fermionic generators of the $\mf{su}(1|1)^{2}$ algebra on these left- and right-moving magnon excitations is
\begin{equation}
\mf{Q}_{\sL} \ket{\phi_{p}} = \sqrt{\alpha} \, \ket{\psi_{p}}, \hspace{0.6cm} 
\mf{S}_{\sL} \ket{\psi_{p}} = \sqrt{\alpha} \, \ket{\phi_{p}}, \hspace{0.6cm}
\mf{Q}_{\sR} \ket{\bar{\phi}_{p}} = \sqrt{\alpha} \, \ket{\bar{\psi}_{p}}, \hspace{0.6cm} 
\mf{S}_{\sR} \ket{\bar{\psi}_{p}} = \sqrt{\alpha} \, \ket{\bar{\phi}_{p}}, 
\end{equation}
with the non-trivial action of the additional $\mf{u}(1)$ generator of the $\mf{u}(1) \ltimes \mf{su}(1|1)^{2}$ algebra, which annihilates the ground state, given by
\[ \mf{X} \ket{\phi_{p}} = \tfrac{1}{2} \hspace{0.05cm} \ket{\phi_{p}}, \hspace{0.85cm}
\mf{X} \ket{\bar{\phi}_{p}} = - \hspace{0.025cm} \tfrac{1}{2} \hspace{0.05cm} \ket{\bar{\phi}_{p}},
\]
and similarly for the primed left- and right-moving magnon excitations with $\alpha \to 1-\alpha$. The generalization to multi-magnon states is again straightforward.

\subsection{$\mf{d}(2,1;\al)^2$ spin-chain with centrally extended $\mf{su}(1|1)^{2}_{\ms{c}}$ excitations}

We have so far considered only the weak-coupling limit of the $\mf{d}(2,1;\al)^2$ spin-chain in which the left- and right-moving excitations decouple.  Beyond this regime, interactions must be taken into account by the introduction of a centrally extended $\mf{su}(1|1)^2$ algebra which links the two otherwise independent $\mf{d}(2,1;\al)_{\sL}$ and $\mf{d}(2,1;\al)_{\sR}$ spin-chains. Let us denote this extended algebra by $\mf{su}(1|1)^2_{\ms c}$.

The algebra $\mf{su}(1|1)^2_{\ms c}$ is generated by the fermionic generators $\mf{Q}_{\ms a}$, $\mf{S}_{\ms a}$, and bosonic generators $\mf{H}_{\ms a}$ and the central elements $\mf{P}$, $\mf{P}^\dag$, where ${\ms a}\in \{\SL,\SR\}$, satisfying%
\footnote{Setting $e_1=\mf{Q}_{\sL}$, $e_2=\mf{S}_{\sR}$, $f_1=\mf{S}_{\sL}$, $f_2=\mf{Q}_{\sR}$, $h_1=\mf{H}_{\sL}$, $h_2=\mf{H}_{\sR}$, $k_1=\mf{P}$, $k_2=\mf{P}^\dagger$, the anti-commutation relations \eqref{extended-algebra} can be written in a more compact form as follows: $\{ e_i,f_j\}=\del_{ij} \, h_i + (1-\del_{ij}) \, k_i$ (see Section 2 in \cite{Re}).
}
\begin{equation} \label{extended-algebra}
\{\mf{Q}_{\ms a},\mf{S}_{\ms b} \} = \delta_{\ms{ab}} \, \mf{H}_{\ms a}, \hspace{0.85cm} \{\mf{Q}_\sL,\mf{Q}_\sR\} = \mf{P}, \hspace{0.85cm} \{\mf{S}_\sL,\mf{S}_\sR\} = \mf{P}^\dag,
\end{equation}
with the remaining relations being trivial. We may further extend this algebra by the inclusion of the element $\mf{X}$, which annihilates the ground state and has non-trivial commutation relations:
\[ 
[\mf{X}, \mf{Q}_{\sL}] = -\tfrac{1}{2} \hspace{0.05cm} \mf{Q}_{\sL}, \hspace{0.85cm} [\mf{X}, \mf{S}_{\sL}] = \tfrac{1}{2} \hspace{0.05cm} \mf{S}_{\sL}, \hspace{0.85cm}
[\mf{X}, \mf{Q}_{\sR}] = \tfrac{1}{2} \hspace{0.05cm} \mf{Q}_{\sR}, \hspace{0.85cm} [\mf{X}, \mf{S}_{\sR}] = - \tfrac{1}{2} \hspace{0.05cm} \mf{S}_{\sR}. 
\]
A dynamic $\mf{d}(2,1;\al)^2$ spin-chain with $\mf{su}(1|1)^2_{\ms c}$--symmetric massive excitations was constructed in \cite{BSS13} and, subsequently, a non-dynamic $\mf{d}(2,1;\al)^2$ spin-chain with an additional Hopf algebra structure was introduced in \cite{BOSST13}.  We now briefly review these constructions.

\subsubsection{Finite spin-chain with length-changing effects}  \label{part3-sec1-2-1}

Here we allow the additional bosonic central elements $\mf{P}$ and $\mf{P}^{\dag}$ of the $\mf{su}(1|1)^2_{\ms c}$ algebra to have a length-changing effect on the spin-chain.
Let us introduce some additional notation: $\mc{Z}^+$ and $\mc{Z}^-$ denote the insertion or removal of a vacuum state (if possible) at the specific spin-chain site in the left-moving magnon excitation $\ket{\varphi^{r}_p}$:
\vspace{-0.15cm}
\[
\ket{\mc{Z}^\pm\varphi^{r}_p} = \sum_{n=1}^J e^{ipn}\left|  {\mc{Z}^{n-1\pm1} \varphi^{r} \mc{Z}^{J-n}} \right\rangle,  \hspace{1.0cm}
\ket{\varphi^{r}_p\mc{Z}^\pm} = \sum_{n=1}^J e^{ipn}\left|  {\mc{Z}^{n-1} \varphi^{r} \mc{Z}^{J-n\pm1}} \right\rangle, 
\]
where we define $|\mc{Z}^{-1}\varphi^{r} \cdots \rangle \equiv |\varphi^{r} \cdots \rangle$ and $|\cdots \varphi^{r} \mc{Z}^{-1} \rangle \equiv | \cdots \varphi^{r} \rangle$ (that is, if there is no vacuum state before or after the field $\varphi^{r}$ to remove, then the state remains unchanged).
Imposing periodic boundary conditions $e^{ipJ}|\mc{Z}^J\varphi\ran = |\varphi \mc{Z}^J\ran$ for a closed spin-chain now gives
\begin{equation}  \label{Z-shift}
\ket{\varphi^{r}_p \hspace{0.025cm} \mc{Z}^\pm} = e^{\pm ip} \, \ket{\mc{Z}^\pm \varphi^{r}_p},
\end{equation}
and similarly for right-moving magnon excitations. For two left-moving magnons, we define
\begin{eqnarray}
&& \nonumber \hspace{-0.65cm} \ket{\mc{Z}^\pm\varphi^{r}_p\varphi^{s}_q} = \sum_{1\leq n<m\leq J}  e^{i(pn + qm)}\left|  {\mc{Z}^{n-1\pm1} \varphi^{r} \mc{Z}^{m-n-1} \varphi^{s} \mc{Z}^{J-m}} \right\rangle,  \\
&& \nonumber \hspace{-0.65cm} \ket{\varphi^{r}_p \mc{Z}^\pm \varphi^{s}_q} = \sum_{1\leq n<m \leq J} e^{i(pn + qm)}\left|  {\mc{Z}^{n-1} \varphi^{r} \mc{Z}^{m-n-1\pm1} \varphi^{s} \mc{Z}^{J-m}} \right\rangle,
\end{eqnarray}
and hence, using the periodic boundary conditions,
\begin{equation}
\ket{\varphi^{r}_{p} \mc{Z}^\pm\varphi^{s}_{q}} = e^{\pm ip} \ket{\mc{Z}^\pm\varphi^{r}_{p} \varphi^{s}_{q}}, \label{Z-shift:2}
\end{equation}
and similarly for two right-moving magnons, or for left- and right-moving magnon excitations.

\paragraph{Single-magnon excitations.}

The action of the fermionic generators of the $\mf{su}(1|1)^2_{\ms c}$ algebra on the left and right-moving magnon excitations was proposed in \cite{BSS13}. Here we will consider a slightly different action,\footnote{This action is equivalent to that of \cite{BSS13} with a rescaling $c_p\to e^{-ip}c_p$ and $d_p\to e^{ip}d_p$ for the left-moving excitations, and a similar rescaling of $\bar{c}_p\to e^{-ip}\bar{c}_p$ and $\bar{d}_p\to e^{ip}\bar{d}_p$ for the right-moving excitations, as can be seen from \eqref{Z-shift}, and will be convenient when we subsequently consider a semi-infinite open spin-chain with a boundary on the right.} with the insertion or removal of $\mc{Z}$ from the left side:
\begin{eqnarray} 
\nonumber && \hspace{-0.75cm} \mf{Q}_\sL \ket{\phi_{p}} = a_{p} \ket{\psi_p}, \hspace{0.6cm} 
\mf{S}_\sL \ket{\psi_{p}} = b_{p} \ket{\phi_p}, \hspace{0.6cm}
\mf{Q}_\sR \ket{\psi_{p}} = c_{p} \ket{\mc{Z}^+ \phi_{p}}, \hspace{0.6cm}
\mf{S}_\sR \ket{\phi_{p}} = d_{p} \ket{\mc{Z}^- \psi_{p}}, \\
&& \hspace{-0.75cm} \mf{Q}_\sR \ket{\bar{\phi}_{p}} = \bar{a}_{p} \ket{\bar \psi_p}, \hspace{0.6cm}
\mf{S}_\sR \ket{\bar{\psi}_{p}} = \bar{b}_{p} \ket{\bar\phi_p}, \hspace{0.6cm} 
\mf{Q}_\sL \ket{\bar{\psi}_{p}} = \bar{c}_{p} \ket{\mc{Z}^+ \bar\phi_p}, \hspace{0.6cm}
\mf{S}_\sL \ket{\bar{\phi}_{p}} = \bar{d}_{p} \ket{\mc{Z}^{-} \bar{\psi}_p}, \hspace{0.8cm} \label{action-one-magnon}
\end{eqnarray}
with the unbarred and barred parameters for left and right-movers. 
The action of the central elements is deduced from the algebra \eqref{extended-algebra}. 
The energy eigenvalues of $\mf{H}=\mf{H}_{\sL}+\mf{H}_{\sR}$ for a left- and right-moving magnon excitation, respectively, are
\[ E_{p} = a_{p} b_{p} + c_{p} d_{p} \hspace{0.75cm} \text{and} \hspace{0.75cm} 
\bar{E}_{p} = \bar{a}_{p} \bar{b}_{p} + \bar{c}_{p} \bar{d}_{p}, \]
while the eigenvalues of $\mf{M} = \mf{H}_{\sL} - \mf{H}_{\sR}$, related to the masses of the magnon excitations, are
\[ m = a_{p} b_{p} - c_{p} d_{p} \hspace{0.75cm} \text{and} \hspace{0.75cm}
-\bar{m} = -(\bar{a}_{p} \bar{b}_{p} - \bar{c}_{p} \bar{d}_{p}),  \]
taken to be independent of the momentum $p$ (but dependent on the unprimed or primed flavour). 
Here $m = \bar{m} = \alpha$ and $m^{\prime} = \bar{m}^{\prime} = 1-\alpha$.
Now, for one physical left-moving magnon excitation, we require the eigenvalues $\mf{P}$ and $\mf{P}^{\dag}$ to vanish to return to a representation of the non-extended $\mf{su}(1|1)^{2}$ symmetry preserved by the vacuum. This leads to the conditions
$a_{p} c_{p}=b_{p} d_{p}=0$ and $\bar{a}_{p} \bar{c}_{p}=\bar{b}_{p} \bar{d}_{p}=0$ with solution $c_{p}=d_{p}=\bar{c}_{p}=\bar{d}_{p}=0$ resulting again in decoupled spin-chains.

\paragraph{Two-magnon excitations.}

Let us now consider two left-moving magnon excitations $\ket{\varphi^{r}_{p} \varphi^{s}_{q}}$ of momenta $p$ and $q$.  The non-trivial action of the fermionic generators of the $\mf{su}(1|1)^2_{\ms c}$ algebra is
\begin{eqnarray}
\nonumber && \hspace{-0.65cm} 
\mf{Q}_{\ms a} \ket{\phi_p \phi_q} = \delta_{\ms a\sL} a_p \ket{\psi_p \phi_q} + \delta_{\ms a\sL} a_q \ket{\phi_p \psi_q}, \hspace{1.99cm} 
\mf{Q}_{\ms a} \ket{\phi_p \psi_q} = \delta_{\ms a\sL} a_p \ket{\psi_p \phi_q} + \delta_{\ms a\sR} e^{ip} c_q \ket{\mc{Z}^+ \phi_p \phi_q}, \\
\nonumber && \hspace{-0.65cm} 
\mf{Q}_{\ms a} \ket{\psi_p \phi_q} = \delta_{\ms a\sR} c_p \ket{\mc{Z}^+ \phi_p \phi_q} - \delta_{\ms a\sL} a_q \ket{\psi_p \psi_q}, \hspace{1.42cm}
\mf{Q}_{\ms a} \ket{\psi_p \psi_q} = \delta_{\ms a\sR} c_p \ket{\mc{Z}^+ \phi_p \psi_q} - \delta_{\ms a\sR} e^{ip} c_q \ket{\mc{Z}^+ \psi_p \phi_q}, \\ 
\nonumber && \hspace{-0.65cm} 
\mf{S}_{\ms a} \ket{\phi_p \phi_q} = \delta_{\ms a\sR} d_p \ket{\mc{Z}^- \psi_p \phi_q} + \delta_{\ms a\sR} e^{-ip} d_q \ket{\mc{Z}^- \phi_p \psi_q}, \hspace{0.15cm} 
\mf{S}_{\ms a} \ket{\phi_p \psi_q} = \delta_{\ms a\sR} d_p \ket{\mc{Z}^- \psi_p \phi_q} + \delta_{\ms a\sL} b_q \ket{\phi_p \phi_q}, \\
&& \hspace{-0.65cm}  
\mf{S}_{\ms a} \ket{\psi_p \phi_q} = \delta_{\ms a\sL} b_p \ket{\phi_p \phi_q} - \delta_{\ms a\sR} e^{-ip} d_q \ket{\mc{Z}^- \psi_p \psi_q}, \hspace{0.74cm}
\mf{S}_{\ms a} \ket{\psi_p \psi_q} = \delta_{\ms a\sL} b_p \ket{\phi_p \psi_q} - \delta_{\ms a\sL} b_q \ket{\psi_p \phi_q},  \label{dynamic-two-magnon}
\end{eqnarray}
and similarly for primed left-moving magnon excitations. The action on two right-moving magnon excitations $\ket{\bar{\varphi}^{r}_{p} \bar{\varphi}^{s}_{q}}$, or a left- and a right-mover, $\ket{\varphi^{r}_{p} \bar{\varphi}^{s}_{q}}$ or $\ket{\bar{\varphi}^{r}_{p} \varphi^{s}_{q}}$, is obtained by interchanging $\SL\leftrightarrow\SR$ indices and using barred notation for the action on the right-movers.  
The energy eigenvalue of the magnon Hamiltonian $\mf{H}$ is $E_{p} + E_{q}$.
Now $\mf{P}$ and $\mf{P}^{\dag}$ must annihilate physical two-magnon states, which implies
\[ a_{p}c_{p} + e^{ip} \, a_{q}c_{q} = h \, (1-e^{i \hspace{0.025cm} (p+q)} ) = 0, \hspace{1.0cm} 
b_{p}d_{p} + e^{-ip} \, b_{q}d_{q} = h \, (1-e^{-i \hspace{0.025cm}(p+q)} ) = 0, \]
and hence $e^{i(p+q)}=1$, with our choice of parameters
\begin{equation} \label{parameters} 
a_{p} = \sqrt{h} \hspace{0.15cm} \eta_{p}, \hspace{0.5cm}  
b_{p} = \sqrt{h} \hspace{0.15cm} \eta_{p}, \hspace{0.5cm} 
c_{p} = -\sqrt{h} \hspace{0.15cm} \frac{i \hspace{0.025cm} \eta_{p}}{x_{p}^{-}}, \hspace{0.5cm} 
d_{p} = \sqrt{h} \hspace{0.15cm}  \frac{i \hspace{0.025cm} \eta_{p}}{x_{p}^{+}}, \hspace{0.5cm} \text{with} \hspace{0.45cm} 
\eta_{p}^{2} = i \, (x_{p}^{-} - x_{p}^{+}). 
\end{equation}
Exactly the same choice of barred parameters $\bar{a}_{p}$, $\bar{b}_{p}$, $\bar{c}_{p}$ and $\bar{d}_{p}$ must apply for the right-moving magnon representation.
Here $x_{p}^{\pm}$ are the Zhukovski variables which satisfy
\begin{equation} 
 \frac{x_{p}^{+}}{x_{p}^{-}} = e^{ip}, \hspace{1.0cm} 
\left( x_{p}^{+} + \frac{1}{x_{p}^{+}} \right) - \left( x_{p}^{-} + \frac{1}{x_{p}^{-}} \right) = \frac{i \hspace{0.05cm} m}{h}. \label{Zhukovski}
\end{equation}
The energy of a single-magnon excitation of momentum $p$ and mass $m$ is given by
\begin{equation} 
E_{p} = \sqrt{m^{2} + 16 \hspace{0.025cm} h^{2}\sin^{2}{\tfrac{p}{2}}}. 
\end{equation}

\subsubsection{Infinite spin-chain with Hopf algebra structure}  \label{part3-sec1-2-2}

In the $J\to\infty$ infinite spin-chain limit, we can drop the $\mc{Z}^{\pm}$ symbols on the left, and thus obtain one- and two-magnon representations of the $\mf{su}(1|1)^2_{\ms c}$ algebra.
The length-changing effects can be encoded in a $\mf{U}$-braided Hopf algebra structure for the $\mf{su}(1|1)^2_{\ms c}$ algebra, similar to that of Appendix B in \cite{BOSST13}. 

\paragraph{Single-magnon representations.}

We want to rewrite the action \eqref{action-one-magnon} in terms of matrix representations of the $\mf{su}(1|1)^2_{\ms c}$ algebra. Let us introduce vector spaces
\[
V_{p}={\rm span}_\C\{ \ket{\phi_{p}}, \ket{\psi_{p}} \},  \hspace{0.85cm} 
V^{\prime}_{p}={\rm span}_\C\{ \ket{\phi^{\prime}_{p}}, \ket{\psi^{\prime}_{p}} \},
\]
for the left-moving magnons, and
\[ 
\bar{V}_{p}={\rm span}_\C\{ \ket{\bar{\phi}_{p}}, \ket{\bar{\psi}_{p}} \}, \hspace{0.85cm} 
\bar{V}^{\prime}_{p}={\rm span}_\C\{ \ket{\bar{\phi}^{\prime}_{p}}, \ket{\bar{\psi}^{\prime}_{p}} \},
\]
for the right-moving magnons. We can identify these vector spaces with $\C^{1|1}$ in the natural way. Now the action \eqref{action-one-magnon} can be defined in terms of the usual supermatrices $\bb{E}_{ij} \in\End(\C^{1|1})$ which span the $\Z_{2}$-graded $\mf{gl}{(1|1)}$ Lie superalgebra,\footnote{The supermatrices $\bb{E}_{ij} \in \End(\C^{1|1})$ have matrix elements $(\bb{E}_{ij})_{ab} = \delta_{ia} \delta_{jb}$, with $\bb{E}_{11}$ and $\bb{E}_{22}$ even elements of degree 0, and $\bb{E}_{12}$ and $\bb{E}_{21}$ odd elements of degree 1. These supermatrices satisfy
\[
\llbracket \bb{E}_{ij}, \bb{E}_{kl}\rrbracket = \delta_{jk} \, \bb{E}_{il} - (-1)^{ \deg{\bb{E}_{ij}} \deg{\bb{E}_{kl}} } \, \delta_{il} \, \bb{E}_{kj}.
\]
with $\llbracket \cdot, \cdot \rrbracket$ the supercommutator. The multiplication of the tensor product of supermatrices is 
\[ (X \ot Y)(Z\ot W) = (-1)^{\deg Z \deg Y} XZ \ot YW \hspace{1.0cm}  \deg(X\ot Y) = \deg X + \deg Y \]
for any $X,Y,Z,W \in \mf{gl}{(1|1)}$ and, in particular, 
\[
(\bb{E}_{ij}\ot \bb{E}_{kl})(\bb{E}_{pr}\ot \bb{E}_{st}) = (-1)^{ \deg{\bb{E}_{kl}} \deg{\bb{E}_{pr}}} \, \delta_{jp} \, \delta_{ls} \,\, \bb{E}_{ir} \ot \bb{E}_{kt}.
\]
The graded permutation operator $\bb{P} \in \End(\C^{1|1}\ot \C^{1|1})$ is
\[
\bb{P} = \sum _{i,j} \, (-1)^{(j-1)} \, \bb{E}_{ij} \ot \bb{E}_{ji}
 =  \bb{E}_{11}\ot \bb{E}_{11} - \bb{E}_{12} \ot \bb{E}_{21} + \bb{E}_{21} \ot \bb{E}_{12} - \bb{E}_{22} \ot \bb{E}_{22}.
\vspace{-0.1cm}
\]
The generalization to supermatrices in $\End(\C^{2|2})$ is straightforward.
}
and the identity matrix is $\bb{I} = \bb{E}_{11} + \bb{E}_{22}$. The left-moving representation $\pi_{p}: \mf{su}(1|1)^2_{\ms c} \to \End{(\C^{1|1})}$ is
\begin{eqnarray} \label{Rep:L}
\nonumber && \hspace{-1.5cm} \pi_p(\mf{Q}_\sL) = a_{p} \, \bb{E}_{21}, \hspace{0.65cm}
 \pi_p(\mf{Q}_\sR) = c_{p} \, \bb{E}_{12},  \hspace{0.65cm}
 \pi_p(\mf{S}_\sL) = b_{p} \, \bb{E}_{12},  \hspace{0.65cm}
 \pi_p(\mf{S}_\sR) = d_{p} \, \bb{E}_{21},  \\
&& \hspace{-1.5cm}  \pi_p(\mf{H}_\sL) = a_p b_p\, \bb{I}, \hspace{0.78cm}
 \pi_p(\mf{H}_\sR) = c_p d_p\, \bb{I}, \hspace{0.75cm}
 \pi_p(\mf{P}) = a_p c_p \, \bb{I}, \hspace{0.82cm}
 \pi_p(\mf{P}^\dag) = b_p d_p \, \bb{I},
\end{eqnarray}
with the parameters \eqref{parameters}.  The right-moving representation $\bar{\pi}_{p} : \mf{su}(1|1)^2_{\ms c} \to \End{(\C^{1|1})}$ is
\begin{eqnarray} \label{Rep:R}
\nonumber && \hspace{-1.5cm} \bar{\pi}_{p}(\mf{Q}_\sR) = a_{p} \, \bb{E}_{\bar{2}\bar{1}}, \hspace{0.65cm}
 \bar{\pi}_{p}(\mf{Q}_\sL) = c_{p} \, \bb{E}_{\bar{1}\bar{2}},  \hspace{0.65cm}
 \bar{\pi}_{p}(\mf{S}_\sR) = b_{p} \, \bb{E}_{\bar{1}\bar{2}},  \hspace{0.64cm}
 \bar{\pi}_{p}(\mf{S}_\sL) = d_{p} \, \bb{E}_{\bar{2}\bar{1}},  \\
&& \hspace{-1.5cm} \bar{\pi}_{p} (\mf{H}_\sR) = a_{p} b_{p} \, \bb{I}, \hspace{0.78cm}
 \bar{\pi}_{p}(\mf{H}_\sL) = c_{p} d_{p} \, \bb{I}, \hspace{0.75cm}
 \bar{\pi}_{p}(\mf{P}) = a_{p} c_{p} \, \bb{I}, \hspace{0.82cm}
 \bar{\pi}_{p}(\mf{P}^\dag) = b_{p} d_{p} \, \bb{I},
\end{eqnarray}
obtained by interchanging $\SL \leftrightarrow \SR$ indices; we have also replaced the indices $1\to\bar{1}$ and $2\to\bar{2}$ to distinguish left and right vector spaces. The primed representations are similarly defined. We will not add subscripts to distinguish the identity matrices; we hope it will always be clear from the context in which space the identity matrix lives.

\paragraph{Hopf algebra.} 

We introduce an additional group-like generator $\mf{U}$, which is central with respect to the $\mf{su}(1|1)^2_{\ms c}$ algebra. The action on any single-magnon excitation is
\begin{equation}
\mf{U} \, | \varphi^{r}_{p} \rangle = e^{i\frac{p}{2}} \hspace{0.025cm} | \varphi^{r}_{p} \rangle,  \hspace{1.0cm}
\mf{U} \, | \bar{\varphi}^{r}_{p} \rangle = e^{i\frac{p}{2}} \hspace{0.025cm} | \bar{\varphi}^{r}_{p} \rangle,
\end{equation}
so that
\begin{equation}
\pi_p(\mf{U}) =  - \hspace{0.05cm} c_{p} d^{-1}_{p} \, \bb{I} = e^{i\frac{p}{2}} \, \bb{I},  \hspace{1.0cm}  
\bar\pi_p(\mf{U}) =  - \hspace{0.05cm} \bar{c}_{p} \bar{d}^{-1}_{p} \, \bb{I} = e^{i\frac{p}{2}} \, \bb{I},
\end{equation}
for our left- and right-moving single-magnon representations. We are now ready to define a Hopf algebra structure on $\mf{su}(1|1)^2_{\ms c}$. We denote this Hopf superalgebra by $\mc{A}$ throughout Part \ref{part3} so that 
\[ L(\mathcal{A})=\mf{su}(1|1)^{2}_{\ms{c}} \]
is the associated Lie superalgebra.

Let $\mathbf{1}$ denote the unit of $\mc{A}$. The coproduct $\Delta_\sL$ corresponding to the action on the left-moving magnon excitations in this spin-chain frame is given by 
\begin{eqnarray}
\nonumber && \hspace{-0.65cm} \Delta_\sL(\mf{Q}_{\ms a}) = \mf{Q}_{\ms a} \ot \mathbf{1} + \mf{U}^{2\delta_{\ms a\sR}} \ot \mf{Q}_{\ms a}, \hspace{1.16cm} \Delta_\sL(\mf{P}) = \mf{P} \ot \mathbf{1} + \mf{U}^2 \ot \mf{P}, \\
\nonumber && \hspace{-0.65cm} \Delta_\sL(\mf{S}_{\ms a}) = \mf{S}_{\ms a} \ot \mathbf{1} + \mf{U}^{-2\delta_{\ms a\sR}}\! \ot \mf{S}_{\ms a},  \hspace{1.0cm} \Delta_\sL(\mf{P}^\dag) = \mf{P}^\dag \ot \mathbf{1} + \mf{U}^{-2} \ot \mf{P}^\dag, \\
\nonumber && \hspace{-0.65cm} \Delta_\sL(\mf{H}_{\ms a}) = \mf{H}_{\ms a} \ot \mathbf{1} + \mathbf{1} \ot \mf{H}_{\ms a}, \hspace{1.89cm} \Delta_\sL(\mf{U}^{\pm1}) = \mf{U}^{\pm1} \ot \mf{U}^{\pm1},
\end{eqnarray}
and the coproduct $\Delta_\sR$ giving the action on the right-moving magnon excitations is obtained by interchanging indices $\sL\leftrightarrow\sR$. In the representation $\pi_{p} \ot \pi_{q}$, this yields \eqref{dynamic-two-magnon} in the infinite spin-chain limit. 
It is convenient to switch to a symmetric frame,\footnote{In the representation $\pi_{p} \otimes \pi_{q}$, we twist the left- and right co-products $\Delta_{\sL} \to T^{-1}_{\sL} \Delta_{\sL} \, T_{\sL}$ and $\Delta_{\sR} \to T^{-1}_{\sR} \Delta_{\sR} \, T_{\sR}$ using the twist matrices
\[ 
\bb{T}_{\sL} = e^{i \frac{p}{2}} \hspace{0.05cm} \bb{I} \ot \bb{E}_{11} +  \bb{I} \ot \bb{E}_{22} \hspace{1.0cm}
\bb{T}_{\sR} = e^{i \frac{p}{2}} \hspace{0.05cm} \bb{I} \ot \bb{E}_{\bar 1 \bar 1} +  \bb{I} \ot \bb{E}_{\bar 2\bar 2}. 
\] \vspace{-0.2cm}} similar to that of \cite{BOSST13}, in which the coproduct $\Delta$ is the same for both left- and right-moving sectors\footnote{This coproduct exhibits a $\Z$-grading (see Remark 2.1 in \cite{Re}).}:
\begin{eqnarray}
&& \nonumber \hspace{-0.75cm} \Delta(\mf{Q}_{\ms a}) = \mf{Q}_{\ms a} \ot \mathbf{1} + \mf{U} \ot \mf{Q}_{\ms a},  \hspace{1.35cm} \Delta(\mf{P}) = \mf{P} \ot \mathbf{1} + \mf{U}^{2} \ot \mf{P},  \\
&& \nonumber \hspace{-0.75cm} \Delta(\mf{S}_{\ms a}) = \mf{S}_{\ms a} \ot \mathbf{1} + \mf{U}^{-1}\! \ot \mf{S}_{\ms a}, \hspace{1.0cm} \Delta(\mf{P}^\dag) = \mf{P}^\dag \ot \mathbf{1} + \mf{U}^{-2} \ot \mf{P}^\dag, \\
&& \hspace{-0.75cm} \Delta(\mf{H}_{\ms a}) = \mf{H}_{\ms a} \ot \mathbf{1} + \mathbf{1} \ot \mf{H}_{\ms a}, \hspace{1.52cm} \Delta(\mf{U}^{\pm1}) = \mf{U}^{\pm1} \ot \mf{U}^{\pm1}. 
\end{eqnarray}
The central elements of the superalgebra  $\mf{C}\in\{\mf{H}_{\ms a},\mf{P},\mf{P}^\dag\}$ must be co-commutative, $\Delta(\mf{C})=\Delta^{\rm{op}}(\mf{C})$, where $\Delta^{\rm op}(\mf{a}) = \mc{P} \hspace{0.05cm} \Delta(\mf{a})$ for $\mf{a} \in \mc{A}$, with $\mc{P}$ the graded permutation operator which permutes the elements of the superalgebra in the coproduct. This is true if\,\footnote{To be precise, we extend $\mf{su}(1|1)^2_{\ms c}$ by $\mf{U}$ and its inverse $\mf{U}^{-1}$, and then consider the Hopf algebra over the quotient of the enveloping algebra of this double-extended algebra by the ideal $\langle \, \mf{P} - \nu_{1} \,(1-\mf{U}^2),\; \mf{P}^\dag-\nu_{2}\,(1-\mf{U}^{-2}) \, \rangle$.}
\begin{equation}
\mf{P} = \nu_{1} \hspace{0.1cm} (1-\mf{U}^2) \qq\text{and}\qq \mf{P}^\dag = \nu_{2} \hspace{0.1cm} (1-\mf{U}^{-2}), \hspace{0.75cm} \text{with} \hspace{0.5cm}  \nu_{1}, \nu_{2} \in \C\backslash\{0\},
\end{equation}
where we choose $\nu_{1} = \nu_{2} = h$ to obtain our previous unitary representation.

Let $\mu: \mc{A} \ot \mc{A} \to \mc{A}$ be the usual associative multiplication of elements of the superalgebra $\mc{A}$. Moreover, let $\imath: \C\to \mc{A}$ be the unit, which maps $\imath(1) =  \mathbf{1}$, and  $\epsilon: \mc{A} \to \C$ the counit, defined by
\begin{equation} 
\epsilon \hspace{0.025cm} (\mf{U}^{\pm1})=1  \hspace{0.75cm} \text{and} \hspace{0.75cm}  \epsilon \hspace{0.025cm} (\mf{J})=0 
\end{equation}
for all $\mf{J}\in\mf{su}(1|1)^2_{\ms c}$.
The antipode $\mathscr{S}: \mc{A} \rightarrow \mc{A}$ must then satisfy $\mu \hspace{0.025cm} (\mathscr{S}\ot id)\,\Delta = \imath \,\epsilon$, which gives
\begin{eqnarray}
&& \nonumber \hspace{-0.65cm} \mathscr{S}(\mf{Q}_{\ms a}) = - \hspace{0.025cm} \mf{U}^{-1} \, \mf{Q}_{\ms a}, \hspace{0.85cm} 
\mathscr{S}(\mf{S}_{\ms a}) = - \hspace{0.025cm} \mf{U} \, \mf{S}_{\ms a}, \hspace{0.97cm}  
\mathscr{S}(\mf{H}_{\ms a}) = - \hspace{0.025cm} \mf{H}_{\ms a}, \\
&& \hspace{-0.65cm} \mathscr{S}(\mf{P}) = - \hspace{0.025cm} \mf{U}^{-2} \, \mf{P},  \hspace{1.2cm} 
\mathscr{S}(\mf{P}^\dag) = - \hspace{0.025cm} \mf{U}^{2} \, \mf{P}^\dag, \hspace{0.8cm} \mathscr{S}(\mf{U}) = \mf{U}^{-1}.  \label{antipode}
\end{eqnarray}
This antipode relates left and right-movers in the representations $\pi_{p}$ and $\pi_{\bar{p}}$:
\begin{equation} \pi_{p}(\mathscr{S}(\mf{a})) =  (\pi_{\bar{p}}(\mf{\bar{a}}))^{\rm str},
\end{equation}
with the charge conjugation matrix trivial. Here $\mf{a} \in \mc{A}$, with $\bar{\mf{a}} \in \mc{A}$ defined by
\begin{eqnarray}
\nonumber 
&& \hspace{-0.65cm}
\bar{\mf{Q}}_{\ms a} = \delta_{\ms a\sL} \, \mf{Q}_{\sR} \, + \, \delta_{\ms a\sR} \, \mf{Q}_{\sL}, \hspace{0.85cm}
 \bar{\mf{P}} = \mf{P}, \hspace{1.14cm}
 \bar{\mf{H}}_{\ms a} \; = \delta_{\ms a\sL} \, \mf{H}_{\sR} \, + \, \delta_{\ms a\sR} \, \mf{H}_{\sL}, \\
 && \hspace{-0.65cm}
\bar{\mf{S}}_{\ms a} = \delta_{\ms a\sL} \, \mf{S}_{\sR} \, + \, \delta_{\ms a\sR} \, \mf{S}_{\sL}, \hspace{0.85cm}
 \bar{\mf{P}}^{\dag} = \mf{P}^{\dag}, \hspace{0.85cm}
 \bar{\mf{U}}^{\pm 1} = \mf{U}^{\pm 1}. \label{bar-algebra}
\end{eqnarray}
The representation $\pi_{\bar{p}}$ has Zhukovski variables 
\begin{equation} \label{bar-Zhukovski} 
x^{\pm}_{\bar{p}} = \frac{1}{x^{\pm}_{p}}. 
\end{equation}

\subsection{Two-magnon scattering and $R$-matrices}

We are interested in the scattering of magnon excitations. Let $\mc{H}_{\rm (in)}$ denote the space of all asymptotic incoming states and let $\mc{H}_{\rm (out)}$ be the space of all asymptotic outgoing states. We consider the limit in which the spin-chain is infinitely long and the number of excitations $n$ is much smaller than the number of spin-chain sites $L$. This allows us to treat the asymptotic states as well separated and non-interacting. Integrability implies that any scattering process factorizes into two-magnon scattering events, in which the only dynamical process allowed is the interchange of magnon momenta and flavours. We need thus only consider such scattering of two-magnon asymptotic states. 

The two-magnon scattering matrix $S(p,q)$ is a map from $\mc{H}_{\rm (in)}$ to $\mc{H}_{\rm (out)}$ which takes an incoming two-magnon state to an outgoing two-magnon state:
\[
S(p,q) \, : \hspace{0.15cm}  \mc{H}_{\rm (in)} \to \mc{H}_{\rm (out)},  \hspace{1.0cm} 
\ket{ \Phi^{\rm 1 \hspace{0.025cm} (in)}_{p} \, \Phi^{\rm 2 \hspace{0.025cm} (in)}_{q} } \mapsto \ket{ \Phi^{\rm 2  \hspace{0.025cm} (out)}_{q} \, \Phi^{\rm 1 \hspace{0.025cm} (out)}_{p} },
\]
where these asymptotic states can be represented by
\[
\ket{ \Phi^{\rm 1 \hspace{0.025cm} (in)}_{p} \, \Phi^{\rm 2 \hspace{0.025cm} (in)}_{q} } \equiv \ket{ \Phi^1_{p} } \ot \ket{ \Phi^2_{q} },  \hspace{1.0cm}
\ket{ \Phi^{\rm 2 \hspace{0.025cm} (out)}_{q} \, \Phi^{\rm 1 \hspace{0.025cm} (out)}_{p} } \equiv \ket{ \Phi^2_{q} } \ot \ket{ \Phi^1_{p} }.
\]
We shall consider integrable two-magnon scattering on the $\mf{d}(2,1;\al)^2$ double-row spin-chain.
Hence the magnons $\ket{\Phi_{p}}$ are vectors in one of the spaces
\[W_p = V_{p} \oplus \bar{V}_{p}={\rm span}_\C \{ \ket{\phi_{p}}, \ket{\psi_{p}}, \ket{\bar{\phi_{p}}}, \ket{\bar{\psi_{p}}} \},  
\hspace{0.6cm}  
W^{\prime}_p = V_{p}^{\prime} \oplus \bar{V}^{\prime}_{p} ={\rm span}_\C  \{ \ket{\phi_{p}^{\prime}}, \ket{\psi_{p}^{\prime}}, \ket{\bar{\phi}_{p}^{\prime}}, \ket{\bar{\psi}_{p}^{\prime}} \}, \]
both isomorphic to $\C^{2|2}$. The scattering matrix then acts as
\[
S(p,q) \, : \hspace{0.15cm} \mc{W}_p \ot \mc{W}_q  \to \mc{W}_q \ot \mc{W}_p,  \hspace{0.75cm} \text{with} \hspace{0.5cm} \mc{W}_{p} \in\{W_{p},W_{p}^{\prime}\},
\]
and also $\mc{V}_{p} \in\{V_{p},V_{p}^{\prime}\}$ and $\bar{\mc{V}}_{p} \in\{\bar{V}_{p},\bar{V}_{p}^{\prime}\}$.  This simplifying notation takes into account both primed and unprimed magnon states.
The Zhukovski variables $x^{\pm}_{p}$ and $x^{\pm}_{q}$ satisfy 
\begin{equation} 
\frac{x_{p}^{+}}{x_{p}^{-}} = e^{ip} \equiv u_{p}^{2}, \hspace{1.2cm} 
\frac{x_{q}^{+}}{x_{q}^{-}} = e^{iq} \equiv u_{q}^{2},
\end{equation}
and the mass-shell constraints
\begin{equation}
\left( x_{p}^{+} + \frac{1}{x_{p}^{+}} \right) - \left( x_{p}^{-} + \frac{1}{x_{p}^{-}} \right) = \frac{i \hspace{0.05cm} m_{p}}{h}, \hspace{0.9cm}
\left( x_{q}^{+} + \frac{1}{x_{q}^{+}} \right) - \left( x_{q}^{-} + \frac{1}{x_{q}^{-}} \right) = \frac{i \hspace{0.05cm} m_{q}}{h}, 
\end{equation}
where $m_{p}, m_{q} \in \{m=\alpha, \, m^{\prime}=1-\alpha\}$ are the masses of the two magnons with momenta $p$ and $q$.

The scattering matrix can be written as
\begin{equation}
S(p,q) = P \, R(p,q), \hspace{0.75cm} \text{with} \hspace{0.5cm}  R(p,q)\in\End(\mc{W}_p \ot \mc{W}_q),
\end{equation}
in terms of the graded permutation operator $P$.  Therefore $\bb{S}(p,q) = \bb{P} \, \bb{R}(p,q)$ with the $R$-matrix $\bb{R}(p,q)\in\End(\C^{2|2}\otimes \C^{2|2})$.
The S-matrix commutes with all the symmetries of magnon excitations:
\[
[ \, \bb{S}(p,q), \, ( (\pi_{p}\oplus\bar{\pi}_{p}) \otimes (\pi_{q}\oplus\bar{\pi}_{q}) )(\Delta(\mf{a})) \, ] = 0, \hspace{0.75cm} \text{for all} \hspace{0.25cm} \mf{a}\in\mc{A}, \] 
which implies the intertwining equations on the $R$-matrix:
\begin{equation} \label{complete-intertwining}
( (\pi_{p}\oplus\bar{\pi}_{p}) \otimes (\pi_{q}\oplus\bar{\pi}_{q}) )(\Delta^{\rm op}(\mf{a})) \, \bb{R}(p,q) = 
\bb{R}(p,q)  \, ( (\pi_{p}\oplus\bar{\pi}_{p}) \otimes (\pi_{q}\oplus\bar{\pi}_{q}) )(\Delta(\mf{a})), \hspace{0.5cm} \text{for all} \hspace{0.25cm} \mf{a}\in\mc{A}. \hspace{0.1cm}
\end{equation}
We change $\pi_{p}$ and $\pi_{q}$ to representations $\pi_{p}^{\prime}$ and $\pi_{q}^{\prime}$ for primed magnon excitations. The S-matrix is unitary,
$\bb{S}(q,p) \,  \bb{S}(p,q) = \bb{I}$, which implies $(\bb{R}(q,p))^{\rm op} \, \bb{R}(p,q) = \bb{I}$ with $(\bb{R}(p,q))^{\rm op} = \bb{P} \, \bb{R}(p,q) \, \bb{P}$.

\subsubsection{Complete and partial $R$-matrices} \label{part3-sec1-3-1}

We now describe the structure of the $R$-matrix $\bb{R}(p,q)$.
Recall that the 4-dimensional vector space $\mc{W}_{p}=\mc{V}_{p}\oplus \bar{\mc{V}}_{p}$ is a direct sum of two $2$-dimensional vector spaces $\mc{V}_{p}, \bar{\mc{V}}_{p} \cong \C^{1|1}$, so that $\mc{W}_{p} \cong \C^{2|2}$.  Thus $\mc{W}_p\ot \mc{W}_q$ can be split into four $4$-dimensional subspaces, and consequently $\bb{R}(p,q)$ is a $16\times 16$ matrix that can be decomposed into 16 sectors.
However, conservation of chirality (the total number of left- and right-moving magnon states) and mass impose additional constraints \cite{BSS13}.  Focusing on the case of pure transmission (rather than pure reflection) for the scattering of left and right states, the {\it complete} $R$-matrix decomposes into a direct sum of four {\it partial} $R$-matrices:
\begin{equation} \label{Rfull}
\bb{R}(p,q) = \bb{R}^{\sL \sL}(p,q) \oplus \bb{R}^{\sL \sR}(p,q) \oplus \bb{R}^{\sR \sL}(p,q) \oplus \bb{R}^{\sR \sR}(p,q),
\end{equation}
where $\bb{R}^{\ms{a} \ms{b}}(p,q) \in \End(\C^{1|1} \otimes \C^{1|1})$.
These four partial $R$-matrices depend on Zhukovski variables satisfying mass-shell constraints which may depend on either primed or unprimed masses.

The partial $R$-matrices satisfy intertwining equations 
which are the result of the corresponding partial $S$-matrices commuting with the generators of the Hopf superalgebra $\mc{A}$. 
These $R$-matrices also satisfy unitarity and crossing symmetry conditions, and a discrete $\SL\SR$ symmetry condition made manifest by our choice of a symmetric Hopf algebra.  These conditions determine the complete $R$-matrix up to overall factors (dependent on dressing phases) and imply the Yang-Baxter equation.

\paragraph{Left-left and right-right sectors.}

We write the partial $R$-matrices in the $\SL\SL$ and $\SR\SR$ sectors as
\[
\bb{R}^{\sL\sL}(p,q)
= \sum_{i,j,k,l=1,2} (R^{\,\sL\sL}(p,q) )^{\hspace{0.05cm} i \hspace{0.05cm} k}_{\hspace{0.05cm} j \hspace{0.05cm} l} 
\hspace{0.2cm} \bb{E}_{ij}\ot \bb{E}_{kl},  \hspace{0.75cm}
\bb{R}^{\sR\sR}(p,q)
= \sum_{i,j,k,l=\bar{1},\bar{2}} (R^{\,\sR\sR}(p,q) )^{\hspace{0.05cm} i \hspace{0.05cm} k}_{\hspace{0.05cm} j \hspace{0.05cm} l} 
\hspace{0.2cm} \bb{E}_{ij}\ot \bb{E}_{kl},  
\] 
which depend on the momenta $p$ and $q$, and on the masses $m_{p}$ and $m_{q}$ of the two magnons. 
The intertwining equations in the $\SL\SL$ and $\SR\SR$ sectors are
\begin{eqnarray}
\nonumber && \hspace{-0.5cm} ( \pi_{p} \otimes \pi_{q} ) (\Delta^{\rm op}(\mf{a})) \,\, \bb{R}^{\sL\sL}(p,q) \, = \, \bb{R}^{\sL\sL}(p,q) \,\, ( \pi_{p} \otimes \pi_{q} )(\Delta(\mf{a})), \\
\nonumber && \hspace{-0.5cm} ( \bar{\pi}_{p} \otimes \bar{\pi}_{q} ) (\Delta^{\rm op}(\mf{a})) \,\, \bb{R}^{\sR\sR}(p,q) \, = \, \bb{R}^{\sR\sR}(p,q) \,\, ( \bar{\pi}_{p} \otimes \bar{\pi}_{q} )(\Delta(\mf{a})),  \hspace{0.75cm} 
\end{eqnarray}
for all $\mf{a}\in\mc{A}$.  Notice that
$( \pi_{p} \otimes \pi_{q}) (\Delta^{\rm op}(\mf{a})) = \bb{P} \, ( \pi_{p} \otimes \pi_{q} ) (\Delta(\mf{a})) \, \bb{P}$.
These intertwining equations are linear equations which are relatively easy to solve. They determine $\bb{R}^{\sL\sL}(p,q)$ and $\bb{R}^{\sR\sR}(p,q)$, each up to one complex factor\footnote{This happens because the tensor product of two atypical, 2-dimensional, irreducible representations is isomorphic to the typical, 4-dimensional, irreducible representation of $\mc{A}$ (see, for example, Section 2.3 in \cite{Re}).} which we call $s^{\sL\sL}(p,q)$ and $s^{\sR\sR}(p,q)$:
\begin{eqnarray} \label{Rll}
&& \hspace{-0.65cm} \bb{R}^{\hspace{0.05cm}\sL\sL}(p,q) = s^{\sL\sL}(p,q) \, \Bigg[ 
\bb{E}_{11}\ot \bb{E}_{11} + \frac{(x^{+}_{p}-x^{+}_{q})}{u_{p} (x^{-}_{p} - x^{+}_{q})} \hspace{0.15cm} \bb{E}_{11}\ot \bb{E}_{22} 
+ \frac{u_{q} (x^{-}_{p}-x^{-}_{q})}{ (x^{-}_{p} - x^{+}_{q})} \hspace{0.15cm} \bb{E}_{22} \ot \bb{E}_{11}  \\
\nonumber && \hspace{-0.65cm} \hspace{3.3cm} + \, \frac{u_{q}(x^{+}_{p} - x^{-}_{q})}{u_{p} (x^{-}_{p} - x^{+}_{q})} \hspace{0.15cm} \bb{E}_{22}\ot \bb{E}_{22} 
+ \frac{i  \, \eta_{p} \eta_{q}}{(x^{-}_{p} - x^{+}_{q})} \hspace{0.15cm} \bb{E}_{12}\ot \bb{E}_{21}
- \frac{i \, u_{q} \, \eta_{p} \eta_{q}}{u_{p} (x^{-}_{p} - x^{+}_{q})} \hspace{0.15cm} \bb{E}_{21}\ot \bb{E}_{12} 
 \Bigg],  \hspace{0.35cm} 
\end{eqnarray}
\vspace{-0.25cm}
\begin{eqnarray} \label{Rrr}
&& \hspace{-0.65cm} \bb{R}^{\hspace{0.05cm}\sR\sR}(p,q) = s^{\sR\sR}(p,q) \, \Bigg[ 
\bb{E}_{\bar{1}\bar{1}}\ot \bb{E}_{\bar{1}\bar{1}} + \frac{(x^{+}_{p}-x^{+}_{q})}{u_{p} (x^{-}_{p} - x^{+}_{q})} \hspace{0.15cm} \bb{E}_{\bar{1}\bar{1}}\ot \bb{E}_{\bar{2}\bar{2}} 
+ \frac{u_{q} (x^{-}_{p}-x^{-}_{q})}{ (x^{-}_{p} - x^{+}_{q})} \hspace{0.15cm} \bb{E}_{\bar{2}\bar{2}} \ot \bb{E}_{\bar{1}\bar{1}}  \\
\nonumber && \hspace{-0.65cm} \hspace{3.3cm} + \, \frac{u_{q}(x^{+}_{p} - x^{-}_{q})}{u_{p} (x^{-}_{p} - x^{+}_{q})} \hspace{0.15cm} \bb{E}_{\bar{2}\bar{2}}\ot \bb{E}_{\bar{2}\bar{2}} 
+ \frac{i  \, \eta_{p} \eta_{q}}{(x^{-}_{p} - x^{+}_{q})} \hspace{0.15cm} \bb{E}_{\bar{1}\bar{2}}\ot \bb{E}_{\bar{2}\bar{1}}
- \frac{i \, u_{q} \, \eta_{p} \eta_{q}}{u_{p} (x^{-}_{p} - x^{+}_{q})} \hspace{0.15cm} \bb{E}_{\bar{2}\bar{1}}\ot \bb{E}_{\bar{1}\bar{2}} 
 \Bigg].  \hspace{0.35cm} 
\end{eqnarray}
The unitarity conditions on the $R$-matrices take the form
\[ 
(\bb{R}^{\sL\sL}(q,p))^{\rm op} \, \bb{R}^{\sL\sL}(p,q) = \bb{I}, \hspace{0.6cm} (\bb{R}^{\sR\sR}(q,p))^{\rm op} \, \bb{R}^{\sR\sR}(p,q) = \bb{I}, \hspace{0.6cm} \text{with} \hspace{0.35cm} (\bb{R}^{\ms{ab}}(p,q))^{\rm op} = \bb{P} \, \bb{R}^{\ms{ab}}(p,q) \, \bb{P},
\]
which imply $s^{\sL\sL}(p,q) \, s^{\sL\sL}(q,p) = 1$ and $s^{\sR\sR}(p,q) \, s^{\sR\sR}(q,p) = 1$.

\paragraph{Left-right and right-left sectors.}

We write the partial $R$-matrices in the $\SL\SR$ and $\SR\SL$ sectors as
\[
\bb{R}^{\sL\sR}(p,q) 
= \sum_{^{i,j=1,2;}_{k,l=\bar{1},\bar{2}}} (R^{\hspace{0.05cm} \sL\sR}(p,q))^{\hspace{0.05cm} i \hspace{0.05cm} k}_{\hspace{0.05cm} j \hspace{0.05cm} l} 
\hspace{0.2cm} \bb{E}_{ij}\ot \bb{E}_{kl}, \hspace{0.85cm}
\bb{R}^{\sR\sL}(p,q) 
= \sum_{^{i,j=\bar{1},\bar{2};}_{k,l=1,2}} (R^{\hspace{0.05cm} \sR\sL}(p,q))^{\hspace{0.05cm} i \hspace{0.05cm} k}_{\hspace{0.05cm} j \hspace{0.05cm} l} 
\hspace{0.2cm} \bb{E}_{ij}\ot \bb{E}_{kl}.
\] 
The intertwining equations are given by
\begin{eqnarray}
\nonumber && ( \pi_{p} \ot \bar{\pi}_{q} ) (\Delta^{\rm op}(\mf{a})) \,\, \bb{R}^{\hspace{0.05cm}\sL\sR}(p,q) = 
\bb{R}^{\hspace{0.05cm}\sL\sR}(p,q) \,\, (  \pi_{p} \ot \bar{\pi}_{q} ) )(\Delta(\mf{a})), \\
\nonumber && ( \bar{\pi}_{p} \ot \pi_{q} ) (\Delta^{\rm op}(\mf{a})) \,\, \bb{R}^{\hspace{0.05cm}\sR\sL}(p,q) = 
\bb{R}^{\hspace{0.05cm}\sR\sL}(p,q) \,\, (  \bar{\pi}_{p} \ot \pi_{q} ) )(\Delta(\mf{a})),  \hspace{0.75cm} 
\end{eqnarray}
for all $\mf{a}\in\mc{A}$, which determine the transmission $R$-matrices in the $\SL\SR$ and $\SR\SL$ sectors as
\eqa{
\bb{R}^{\sL\sR}(p,q) &= s^{\sL\sR}(p,q) \, \big[ (x^{+}_{p}x^{+}_{q}-1) (x^{-}_{p}x^{-}_{q} - 1) \big]^{-\frac{1}{2}}\\
& \times \Big[ \,  (x^{+}_{p}x^{-}_{q} - 1) \hspace{0.15cm} \bb{E}_{11} \ot \bb{E}_{\bar{1}\bar{1}} 
+ \,u_{p} \, (x^{-}_{p}x^{-}_{q} - 1 ) \hspace{0.15cm} \bb{E}_{11} \ot \bb{E}_{\bar{2}\bar{2}}
\, + \, u_{q}^{-1} \, (x^{+}_{p}x^{+}_{q} - 1 ) \hspace{0.15cm} \bb{E}_{22} \ot \bb{E}_{\bar{1}\bar{1}} \qq \nonumber\\
& \qu\; + \, u_{p} u_{q}^{-1} \, (x^{-}_{p}x^{+}_{q} - 1)  \hspace{0.15cm} \bb{E}_{22} \ot \bb{E}_{\bar{2}\bar{2}}  
\, + \, u_{p}  \, \eta_{p}\eta_{q} \hspace{0.15cm} \bb{E}_{12}\ot \bb{E}_{\bar{1}\bar{2}} 
\, + \, \eta_{p}\eta_{q} \, u_{q}^{-1} \hspace{0.15cm} \bb{E}_{21}\ot \bb{E}_{\bar{2}\bar{1}} \Big], \nonumber
}
\vspace{-0.5cm}
\eqa{
\bb{R}^{\sR\sL}(p,q) &= s^{\sR\sL}(p,q) \, \big[ (x^{+}_{p}x^{+}_{q}-1) (x^{-}_{p}x^{-}_{q} - 1) \big]^{-\frac{1}{2}}\\
& \times \Big[ \,  (x^{+}_{p}x^{-}_{q} - 1) \hspace{0.15cm} \bb{E}_{\bar{1}\bar{1}} \ot \bb{E}_{11} 
+ \,u_{p} \, (x^{-}_{p}x^{-}_{q} - 1 ) \hspace{0.15cm} \bb{E}_{\bar{1}\bar{1}} \ot \bb{E}_{22}
\, + \, u_{q}^{-1} \, (x^{+}_{p}x^{+}_{q} - 1 ) \hspace{0.15cm} \bb{E}_{\bar{2}\bar{2}} \ot \bb{E}_{11} \qq \nonumber\\
& \qu\; + \, u_{p} u_{q}^{-1} \, (x^{-}_{p}x^{+}_{q} - 1)  \hspace{0.15cm} \bb{E}_{\bar{2}\bar{2}} \ot \bb{E}_{22}  
\, + \, u_{p}  \, \eta_{p}\eta_{q} \hspace{0.15cm} \bb{E}_{\bar{1}\bar{2}}\ot \bb{E}_{12} 
\, + \, \eta_{p}\eta_{q} \, u_{q}^{-1} \hspace{0.15cm} \bb{E}_{\bar{2}\bar{1}}\ot \bb{E}_{21} \Big], \nonumber
}
up to the overall factors $s^{\sL\sR}(p,q)$ and $s^{\sR\sL}(p,q)$.
The unitarity conditions are
\[ 
(\bb{R}^{\sL\sR}(q,p))^{\rm op} \, \bb{R}^{\sR\sL}(p,q) = \bb{I}, \hspace{0.85cm} 
(\bb{R}^{\sR\sL}(q,p))^{\rm op} \, \bb{R}^{\sL\sR}(p,q) = \bb{I},  
 \] 
which imply $s^{\sL\sR}(p,q) \, s^{\sR\sL}(q,p) = 1$ and $s^{\sR\sL}(p,q) \, s^{\sL\sR}(q,p) = 1$.

\smallskip

We require the parity relation $s^{\ms{ba}}(-q,-p) = s^{\ms{ab}}(p,q)$ for the scale factors in the $R$-matrix, and also impose a discrete $\SL\SR$ symmetry  
$s^{\sL\sL}(p,q) = s^{\sR\sR}(p,q)$ and $s^{\sL\sR}(p,q) = s^{\sR\sL}(p,q)$ as in \cite{BSS13}.
These conditions will be necessary to derive reflection matrix solutions of the boundary Yang-Baxter equation in Chapter \ref{part3-sec2}.
We note, finally, that additional {\it bulk crossing symmetry conditions} must be imposed which further constrain the scale factors $s^{\ms{ab}}(p,q)$.  A proposal for the solution to these crossing symmetry conditions has been advanced in \cite{Borsato:2013hoa}.

\subsubsection{Yang-Baxter equation.}

\begin{figure}[htb!]
\begin{center}
\includegraphics[scale=1.6]{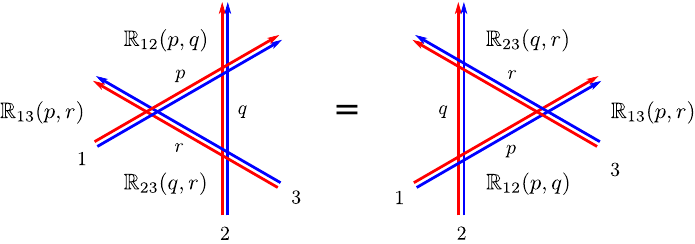}
\caption{Three-magnon scattering factorizes into a succession of two-magnon scattering events. The double red and blue line indicates the {\it direct sum of left and right magnons states} with scattering described by the complete $R$-matrix.  The scattering of individual red or blue lines (left or right magnons) is described by partial $R$-matrices.  This diagram gives both the Yang-Baxter equation for the complete $R$-matrix (treating double lines as a single composite line) and the Yang-Baxter equation for the partial $R$-matrices (choosing a red or blue line from each pair to give 8 possibilities).}  \label{figure-YBE}
\vspace{-0.2cm}
\end{center}
\end{figure}

The complete $R$-matrix $\bb{R}(p,q)$ must satisfy
\begin{equation}  \label{YBE-complete} 
\bb{R}_{12}(p,q) \; \bb{R}_{13}(p,r) \; \bb{R}_{23}(q,r) \, = \, \bb{R}_{23}(q,r) \; \bb{R}_{13}(p,r) \; \bb{R}_{12}(p,q),  
\end{equation}
which is the Yang-Baxter equation of an integrable system.  
The partial $R$-matrices then satisfy
\begin{equation}  \label{YBE-partial} 
\bb{R}^{\ms{a} \ms{b}}_{12}(p,q) \; \bb{R}^{\ms{a} \ms{c}}_{13}(p,r) \; \bb{R}^{\ms{b} \ms{c}}_{23}(q,r) \, 
= \, \bb{R}^{\ms{b} \ms{c}}_{23}(q,r) \; \bb{R}^{\ms{a} \ms{c}}_{13}(p,r) \; \bb{R}^{\ms{a} \ms{c}}_{12}(p,q)  
\end{equation}
for all $\ms{a}, \ms{b}, \ms{c} \in \{ \SL,\SR \}$ (see Figure \ref{figure-YBE}).  Here we define
\[ 
\bb{R}^{\ms{a} \ms{b}}_{12}(p,q) = \bb{R}^{\ms{a}\ms{b}}(p,q) \ot \bb{I}, \hspace{0.5cm}
\bb{R}^{\ms{a} \ms{b}}_{23}(p,q) = \bb{I} \ot \bb{R}^{\ms{a}\ms{b}}(p,q),  \hspace{0.5cm}
\bb{R}^{\ms{a} \ms{b}}_{13}(p,q) = ( \bb{I} \ot \bb{P})( \bb{R}^{\ms{a} \ms{b}}(p,q) \ot \bb{I} )(\bb{I} \ot \bb{P} ), 
\]
and similarly for $\bb{R}_{12}(p,q)$, $\bb{R}_{13}(p,q)$ and $\bb{R}_{23}(p,q)$ in terms of the complete $R$-matrices, with a change to the identity matrix in $\End(\C^{2|2})$ and the graded permutation matrix in $\End(\C^{2|2} \ot \C^{2|2})$.

\vspace{0.2cm}

We can check that the $R$-matrices above satisfy \eqref{YBE-partial} for  $(\ms{a} \ms{b} \ms{c})$ in the homogeneous ($\SL\SL\SL$, $\SR\SR\SR$) and mixed ($\SL\SL\SR$, $\SL\SR\SL$, $\SR\SL\SL$, $\SL\SR\SR$, $\SR\SL\SR$, $\SR\SR\SL$) sectors. Note that the discrete $\SL\SR$  symmetry means we need only check half of these equations. This ensures that the complete $R$-matrix satisfies \eqref{YBE-complete}.

\section{Integrable open $\mf{d}(2,1;\alpha)^2$ spin-chain and reflection matrices}  \label{part3-sec2}

In this chapter we consider the boundary scattering of magnon excitations of a $\mf{d}(2,1;\alpha)^2$ open spin-chain off an integrable boundary. These correspond to massive excitations of an open superstring ending on a D-brane in $AdS_{3}\times S^{3} \times S^{3\prime} \times S^{1}$, such as one of the maximal D1- or D5-brane giant gravitons described in Chapter \ref{part2-sec1}.

\subsection{Open spin-chains and boundary scattering} \label{part3-sec2-1}

\subsubsection{Double-row $\mf{d}(2,1;\alpha)^2$ open spin-chain with $\mf{su}(1|1)^{2}_{\ms{c}}$ excitations} \label{part3-sec2-1-1}

\paragraph{Semi-infinite open spin-chain.}  

In the infinite $J\to\infty$ spin-chain limit we can consider one end of the spin-chain at a time -- we choose the right end of the open spin-chain. Thus we obtain a semi-infinite spin-chain with a distinguished site at the right end which we call the boundary site. The vector occupying this site transforms in a representation of a boundary subalgebra, dictated by boundary conditions.  The ground state of the semi-infinite open spin-chain is
\eq{
\ket{0} = \ket{\mc{Z}^J \mc{F}_\sB},
}
where $\mc{F}_\sB$ is the right boundary field. This infinitely heavy state represents the whole D-brane to which the open superstring is attached (to account for conservation of momentum).\footnote{A similar construction is presented in Section 2.2 of \cite{HM:2007}, in which open superstrings on the $AdS_5\times S^5$ background are attached to $Y=0$ and $Z=0$ maximal giant gravitons. In the dual CFT$_{4}$ description, an open superstring corresponds to a gauge invariant local operator ${\rm Tr}(Z\ldots Z)$, with impurities, attached to a ${\rm det}(Y)$ or ${\rm det}(Z)$ operator dual to a maximal D3-brane giant graviton.  The full determinant plays the role of the boundary field $\mc{F}_\sB$. Here the CFT$_{2}$ dual of IIB superstring theory on $AdS_3 \times S^3 \times S^{3\prime} \times S^1$ is not known, although it is conjectured to arise as the IR limit of the D1-D5-D5$^{\prime}$ worldvolume gauge theory of \cite{Tong:2014}. We expect, however, that an open superstring attached to our maximal D1 and D5-brane giant gravitons will be dual to similar determinant-like CFT$_{2}$ operators. 
}

We will enumerate the sites of the semi-infinite open spin-chain by $-J,-J+1,\ldots,-1,0$, with $0$ denoting the boundary field site. In this notation the fundamental excitations are described by the spin-chain state vectors
\begin{equation}
\big|\varphi^r_{(n)}\big\rangle_{\sB} = \ket{\mc{Z}^{J-n} \varphi^{r} \mc{Z}^{n-1}\mc{F}_\sB}, \hspace{1.0cm}
\big|\bar\varphi^{r}_{(n)}\big\rangle_{\sB} = \ket{\mc{Z}^{n-1} \bar{\varphi}^{r} \mc{Z}^{J-n}\mc{F}_\sB}, 
\end{equation}
where $\varphi^r$, $\bar\varphi^r$ represent the excitations discussed in Section \ref{part3-sec3-1-2}. The low-lying left- and right-moving single-magnon excitations are now given by
\eq{ \label{in-states}
\big|\varphi^r_p\big\rangle_{\sB} = \sum_{n=1}^J e^{-ipn}\, \big|\varphi^r_{(n)}\big\rangle_{\sB}, \hspace{1.0cm} 
\big|\bar{\varphi}^r_p\big\rangle_{\sB} = \sum_{n=1}^J e^{-ipn}\, \big|\bar{\varphi}^r_{(n)}\big\rangle_{\sB}.
}
The supercharges acting on these states insert or remove a $\mc{Z}$ field from the left side. In the $J\to\infty$ limit, in which the state has infinite length in the left direction, this does not change the length of the spin-chain or the location of the excitation.
We can identify magnon states of the semi-infinite open spin-chain with magnon states of the infinite closed spin-chain, with an extra boundary state,  
\begin{equation}
\big|\varphi^r_p\big\rangle_{\sB} = \ket{\varphi^r_p} \otimes \ket{0}_{\sB}, \hspace{1.0cm}
\big|\bar{\varphi}^r_p\big\rangle_{\sB} = \ket{\bar{\varphi}^r_p} \otimes \ket{0}_{\sB},
\end{equation}
where $\ket{\varphi^r_p}$ and $\ket{\bar{\varphi}^r_p}$ are bulk magnon excitations in the vector spaces $\mc{V}_p$ and $\bar{\mc{V}}_p$, with the action of the Hopf superalgebra $\mc{A}$ defined in the usual way.  The boundary field $\mc{F}_{\sB}$ is represented by the boundary vacuum state $\ket{0}_{\sB}$. The generalization for the multi-magnon states is straightforward.

\paragraph{Boundary algebra.}  

The symmetries of the boundary site are related to those symmetries of the D-brane which survive the choice of vacuum $\mc{Z}$.   These boundary symmetries are generated by a subalgebra of the 
bulk symmetry algebra, which in our case is Hopf superalgebra $\mc{A}$. We will denote the boundary symmetry algebra by $\mc{B}$ with elements $\mf{b}\in\mc{B}$. 

Since $\mc{B}\subset\mc{A}$, the elements $\mf{b}$ inherit all the additional algebraic structures (coproduct, counit, etc.) defined on $\mc{A}$. However, $\mc{B}$ is not a Hopf subalgebra of $\mc{A}$. The integrability assumption requires boundary algebra to be a coideal subalgebra of the bulk symmetry algebra:
\begin{equation}
\Delta(\mf{b}) \in \mc{A} \ot \mc{B} \hspace{0.75cm} \text{for all} \hspace{0.5cm}  \mf{b} \in \mc{B}.
\end{equation}
The boundary vacuum state $\ket{0}_{\sB}$ is taken to be a vector in the trivial representation of the coideal subalgebra $\mc{B}$, defined by the counit map $\eps$. We call this the {\it singlet boundary}. We will later extend this construction to accommodate a vector representation of $\mc{B}$ -- called the {\it vector boundary}.  

\paragraph{Boundary integrability.}

It is known that integrability is preserved if the bulk and boundary Lie algebras form a symmetric pair (see \cite{Mackay04} and references therein for a review of boundary integrability).  
Recall that a symmetric pair of Lie algebras is a pair $(\mfg,\mf{h})$ such that $\mfg=\mf{h}\op\mf{m}$, where $\mf{h}$ is a subalgebra of $\mfg$ and the following relations hold:
\eq{ 
[\mf{h},\mf{h}] \subset \mf{h},  \hspace{1.0cm} [\mf{h},\mf{m}] \subset \mf{m},  \hspace{1.0cm}  [\mf{m},\mf{m}] \subset \mf{h} .  \label{symm-pair}
}
The statement above is true if $\mfg$ is a simple complex Lie algebra.  Integrability is then ensured by the existence of a twisted Yangian associated with the symmetric pair $(\mfg,\mf{h})$, the Cartan subalgebra of which is an infinite-dimensional abelian algebra. By a quantum extension of the classical Liouville integrability theorem, this is enough to ensure that the system is integrable.%
\footnote{There is no analogue of the Liouville-Arnold theorem for infinite-dimensional quantum systems; thus we consider integrability as the set of constraints that are necessary to ensure factorized scattering.}
The question of boundary integrability becomes much more complicated if $\mfg$ is non-simple, since both the Yangian and twisted Yangian (for a given symmetric pair) are not well-defined. Integrability then needs to be examined on a case-by-case basis. 
Likewise, the integrable structures associated with symmetric pairs of Lie superalgebras often require further investigation.
An alternative way of verifying the integrability of boundary scattering is by finding the reflection $K$-matrix which is a solution of the reflection equation and intertwines the boundary symmetries. 

\subsubsection{Boundary scattering and $K$-matrices} \label{part3-sec2-1-2}

\paragraph{Outgoing single magnon representations $\pi_{-p}$ and $\bar{\pi}_{-p}$.}  

Let us take the states \eqref{in-states} to be {\it incoming} magnons with momentum $p$. To define the boundary scattering theory, we need a notion of {\it outgoing} states with opposite momentum $-p$. We will denote the corresponding vector spaces by 
$\mc{V}_{-p}$ and $\bar{\mc{V}}_{-p}$. The left- and right-moving outgoing magnon representations $\pi_{-p}:\mc{A} \to \End(\C^{1|1})$ and $\bar{\pi}_{-p}:\mc{A} \to \End(\C^{1|1})$ are parametrized by the labels $a_{-p}$, $b_{-p}$, $c_{-p}$ and $d_{-p}$, with Zhukovski variables $x^\pm_{-p}$. Recall that $x^\pm_{p}$ satisfy \eqref{Zhukovski}: the first equality implies $x^\pm_{-p}=-f_p x^\mp_p$, for some phase factor $f_p$, while the second equality sets $f_p=1$.  Thus we obtain 
\begin{equation}
x^\pm_{-p} = - \hspace{0.05cm} x^\mp_p,   \hspace{1.2cm} \eta_{-p} = \eta_{p}. 
\end{equation}
The outgoing left-moving representation $\pi_{-p}$ is \eqref{Rep:L} with the parameters \eqref{parameters} replaced by
\begin{equation} \label{parameters-reflected}
a_{-p} = \sqrt{h} \hspace{0.15cm} \eta_{p}, \hspace{0.65cm}  
b_{-p} = \sqrt{h} \hspace{0.15cm} \eta_{p},  \hspace{0.65cm}
c_{-p} = \sqrt{h} \hspace{0.15cm} \frac{i \hspace{0.025cm} \eta_{p}}{x_{p}^{+}},   \hspace{0.65cm}
d_{-p} = -\sqrt{h} \hspace{0.15cm} \frac{i \hspace{0.025cm} \eta_{p}}{x_{p}^{-}}, \hspace{0.65cm} 
u^{2}_{-p} = \frac{x_{p}^{-}}{x_{p}^{+}}.
\end{equation}
The outgoing right-moving representation $\bar\pi_{-p}$ is given by \eqref{Rep:R} with an identical replacement.

\paragraph{Singlet boundary scattering.}

Boundary scattering on the semi-infinite open spin-chain is described by a boundary scattering matrix $S_{\rm boundary}(p)$ which maps incoming states to outgoing states, while keeping the boundary fixed. As for bulk scattering, we denote the space of all asymptotic incoming magnon states of the open spin-chain by $\mc{H}_{\rm (in)}$, and the space of all outgoing states by $\mc{H}_{\rm (out)}$.  By the integrability hypothesis, we need only consider the reflection of single magnons off the boundary.
The boundary scattering matrix maps $\mc{H}_{\rm (in)}$ to $\mc{H}_{\rm (out)}$:    
$$
S_{\rm boundary}(p) \, : \hspace{0.15cm} \mc{H}_{\rm (in)} \to \mc{H}_{\rm (out)}, \hspace{0.75cm} \ket{ \Phi^{\rm \hspace{0.025cm} (in)}_{p} } \mapsto \ket{ \Phi^{\rm \hspace{0.025cm} (out)}_{-p} }.
$$
We can write the incoming and outgoing states as
$$
\ket{ \Phi^{\rm \hspace{0.025cm} (in)}_{p} } = \ket{ \Phi_{p} } \ot \ket{0}_{\sB},  \hspace{0.75cm}
\ket{ \Phi^{\rm \hspace{0.025cm} (out)}_{-p} } = \ket{ \Phi_{-p} } \ot \ket{0}_{\sB},
$$
where $\ket{ \Phi_{p} }\in\mc{W}_p$ and $\ket{ \Phi_{-p} }\in\mc{W}_{-p}$ are bulk magnons, and $\ket{0}_\sB$ is the singlet boundary state.

For our purposes it will be convenient to introduce the boundary intertwining $K$-matrix.  Let $\kappa$ denote the natural {\it reflection} map which acts as the identity map on $|0\ran_\sB$ and
$$
\kappa \, :\hspace{0.15cm} \mc{W}_p \to \mc{W}_{-p}, \hspace{0.75cm}  \ket{ \Phi_{p} } \mapsto \ket{ \Phi_{-p} }, 
$$
which is the canonical isomorphism $\mc{W}_p\cong \mc{W}_{-p}$. The boundary scattering matrix $S_{\rm boundary}(p)$ is  the composition\footnote{This composition is well-defined, since $|0\ran_\sB$ is a state in a one-dimensional vector space ($\cong\C$) and $\mc{W}_p\ot\C \cong \mc{W}_p$.}
\begin{equation}
S_{\rm boundary}(p) = \kappa \; K(p), \hspace{0.75cm} \text{with} \hspace{0.35cm}  K(p) \in\End (\mc{W}_p).
\end{equation}
Now $\mc{W}_{p} \cong \C^{2|2}$, so the $K$-matrix corresponds to $\bb{K}(p) \in \End(\C^{2|2})$.  The boundary $S$-matrix must commute with all the boundary symmetries, which yields the boundary intertwining equations for the $K$-matrix.  There is also a unitary condition on the boundary $S$-matrix.

\paragraph{Vector boundary scattering.} 

Let us consider a vector representation of the boundary algebra.
A boundary vector state is anticipated to have an interpretation as a magnon state\footnote{A boundary state is created when the boundary absorbs magnon excitations via the so-called boundary {\it bootstrap} procedure (see, for example, Section 3 in \cite{GZ93}). Boundary states for open superstrings in $AdS_5\times S^5$ were considered in \cite{ALT:2009,MR:2012,ABR07}. We will give a detailed description of similar boundary states for open superstrings in $AdS_{3} \times S^{3} \times S^{3\prime} \times S^{1}$ in a forthcoming publication.} $\ket{\Phi_\sB} \in \mc{W}_\sB$, with maximum total momentum $\pi$, absorbed by a singlet boundary state $|0\ran_\sB$. The boundary Lie algebra associated with the coideal subalgebra, in this case denoted $\mc{B}_{\sT} \subset \mc{A}$, is thus the totally supersymmetric $\mf{su}(1|1)^{2}_{\ms{c}}$ symmetry of bulk magnon excitations.

In the case of this vector boundary state, the incoming and outgoing states in $\mc{H}_{\rm (in)}$ and $\mc{H}_{\rm (out)}$ have the following tensor product decomposition:
$$
\ket{ \Phi^{\rm \hspace{0.025cm} (in)}_{p} } = \ket{ \Phi_{p} } \ot \ket{\Phi_\sB} \ot |0\ran_\sB, \hspace{0.75cm}
\ket{ \Phi^{\rm \hspace{0.025cm} (out)}_{-p} } = \ket{ \Phi_{-p} } \ot \ket{\Phi_\sB} \ot |0\ran_\sB,
$$
where $\ket{\Phi_\sB}\in\mc{W}_\sB$ is a vector in the boundary space. We will now denote the boundary scattering matrix by $S_{\rm boundary}(p,\SB)$. As before, we write this as a composition of the reflection map $\kappa$, which now acts as the identity map on the boundary vector state, and a reflection $K$-matrix:
\begin{equation}
S_{\rm boundary}(p,\SB) = \kappa \; K(p,\SB), \hspace{0.75cm} \text{with}\hspace{0.5cm} K(p,\SB) \in \End( \mc{W}_p \ot \mc{W}_\sB) .
\end{equation}
Now  $\mc{W}_{p}, \mc{W}_{\sB}\cong \C^{2|2}$, so the $K$-matrix $K(p,\SB)$ corresponds to $\bb{K}(p,\SB) \in \End(\C^{2|2}\otimes \C^{2|2})$.

\smallskip

We will discuss singlet boundaries in Section \ref{sec:4.2} and the vector boundary in Section \ref{sec:4.3}.
Note that only a vector representation (not a singlet representation) of the boundary algebra $\mc{B}_{\sT}$ is possible due to the inclusion of the central elements $\mf{P}$ and $\mf{P}^{\dag}$. We will explain this argument in detail in subsequent sections.

\subsection{Singlet boundaries} \label{sec:4.2}

\paragraph{Boundary algebras.}

There are four boundary subalgebras $\mc{B}$ of the bulk Hopf superalgebra $\mathcal{A}$ which describe the scattering of magnons off singlet boundaries. The associated Lie algebras $L(\mc{B})$ can be compared with the boundary algebras in Tables \ref{table-D5} and \ref{table-D1}. We will denote them as follows:
\vspace{-0.75cm}
\begin{itemize}
\item[$\ast$]{{\bf left \& right half-supersymmetric boundary algebras} $\mc{B}_{\sL}$, $\mc{B}_{\sR}$, corresponding to D-branes preserving half of the bulk supersymmetries, and the magnon Hamiltonian $\mf{H}$ and $\mf{M}$, implying chiral boundary scattering.  The boundary Lie algebras are $\mf{su}(1|1)_{\sL} \oplus \mf{u}(1)_{\sR}$ and $\mf{u}(1)_{\sL} \oplus \mf{su}(1|1)_{\sR}$.}
\vspace{-0.68cm}
\item[$\ast$]{{\bf non-supersymmetric chiral boundary algebra} $\mc{B}_{\sN\sC}$, corresponding to D-branes which do not respect any bulk supersymmetries, but preserve both  $\mf{H}$ and $\mf{M}$.  The boundary Lie algebra is $\mf{u}(1)_{\sL} \oplus \mf{u}(1)_{\sR}$. We will show that scattering off the non-supersymmetric chiral boundary has a {\it hidden symmetry}, denoted $\mc{B}_{\sD}$, at the level of the Hopf superalgebra.}
\vspace{-0.2cm}
\item[$\ast$]{{\bf non-supersymmetric achiral boundary algebra} $\mc{B}_{\sN\sA}$, corresponding to D-branes which preserve $\mf{H}$ and no bulk supersymmetries. The associated boundary Lie algebra is $\mf{u}(1)_{+}$.}
\end{itemize}
\vspace{-0.2cm}

\paragraph{Boundary intertwining equations.}  

The $K$-matrix $\bb{K}(p)\in\End(\C^{2|2})$ is the boundary analogue of the bulk $R$-matrix and is required to satisfy the boundary intertwining equations
\begin{equation} \label{singlet-boundary-intertwining}
((\pi_{-p} \oplus \bar{\pi}_{-p}) \otimes \epsilon)(\Delta(\mf{b})) \hspace{0.15cm} \bb{K}(p) \, = \, 
\bb{K}(p) \hspace{0.15cm} ((\pi_{p} \oplus \bar{\pi}_{p}) \otimes \epsilon)(\Delta(\mf{b}))
\end{equation}
for all $\mf{b}\in\mc{B}$ for a given boundary subalgebra $\mc{B}$.  For those $\mf{b} \in \mf{su}(1|1)^{2}_{\ms{c}}$, this simplifies to the form
\begin{equation} \label{singlet-boundary-intertwining-simplified}
(\pi_{-p} \oplus \bar{\pi}_{-p})(\mf{b}) \hspace{0.15cm} \bb{K}(p) \, = \, \bb{K}(p) \hspace{0.15cm} (\pi_{p} \oplus \bar{\pi}_{p})(\mf{b}),
\end{equation}
where we have dropped the trivial boundary representation $\eps$, since $\eps(\mf{b})=0$.

The {\it complete} K-matrix $\bb{K}(p)$ can have four sectors which correspond to {\it chiral} reflections (left-to-left $\bb{K}^{\sL}(p)$ and right-to-right $\bb{K}^{\sR}(p)$) and {\it achiral} reflections (left-to-right $\bb{A}^{\sL}(p)$ and right-to-left $\bb{A}^{\sR}(p)$). We denote these {\it partial} $K$-matrices  by $\bb{K}^{\ms{a}}(p), \, \bb{A}^{\ms{a}}(p)\in\End(\C^{1|1})$. 
Here the superscript $\ms{a}$ denotes the chirality of the  incoming magnon before the reflection. 
The complete $K$-matrix is then 
\begin{equation} \label{singlet-complete-K}
\bb{K}(p) \, = \, 
\begin{pmatrix} \bb{K}^{\sL}(p) &  \bb{A}^{\sR}(p) \\ 
\bb{A}^{\sL}(p) &  \bb{K}^{\sR}(p)
\end{pmatrix} \, = \, 
\big(\bb{K}^{\sL}(p) \oplus \bb{K}^{\sR}(p) \big) \, + \, 
\begin{pmatrix} 0 & \bb{I} \\ \bb{I} & 0 \end{pmatrix} \big(\bb{A}^{\sL}(p) \oplus \bb{A}^{\sR}(p) \big),
\end{equation}
which is also required to satisfy the unitarity condition $\bb{K}(-p) \, \bb{K}(p) = \bb{I}$.

\paragraph{Constraints from the central elements.}

The central elements of the boundary subalgebra $\mc{B}$ play a crucial role in boundary scattering. Let us explain why. A central element $\mf{C}\in\mc{B}$ is required to commute with the boundary scattering matrix and thus must intertwine the $K$-matrix $\bb{K}(p)$ trivially.
Since $\mc{B}\subset\mc{A}$, there are five candidates for central elements in $\mc{B}$. Consider $\mf{H}_{\ms{a}}$ with $\ms{a} \in \{\SL,\SR\}$. Note that
\[
\pi_{p}(\mf{H}_{\ms{a}}) \neq \bar{\pi}_{-p} (\mf{H}_{\ms{a}}), \hspace{1.0cm} 
\bar{\pi}_{p}(\mf{H}_{\ms{a}}) \neq \pi_{-p}(\mf{H}_{\ms{a}}), 
\]
and thus chirality is conserved if either (or both) $\mf{H}_{\sL}$ or $\mf{H}_{\sR}$ are in $\mc{B}$. However, chirality is not conserved if only the linear combination $\mf{H} = \mf{H}_\sL+\mf{H}_\sR$, but not $\mf{M} = \mf{H}_\sL-\mf{H}_\sR$, is in $\mc{B}$, since
\[
\pi_{p}(\mf{H}) = \bar{\pi}_{-p}(\mf{H}), \hspace{1.0cm} 
\pi_{p}(\mf{M}) \neq \bar{\pi}_{-p}(\mf{M}).
\]
Now let us consider $\mf{C}\in \{\mf{P}, \mf{P}^\dagger\}$. Then
\[
\pi_{p}(\mf{C}) \neq \pi_{-p}(\mf{C}), \hspace{1.0cm} 
\bar\pi_p(\mf{C}) \neq \bar\pi_{-p}(\mf{C}),
\]
so, if the boundary representation is trivial, then the central elements $\mf{P}$ and $\mf{P}^\dagger$ \emph{cannot} be in the boundary subalgebra. However, suitable linear and quadratic combinations of  $\mf{P}$ and $\mf{P}^{\dag}$ are allowed.  In particular, 
$\mf{P}^\pm\equiv \mf{P}^\dag \pm \mf{P}$ and $\mf{K}\equiv \mf{P}^\dag \mf{P}$ satisfy
\begin{eqnarray}
\nonumber && \hspace{-0.65cm} \pi_{p}(\mf{P}^+) = \pi_{-p}(\mf{P}^+), \hspace{0.85cm} 
\pi_{p}(\mf{P}^-) \neq \pi_{-p}(\mf{P}^-), \hspace{0.85cm}
\pi_p(\mf{K}) = \pi_{-p}(\mf{K}), \\
\nonumber && \hspace{-0.65cm} \bar{\pi}_{p}(\mf{P}^+) = \bar{\pi}_{-p}(\mf{P}^+), \hspace{0.85cm} 
\bar{\pi}_{p}(\mf{P}^-) \neq \bar{\pi}_{-p}(\mf{P}^-), \hspace{0.85cm}
\bar{\pi}_p(\mf{K}) = \bar{\pi}_{-p}(\mf{K}), 
\end{eqnarray}
and thus we may have $\mf{P}^+,\mf{K}\in\mc{B}$, but $\mf{P}^-\notin\mc{B}$, if the boundary representation is trivial. We also note that $\mf{P}^+,\mf{K}\in\mc{B}$ does not imply any constraints on the chirality of the reflection.

The last central element we need to consider is $\mf{U}$. Since $\eps(\mf{U})=1$, we should note that
$$
((\pi_p \oplus \bar{\pi}_{p})\ot \eps)(\Delta(\mf{U})) \neq ((\pi_{-p} \oplus \bar{\pi}_{-p}) \ot \eps)(\Delta(\mf{U})) ,
$$ 
which implies that $\mf{U}$ cannot be in the boundary algebra $\mc{B}$, if boundary representation is given by the counit $\eps$.  We will show in the next section that this is also true for the vector boundary.

\subsubsection{Boundary subalgebras and $K$-matrices} \label{part3-sec2-2-1}

\paragraph{Left and right half-supersymmetric boundary algebras.}

We define the left and right half-supersymmetric boundary superalgebras, $\mc{B}_{\sL}$ and $\mc{B}_{\sR}$, to be coideal subalgebras of $\mc{A}$ generated as
\eq{
\mc{B}_\sL = \big\lan \mf{Q}_{\sL}, \, \mf{S}_{\sL}, \, \mf{H}_{\ms{\sL}}, \, \mf{H}_{\ms{\sR}} \big\ran, \hspace{0.85cm} 
\mc{B}_\sR = \big\lan \mf{Q}_{\sR}, \, \mf{S}_{\sR}, \, \mf{H}_{\ms{\sL}}, \, \mf{H}_{\ms{\sR}}  \big\ran.  
}
It remains to check the symmetric pair property. Let 
\begin{equation} 
\mf{g} \equiv L(\mc{A})=\mf{su}(1|1)^2_\ms{c}, \hspace{0.6cm} 
\mf{h}_{\sL} \equiv L(\mc{B}_\sL)=\mf{su}(1|1)_\sL \op \mf{u}(1)_\sR, \hspace{0.6cm} 
\mf{h}_{\sR} \equiv L(\mc{B}_\sR)=\mf{u}(1)_\sL \op \mf{su}(1|1)_\sR
\end{equation}
denote the associated Lie superalgebras, and $\mf{m}_{\sL}$ and $\mf{m}_{\sR}$ denote spaces generated by 
$\{\mf{Q}_{\sR}, \, \mf{S}_{\sR}, \, \mf{P},\mf{P}^\dag\}$ and
$\{\mf{Q}_{\sL}, \, \mf{S}_{\sL}, \, \mf{P},\mf{P}^\dag\}$, respectively. Thus $\mf{g}=\mf{h}_{\sL}\op\mf{m}_{\sL}$ and $\mf{g}=\mf{h}_{\sR}\op\mf{m}_{\sR}$ as vector spaces, and an easy computation shows that the property \eqref{symm-pair} indeed holds.
The boundary Lie superalgebra $\mf{h}_{\sL}$ is that of an open superstring on $AdS_{3} \times S^{3} \times S^{3\prime} \times S^{1}$ ending on the $\bar{Y}=0$ or $\bar{Y}^{\prime}=0$ half of the D5-brane maximal giant graviton, or the $\bar{Y}=\bar{Y}^{\prime}=0$ D1-brane giant (which is the intersection of $\bar{Y}=0$ and $\bar{Y}^{\prime}=0$ giants). 
The boundary Lie superalgebra $\mf{h}_{\sR}$  is that of an open superstring ending on the $Y=0$ or $Y^{\prime}=0$ half of the D5-brane maximal giant, or the $Y=Y^{\prime}=0$ D1-brane giant. 
These are analogues of open superstrings attached to $\bar{Y}=0$ and $Y=0$ giant gravitons in $AdS_{5}\times S^{5}$ \cite{HM:2007}.

Let us construct the $K$-matrices $\bb{K}_{\mc{B}_{\ms{a}}}(p)$ for both $\ms{a} \in \{\SL,\SR\}$. Since $\mf{H}_{\sL}, \mf{H}_{\sR} \in\mc{B}_{\ms{a}}$, these $K$-matrices are chiral:
\begin{equation} \label{K-matrix-singlet-a}
\bb{K}_{\mc{B}_{\ms{a}}}(p) = \bb{K}_{\mc{B}_{\ms{a}}}^{\sL}(p) \op \bb{K}_{\mc{B}_{\ms{a}}}^{\sR}(p),
\end{equation}
where
\[
\bb{K}_{\mc{B}_{\ms{a}}}^{\sL}(p)
= \sum_{i,j=1,2} (K^{\sL}_{\mc{B}_{\ms{a}}}(p) )^{\hspace{0.05cm} i }_{\hspace{0.05cm} j } \hspace{0.15cm} \bb{E}_{ij},  \hspace{1.0cm}
\bb{K}_{\mc{B}_{\sL}}^{\sR}(p)
= \sum_{i,j=\bar{1},\bar{2}} (K^{\sR}_{\mc{B}_{\ms{a}}}(p) )^{\hspace{0.05cm} i }_{\hspace{0.05cm} j }  \hspace{0.15cm} \bb{E}_{ij} . 
\] 
The boundary intertwining equations are given by
\begin{equation} \label{a-boundary-intertwining}
\pi_{-p} (\mf{b}) \, \bb{K}_{\mc{B}_{\ms{a}}}^{\sL}(p) = \bb{K}_{\mc{B}_{\ms{a}}}^{\sL}(p) \, \pi_{p} (\mf{b}),  \hspace{1.0cm}
\bar{\pi}_{-p} (\mf{b}) \, \bb{K}_{\mc{B}_{\ms{a}}}^{\sR}(p) = \bb{K}_{\mc{B}_{\ms{a}}}^{\sR}(p) \, \bar{\pi}_{p}(\mf{b}), 
\end{equation}
for all $\mf{b} \in \mc{B}_{\ms{a}}$. The unitarity conditions are 
$\bb{K}_{\mc{B}_{\ms{a}}}^{\sL}(-p) \, \bb{K}_{\mc{B}_{\ms{a}}}^{\sL}(p) = \bb{I}$ and 
$\bb{K}_{\mc{B}_{\ms{a}}}^{\sR}(-p) \, \bb{K}_{\mc{B}_{\ms{a}}}^{\sR}(p) = \bb{I}$.
The solutions to these intertwining equations take the form
\begin{eqnarray}
& \hspace{-0.75cm} \bb{K}^{\sL}_{\mc{B}_{\sL}}(p) = k^{\sL}_{\mc{B}_{\sL}}(p) \hspace{0.15cm}  \bb{I}, \hspace{2.25cm} & \hspace{0.5cm}
\bb{K}^{\sR}_{\mc{B}_{\sL}}(p) = k^{\sR}_{\mc{B}_{\sL}}(p) \hspace{0.15cm} (\bb{E}_{\bar{1}\bar{1}} - u^{2}_p\,\bb{E}_{\bar{2}\bar{2}} ), \\
& \hspace{-0.75cm} \bb{K}_{\mc{B}_{\sR}}^{\sL}(p) = k^{\sL}_{\mc{B}_{\sR}}(p) \hspace{0.15cm} (\bb{E}_{11} - u^{2}_p \, \bb{E}_{22} ), & \hspace{0.5cm}
\bb{K}_{\mc{B}_{\sR}}^{\sR}(p) = k^{\sR}_{\mc{B}_{\sR}}(p) \hspace{0.15cm} \bb{I},
\end{eqnarray}
with both $k^{\sL}_{\mc{B}_{\ms{a}}}(-p) \, k^{\sL}_{\mc{B}_{\ms{a}}}(p) = 1$ and $k^{\sR}_{\mc{B}_{\ms{a}}}(-p) \, k^{\sR}_{\mc{B}_{\ms{a}}}(p) = 1$ for unitarity.

\paragraph{Non-supersymmetric chiral boundary algebra.}

Let us now consider a boundary algebra containing no supercharges. We set
\eq{
\mc{B}_{\sN\sC} = \big\lan \mf{H}_\sL , \mf{H}_\sR \big\ran.
}
The boundary Lie algebra $L(\mc{B}_{\sN\sC}) = \mf{u}(1)_{\sL} \oplus \mf{u}(1)_{\sR}$ is from an open superstring on $AdS_{3} \times S^{3} \times S^{3\prime} \times S^{1}$ attached to the $\bar{Z}=0$ or $\bar{Z}^{\prime}=0$ half of the D5-brane maximal giant or to the $Z=\bar{Z}^{\prime}=0$  D1-brane  giant. 
This is analogous to an open superstring ending on the $\bar{Z}=0$ giant graviton in $AdS_{5}\times S^{5}$ \cite{HM:2007}.

Now the coideal boundary algebra $\mc{B}_{\sN\sC}$ is a subalgebra of $\mc{A}$, and the constraints from the central elements imply that boundary reflections must be chiral. The $K$-matrix is thus
\begin{equation}
\bb{K}_{\mc{B}_{\sN\sC}}(p) = \bb{K}_{\mc{B}_{\sN\sC}}^{\sL}(p) \op \bb{K}_{\mc{B}_{\sN\sC}}^{\sR}(p).
\end{equation}
Since there are no additional constraints coming from the intertwining equations of $\mc{B}_{\sN\sC}$, we must solve the boundary Yang-Baxter equation or reflection equation \eqref{RE-complete-singlet}, discussed in the next subsection, directly.
The solution to this reflection equation has
\begin{equation} 
\bb{K}_{\mc{B}_{\sN\sC}}^{\sL}(p) = k^{\sL}_{\mc{B}_{\sN\sC}}(p) \, \bigg[ \bb{E}_{11} + \frac{(c - x^{+}_{p})}{(c + x^{-}_{p})} \, \bb{E}_{22} \bigg],  \hspace{0.75cm}
\bb{K}_{\mc{B}_{\sN\sC}}^{\sR}(p) = k^{\sR}_{\mc{B}_{\sN\sC}}(p) \, \bigg[ \bb{E}_{\bar{1}\bar{1}} + \frac{(1 + c \, x^{+}_{p})}{(1 - c \, x^{-}_{p})} \, \bb{E}_{\bar{2}\bar{2}} \bigg], \label{KD}
\end{equation}
where $k^{\sL}_{\mc{B}_{\sN\sC}}(-p)\,k^{\sL}_{\mc{B}_{\sN\sC}}(p) = 1$ and $k^{\sR}_{\mc{B}_{\sN\sC}}(-p)\, k^{\sR}_{\mc{B}_{\sN\sC}}(p) = 1$ for unitarity. 
The parameter $c \in \C$, which is interpreted as a free boundary parameter, has several interesting values. Setting $c=\tan \theta$ gives
\[
\underset{\theta \to \frac{\pi}{2}}{\rm lim} \hspace{0.15cm} \bb{K}_{\mc{B}_{\sN\sC}}(p) = \bb{K}_{\mc{B}_{\sL}}(p), \hspace{1.0cm}
\underset{\theta \to 0}{\rm lim} \hspace{0.15cm} \bb{K}_{\mc{B}_{\sN\sC}}(p) = \bb{K}_{\mc{B}_{\sR}}(p),
\]
whereas, for $c^{2} = -1$,  the partial $K$-matrices (\ref{KD}) become identical, since now
\[ \frac{(c - x^{+}_{p})}{(c + x^{-}_{p})} = \frac{(1 + c \, x^{+}_{p})}{(1 - c \, x^{-}_{p})}. \]

Now we want to ask: Does there exist a larger supersymmetric subalgebra of $\mc{A}$ which yields this $K$-matrix as a solution of the intertwining equations? The answer is yes. 
Let us introduce the diagonally supersymmetric boundary algebra
\begin{equation}
\mc{B}_{\sD} = \big\lan \, \mf{q}_{+}, \, \mf{q}_{-}, \, \mf{d} , \, \tilde{\mf{d}} \, \big\ran,  \label{BD}
\end{equation}
which is a coideal subalgebra of $\mc{A}$.  Here
\begin{eqnarray}
\nonumber && \hspace{-0.75cm} \mf{q}_{+} = \mf{P}^{\dag} \mf{Q}_{\sL} + i c \, \mf{P} \, \mf{S}_{\sR}, \hspace{1.0cm} 
\mf{d} = \big( \mf{H}_{\sL} - c^2 \mf{H}_\sR + i c \, (\mf{P}+\mf{P}^\dag) \big) \, \mf{K}, \\
&& \hspace{-0.75cm} \mf{q}_{-} = \mf{P} \, \mf{S}_{\sL} + i c \, \mf{P}^{\dag} \mf{Q}_{\sR}, \hspace{1.0cm} 
\tilde{\mf{d}} = \big( \mf{H}_{\sL} - c^2 \mf{H}_\sR - i c \, (\mf{P}+\mf{P}^\dag) \big) \, \mf{K},
\end{eqnarray}
with $\mf{K}=\mf{P}\,\mf{P}^\dag$. We also introduce the space $\mc{M}_{\sD}$ generated by $\{ \mf{s}_{+}, \, \mf{s}_{-}, \, \mf{n}, \, \tilde{\mf{n}} \, \}$, where
\begin{eqnarray}
\nonumber && \hspace{-0.75cm} \mf{s}_{+} = \mf{P}^{\dag} \mf{Q}_{\sL} - i c \, \mf{P}\,\mf{S}_{\sR}, \hspace{1.0cm} 
\mf{n} = \big(\mf{H}_\sL + c^2 \mf{H}_\sR + i c \, (\mf{P}-\mf{P}^\dag)\big)\,\mf{K}, \\
&& \hspace{-0.75cm} \mf{s}_{-} = \mf{P} \, \mf{S}_{\sL} - i c \, \mf{P}^{\dag} \mf{Q}_{\sR}, \hspace{1.0cm} 
\tilde{\mf{n}} = \big(\mf{H}_\sL + c^2 \mf{H}_\sR - i c \, (\mf{P}-\mf{P}^\dag)\big)\,\mf{K}.
\end{eqnarray} 
The generators of $\mc{B}_\sD$ and $\mc{M}_{\sD}$ satisfy the following non-trivial identities
\spl{
\{ \mf{q}_{+}, \mf{q}_{-} \} &= \mf{d},  \hspace{0.85cm}
\{ \mf{q}_{+}, \mf{s}_{-} \} = \mf{n},  \hspace{0.85cm}
\{ \mf{q}_{+}, \mf{s}_{+} \} = 0, \\
\{ \mf{s}_{+}, \mf{s}_{-} \} & = \tilde{\mf{d}},  \hspace{0.85cm}
\{ \mf{q}_{-}, \mf{s}_{+} \} = \tilde{\mf{n}},  \hspace{0.85cm}
\{ \mf{q}_{-}, \mf{s}_{-} \} = 0,
}
where $\mf{d}$, $\tilde{\mf{d}}$ and $\mf{n}$, $\tilde{\mf{n}}$ are central elements. Notice these relations are identical to those in \eqref{extended-algebra}. Thus the associated Lie superalgebra 
$L(\mc{B}_{\sD}\oplus \mc{M}_{\sD})$ is isomorphic to $L(\mc{A})$, while $L(\mc{B}_\sD)=\mf{su}(1|1)_\sD\op \mf{u}(1)_\sD$ consists of the Lie superalgebra 
$\mf{su}(1|1)_\sD$ generated by the triple $\{\mf{q}_{+},\,\mf{q}_{-},\,\mf{d}\}$, and $\mf{u}(1)_\sD$ generated by $\tilde{\mf{d}}$. Moreover, 
\begin{equation} 
\mf{g} \equiv L(\mc{B}_{\sD}\oplus \mc{M}_{\sD}) \cong L(\mc{A}) = \mf{su}(1|1)^{2}_{\ms{c}}, \hspace{0.85cm} 
\mf{h}_{\sD} \equiv L(\mc{B}_\sD)=\mf{su}(1|1)_\sD\op \mf{u}(1)_\sD 
\end{equation}
gives a symmetric pair  $(\mf{g},\mf{h}_{\sD})$ of Lie superalgebras. 

Solving the intertwining equations of the superalgebra $\mc{B}_{\sD}$, with $c\in\C$ an arbitrary parameter such that $c^2\neq-1$, gives precisely the $K$-matrix $\bb{K}_{\mc{B}_{\sN\sC}}(p)$ obtained above. Hence the algebra $\mc{B}_{\sD}$ can be understood as a {\it hidden symmetry} of the non-supersymmetric chiral boundary for $c^2\neq-1$.

We note, however, that setting $c^2=-1$ yields a solution of the intertwining equations for $\mc{B}_\sD$ which consists of both chiral parts, 
$\bb{K}_{\mc{B}_{\sN\sC}}^{\sL}(p)$ and $\bb{K}_{\mc{B}_{\sN\sC}}^{\sR}(p)$, and achiral parts $\bb{A}_{\mc{B}_{\sN\sC}}^{\sR}(p)$ and $\bb{A}_{\mc{B}_{\sN\sC}}^{\sL}(p)$.  This solution does not satisfy the reflection equation. There is no obvious hidden symmetry for $c^{2} = -1$.

\paragraph{Non-supersymmetric achiral boundary algebra.}

Let us now consider the possibility of a boundary algebra containing only the magnon Hamiltonian. Let us assume that
\begin{equation}
\mc{B}_{\sN\sA} = \lan \mf{H} \ran, \hspace{0.75cm} \text{with} \hspace{0.35cm} \mf{H} =\mf{H}_\sL + \mf{H}_\sR,
\end{equation}
which allows for both chiral and achiral reflections. The associated boundary Lie algebra, given by $L(\mc{B}_{\sN\sA}) = \mf{u}(1)_{+}$, is that of an open superstring on 
$AdS_{3}\times S^{3} \times S^{3\prime} \times S^{1}$ ending on the $\bar{Y}=Y^{\prime}=0$ or $Y=\bar{Y}^{\prime}=0$ D1-brane maximal giant graviton.
We choose an achiral ansatz for the $K$-matrix 
\begin{equation}
\bb{K}_{\mc{B}_{\sN\sA}}(p) \, = \, 
\begin{pmatrix} 0 &  \bb{A}^{\sR}_{\mc{B}_{\sN\sA}}(p) \\ 
\bb{A}^{\sL}_{\mc{B}_{\sN\sA}}(p) &  0
\end{pmatrix} \, = \, 
\begin{pmatrix} 0 & \bb{I} \\ \bb{I} & 0 \end{pmatrix} \big(\bb{A}_{\mc{B}_{\sN\sA}}^{\sL}(p) \oplus \bb{A}_{\mc{B}_{\sN\sA}}^{\sR}(p) \big),
\end{equation}
where
$$
\bb{A}_{\mc{B}_{\sN\sA}}^{\sR}(p) = \sum_{^{i=1,2}_{j=\bar{1},\bar{2}}} (A^{\sR}_{\mc{B}_{\sN\sA}}(p))^{i}_{j} \hspace{0.15cm} \bb{E}_{ij},  \hspace{0.85cm}
\bb{A}_{\mc{B}_{\sN\sA}}^{\sL}(p) = \sum_{^{i=\bar{1},\bar{2}}_{j=1,2}} (A^{\sL}_{\mc{B}_{\sN\sA}}(p))^{i}_{j} \hspace{0.15cm} \bb{E}_{ij}.
$$
There are no constraints from the intertwining equations of $\mc{B}_{\sN\sA}$.  We must therefore proceed by solving the reflection equation \eqref{RE-complete-singlet} directly -- which eventually yields a solution of the form
\begin{eqnarray}
\nonumber && \hspace{-0.65cm} \bb{A}_{\mc{B}_{\sN\sA}}^{\sR}(p) = a_{\mc{B}_{\sN\sA}}^{\sR}(p) \hspace{0.15cm} \big(1 + x^{+}_{p} x^{-}_{p}\big)^{-\frac{1}{2}} \hspace{0.1cm}
\bigg[- c\,u_p^{-1} \, (i+x^{+}_p)\,\bb{E}_{1\bar{1}} + i \eta_p (\bb{E}_{1\bar{2}} + \bb{E}_{2\bar{1}}) - c^{-1} \, u_p \, (i-x^-_p) \, \bb{E}_{2\bar{2}} \bigg], \\
&& \hspace{-0.65cm} \bb{A}_{\mc{B}_{\sN\sA}}^{\sL}(p) = a_{\mc{B}_{\sN\sA}}^{\sL}(p) \hspace{0.15cm} \big(1 + x^{+}_{p} x^{-}_{p}\big)^{-\frac{1}{2}} \hspace{0.1cm}
\bigg[ c^{-1} \, u_{p}^{-1} \, (i + x^{+}_{p}) \, \bb{E}_{\bar{1}1} + i \eta_{p} \, (\bb{E}_{\bar{1}2} + \bb{E}_{\bar{2}1}) + c\,u_p(i-x^{-}_{p}) \, \bb{E}_{\bar{2}2} \bigg] \hspace{1.2cm}
\end{eqnarray}
for any $c\in\C$.  Here $a_{\mc{B}_{\sN\sA}}^{\sR}(-p) \, a_{\mc{B}_{\sN\sA}}^{\sL}(p) = 1$ and 
$a^{\sL}_{\mc{B}_{\sN\sA}}(-p) \, a^{\sR}_{\mc{B}_{\sN\sA}}(p) = 1$ for unitarity.  

We can ask the same question as before: does there exist a coideal subalgebra of $\mc{A}$ which yields this $K$-matrix as a solution of the intertwining equations? We were not able to find such an algebra, but this does not exclude its existence.

\subsubsection{Reflection equation}

\begin{figure}[htb!]
\begin{center}
\includegraphics[scale=1.3]{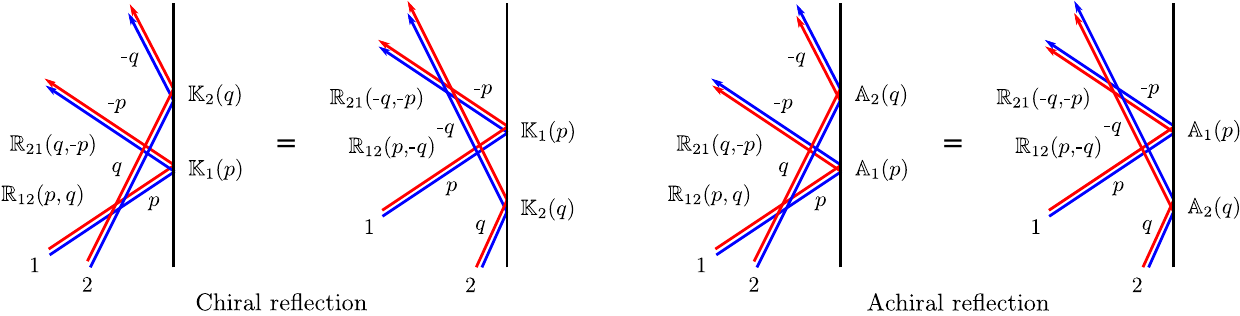}
\caption{A two-magnon reflection off a singlet boundary factorizes into a succession of single-magnon reflections and two-magnon scattering events.  The double red and blue line again indices the {\it direct sum of left and right magnon states} which scatter by complete $R$ and $K$-matrices.} \label{figure-RE-singlet}
\vspace{-0.35cm}
\end{center}
\end{figure}

A reflection $K$-matrix must satisfy the boundary Yang-Baxter equation, also called the reflection equation \cite{Sk:1988}, 
\begin{equation} \label{RE-complete-singlet} 
\bb{K}_{2}(q) \; \bb{R}_{21}(q,-p) \; \bb{K}_{1}(p) \; \bb{R}_{12}(p,q) \, = \, 
\bb{R}_{21}(-q,-p) \; \bb{K}_1(p) \; \bb{R}_{12}(p,-q) \; \bb{K}_2(q). 
\end{equation}
Let us now consider chiral and achiral reflections separately.
This reflection equation is equivalent, for the partial $R$-matrices and partial chiral $K$-matrices, to
\begin{equation} \label{RE-partial-singlet-chiral}
\bb{K}^{\ms{b}}_{2}(q) \; \bb{R}^{\ms{ba}}_{21}(q,-p) \; \bb{K}^{\ms{a}}_{1}(p) \; \bb{R}^{\ms{ab}}_{12}(p,q) \, 
= \bb{R}^{\ms{ba}}_{21}(-q,-p) \; \bb{K}^{\ms{a}}_{1}(p) \; \bb{R}^{\ms{ab}}_{12}(p,-q) \; \bb{K}^{\ms{b}}_{2}(q) ,
\end{equation}
and, for the partial $R$-matrices and the partial achiral $K$-matrices, to
\begin{equation} \label{RE-partial-singlet-achiral}
\bb{A}^{\ms{b}}_{2}(q) \; \bb{R}^{\ms{b\bar{a}}}_{21}(q,-p) \; \bb{A}^{\ms{a}}_{1}(p) \; \bb{R}^{\ms{ab}}_{12}(p,q) \, 
= \bb{R}^{\ms{\bar{b}a}}_{21}(-q,-p) \; \bb{A}^{\ms{a}}_{1}(p) \; \bb{R}^{\ms{a\bar{b}}}_{12}(p,-q) \; \bb{A}^{\ms{b}}_{2}(q) ,
\end{equation}
for all $\ms{a}, \ms{b}, \ms{c} \in \{ \SL,\SR \}$, with $\ms{\bar \SL}=\SR$ and $\ms{\bar \SR}=\SL$ (see Figure \ref{figure-RE-singlet}).
Here we define
\begin{eqnarray}
\nonumber && \hspace{-0.75cm}
\bb{R}^{\ms{a} \ms{b}}_{12}(p,q) = \bb{R}^{\ms{a}\ms{b}}(p,q), \hspace{1.45cm} 
\bb{K}^{\ms{a}}_1(p) = \bb{K}^{\ms{a}}(p) \ot \bb{I},  \hspace{0.85cm}
\bb{A}^{\ms{a}}_1(p) = \bb{A}^{\ms{a}}(p) \ot \bb{I},  \\
\nonumber && \hspace{-0.75cm} \bb{R}^{\ms{a} \ms{b}}_{21}(p,q) = \bb{P} \, \bb{R}^{\ms{a}\ms{b}}(p,q) \, \bb{P} , \hspace{0.85cm}
\bb{K}^{\ms{a}}_2(p) = \bb{I} \ot \bb{K}^{\ms{a}}(p), \hspace{0.85cm}
\bb{A}^{\ms{a}}_2(p) = \bb{I} \ot \bb{A}^{\ms{a}}(p),  \hspace{0.35cm}
\end{eqnarray}
and similarly for $\bb{R}_{12}(p,q)$ and $\bb{R}_{21}(p,q)$, and $\bb{K}_1(p)$ and $\bb{K}_2(q)$ in terms of the complete $R$-matrices and $K$-matrices. 
All the $K$-matrices constructed above are found to satisfy the reflection equation, if we impose the parity and discrete $\SL\SR$ symmetry constraints on the scale factors $s^{\ms{ab}}(p,q)$ in the partial $R$-matrices, which were described in Section \ref{part3-sec1-3-1}.  

\subsection{Vector boundary} \label{sec:4.3}

The totally supersymmetric boundary superalgebra $\mc{B}_{\sT}$ is the coideal subalgebra of $\mc{A}$ generated as
\begin{equation}
\mc{B}_{\sT} = \big\lan \mf{Q}_\sL,\mf{S}_\sL,\mf{H}_\sL, \mf{Q}_\sR,\mf{S}_\sR,\mf{H}_\sR,\mf{P}, \mf{P}^\dag \big\ran. \label{BT}
\end{equation}
Note that elements $\mf{U},\mf{U}^{-1}\notin\mc{B}_\sT$.  Since $\mf{U}$ and $\mf{U}^{-1}$ appear in the left tensor factor of the coproduct only, the coideal property $\Delta(\mf{b}) \in \mc{A} \ot \mc{B}_\sT$ for all $\mf{b}\in\mc{B}_\sT$ is satisfied.

The vector boundary state transforms in the left- or right-moving representation, $\pi_{\sB}$ or $\bar{\pi}_{\sB}$, of the boundary Lie superalgebra 
$L(\mc{B}_{\sT})=\mf{su}(1|1)^{2}_{\ms{c}}$. This boundary superalgebra arises from an open superstring ending on the $Z=0$ or $Z^{\prime}=0$ half of the D5-brane maximal giant graviton, or the $Z=Z^{\prime}=0$ D1-brane giant (the intersection of $Z=0$ and $Z^{\prime}=0$ 
giants), shown in Tables \ref{table-D5} and \ref{table-D1}. This is analogous to an open superstring attached to the $Z=0$ giant graviton in $AdS_{5}\times S^{5}$ \cite{HM:2007}. 

\paragraph{Boundary representations $\pi_\sB$ and $\bar{\pi}_\sB$.}

We define the vector spaces associated with left- and right-moving boundary vector states in the same way as for the magnons in the bulk:
\[
V_{\sB}={\rm span}_{\C}\{ | \phi_\sB\rangle, | \psi_\sB\rangle \} \cong \C^{1|1}, \hspace{1.0cm} 
\bar{V}_{\sB}={\rm span}_{\C}\{ | \bar\phi_\sB\rangle, | \bar\psi_\sB\rangle \} \cong \C^{1|1},
\]  
and set $W_{\sB} = V_{\sB} \oplus \bar{V}_{\sB} \cong \C^{2|2}$. The left boundary representation $\pi_\sB : \mc{B}_{\sT} \to \End(\C^{1|1})$ is given  by \eqref{Rep:L}, with the subindex $p$ replaced by $\SB$. The right boundary representation $\bar\pi_\sB : \mc{B}_{\sT} \to \End(\C^{1|1})$ is analogous to \eqref{Rep:R} subject to the $p\to\SB$ replacement. The primed space $W'_\sB = V^{\prime}_{\sB} \oplus \bar{V}^{\prime}_{\sB}$, and representations $\pi^{\prime}_{\sB}$ and $\bar{\pi}^{\prime}_{\sB}$ are defined in a similar way. We will use the notation $\mc{W}_\sB \in\{W_\sB, W'_\sB\}$.

We may choose the following parametrization of the boundary parameters:
\begin{equation}
a_{\sB} = \sqrt{h} \hspace{0.1cm} \eta_{\sB}, \hspace{0.65cm}
b_{\sB} = \sqrt{h} \hspace{0.1cm} \eta_{\sB}, \hspace{0.65cm}
c_{\sB} = \sqrt{h} \hspace{0.1cm} \frac{i \eta_{\sB}}{x_{\sB}}, \hspace{0.65cm}
d_{\sB} = \sqrt{h} \hspace{0.1cm} \frac{i \eta_{\sB}}{x_{\sB}}, \hspace{0.65cm} \eta_\sB^2 = - i x_{\sB}, \label{abcd:boundary}
\end{equation}
where $x_\sB$ is the boundary Zhukovski variable satisfying the boundary mass-shell identity
\begin{equation}
x_{\sB} + \frac{1}{x_\sB} = \frac{i m_\sB}{h},
\end{equation}
with $m_\sB$ the boundary mass parameter.
This boundary representation can be obtained from $\pi_{p}$ by setting the momentum to $p=\pi$ (so that $x_{\sB} \equiv x_{p=\pi}^{+} = - \hspace{0.025cm} x_{p=\pi}^{-}$), defining the boundary mass parameter to be $m_\sB = m_{p=\pi}$ and rescaling $h \to h/2$.  We expect this representation to describe a magnon state absorbed by the boundary.
Notice that the total bulk and boundary momentum before and after the reflection sum to $p+\pi-p+\pi=2\pi\sim 0$ due to periodicity -- which is conservation of momentum for an elastic reflection off an infinitely massive boundary. 
We will further justify these boundary parameters when we discuss the constraints from the central elements in $\mc{B}_\sT$.

\paragraph{Boundary intertwining equations.}

For this case of a vector boundary, the intertwining equations for the K-matrix $K_{\sT}(p,\SB)\in\End(\mc{W}_p\ot \mc{W}_\sB)$, corresponding to 
$\bb{K}_{\sT}(p,\SB) \in \End(\C^{2|2}\otimes\C^{2|2})$, are very similar to those imposed on the bulk $R$-matrix in Chapter \ref{part3-sec1}. The boundary scattering involves three states: bulk magnon, boundary vector state and the boundary singlet vacuum state. Since the boundary vacuum state is described by the trival representation $\eps$ satisfying $\eps(\mf{b})=0$ for all $\mf{b}\in\mc{B}_\sT$, we can drop it to obtain the boundary intertwining equations for the vector boundary:
\begin{equation} \label{intw:b2} 
(({\pi}_{-p} \oplus \bar{\pi}_{-p}) \ot (\pi_\sB \oplus \bar{\pi}_{\sB}))(\Delta(\mf{b})) \hspace{0.15cm} \bb{K}_{\mc{B}_{\sT}}(p,\SB) \, = \,  
\bb{K}_{\mc{B}_{\sT}}(p,\SB) \,((\pi_{p} \oplus \bar{\pi}_{p}) \ot (\pi_\sB \oplus \bar{\pi}_{\sB}))(\Delta(\mf{b})) 
\end{equation}
for all $\mf{b}\in\mc{B}_{\sT}$. 

\paragraph{Constraints from the central elements for vector boundary.}

Central elements of $\mc{B}_{\sT}$ imply important constraints for reflections off a vector boundary and for the parameters of the boundary representation $\pi_{\sB}$. Recall that central elements must intertwine the reflection matrix trivially, which means that, for a chiral reflection,
\[
(\pi_p\ot\pi_{\sB})(\Delta(\mathfrak{C})) = (\pi_{-p}\ot\pi_{\sB})(\Delta(\mathfrak{C})), \hspace{0.85cm} 
(\bar\pi_p\ot\bar\pi_{\sB})(\Delta(\mathfrak{C})) = (\bar\pi_{-p}\ot\bar\pi_{\sB})(\Delta(\mathfrak{C}))   
\]
for all $\mathfrak{C} \in\{\mf{H}_{\ms{a}}, \mf{P}, \mf{P}^\dagger\}$. Suppose that we do not know the boundary representation $\pi_\sB$. Since the elements $\mf{P}$ and $\mf{P}^\dag$ are central, we must have $\pi_\sB(\mf{P})=f_\sB \,\bb{I}$ and $\pi_\sB(\mf{P}^\dag) = f^\dag_\sB\, \bb{I}$ for some $f_\sB, f^\dag_\sB \in \C$. Using $\pi_p(\mf{P})=h \, (1-u_p^2)\, \bb{I}$ and $\pi_p \, (\mf{P}^\dag)=h \, (1-u_p^{-2})\, \bb{I}$, and the boundary intertwining equations, we obtain
\[
h \, (1-u_p^2) + u_p^2 \, f_\sB \, = \, h \, (1-u_p^{-2}) + u_p^{-2} \, f_\sB,  \hspace{0.85cm} 
h \, (1-u_p^{-2}) + u_p^{-2} \, f^\dag_\sB \, =  \, h \, (1-u^{2}_p) + u_p^{2} \, f^\dag_\sB,
\]
giving $f_\sB= f^\dag_\sB= h$. Now notice that $\pi_{p}(\mf{P})|_{p=\pi, \, h\to h/2} = \pi(\mf{P}^\dag)|_{p=\pi, \, h\to h/2} =h \, \bb{I}$, which justifies our interpretation of the vector boundary as a bulk magnon state with momentum $p=\pi$. 
A vector state at the boundary is always necessary for the boundary algebra $\mc{B}_{\sT}$, since the central elements $\mf{P}$ and $\mf{P}^{\dag}$ cannot be preserved by a singlet boundary.

The next step is to check the intertwining equations for achiral reflections. For example,
\[
(\pi_{p} \ot \pi_{\sB})(\Delta(\mf{H}_{\ms{a}})) \neq (\bar\pi_{-p}\ot\pi_{\sB})(\Delta(\mf{H}_{\ms{a}})), \hspace{0.85cm} 
(\bar{\pi_{p}}\ot\bar\pi_{\sB})(\Delta(\mf{H}_{\ms{a}}))) \neq (\pi_{-p}\ot\bar\pi_B)(\Delta(\mf{H}_{\ms{a}})).
\]
However,
\[
(\pi_p\ot\bar\pi_{\sB})(\Delta(\mathfrak{C})) = (\bar\pi_{-p}\ot\pi_{\sB})(\Delta(\mathfrak{C})), \hspace{0.85cm} 
(\bar{\pi_{p}}\ot\pi_{\sB})(\Delta(\mathfrak{C})) = (\pi_{-p}\ot\bar\pi_{\sB})(\Delta(\mathfrak{C})),
\]
for all $\mathfrak{C} \in\{\mf{H}_{\ms{a}}, \mf{P}, \mf{P}^\dagger\}$. 
This implies that the total number of left or right states is a conserved quantum number as a result of the central elements in $\mc{B}_{\sT}$. 
Scattering off a vector boundary, which intertwines the representations $\pi_{p} \ot \bar\pi_{\sB}$ and $\bar\pi_{-p}\ot\pi_{\sB}$ or $\bar\pi_{p}\ot\pi_{\sB}$ and $\pi_{-p}\ot\bar\pi_{\sB}$, is forbidden rather by the intertwining equations for the supercharges.  Thus the $K$-matrix $\bb{K}_{\mc{B}_{\sT}}(p)$ decomposes into the four sectors (left-from-left, right-from-right, left-from-right and right-from-left) with $\bb{K}^{\ms{ab}}_{\mc{B}_{\sT}}(p)$ describing the chiral reflection of a magnon of chirality $\ms{a}$ from a boundary of chirality $\ms{b}$.

\subsubsection{Complete and partial $K$-matrices}

The {\it complete} $K$-matrix decomposes into the direct sum
\begin{equation}
\bb{K}_{\mc{B}_{\sT}}(p,\SB) = \bb{K}^{\sL\sL}_{\mc{B}_{\sT}}(p,\SB)\op \bb{K}^{\sR\sR}_{\mc{B}_{\sT}}(p,\SB)\op \bb{K}^{\sL\sR}_{\mc{B}_{\sT}}(p,\SB)\op \bb{K}^{\sR\sL}_{\mc{B}_{\sT}}(p,\SB) \label{KF-dec}
\end{equation}
of {\it partial} $K$-matrices in the $\SL\SL$, $\SR\SR$, $\SL\SR$ and $\SR\SL$ decoupled sectors. The partial $K$-matrices are solutions of the boundary intertwining and reflection equations, and the unitarity condition.

\paragraph{Left-left and right-right sectors.}

We write the partial $K$-matrices in the $\SL\SL$ and $\SR\SR$ sectors as
\[
\bb{K}^{\sL\sL}_{\mc{B}_{\sT}}(p,\SB)
= \sum_{i,j,k,l=1,2} (K^{\sL\sL}(p,\SB))^{\hspace{0.05cm} i \hspace{0.05cm} k}_{\hspace{0.05cm} j \hspace{0.05cm} l} \hspace{0.15cm} (\bb{E}_{ij}\ot \bb{E}_{kl}), \hspace{0.6cm}
\bb{K}^{\sR\sR}_{\mc{B}_{\sT}}(p,\SB)
= \sum_{i,j,k,l=\bar{1},\bar{2}} (K^{\sR\sR}(p,\SB))^{\hspace{0.05cm} i \hspace{0.05cm} k}_{\hspace{0.05cm} j \hspace{0.05cm} l} \hspace{0.15cm} (\bb{E}_{ij} \ot \bb{E}_{kl} ) ,
\] 
which depend on the magnon momentum $p$ and its mass $m_{p}$ through the Zhukovski variables $x_{p}^{\pm}$, and the boundary mass parameter $m_{\sB}$ through $x_{\sB}$.
The boundary intertwining equations are
\begin{eqnarray}
\nonumber && \hspace{-0.65cm} 
( \pi_{-p} \ot \pi_{\sB} ) (\Delta(\frak{b})) \hspace{0.15cm} \bb{K}^{\sL\sL}_{\mc{B}_{\sT}}(p,\SB) \, = \,
\bb{K}^{\sL\sL}_{\mc{B}_{\sT}}(p,\SB) \hspace{0.15cm} ( \pi_{p} \ot \pi_{\sB} )(\Delta(\frak{b})), \\
\nonumber && \hspace{-0.65cm} 
( \bar{\pi}_{-p} \ot \bar{\pi}_{\sB} ) (\Delta(\frak{b})) \hspace{0.15cm} \bb{K}^{\sR\sR}_{\mc{B}_{\sT}}(p,\SB) \, = \,
\bb{K}^{\sR\sR}_{\mc{B}_{\sT}}(p,\SB) \hspace{0.15cm} ( \bar{\pi}_{p} \ot \bar{\pi}_{\sB} )(\Delta(\frak{b})),  
\end{eqnarray}
for all $\frak{b} \in \mc{B}_\sT$. The unitarity condition is  $\bb{K}^{\sL\sL}_{\mc{B}_{\sT}}(-p,\SB) \, \bb{K}^{\sL\sL}_{\mc{B}_{\sT}}(p,\SB) = \bb{I}$ and $\bb{K}^{\sR\sR}_{\mc{B}_{\sT}}(-p,\SB) \, \bb{K}^{\sR\sR}_{\mc{B}_{\sT}}(p,\SB) = \bb{I}$. 
We find the solutions of these boundary intertwining equations to be
\begin{eqnarray}
&& \hspace{-0.65cm}
\bb{K}^{\sL\sL}_{\mc{B}_{\sT}}(p,\SB) = k^{\sL\sL}_{\mc{B}_{\sT}}(p, \SB) \, \bigg[ \bb{E}_{11}\ot \bb{E}_{11} 
+ \frac{(x^{+}_{p} - u^{-2}_{p}\, x_{\sB})}{(x^{-}_{p} - x_{\sB})} \, \bb{E}_{11}\ot \bb{E}_{22} 
+ \frac{(x^{-}_{p} + u_{p}^{2} \, x_{\sB})}{(x^{-}_{p} - x_{\sB})} \, \bb{E}_{22} \ot \bb{E}_{11} \\
\nonumber && \hspace{-0.65cm} \hspace{3.5cm}
+ \, \frac{(x^{+}_{p} + x_{\sB})}{(x^{-}_{p} - x_{\sB})} \, \bb{E}_{22}\ot \bb{E}_{22}
+ \frac{i \, (u_{p} + u_{p}^{-1} ) \, \eta_p \eta_\sB }{(x^{-}_{p} - x_{\sB})} \,\bb{E}_{12}\ot \bb{E}_{21}
- \frac{i \, (u_{p} + u_{p}^{-1} ) \, \eta_p \eta_\sB }{(x^{-}_{p} - x_{\sB})} \, \bb{E}_{21}\ot \bb{E}_{12} \bigg], 
\end{eqnarray}
\begin{eqnarray}
&& \hspace{-0.65cm}
\bb{K}^{\sR\sR}_{\mc{B}_{\sT}}(p,\SB) = k^{\sR\sR}_{\mc{B}_{\sT}}(p, \SB) \, \bigg[ \bb{E}_{\bar{1}\bar{1}}\ot \bb{E}_{\bar{1}\bar{1}} 
+ \frac{(x^{+}_{p} - u^{-2}_{p}\, x_{\sB})}{(x^{-}_{p} - x_{\sB})} \, \bb{E}_{\bar{1}\bar{1}}\ot \bb{E}_{\bar{2}\bar{2}} 
+ \frac{(x^{-}_{p} + u_{p}^{2} \, x_{\sB})}{(x^{-}_{p} - x_{\sB})} \, \bb{E}_{\bar{2}\bar{2}} \ot \bb{E}_{\bar{1}\bar{1}}  \\
\nonumber  && \hspace{-0.65cm} \hspace{3.5cm}
+ \, \frac{(x^{+}_{p} + x_{\sB})}{(x^{-}_{p} - x_{\sB})} \, \bb{E}_{\bar{2}\bar{2}}\ot \bb{E}_{\bar{2}\bar{2}}
+  \frac{i \, (u_{p} + u_{p}^{-1} ) \, \eta_p \eta_\sB }{(x^{-}_{p} - x_{\sB})} \,\bb{E}_{\bar{1}\bar{2}}\ot \bb{E}_{\bar{2}\bar{1}}
-  \frac{i \, (u_{p} + u_{p}^{-1} ) \, \eta_p \eta_\sB }{(x^{-}_{p} - x_{\sB})} \, \bb{E}_{\bar{2}\bar{1}}\ot \bb{E}_{\bar{1}\bar{2}} \bigg], 
\end{eqnarray}
where $ k^{\sL\sL}_{\mc{B}_{\sT}}(-p, \SB) \, k^{\sL\sL}_{\mc{B}_{\sT}}(p, \SB) = 1$ and 
$ k^{\sR\sR}_{\mc{B}_{\sT}}(-p, \SB) \, k^{\sR\sR}_{\mc{B}_{\sT}}(p, \SB) = 1$ for unitarity.

\paragraph{Left-right and right-left sectors.}

We write the partial $K$-matrices in the $\SL\SR$ and $\SR\SL$ sectors as
\[
\bb{K}^{\sL\sR}_{\mc{B}_{\sT}}(p,\SB)
= \!\! \sum_{^{i,j=1,2}_{k,l=\bar{1},\bar{2}}} \!\! (K^{\sL\sR}(p,\SB))^{\hspace{0.05cm} i \hspace{0.05cm} k}_{\hspace{0.05cm} j \hspace{0.05cm} l} \hspace{0.15cm} (\bb{E}_{ij}\ot \bb{E}_{kl}), \hspace{0.75cm}
\bb{K}^{\sR\sL}_{\mc{B}_{\sT}}(p,\SB)
= \!\! \sum_{^{i,j=\bar{1},\bar{2}}_{k,l=1,2}} \!\! (K^{\sR\sL}(p,\SB))^{\hspace{0.05cm} i \hspace{0.05cm} k}_{\hspace{0.05cm} j \hspace{0.05cm} l} \, (\bb{E}_{ij} \ot \bb{E}_{kl}). 
\] 
The left-right $K$-matrix is a solution of the boundary intertwining equations
\begin{eqnarray}
\nonumber && \hspace{-0.65cm} 
( \pi_{-p} \ot \bar{\pi}_{\sB} ) (\Delta(\frak{b})) \hspace{0.15cm} \bb{K}^{\sL\sR}_{\mc{B}_{\sT}}(p,\SB) \, = \,
\bb{K}^{\sL\sR}(p,\SB) \hspace{0.15cm} ( \pi_{p} \ot \bar{\pi}_{\sB} )(\Delta(\frak{b})), \\
\nonumber && \hspace{-0.65cm} 
( \bar{\pi}_{-p} \ot \pi_{\sB} ) (\Delta(\frak{b})) \hspace{0.15cm} \bb{K}^{\sR\sL}_{\mc{B}_{\sT}}(p,\SB) \, = \,
\bb{K}^{\sR\sL}(p,\SB) \hspace{0.15cm} ( \bar{\pi}_{p} \ot \pi_{\sB} )(\Delta(\frak{b})),
\end{eqnarray}
for all $\frak{b} \in \mc{B}_\sT$. The unitarity condition is
$\bb{K}^{\sL\sR}_{\mc{B}_{\sT}}(-p,\SB) \, \bb{K}^{\sL\sR}_{\mc{B}_{\sT}}(p,\SB) = \bb{I}$ and 
$\bb{K}^{\sR\sL}_{\mc{B}_{\sT}}(-p,\SB) \, \bb{K}^{\sR\sL}_{\mc{B}_{\sT}}(p,\SB) = \bb{I}$. 
The solutions of the boundary intertwining equations take the form
\begin{align}
\bb{K}^{\sL\sR}_{\mc{B}_{\sT}}(p,\SB)
\nonumber &= k^{\sL\sR}_{\mc{B}_{\sT}}(p,\SB) \,\, \big[(1 - x^{+}_{p} x_{\sB})(1 + x^{-}_{p} x_{\sB})\big]^{-\frac{1}{2}}
  \\
& \times \bigg[ \big(x^{+}_{p} x_{\sB} + u_{p}^{-2} \big) \hspace{0.15cm} \bb{E}_{11}\ot \bb{E}_{\bar{1}\bar{1}} + 
\big(x^{-}_{p} x_{\sB} + 1\big) \hspace{0.15cm} \bb{E}_{11}\ot \bb{E}_{\bar{2}\bar{2}}  
+ \big(x^+_p x_\sB -1\big) \hspace{0.15cm} \bb{E}_{22} \ot \bb{E}_{\bar{1}\bar{1}} \\ 
& \hspace{0.35cm} + \big(x^-_p x_\sB - u_p^{2}\big) \hspace{0.15cm} \bb{E}_{22}\ot \bb{E}_{\bar{2}\bar{2}}
- \big(u_{p} + u^{-1}_{p}\big) \, \eta_{p} \eta_{\sB} \hspace{0.15cm} \bb{E}_{12}\ot \bb{E}_{\bar{1}\bar{2}} 
+ \big(u_p + u^{-1}_p\big) \, \eta_p \eta_\sB \hspace{0.15cm} \bb{E}_{21}\ot \bb{E}_{\bar{2}\bar{1}} \bigg], \nonumber
\end{align}
\vspace{-0.5cm}
\begin{align}
\nonumber \bb{K}^{\sR\sL}_{\mc{B}_{\sT}}(p,\SB)
&= k^{\sL\sR}_{\mc{B}_{\sT}}(p, \SB) \,\, \big[(1 - x^{+}_{p} x_{\sB})(1 + x^{-}_{p} x_{\sB})\big]^{-\frac{1}{2}}
  \\
& \times \bigg[ \big(x^{+}_{p} x_{\sB} + u_{p}^{-2} \big) \hspace{0.15cm} \bb{E}_{\bar{1}\bar{1}}\ot \bb{E}_{11} + 
\big(x^{-}_{p} x_{\sB} + 1\big) \hspace{0.15cm} \bb{E}_{\bar{1}\bar{1}}\ot \bb{E}_{22}  
+ \big(x^+_p x_\sB -1\big) \hspace{0.15cm} \bb{E}_{\bar{2}\bar{2}} \ot \bb{E}_{11} \\ 
& \hspace{0.35cm} + \big(x^-_p x_\sB - u_p^{2}\big) \hspace{0.15cm} \bb{E}_{\bar{2}\bar{2}}\ot \bb{E}_{22}
- \big(u_{p} + u^{-1}_{p}\big) \, \eta_{p} \eta_{\sB} \hspace{0.15cm} \bb{E}_{\bar{1}\bar{2}}\ot \bb{E}_{12} 
+ \big(u_p + u^{-1}_p\big) \, \eta_p \eta_\sB \hspace{0.15cm} \bb{E}_{\bar{2}\bar{1}}\ot \bb{E}_{21} \bigg], \nonumber
\end{align}
where $ k^{\sL\sR}_{\mc{B}_{\sT}}(-p, \SB) \, k^{\sL\sR}_{\mc{B}_{\sT}}(p, \SB) = 1$ and 
$ k^{\sR\sL}_{\mc{B}_{\sT}}(-p, \SB) \, k^{\sR\sL}_{\mc{B}_{\sT}}(p, \SB) = 1$ for unitarity.

\smallskip

We note that it is necessary to impose {\it boundary crossing symmetry conditions} on all our $K$-matrices, which constrain the scale factors $k^{\ms{a}}(p)$ and $a^{\ms{a}}(p)$. We anticipate that these will be related to the dressing phases of \cite{Borsato:2013hoa} in the bulk $R$-matrix. We leave this for future research.

\subsubsection{Reflection equation}

\begin{figure}[htbb!]
\begin{center}
\includegraphics[scale=1.5]{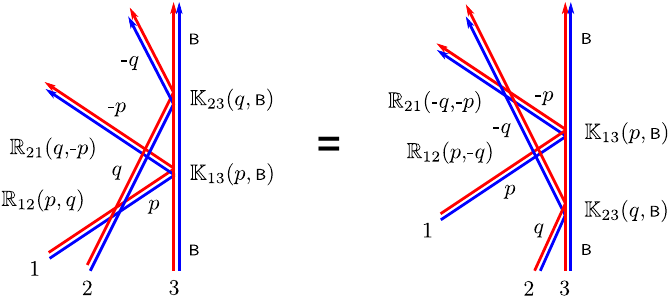} 
\caption{A two-magnon reflection off a vector boundary factorizes into a succession of single magnon reflections and two-magnon scattering events. The double red and blue line again indicates a {\it direct sum of left and right magnons or boundary vector states}.} \label{figure-RE-vector}
\vspace{-0.35cm}
\end{center}
\end{figure}

The reflection equation for the complete $K$-matrix is
\begin{equation} \label{RE-complete} 
\bb{K}_{23}(q,\SB)  \; \bb{R}_{21}(q,-p) \; \bb{K}_{13}(p,\SB) \; \bb{R}_{12}(p,q) \, = \, 
\bb{R}_{21}(-q,-p) \; \bb{K}_{13}(p,\SB) \; \bb{R}_{12}(p,-q) \; \bb{K}_{23}(q,\SB), 
\end{equation}
which is equivalent, for the partial $R$-matrices and $K$-matrices, to
\begin{equation} \label{RE-partial}
\bb{K}^{\ms{bc}}_{23}(q,\SB)  \; \bb{R}^{\ms{ba}}_{21}(q,-p) \; \bb{K}^{\ms{ac}}_{13}(p,\SB) \; \bb{R}^{\ms{ab}}_{12}(p,q) \;
= \bb{R}^{\ms{ba}}_{21}(-q,-p) \; \bb{K}^{\ms{ac}}_{13}(p,\SB) \; \bb{R}^{\ms{ab}}_{12}(p,-q) \; \bb{K}^{\ms{bc}}_{23}(q,\SB),
\end{equation}
for all $\ms{a}, \ms{b}, \ms{c} \in \{ \SL,\SR \}$ (see Figure \ref{figure-RE-vector}). 
Here we define
\begin{eqnarray}
\nonumber && \hspace{-0.65cm} \bb{R}^{\ms{a} \ms{b}}_{12}(p,q) = \bb{R}^{\ms{a}\ms{b}}(p,q)\ot \bb{I}, \hspace{1.9cm}
\bb{K}^{\ms{a} \ms{b}}_{13}(p,\SB) = (\bb{I} \ot \bb{P} )(\bb{K}^{\ms{a} \ms{b}}(p,\SB) \ot \bb{I})(\bb{I} \ot \bb{P}),  \\
\nonumber && \hspace{-0.65cm} 
\bb{R}^{\ms{a} \ms{b}}_{21}(p,q) = (\bb{P} \, \bb{R}^{\ms{a}\ms{b}}(p,q) \, \bb{P}) \ot \bb{I}, \hspace{1.0cm}
\bb{K}^{\ms{a} \ms{b}}_{23}(p,\SB) = \bb{I} \otimes \bb{K}^{\ms{a}\ms{b}}(p,\SB)
\end{eqnarray}
and similarly for $\bb{R}_{12}(p,q)$, $\bb{R}_{21}(q,p)$, $\bb{K}_{13}(p,\SB)$ and $\bb{K}_{23}(q,\SB)$ in terms of complete $R$ and $K$-matrices.  
\vspace{-0.2cm}

A direct computation shows that the $R$-matrices and $K$-matrices for the vector boundary satisfy \eqref{RE-partial} for $(\ms{a} \ms{b} \ms{c})$ in the homogeneous ($\SL\SL\SL$, $\SR\SR\SR$) and mixed ($\SL\SL\SR$, $\SL\SR\SL$, $\SR\SL\SL$, $\SL\SR\SR$, $\SR\SL\SR$, $\SR\SR\SL$) sectors. This ensures that the complete $R$-matrix and $K$-matrix satisfy \eqref{RE-complete}.


\part{$\mf{psu}(1,1|2)^2$ spin-chains in $AdS_{3}\times S^{3} \times T^{4}$}  \label{part4}


\section{Integrable closed $\mf{psu}(1,1|2)^2$ spin-chain and scattering matrices}   \label{part4-sec1} 

The bosonic isometry group of the $AdS_3 \times S^3 \times T^4$ supergravity background is
\[
SO(2,2) \times SO(4) \times U(1)^{4},
\] 
with the Lie algebra splitting into left- and right-movers
\[
\mf{so}(2,2) \sim \mf{su}(1,1)_{\sL} \op \mf{su}(1,1)_{\sR}, \hspace{0.75cm} \mf{so}(4) \sim \mf{su}(2)_{\sL} \op \mf{su}(2)_{\sR}.
\]
The bosonic isometries can thus be rearranged into 
\[
[\mf{su}(1,1) \op \mf{su}(2)]_{\sL} \oplus [\mf{su}(1,1) \op \mf{su}(2)]_{\sR} \oplus \mf{u}(1)^{4},
\]
which is the bosonic part of the superisometry algebra 
\[
\mf{psu}(1,1|2)_{\sL} \oplus \mf{psu}(1,1|2)_{\sR} \oplus \mf{u}(1)^{4}.
\]

Massive excitations of the worldsheet of a closed superstring on $AdS_3 \times S^3 \times T^4$ can be identified with the magnon excitations of a homogeneous double-row $\mf{psu}(1,1|2)^{2}$ closed spin-chain. These magnons transform under a centrally extended $[\mf{psu}(1|1)^{2} \oplus \mf{u}(1)]^{2}_{\ms{c}}$ algebra\footnote{Note that here we use $\mf{psu}(1|1)^{2} \oplus \mf{u}(1)$ to denote the direct sum of Lie algebras as vector spaces.} (two copies of 
$\mf{su}(1|1)^{2}_{\ms{c}}$ with the Cartan and central elements identified, respectively). The left and right-moving excitations decouple in the weak coupling limit.
This chapter contains a review  based on \cite{BOSST13} of this integrable closed $\mf{psu}(1,1|2)^{2}$ spin-chain and the $S$-matrix describing the scattering of magnon excitations.

\subsection{$\mf{psu}(1,1|2)^2$ closed spin-chain with $[\mf{psu}(1|1)^{2} \oplus \mf{u}(1)]^{2}$ excitations} \label{part4-sec1-1}

\subsubsection{Single-row $\mf{psu}(1,1|2)$ spin-chain with $\mf{psu}(1|1)^{2} \oplus \mf{u}(1)$ excitations} \label{part4-sec1-1-1}

\paragraph{Symmetry generators.}  The $\mf{psu}(1,1|2)$ superalgebra shown in Appendix \ref{app:B} has bosonic generators
\[ \mf{J}_{0},\mf{J}_{b} \in \mf{su}(1,1), \hspace{0.85cm} \mf{L}_{\beta},\mf{L}_{5} \in \mf{su}(2) \]
of $\mf{su}(1,1) \op \mf{su}(2) $, and fermionic generators $\mf{Q}_{b\hspace{0.035cm}\beta\dot{\beta}}$ labeled by $b, \beta, \dot{\beta} = {\pm}$ indices.
There is also a $\mf{u}(1)$ automorphism $\mf{R}_{8}$.

\paragraph{Sites.}  A site in this homogeneous single-row $\mf{psu}(1,1|2)$ spin-chain is the module
\[M \equiv M(-\tfrac{1}{2},\tfrac{1}{2},0) =  \text{span}_{\C}\{ \ket{\phi^{(n)}_{\dot{\beta}}},  \ket{\psi^{(n)}_{\beta}} \}. \]
A vector at this site transforms in the half-BPS representation of the $\mf{psu}(1,1|2)$ superalgebra shown in Appendix \ref{app:B}.  The vacuum state is
\begin{equation} 
\ket{\mc{Z}} = \ket{\phi_{+}^{(0)}},
\end{equation}
and the four fundamental excitations $\varphi^{\beta\dot{\beta}}$ are
\begin{equation} 
\hspace{-0.25cm} \ket{\varphi^{+\dot{+}}} = \ket{\phi^{(0)}_{-}}, \hspace{0.75cm} \ket{\varphi^{-\dot{-}}} = \ket{\phi^{(1)}_{+}},  \hspace{0.75cm}
\ket{\varphi^{+\dot{-}}} = - \ket{\psi^{(0)}_{-}}, \hspace{0.75cm} \ket{\varphi^{-\dot{+}}} = \ket{\psi^{(0)}_{+}},  
\end{equation}
which transform under a $\mf{psu}(1|1)^{2} \oplus \mf{u}(1)$ algebra with fermionic $\mf{psu}(1|1)^{2}$ generators
\[ \mf{Q}_{1} \equiv \mf{Q}_{-++}, \hspace{0.75cm} \mf{Q}_{2} \equiv - \hspace{0.05cm} \mf{Q}_{-+-}, \hspace{0.75cm} 
\mf{S}_{1} \equiv \mf{Q}_{+--}, \hspace{0.75cm} \mf{S}_{2} \equiv \mf{Q}_{+-+} \]
satisfying $ \{\mf{Q}_{i}, \mf{S}_{j} \} = \mf{H} \hspace{0.1cm} \delta_{ij}$, for $i,j \in \{1,2\}$.
The additional bosonic $\mf{u}(1)$ generator
$\mf{H} = - \hspace{0.05cm} \mf{J}_{0} - \mf{L}_{5}$ is the magnon Hamiltonian. 
We notice that this $\mf{psu}(1|1)^{2} \oplus \mf{u}(1)$ algebra can, alternatively, be viewed as two copies of $\mf{su}(1|1)^{2}$ with the Cartan elements identified.  We can extend this algebra to $\mf{u}(1)^{2} \ltimes [\mf{psu}(1|1)^{2} \oplus \mf{u}(1)]$ by introducing 
$\mf{X}_{1} = - \hspace{0.05cm} \tfrac{1}{2} \hspace{0.05cm} \mf{L}_{5} - \tfrac{1}{2} \hspace{0.05cm} \mf{R}_{8}$ and 
$\mf{X}_{2} = - \hspace{0.05cm} \tfrac{1}{2} \hspace{0.05cm} \mf{L}_{5} + \tfrac{1}{2} \hspace{0.05cm} \mf{R}_{8}$,  which satisfy
\begin{equation} 
\{ \mf{Q}_{i}, \mf{S}_{j} \} = \mf{H} \hspace{0.1cm} \delta_{ij}, \hspace{0.85cm} 
[\mf{X}_{i},\mf{Q}_{j}] = - \hspace{0.05cm} \tfrac{1}{2} \hspace{0.05cm} \delta_{ij} \mf{Q}_{i}, \hspace{0.85cm} 
[\mf{X}_{i},\mf{S}_{j}] =  \tfrac{1}{2} \hspace{0.05cm} \delta_{ij} \mf{S}_{i},
\end{equation}
but do not annihilate the vacuum state. 

\paragraph{Spin-chain.}

The homogeneous single-row spin-chain with $J$ sites is now identified with the module $M^{\otimes J}$.  The spin-chain vacuum and fundamental excitations are
\begin{equation} 
\ket{0}  = \ket{\mc{Z}^{J}}, \hspace{1.0cm} \ket{\varphi^{\beta\dot{\beta}}_{(n)}}  = \ket{\mc{Z}^{n-1} \varphi^{\beta\dot{\beta}} \mc{Z}^{J-n}}. 
\end{equation}
Single magnon excitations are constructed as vectors in momentum space:
\begin{equation} 
\ket{\varphi^{\beta\dot{\beta}}_{p}} = \sum_{n=1}^{J} e^{ipn} \ket{\varphi^{\beta\dot{\beta}}_{(n)}}. 
\end{equation}
The action of the fermionic $\mf{psu}(1|1)^{2}$ generators  on
a magnon state is
\begin{eqnarray}
\nonumber && \hspace{-0.75cm} \mf{Q}_{1} \ket{\varphi^{+\dot{\beta}}_{p}} = \ket{\varphi^{-\dot{\beta}}_{p}}, \hspace{1.82cm} 
\mf{S}_{1} \ket{\varphi^{+ \dot{\beta}}_{p}} = 0, \hspace{0.5cm}
\mf{Q}_{1} \ket{\varphi^{- \dot{\beta}}_{p}} = 0, \hspace{0.5cm}
\mf{S}_{1} \ket{\varphi^{- \dot{\beta}}_{p}} = \ket{\varphi^{+ \dot{\beta}}_{p}}, \\
&&  \hspace{-0.75cm}
\mf{Q}_{2} \ket{\varphi^{\beta +}_{p}} =  (-1)^{\delta_{\beta -}} \ket{\varphi^{\beta -}_{p}}, \hspace{0.5cm} 
\mf{S}_{2} \ket{\varphi^{\beta +}_{p}} = 0, \hspace{0.5cm}
\mf{Q}_{2} \ket{\varphi^{\beta -}_{p}} = 0, \hspace{0.5cm}
\mf{S}_{2} \ket{\varphi^{\beta -}_{p}} =  (-1)^{\delta_{\beta -}} \ket{\varphi^{\beta +}_{p}}. \hspace{0.75cm}
\end{eqnarray}
These low-lying magnon excitations have energy 1 with respect to the magnon Hamiltonian $\mf{H}$.  Here also
\begin{eqnarray} 
\nonumber & \hspace{-0.75cm} 
\mf{X}_{i} \ket{0} = -\tfrac{J}{2} \ket{0},  \hspace{0.5cm} &
\mf{X}_{i} \ket{\varphi^{+\dot{+}}_{p}} = (-\tfrac{J}{2} + \tfrac{1}{2}) \hspace{0.05cm} \ket{\varphi^{+\dot{+}}_{p}},  \hspace{1.0cm}
\mf{X}_{i} \ket{\varphi^{-\dot{-}}_{p}} = -\tfrac{J}{2} \hspace{0.05cm} \ket{\varphi^{-\dot{-}}_{p}},  \\
\nonumber &&  
\mf{X}_{1} \ket{\varphi^{+\dot{-}}_{p}} = (-\tfrac{J}{2} + \tfrac{1}{2}) \ket{\varphi^{+\dot{-}}_{p}},  \hspace{1.0cm}
\mf{X}_{2} \ket{\varphi^{+\dot{-}}_{p}} = -\tfrac{J}{2} \ket{\varphi^{+\dot{-}}_{p}},  \\
\nonumber &&  
\mf{X}_{1} \ket{\varphi^{-\dot{+}}_{p}} = -\tfrac{J}{2} \ket{\varphi^{-\dot{+}}_{p}},  \hspace{2.0cm}
\mf{X}_{2} \ket{\varphi^{-\dot{+}}_{p}} = (-\tfrac{J}{2} + \tfrac{1}{2}) \ket{\varphi^{-\dot{+}}_{p}}. \hspace{0.5cm}
\end{eqnarray}
The standard procedure by which single-magnon excitations may be generalized to multi-magnon excitations was described in Section \ref{part3-sec1-1-1}.

\subsubsection{Double-row $\mf{psu}(1,1|2)^2$ closed spin-chain with $[\mf{psu}(1|1)^{2} \oplus \mf{u}(1)]^{2}$ excitations} \label{part4-sec1-1-2}

The homogeneous double-row $\mf{psu}(1,1|2)^2$ spin-chain consists of left and right-moving $\mf{psu}(1,1|2)_{\sL}$ and $\mf{psu}(1,1|2)_{\sR}$ spin-chains which decouple at weak coupling.

\paragraph{Sites.}  

Sites of the left and right-moving spin-chains form the module $M_{\sL} \otimes M_{\sR}$, with $M_{\sL}$ and $M_{\sR}$ the left and right copies of the module $M$.  The vacuum state and fundamental excitations are
\begin{equation} 
\ket{\mc{Z}} = \left|{\small\begin{pmatrix} \mc{Z}_{\sL} \\ \mc{Z}_{\sR} \end{pmatrix}} \right\rangle, \hspace{0.85cm}
\ket{\varphi^{\beta\dot{\beta}}} = \left|{\small\begin{pmatrix} \varphi^{\beta\dot{\beta}}_{\sL} \\ \mc{Z}_{\sR} \end{pmatrix}} \right\rangle, \hspace{0.85cm} 
\ket{\bar{\varphi}^{\beta\dot{\beta}}} = \left|{\small\begin{pmatrix} \mc{Z}_{\sL} \\ \varphi^{\beta\dot{\beta}}_{\sR} \end{pmatrix}} \right\rangle,
\end{equation}
which transform under the $[\mf{psu}(1|1)^{2} \oplus \mf{u}(1)]^{2}$ algebra
\[ \{ \mf{Q}_{\ms{a}i}, \mf{S}_{\ms{b}j} \} = \mf{H}_{\ms{a}} \, \delta_{\ms{ab}} \, \delta_{ij}, \hspace{0.85cm} 
[ \mf{X}_{\ms{a}i}, \mf{Q}_{\ms{b}j} ] = - \hspace{0.05cm} \tfrac{1}{2} \hspace{0.05cm} \mf{Q}_{\ms{a} i} \, \delta_{\ms{ab}} \, \delta_{ij}, \hspace{0.85cm}
[ \mf{X}_{\ms{a} i}, \mf{S}_{\ms{b} j} ] = \tfrac{1}{2} \hspace{0.05cm} \mf{S}_{\ms{a} i} \, \delta_{\ms{ab}} \, \delta_{ij}, \]
with $\ms{a},\ms{b} \in \{\SL, \SR\}$ and $i,j \in \{1,2\}$. Notice that $\mf{X}_{i} = \mf{X}_{\sL i} - \mf{X}_{\sR i}$ do annihilate the vacuum state, although $\mf{X}_{\sL i}$ and $\mf{X}_{\sR i}$ individually do not.  We define $\mf{H} = \mf{H}_{\sL} + \mf{H}_{\sR}$ and $\mf{M} = \mf{H}_{\sL} - \mf{H}_{\sR}$, with $\mf{H}$ the magnon Hamiltonian. Again, we will focus on well-separated magnon excitations in the $J \to \infty$ limit so that the left and right-moving excitations $\varphi_{\sL}^{\beta\dot{\beta}}$ and $\varphi_{\sR}^{\beta\dot{\beta}}$ do not coincide.

\paragraph{Spin-chain.}  

The homogeneous double-row spin-chain is identified with the module $(M_{\sL} \otimes M_{\sR})^{\otimes J}$.  The ground state is
\begin{equation}
\ket{0} = \ket{\mc{Z}^{J}} = \left| {\small\begin{pmatrix}\mc{Z}_{\sL} \\ \mc{Z}_{\sR} \end{pmatrix}^{\!J}} \right\rangle, \hspace{0.75cm}
\end{equation}
and left- and right-moving fundamental excitations are
\begin{eqnarray}
\nonumber && \ket{\varphi^{\beta\dot{\beta}}_{(n)}} = \ket{\mc{Z}^{n-1} \varphi^{\beta\dot{\beta}} \mc{Z}^{J-n}} = \left|  {\small\begin{pmatrix} \mc{Z}_{\sL} \\ \mc{Z}_{\sR} \end{pmatrix}^{\!n-1}\! \begin{pmatrix}\varphi_{\sL}^{\beta\dot{\beta}} \\ \mc{Z}_{\sR} \end{pmatrix}\! \, \begin{pmatrix}\mc{Z}_{\sL} \\ \mc{Z}_{\sR} \end{pmatrix}^{\!J-n}}\right\rangle, \\
&& \ket{\bar\varphi^{\beta\dot{\beta}}_{(n)}} = \ket{\mc{Z}^{n-1} \bar{\varphi}^{\beta\dot{\beta}} \mc{Z}^{J-n}} = \left|  {\small\begin{pmatrix} \mc{Z}_{\sL} \\ \mc{Z}_{\sR} \end{pmatrix}^{\!n-1}\! \begin{pmatrix}\mc{Z}_{\sL} \\ 
\varphi_{\sR}^{\beta\dot{\beta}}\end{pmatrix}\! \, \begin{pmatrix} \mc{Z}_{\sL} \\ \mc{Z}_{\sR} \end{pmatrix}^{\!J-n}}\right\rangle,
\end{eqnarray}
with low-lying left- and right-moving single magnon excitations given by
\begin{equation}
\ket{\varphi^{\beta\dot{\beta}}_p} = \sum_{n=1}^J e^{ipn}\, \ket{\varphi^{\beta\dot{\beta}}_{(n)}}, \hspace{1.0cm} 
\ket{\bar{\varphi}^{\beta\dot{\beta}}_p} = \sum_{n=1}^J e^{ipn}\, \ket{\bar{\varphi}^{\beta\dot{\beta}}_{(n)}}.
\end{equation}
The left- and right-moving magnon excitations $ \ket{\varphi^{\beta \dot{\beta}}_{p}}$ and $ \ket{\bar{\varphi}^{\beta\dot{\beta}}_{p}}$ have energy eigenvalue 1 of the magnon Hamiltonian $\mf{H}$, whereas the left/right-movers have mass eigenvalues $\pm 1$, with thus a particles/antiparticle interpretation.
The non-trivial action of the fermionic generators of the $[\mf{psu}(1|1)^{2} \oplus \mf{u}(1)]^{2}$ algebra on these magnon states is
\begin{eqnarray}
\nonumber && \hspace{-0.65cm} \mf{Q}_{\sL 1} \ket{\varphi^{+\dot{\beta}}_{p}} = \ket{\varphi^{-\dot{\beta}}_{p}}, \hspace{0.3cm} 
\mf{S}_{\sL 1} \ket{\varphi^{- \dot{\beta}}_{p}} = \ket{\varphi^{+ \dot{\beta}}_{p}}, \hspace{0.3cm}
\mf{Q}_{\sL 2} \ket{\varphi^{\beta +}_{p}} = (-1)^{\delta_{\beta -}} \ket{\varphi^{\beta -}_{p}}, \hspace{0.3cm} 
\mf{S}_{\sL 2} \ket{\varphi^{\beta -}_{p}} = (-1)^{\delta_{\beta -}} \ket{\varphi^{\beta +}_{p}}, \\
\nonumber &&  \hspace{-0.65cm}
\mf{Q}_{\sR 1} \ket{\bar{\varphi}^{+\dot{\beta}}_{p}} = \ket{\bar{\varphi}^{-\dot{\beta}}_{p}}, \hspace{0.275cm} 
\mf{S}_{\sR 1} \ket{\bar{\varphi}^{- \dot{\beta}}_{p}} = \ket{\bar{\varphi}^{+ \dot{\beta}}_{p}}, \hspace{0.275cm}
\mf{Q}_{\sR 2} \ket{\bar{\varphi}^{\beta +}_{p}} = (-1)^{\delta_{\beta -}} \ket{\bar{\varphi}^{\beta -}_{p}}, \hspace{0.275cm} 
\mf{S}_{\sR 2} \ket{\bar{\varphi}^{\beta -}_{p}} = (-1)^{\delta_{\beta -}} \ket{\bar{\varphi}^{\beta +}_{p}}, \\
\end{eqnarray}
with the non-trivial action of the $\mf{u}(1)$ generators $\mf{X}_{i}$ of the $\mf{u}(1)^{2} \ltimes [\mf{psu}(1|1)^{2} \oplus \mf{u}(1)]^{2}$ algebra, which annihilate the ground state, given by
\begin{eqnarray} 
\nonumber && \hspace{-0.65cm} 
\mf{X}_{i} \ket{\varphi^{+\dot{+}}_{p}} = \tfrac{1}{2} \hspace{0.05cm} \ket{\varphi^{+\dot{+}}_{p}},  \hspace{1.21cm}
\mf{X}_{1} \ket{\varphi^{+\dot{-}}_{p}} = \tfrac{1}{2} \hspace{0.05cm} \ket{\varphi^{+\dot{-}}_{p}},  \hspace{1.15cm}
\mf{X}_{2} \ket{\varphi^{-\dot{+}}_{p}} = \tfrac{1}{2} \hspace{0.05cm} \ket{\varphi^{-\dot{+}}_{p}},  \\
\nonumber && \hspace{-0.65cm} 
\mf{X}_{i} \ket{\bar{\varphi}^{+\dot{+}}_{p}} = - \hspace{0.05cm} \tfrac{1}{2} \hspace{0.05cm}  \ket{\bar{\varphi}^{+\dot{+}}_{p}}, \hspace{0.85cm}
\mf{X}_{1} \ket{\bar{\varphi}^{+\dot{-}}_{p}} = - \tfrac{1}{2} \hspace{0.05cm} \ket{\bar{\varphi}^{+\dot{-}}_{p}}, \hspace{0.85cm}
\mf{X}_{2} \ket{\bar{\varphi}^{-\dot{+}}_{p}} = - \tfrac{1}{2} \hspace{0.05cm} \ket{\bar{\varphi}^{-\dot{+}}_{p}}.
\end{eqnarray}
We again make use of the standard generalization to multi-magnon excitations.

\subsection{$\mf{psu}(1,1|2)^2$ spin-chain with centrally extended $[\mf{psu}(1|1)^{2} \oplus \mf{u}(1)]^{2}_{\ms{c}}$ excitations} \label{part4-sec1-2}

Beyond the weak-coupling limit of the $\mf{psu}(1,1|2)^2$ spin-chain in which the left- and right-moving excitations decouple, we must centrally extend the subalgebra of the massive magnon excitations to $[\mf{psu}(1|1)^{2} \oplus \mf{u}(1)]^{2}_{\ms{c}}$.  
This centrally extended algebra has fermionic generators $\mf{Q}_{\ms{a} i}$, $\mf{S}_{\ms{a} i}$ and bosonic generators $\mf{H}_{\ms{a}}$, $\mf{P}$, $\mf{P}^{\dag}$, with $\ms{a} \in \{\SL,\SR\}$ and $i \in \{1,2\}$.  These generators satisfy
\begin{equation} \label{extended-algebra-psu} 
\{\mf{Q}_{\ms{a} i }, \mf{S}_{\ms{b} j}\} = \mf{H}_{\ms{a}} \, \delta_{\ms{ab}} \, \delta_{ij}, \hspace{0.85cm}
\{ \mf{Q}_{\sL i}, \mf{Q}_{\sR j} \} = \mf{P} \, \delta_{ij}, \hspace{0.85cm}
\{ \mf{S}_{\sL i}, \mf{S}_{\sR j} \} = \mf{P}^{\dag} \, \delta_{ij}. 
\end{equation}
Here $\mf{P}$ and $\mf{P}^{\dag}$ are the new central elements.  The dynamic and non-dynamic $\mf{psu}(1,1|2)^{2}$ spin-chains were studied in detail in \cite{BOSST13}.  Let us briefly review these constructions.

\subsubsection{Finite spin-chain with length-changing effects}  \label{part4-sec1-2-1}

The bosonic central elements $\mf{P}$ and $\mf{P}^{\dag}$  have length-changing effects on the closed finite spin-chain above.  Here $\mc{Z}^{+}$ and $\mc{Z}^{-}$ insert or remove a vacuum state, as described in Section \ref{part3-sec1-2-1}.

\paragraph{Single-magnon excitations.}

The action of the fermionic generators of the $[\mf{psu}(1|1)^{2} \oplus \mf{u}(1)]^{2}_{\ms{c}}$ algebra on the left-moving magnon excitations is \cite{BOSST13}:
\begin{eqnarray} 
\nonumber && \hspace{-0.65cm} \mf{Q}_{\sL 1} \ket{\varphi^{+\dot{\beta}}_{p}} = a_{p} \ket{\varphi^{-\dot{\beta}}_{p}}, \hspace{2.84cm} 
\mf{S}_{\sL 1} \ket{\varphi^{- \dot{\beta}}_{p}} = b_{p} \ket{\varphi^{+ \dot{\beta}}_{p}}, \\
\nonumber && \hspace{-0.65cm} \mf{Q}_{\sL 2} \ket{\varphi^{\beta +}_{p}} = (-1)^{\delta_{\beta -}} a_{p} \ket{\varphi^{\beta -}_{p}}, \hspace{1.52cm} 
\mf{S}_{\sL 2} \ket{\varphi^{\beta -}_{p}} = (-1)^{\delta_{\beta -}} b_{p} \ket{\varphi^{\beta +}_{p}}, \\
\nonumber && \hspace{-0.65cm} \mf{Q}_{\sR 1} \ket{\varphi^{-\dot{\beta}}_{p}} = c_{p} \ket{\mc{Z}^{+} \varphi^{+\dot{\beta}}_{p}}, \hspace{2.3cm} 
\mf{S}_{\sR 1} \ket{\varphi^{+ \dot{\beta}}_{p}} = d_{p} \ket{\mc{Z}^{-} \varphi^{- \dot{\beta}}_{p}},  \\
&& \hspace{-0.65cm} \mf{Q}_{\sR 2} \ket{\varphi^{\beta -}_{p}} = (-1)^{\delta_{\beta -}} c_{p} \ket{\mc{Z}^{+} \varphi^{\beta +}_{p}}, \hspace{1.0cm}
\mf{S}_{\sR 2} \ket{\varphi^{\beta +}_{p}} = (-1)^{\delta_{\beta -}} d_{p} \ket{\mc{Z}^{-} \varphi^{\beta -}_{p}},  \label{action-one-left-magnon-T4}
\end{eqnarray}
and, similarly, on the right-moving magnon excitations
\begin{eqnarray} 
\nonumber && \hspace{-0.65cm} \mf{Q}_{\sR 1} \ket{\bar{\varphi}^{+\dot{\beta}}_{p}} = \bar{a}_{p} \ket{\bar{\varphi}^{-\dot{\beta}}_{p}}, \hspace{2.84cm} 
\mf{S}_{\sR 1} \ket{\bar{\varphi}^{- \dot{\beta}}_{p}} = \bar{b}_{p} \ket{\bar{\varphi}^{+ \dot{\beta}}_{p}}, \\
\nonumber && \hspace{-0.65cm} \mf{Q}_{\sR 2} \ket{\bar{\varphi}^{\beta +}_{p}} = (-1)^{\delta_{\beta -}} \bar{a}_{p} \ket{\bar{\varphi}^{\beta -}_{p}}, \hspace{1.52cm} 
\mf{S}_{\sR 2} \ket{\bar{\varphi}^{\beta -}_{p}} = (-1)^{\delta_{\beta -}} \bar{b}_{p} \ket{\bar{\varphi}^{\beta +}_{p}}, \\
\nonumber && \hspace{-0.65cm} \mf{Q}_{\sL 1} \ket{\bar{\varphi}^{-\dot{\beta}}_{p}} = \bar{c}_{p} \ket{\mc{Z}^{+} \bar{\varphi}^{+\dot{\beta}}_{p}}, \hspace{2.3cm} 
\mf{S}_{\sL 1} \ket{\bar{\varphi}^{+ \dot{\beta}}_{p}} = \bar{d}_{p} \ket{\mc{Z}^{-} \bar{\varphi}^{- \dot{\beta}}_{p}},   \\
 && \hspace{-0.65cm} \mf{Q}_{\sL 2} \ket{\bar{\varphi}^{\beta -}_{p}} = (-1)^{\delta_{\beta -}} \bar{c}_{p} \ket{\mc{Z}^{+} \bar{\varphi}^{\beta +}_{p}}, \hspace{1.0cm}
\mf{S}_{\sL 2} \ket{\bar{\varphi}^{\beta +}_{p}} = (-1)^{\delta_{\beta -}} \bar{d}_{p} \ket{\mc{Z}^{-} \bar{\varphi}^{\beta -}_{p}}.
\label{action-one-right-magnon-T4}
\end{eqnarray}
The energy eigenvalues of $\mf{H} = \mf{H}_{\sL} + \mf{H}_{\sR}$ for the left- and right-moving magnons are
\[ E_{p} = a_{p}b_{p} + c_{p}d_{p} \hspace{0.75cm} \text{and} \hspace{0.75cm} \bar{E}_{p} = \bar{a}_{p}\bar{b}_{p} + \bar{c}_{p}\bar{d}_{p},  \]
and the eigenvalues of the mass operator $\mf{M} = \mf{H}_{\sL} - \mf{H}_{\sR}$ are
\[ m = a_{p}b_{p} - c_{p}d_{p} = 1 \hspace{0.75cm} \text{and} \hspace{0.75cm} -\bar{m} = -(\bar{a}_{p}\bar{b}_{p} - \bar{c}_{p}\bar{d}_{p}) = -1. \]
Physical single magnon states should again be annihilated by the central elements $\mf{P}$ and $\mf{P}^{\dag}$, which would imply $c_{p}=d_{p}=\bar{c}_{p}=\bar{d}_{p}=0$ so we revert to a magnon state of the decoupled spin-chain.

\paragraph{Two-magnon excitations.}

We can write similar excitations for two left-moving magnons of momenta $p$ and $q$ to those in Section \ref{part3-sec1-2-1}, as well as for two right-moving magnons, and for left- and right-moving magnons
$\ket{\varphi^{\beta\dot{\beta}}_{p} \varphi^{\gamma\dot{\gamma}}_{q}}$,
$\ket{\bar{\varphi}^{\beta\dot{\beta}}_{p} \bar{\varphi}^{\gamma\dot{\gamma}}_{q}}$,
$\ket{\varphi^{\beta\dot{\beta}}_{p} \bar{\varphi}^{\gamma\dot{\gamma}}_{q}}$,
$\ket{\bar{\varphi}^{\beta\dot{\beta}}_{p} \varphi^{\gamma\dot{\gamma}}_{q}}$
in terms of these length-changing effects.  We find that, for these two-magnon states to be annihilated by the central elements $\mf{P}$ and $\mf{P}^{\dag}$, we must make use of the parameterization (\ref{parameters}) for both $a_{p}$, $b_{p}$, $c_{p}$, $d_{p}$ and $\bar{a}_{p}$, $\bar{b}_{p}$, $\bar{c}_{p}$, $\bar{d}_{p}$, satisfying the same constraints (\ref{Zhukovski}), with now unit mass $m=\bar{m}=1$.

\subsubsection{Infinite spin-chain with Hopf algebra structure} \label{part4-sec1-2-2}
Now, in the $J\to\infty$ infinite spin-chain limit, we can encode the length-changing effects rather in a $\mf{U}$-braided Hopf algebra structure for the $[\mf{psu}(1|1)^{2}\oplus \mf{u}(1)]^{2}_{\ms c}$ superalgebra, as in Appendix B of \cite{BOSST13}.

\paragraph{Single-magnon representations.}

We will write the actions \eqref{action-one-left-magnon-T4} and \eqref{action-one-right-magnon-T4} in terms of matrix representations of $[\mf{psu}(1|1)^2 \oplus \mf{u}(1)]^{2}_{\ms c}$.  We must first introduce the vector spaces
\[
\mc{V}_{p}={\rm span}_\C\{ \ket{\varphi^{+\dot{+}}_{p}}, \ket{\varphi^{+\dot{-}}_{p}}, \ket{\varphi^{-\dot{+}}_{p}}, \ket{\varphi^{-\dot{-}}_{p}} \},  \hspace{0.85cm} 
\mc{\bar{V}}_{p}={\rm span}_\C\{ \ket{\bar{\varphi}^{+\dot{+}}_{p}}, \ket{\bar{\varphi}^{+\dot{-}}_{p}}, \ket{\bar{\varphi}^{-\dot{+}}_{p}}, \ket{\bar{\varphi}^{-\dot{-}}_{p}} \},
\]
for left- and right-moving magnons. We can identify these vector spaces with $\C^{1|1}\otimes\dot{\C}^{1|1}$.  The action \eqref{action-one-left-magnon-T4} can be encoded in the left-moving representation $\pi_{p}: [\mf{psu}(1|1)^2\oplus \mf{u}(1)]^{2}_{\ms c} \to \End{(\C^{1|1}\otimes \dot{\C}^{1|1} )}$:
\begin{eqnarray} \label{Rep:L-T4}
\nonumber && \hspace{-0.75cm} \pi_p(\mf{Q}_{\sL 1}) = a_{p} \, \bb{E}_{21} \otimes \bb{I}, \hspace{0.5cm}
 \pi_p(\mf{Q}_{\sL 2}) = a_{p} \, \bb{I} \otimes \bb{E}_{21}, \hspace{0.5cm}
 \pi_p(\mf{Q}_{\sR 1}) = c_{p} \, \bb{E}_{12} \otimes \bb{I},  \hspace{0.5cm}
 \pi_p(\mf{Q}_{\sR 1}) = c_{p} \, \bb{I} \otimes \bb{E}_{12},   \\
\nonumber && \hspace{-0.75cm}  \pi_p(\mf{S}_{\sL 1}) = b_{p} \, \bb{E}_{12} \otimes \bb{I}, \hspace{0.53cm}
 \pi_p(\mf{S}_{\sL 2}) = b_{p} \, \bb{I} \otimes \bb{E}_{12}, \hspace{0.54cm}
 \pi_p(\mf{S}_{\sR 1}) = d_{p} \, \bb{E}_{21} \otimes \bb{I}, \hspace{0.47cm} 
 \pi_p(\mf{S}_{\sR 2}) = d_{p} \, \bb{I} \otimes \bb{E}_{21},  \\
&& \nonumber  \hspace{-0.75cm}  \pi_p(\mf{H}_\sL) = a_p b_p \, \bb{I} \otimes \bb{I}, \hspace{0.79cm}
 \pi_p(\mf{H}_\sR) = c_p d_p\, \bb{I} \otimes \bb{I},  \hspace{0.75cm}
 \pi_p(\mf{P}) = a_p c_p \, \bb{I} \otimes \bb{I},  \hspace{0.85cm}
 \pi_p(\mf{P}^\dag) = b_p d_p \, \bb{I} \otimes \bb{I}, \\
\end{eqnarray}
and, similarly, the action of the generators \eqref{action-one-right-magnon-T4} can be encoded in the right-moving magnon representation $\bar{\pi}_{p} : [\mf{psu}(1|1)^2\oplus \mf{u}(1)]^{2}_{\ms c} \to \End{(\C^{1|1}\otimes \C^{1|1})}$:
\begin{eqnarray} \label{Rep:R-T4}
\nonumber && \hspace{-0.75cm} \pi_p(\mf{Q}_{\sR 1}) = a_{p} \, \bb{E}_{\bar{2}\bar{1}} \otimes \bb{I}, \hspace{0.5cm}
 \pi_p(\mf{Q}_{\sR 2}) = a_{p} \, \bb{I} \otimes \bb{E}_{\bar{2}\bar{1}}, \hspace{0.5cm}
 \pi_p(\mf{Q}_{\sL 1}) = c_{p} \, \bb{E}_{\bar{1}\bar{2}} \otimes \bb{I},  \hspace{0.5cm}
 \pi_p(\mf{Q}_{\sL 1}) = c_{p} \, \bb{I} \otimes \bb{E}_{\bar{1}\bar{2}},   \\
\nonumber && \hspace{-0.75cm}  \pi_p(\mf{S}_{\sR 1}) = b_{p} \, \bb{E}_{\bar{1}\bar{2}} \otimes \bb{I}, \hspace{0.53cm}
 \pi_p(\mf{S}_{\sR 2}) = b_{p} \, \bb{I} \otimes \bb{E}_{\bar{1}\bar{2}}, \hspace{0.54cm}
 \pi_p(\mf{S}_{\sL 1}) = d_{p} \, \bb{E}_{\bar{2}\bar{1}} \otimes \bb{I}, \hspace{0.47cm} 
 \pi_p(\mf{S}_{\sL 2}) = d_{p} \, \bb{I} \otimes \bb{E}_{\bar{2}\bar{1}},  \\
&& \nonumber \hspace{-0.75cm}  \pi_p(\mf{H}_\sR) = a_p b_p \, \bb{I} \otimes \bb{I}, \hspace{0.79cm}
 \pi_p(\mf{H}_\sL) = c_p d_p\, \bb{I} \otimes \bb{I},  \hspace{0.8cm}
 \pi_p(\mf{P}) = a_p c_p \, \bb{I} \otimes \bb{I},  \hspace{0.9cm}
 \pi_p(\mf{P}^\dag) = b_p d_p \, \bb{I} \otimes \bb{I}, \\
\end{eqnarray}
both with parameters \eqref{parameters}.

\paragraph{Hopf algebra.} 

Again we introduce an additional generator $\mf{U}$, which is central with respect to the $[\mf{psu}(1|1)^2\oplus \mf{u}(1)]^{2}_{\ms c}$ superalgebra.  The action on any single-magnon excitation is
\begin{equation}
\mf{U} \, \ket{\varphi^{\beta\dot{\beta}}_{p}}= e^{i\frac{p}{2}} \hspace{0.025cm} \ket{ \varphi^{\beta\dot{\beta}}_{p} },  \hspace{1.0cm}
\mf{U} \, \ket{\bar{\varphi}^{\beta\dot{\beta}}_{p}}= e^{i\frac{p}{2}} \hspace{0.025cm} \ket{ \bar{\varphi}^{\beta\dot{\beta}}_{p} }, 
\end{equation}
and hence
\begin{equation}
\pi_p(\mf{U}) =  - \hspace{0.05cm} c_{p} d^{-1}_{p} \, \bb{I} \otimes \bb{I} = e^{i\frac{p}{2}} \, \bb{I} \otimes \bb{I}, \hspace{1.0cm}  
\bar\pi_p(\mf{U}) =  - \hspace{0.05cm} \bar{c}_{p} \bar{d}^{-1}_{p} \, \bb{I} \otimes \bb{I} = e^{i\frac{p}{2}} \, \bb{I} \otimes \bb{I},
\end{equation}
in the left- and right-moving single-magnon representations. As in Section \ref{part3-sec1-2-2}, we define a Hopf algebra structure on $[\mf{psu}(1|1)^2\oplus \mf{u}(1)]^{2}_{\ms c}$, denoting this Hopf superalgebra by $\mc{A}$ throughout Part \ref{part4}.  
\[ L(\mathcal{A})=[\mf{psu}(1|1)^2\oplus \mf{u}(1)]^{2}_{\ms c} \]
is the associated Lie superalgebra.

We again choose a symmetric frame in which the coproduct takes a form similar to that of \cite{BOSST13}:
\begin{eqnarray}
&& \nonumber \hspace{-0.65cm} \Delta(\mf{Q}_{\ms{a} i}) = \mf{Q}_{\ms{a} i} \ot \mathbf{1} + \mf{U} \ot \mf{Q}_{\ms{a} i},  \hspace{1.34cm} \Delta(\mf{P}) = \mf{P} \ot \mathbf{1} + \mf{U}^{2} \ot \mf{P},  \\
&& \nonumber \hspace{-0.65cm} \Delta(\mf{S}_{\ms{a} i}) = \mf{S}_{\ms{a} i} \ot \mathbf{1} + \mf{U}^{-1}\! \ot \mf{S}_{\ms{a} i}, \hspace{1.0cm} \Delta(\mf{P}^\dag) = \mf{P}^\dag \ot \mathbf{1} + \mf{U}^{-2} \ot \mf{P}^\dag, \\
&& \hspace{-0.65cm} \Delta(\mf{H}_{\ms a}) = \mf{H}_{\ms{a}} \ot \mathbf{1} + \mathbf{1} \ot \mf{H}_{\ms{a}}, \hspace{1.79cm} \Delta(\mf{U}^{\pm1}) = \mf{U}^{\pm1} \ot \mf{U}^{\pm1},
\end{eqnarray}
with opposite coproduct $\Delta^{\rm op}(\mf{a}) = \mc{P} \hspace{0.05cm} \Delta(\mf{a})$.
The central elements co-commute, $\Delta(\mf{C})=\Delta^{\rm{op}}(\mf{C})$ for 
$\mf{C}\in\{\mf{H}_{\ms a},\mf{P},\mf{P}^\dag\}$, which implies
$ \mf{P} = \nu_{1} \hspace{0.1cm} (1-\mf{U}^2)$ and $\mf{P}^\dag = \nu_{2} \hspace{0.1cm} (1-\mf{U}^{-2})$, again using $\nu_{1}=\nu_{2}=h$ to obtain our representations $\pi_{p}$ and $\bar{\pi}_{p}$.  All the other Hopf algebra structures of Section \ref{part3-sec1-2-2} generalize in the obvious manner. In particular,
the antipode $\mathscr{S}: \mc{A} \rightarrow \mc{A}$ is
\begin{eqnarray}
&& \nonumber \hspace{-0.65cm} \mathscr{S}(\mf{Q}_{\ms {a} i}) = - \hspace{0.025cm} \mf{U}^{-1} \, \mf{Q}_{\ms {a} i}, \hspace{0.85cm} 
\mathscr{S}(\mf{S}_{\ms{a} i}) = - \hspace{0.025cm} \mf{U} \, \mf{S}_{\ms {a} i}, \hspace{0.85cm}  
\mathscr{S}(\mf{H}_{\ms{a}}) = - \hspace{0.025cm} \mf{H}_{\ms{a}}, \\
&& \hspace{-0.65cm} \mathscr{S}(\mf{P}) = - \hspace{0.025cm} \mf{U}^{-2} \, \mf{P},  \hspace{1.33cm} 
\mathscr{S}(\mf{P}^\dag) = - \hspace{0.025cm} \mf{U}^{2} \, \mf{P}^\dag, \hspace{0.95cm} \mathscr{S}(\mf{U}) = \mf{U}^{-1},
\end{eqnarray}
which relates left- and right-movers in the representations $\pi_{p}$ and $\pi_{\bar{p}}$ through
\begin{equation} 
\pi_{p}(\mathscr{S}(\mf{a})) =  (\pi_{\bar{p}}(\mf{\bar{a}}))^{\rm str},
\end{equation}
with the charge conjugation matrix trivial. Here $\mf{a} \in \mc{A}$, with $\bar{\mf{a}} \in \mc{A}$ defined by
\begin{eqnarray}
\nonumber 
&& \hspace{-0.65cm}
\bar{\mf{Q}}_{\ms{a} i} = \delta_{\ms a\sL} \, \mf{Q}_{\sR i} \, + \, \delta_{\ms a\sR} \, \mf{Q}_{\sL i}, \hspace{0.85cm}
 \bar{\mf{P}} = \mf{P}, \hspace{1.14cm}
 \bar{\mf{H}}_{\ms{a}} \; = \delta_{\ms a\sL} \, \mf{H}_{\sR} \, + \, \delta_{\ms a\sR} \, \mf{H}_{\sL}, \\
&& \hspace{-0.65cm}
\bar{\mf{S}}_{\ms{a} i} = \delta_{\ms{a} \sL} \, \mf{S}_{\sR i} \, + \, \delta_{\ms{a} \sR} \, \mf{S}_{\sL i}, \hspace{0.85cm}
 \bar{\mf{P}}^{\dag} = \mf{P}^{\dag}, \hspace{0.85cm}
 \bar{\mf{U}}^{\pm 1} = \mf{U}^{\pm 1}.
\end{eqnarray}
%

\subsection{Two-magnon scattering and R-matrices} \label{part4-sec1-3} 

The two-magnon scattering matrix is
\begin{equation} 
S(p,q) = P \, R(p,q), \hspace{0.75cm} \text{with} \hspace{0.5cm}  R(p,q) \in \End(\mc{W}_{p} \otimes \mc{W}_{q}),
\end{equation} 
where now
\[\mc{W}_{p} = \mc{V}_{p} \oplus \mc{\bar{V}}_{p} = {\rm span}_\C \{\ket{\varphi^{+\dot{+}}_{p}}, \ket{\varphi^{+\dot{-}}_{p}}, \ket{\varphi^{-\dot{+}}_{p}}, \ket{\varphi^{-\dot{-}}_{p}}, \ket{\bar{\varphi}^{+\dot{+}}_{p}}, \ket{\bar{\varphi}^{+\dot{-}}_{p}}, \ket{\bar{\varphi}^{-\dot{+}}_{p}}, \ket{\bar{\varphi}^{-\dot{-}}_{p}}  \}.  \]
Again, $R(p,q)$ can be decomposed into a direct sum of four sectors  $R^{\ms{ab}}(p,q)$ corresponding to the partial $R$-matrices 
$\bb{R}^{\ms{ab}}(p,q) \in \End((\C^{1|1} \otimes \dot{\C}^{1|1})\otimes (\C^{1|1} \otimes \dot{\C}^{1|1}))$.
Here the complete $R$-matrix 
\begin{equation}
\bb{R}(p,q) = \bb{R}^{\sL \sL}(p,q) \oplus \bb{R}^{\sL \sR}(p,q) \oplus \bb{R}^{\sR \sL}(p,q) \oplus \bb{R}^{\sR \sR}(p,q)
\end{equation}
satisfies the intertwining equations (\ref{complete-intertwining}), with $\pi_{p}$ and $\bar{\pi}_{p}$ now the representations 
(\ref{Rep:L-T4}) and (\ref{Rep:R-T4}), and a similar unitarity condition.

\pagebreak

The intertwining equations of $\mf{Q}_{\ms{a} 1}$ and $\mf{S}_{\ms{a} 1}$ intertwine only the $\C^{1|1}$ space non-trivially, and similarly those of 
$\mf{Q}_{\ms{a} 2}$ and $\mf{S}_{\ms{a} 2}$ intertwine only $\dot{\C}^{1|1}$. The intertwining equations of $\mf{H}_{\ms{a}}$ may be thought of as acting non-trivially on either space and trivially on the other, and just ensure the decomposition of the $R$-matrix into block-diagonal form.  Thus two copies of the partial $R$-matrices of the $\mf{d}(2,1;\alpha)^{2}$ spin-chain given in Section \ref{part3-sec1-3-1},
\[
\bb{R}^{\ms{ab}}(p,q) 
= \sum_{i, \hspace{0.03cm} j, \hspace{0.03cm} k, \hspace{0.03cm} l} (R^{\hspace{0.025cm}\ms{ab}}(p,q))^{\hspace{0.05cm} i \hspace{0.05cm} k}_{\hspace{0.05cm} j \hspace{0.05cm} l} 
\hspace{0.2cm} \bb{E}_{ij}\ot \bb{E}_{kl}, \hspace{0.75cm}
\dot{\bb{R}}^{\ms{ab}}(p,q) 
= \sum_{\dot{i}, \hspace{0.03cm} \dot{j}, \hspace{0.03cm} \dot{k}, \hspace{0.03cm} \dot{l}} (R^{\hspace{0.025cm} \ms{ab}}(p,q))^{\hspace{0.05cm} \dot{i} \hspace{0.05cm} \dot{k}}_{\hspace{0.05cm} \dot{j} \hspace{0.05cm} \dot{l}} 
\hspace{0.2cm} \bb{E}_{\dot{i} \dot{j}}\ot \bb{E}_{\dot{k} \dot{l}},
\]
can be used to build an $R$-matrix $\bb{R}(p,q)$ for the $\mf{psu}(1,1|2)^{2}$ spin-chain \cite{BOSST13}. This decomposes into the partial $R$-matrices given by
\begin{equation} 
\bb{R}^{\ms{ab}}(p,q) \, = \sum_{^{i,\dot{i},j,\dot{j},}_{k ,\dot{k},\ell,\dot{\ell}}} 
(R^{\ms{ab}}(p,q) )^{i \hspace{0.03cm} \dot{i}, \, k \dot{k}}_{j \dot{j}, \, \ell \hspace{0.01cm} \dot{\ell}} 
\hspace{0.2cm} (\bb{E}_{ij}\ot \bb{E}_{\dot{i}\dot{j}}) \otimes  (\bb{E}_{k\ell}\ot \bb{E}_{\dot{k} \dot{\ell}}),  
\end{equation}
with\footnote{We make use of the isomorphism $\bb{I} \otimes \bb{P} \otimes \bb{I}$ between $\C^{1|1} \otimes \C^{1|1} \otimes \dot{\C}^{1|1} \otimes \dot{\C}^{1|1}$ and $\C^{1|1} \otimes \dot{\C}^{1|1}\otimes \C^{1|1} \otimes \dot{\C}^{1|1}$ which maps the graded tensor product
$R^{\ms{ab}}(p,q) \ot \dot{R}^{\ms{ab}}(p,q)$ of the $R$-matrices in Section \ref{part3-sec1-3-1} to the $R$-matrix $R^{\ms{ab}}(p,q)$ here.}
\begin{equation} 
(R^{\ms{ab}}(p,q) )^{\hspace{0.05cm} i \hspace{0.04cm} \dot{i}, \hspace{0.05cm} k  \dot{k}}_{j \dot{j}, \hspace{0.05cm} \ell \hspace{0.01cm} \dot{\ell}}  
\, = \, (-1)^{(k+\ell)(\dot{i}+\dot{j})} \hspace{0.15cm} (R^{\ms{ab}}(p,q) )^{i \hspace{0.03cm} k}_{j \hspace{0.01cm} \ell} \hspace{0.15cm} 
(R^{\ms{ab}}(p,q) )^{\dot{i} \hspace{0.04cm} \dot{k}}_{\dot{j} \hspace{0.01cm} \dot{\ell}}, 
\end{equation}
where $i,\dot{i},j,\dot{j} \in \{1,2\}$ and $\{\bar{1},\bar{2}\}$ for $\ms{a} = \SL$ and $\ms{a} = \SR$, respectively, and
$k,\dot{k},\ell,\dot{\ell} \in \{1,2\}$  and $\{\bar{1},\bar{2}\}$  for $\ms{b} = \SL$ and $\ms{b}=\SR$.  
Now the Zhukovski variables satisfy a mass shell constraint (\ref{Zhukovski}) with unit mass $m=1$.  
This $R$-matrix satisfies the Yang-Baxter equation (\ref{YBE-complete}) and a unitarity condition.


\section{Integrable open $\mf{psu}(1,1|2)^2$ spin-chain and reflection matrices}   \label{part4-sec2} 

Let us now consider the boundary scattering of magnon excitations of a $\mf{psu}(1,1|2)^2$ open spin-chain off an integrable boundary.  These correspond to massive excitations of an open superstring ending on D-branes in $AdS_{3}\times S^{3}\times T^{4}$, such as one of the maximal giant gravitons discussed in Chapter \ref{part2-sec2}.

\subsection{Open spin-chains and boundary scattering} \label{part4-sec2-1} 

\subsubsection{Double-row $\mf{psu}(1,1|2)^2$ open spin-chain with $[\mf{psu}(1|1)^{2}\oplus \mf{u}(1)]^{2}_{\ms{c}}$ excitations}

\paragraph{Semi-infinite open spin-chain.}  Again we consider a semi-infinite open spin-chain with $J \to \infty$ which has a boundary site on the right side.  The ground state is
\eq{
\ket{0} = \ket{\mc{Z}^J \mc{F}_\sB},
}
with $\mc{F}_\sB$ an infinitely heavy boundary field.  Fundamental excitations now take the form
\begin{equation}
\big|\varphi^{\beta\dot{\beta}}_{(n)}\big\rangle_{\sB} = \ket{\mc{Z}^{J-n} \varphi^{\beta\dot{\beta}} \mc{Z}^{n-1}\mc{F}_\sB}, \hspace{1.0cm}
\big|\bar\varphi^{\beta\dot{\beta}}_{(n)}\rangle_{\sB} = \ket{\mc{Z}^{n-1} \bar{\varphi}^{\beta\dot{\beta}} \mc{Z}^{J-n}\mc{F}_\sB} 
\end{equation}
in terms of the left- and right-moving excitations $\varphi^{\beta\dot{\beta}}$ and $\bar{\varphi}^{\beta\dot{\beta}}$ of Section \ref{part4-sec1-1-2}.  The low-lying left- and right single-magnon excitations of the double-row open spin-chain are thus
\eq{ \label{in-states-T4}
\big|\varphi^{\beta\dot{\beta}}_p\big\rangle_{\sB} = \sum_{n=1}^J \, e^{-ipn}\, \big|\varphi^{\beta\dot{\beta}}_{(n)}\big\rangle_{\sB}, \hspace{1.0cm} 
\big|\bar{\varphi}^{\beta\dot{\beta}}_p\big\rangle_{\sB} = \sum_{n=1}^J \, e^{-ipn}\, \big|\bar{\varphi}^{\beta\dot{\beta}}_{(n)}\big\rangle_{\sB}.
}
In the $J\to\infty$ limit, magnon states of the semi-infinite open spin-chain can again be identified with magnon states of the closed spin-chain, with an additional boundary state,   
\begin{equation}
\big|\varphi^{\beta\dot{\beta}}_p\big\rangle_{\sB} = \big|\varphi^{\beta\dot{\beta}}_p\big\rangle \otimes \ket{0}_{\sB}, \hspace{1.0cm}
\big|\bar{\varphi}^{\beta\dot{\beta}}_p\big\rangle_{\sB} = \big|\bar{\varphi}^{\beta\dot{\beta}}_p \big\rangle \otimes \ket{0}_{\sB}
\end{equation}
with $\ket{\varphi^{\beta\dot{\beta}}_p}$ and $\ket{\bar{\varphi}^{\beta\dot{\beta}}_p}$ bulk magnon excitations in $\mc{V}_p$ and $\bar{\mc{V}}_p$. 
The length changing effects of the dynamic spin-chain are encoded in the Hopf algebra $\mc{A}$ of Section \ref{part4-sec1-2-2}.
The boundary field $\mc{F}_{\sB}$ is represented by the boundary vacuum state $\ket{0}_{\sB}$.   
We can generalize to multi-magnon states. 

\paragraph{Boundary algebra.}  
 
The boundary algebra $\mc{B} \subset \mc{A}$ must be a coideal subalgebra of the bulk Hopf superalgebra $\mc{A}$.  The singlet boundary state $\ket{0}_{\sB}$ transforms in the trivial representation of $\mc{B}$ defined by the counit map $\epsilon$. A vector state $\ket{\Phi}_{\sB}$ at the boundary $\ket{0}_{\sB}$ is also possible.

\subsubsection{Boundary scattering and $K$-matrices}

\paragraph{Outgoing single-magnon representations $\pi_{-p}$ and $\bar{\pi}_{-p}$.}  

Incoming magnons are states in $\mc{V}_{p}$ and $\bar{\mc{V}}_{p}$, whereas outgoing magnons are states in the vector spaces $\mc{V}_{-p}$ and $\bar{\mc{V}}_{-p}$.  
The incoming single-magnon representations $\pi_{p}$ and $\bar{\pi}_{p}$ are shown in (\ref{Rep:L-T4}) and (\ref{Rep:R-T4}). The outgoing single-magnon representations  $\pi_{-p}:\mc{A} \to \End(\C^{1|1} \otimes \dot{\C}^{1|1})$ and $\bar{\pi}_{-p}:\mc{A} \to \End(\C^{1|1} \otimes \dot{\C}^{1|1})$ are obtained by replacing the parameters $a_{p}$, $b_{p}$, $c_{p}$, $d_{p}$ with the parameters $a_{-p}$, $b_{-p}$, $c_{-p}$, $d_{-p}$ given in (\ref{parameters-reflected}). 

\paragraph{Singlet boundary scattering.}  The boundary scattering matrix is
\begin{equation} 
S_{\rm boundary}(p) = \kappa \; K(p), \hspace{0.75cm} \text{with} \hspace{0.5cm}  K(p) \in\End (\mc{W}_p),
\end{equation}
with $\kappa$ the reflection map.  The scattering of magnons off a singlet boundary $|0\ran_\sB$ was described in Section \ref{part3-sec2-1-2}.  The vector space $\mc{W}_p = \mc{V}_{p} \oplus \mc{V}_{p}$ with $\mc{V}_{p}, \bar{\mc{V}}_{p} \cong \C^{1|1} \otimes \dot{\C}^{1|1}$ is discussed in Section \ref{part4-sec1-3}.  

\paragraph{Vector boundary scattering.} 

The boundary scattering matrix is now
\begin{equation}
S_{\rm boundary}(p,\SB) = \kappa \; K(p,\SB), \hspace{0.75cm} \text{with}\hspace{0.5cm} K(p,\SB) \in \End( \mc{W}_p \ot \mc{W}_\sB).
\end{equation}
Here $\mc{W}_{p}$ and $\mc{W}_{\sB}$ are both modules of  $\mc{B}_{(\sT,\sT)}=[\mf{psu}(1|1)^{2}\oplus \mf{u}(1)]^{2}_{\ms{c}}$.  The scattering of magnon excitations of an open spin-chain off a vector boundary $\ket{\Phi_\sB} \ot |0\ran_\sB$ was described in Section \ref{part3-sec2-1-2}. 

\smallskip

We will discuss singlet boundaries in Section \ref{part4-sec2-2}  and the vector boundary in Section \ref{part4-sec2-3}.  

\subsection{Singlet boundaries} \label{part4-sec2-2} 

\paragraph{Boundary algebras.}

There are now more possible coideal boundary subalgebras $\mc{B}$ of the bulk Hopf superalgebra $\mc{A}$, describing magnons scattering off singlet boundaries, than in Section \ref{sec:4.2}.
The associated Lie algebras $L(\mc{B})$ can be compared with the boundary algebras shown in Table \ref{table-D1D5}.
\vspace{-0.75cm}
\begin{itemize}
\item[$\ast$]{{ \bf left, right \& mixed half-supersymmetric boundary algebras} $\mc{B}_{(\sL,\sL)}$, $\mc{B}_{(\sR,\sR)}$, $\mc{B}_{(\sL,\sR)}$, $\mc{B}_{(\sR,\sL)}$, corresponding to D-branes preserving half the bulk supersymmetries, and $\mf{H}$ and $\mf{M}$, implying chiral boundary scattering.  The boundary Lie algebras associated with these coideal boundary subalgebras are $\mf{psu}(1|1)^{2}_{\sL} \oplus \mf{u}(1)_{\sL} \oplus \mf{u}(1)_{\sR}$, $ \mf{u}(1)_{\sL} \oplus \mf{psu}(1|1)^{2}_{\sR} \oplus \mf{u}(1)_{\sR}$ and 
$\mf{su}(1|1)_{\sL} \oplus \mf{su}(1|1)_{\sR}$.}
\vspace{-0.2cm}
\item[$\ast$]{{\bf non-supersymmetric chiral boundary algebra $\mc{B}_{(\sN\sC,\sN\sC)}$}, corresponding to D-branes which preserve none of the bulk supersymmetries, but do preserve $\mf{H}$ and $\mf{M}$. The associated boundary Lie algebra is $\mf{u}(1)_{\sL} \oplus \mf{u}(1)_{\sR}$. We will show that scattering off this boundary has a {\it hidden symmetry}, denoted $\mc{B}_{(\sD,\sD)}$, contained in the bulk Hopf superalgebra.}
\vspace{-0.2cm}
\item[$\ast$]{{\bf left \& right quarter-supersymmetric boundary algebras} $\mc{B}_{(\sL,\sN\sC)}$, $\mc{B}_{(\sN\sC,\sL)}$, $\mc{B}_{(\sR,\sN\sC)}$, $\mc{B}_{(\sN\sC,\sR)}$, corresponding to D-branes preserving a quarter of the bulk supersymmetries, and $\mf{H}$ and $\mf{M}$. 
The associated boundary Lie algebras are $\mf{su}(1|1)_{\sL} \oplus \mf{u}(1)_{\sR}$ and $\mf{u}(1)_{\sL} \oplus \mf{su}(1|1)_{\sR}$. We will show that these boundary scattering processes preserve {\it hidden symmetries}, denoted $\mc{B}_{(\sL,\sD)}$, $\mc{B}_{(\sD,\sL)}$, $\mc{B}_{(\sR,\sD)}$, $\mc{B}_{(\sD,\sR)}$, at the level of the Hopf superalgebra.}
\vspace{-0.2cm}
\item[$\ast$]{{ \bf non-supersymmetric achiral boundary algebra} $\mc{B}_{(\sN\sA,\sN\sA)}$, corresponding to D-branes  
preserving $\mf{H}$, but no bulk supersymmetries. The associated boundary Lie algebra is $\mf{u}(1)_{+}$.}
\end{itemize}
\vspace{-0.2cm}

\paragraph{Boundary intertwining equations.}  

The $K$-matrix $K(p) \in \End (\mc{W}_p)$ again has four sectors: $\bb{K}^{\ms{a}}(p)$ are chiral reflections and $\bb{A}^{\ms{a}}(p)$ are achiral reflections.
The partial $K$-matrices $\bb{K}^{\ms{a}}(p),\,  \bb{A}^{\ms{a}}(p) \in \End(\C^{1|1} \otimes \dot{\C}^{1|1})$. 
The complete $K$-matrix $\bb{K}(p)$ takes the form shown in (\ref{singlet-complete-K}).
The boundary intertwining equations (\ref{singlet-boundary-intertwining}) for all $\mf{b} \in \mc{B}$ simplify to
(\ref{singlet-boundary-intertwining-simplified}) for those $\mf{b} \in [\mf{psu}(1|1)^{2}\oplus \mf{u}(1)]^{2}_{\ms{c}}$.
Here $\pi_{p}$ and $\bar{\pi}_{p}$ are now the representations (\ref{Rep:L-T4}) and (\ref{Rep:R-T4}), and similarly for the reflected representations.

Integrable $K$-matrices satisfying the reflection equation (\ref{RE-complete}) can be built from two of the $K$-matrices $\bb{K}_{\mc{B}_{1}}(p)$ and $\dot{\bb{K}}_{\mc{B}_{2}}(p)$ given in Section \ref{part3-sec2-2-1} -- which have partial $K$-matrices
\begin{eqnarray} 
&& \nonumber \hspace{-0.65cm} \bb{K}^{\ms{a}}_{\mc{B}_{1}}(p) =  \sum_{i, \hspace{0.03cm} j} (K^{\ms{a}}_{\mc{B}_{1}}(p))^{\hspace{0.05cm} i}_{\hspace{0.03cm} j} \hspace{0.15cm} \bb{E}_{ij}, \hspace{0.85cm}
\bb{A}^{\ms{a}}_{\mc{B}_{1}}(p) =  \sum_{i, \hspace{0.03cm} j} (A^{\ms{a}}_{\mc{B}_{1}}(p))^{\hspace{0.05cm} i}_{\hspace{0.03cm} j} \hspace{0.15cm} \bb{E}_{ij}, \\
&& \hspace{-0.65cm} 
\dot{\bb{K}}^{\ms{a} }_{\mc{B}_{2}}(p) =  \sum_{\dot{i}, \hspace{0.03cm} \dot{j}} (K^{\ms{a}}_{\mc{B}_{2}}(p))^{\hspace{0.05cm} \dot{i} }_{\hspace{0.03cm} \dot{j}} \hspace{0.15cm} \bb{E}_{\dot{i}\dot{j}},
\hspace{0.85cm}
\dot{\bb{A}}^{\ms{a} }_{\mc{B}_{2}}(p) =  \sum_{\dot{i}, \hspace{0.03cm} \dot{j}} (A^{\ms{a}}_{\mc{B}_{2}}(p))^{\hspace{0.05cm} \dot{i} }_{\hspace{0.03cm} \dot{j}} \hspace{0.15cm} \bb{E}_{\dot{i}\dot{j}}. 
\end{eqnarray}
The result is a complete $K$-matrix solution of the boundary intertwining equations associated with a boundary coideal subalgebra, denoted $\mc{B}_{(1,2)}$.  The partial $K$-matrices are given by
\begin{eqnarray}
&& \nonumber \hspace{-0.65cm} 
\bb{K}^{\ms{a}}_{\mc{B}_{(1,2)}}(p) = \bb{K}^{\ms{a}}_{\mc{B}_{1}}(p) \otimes \dot{\bb{K}}^{\ms{a}}_{\mc{B}_{2}}(p) = \sum_{i, \hspace{0.03cm} j, \hspace{0.03cm} \dot{i}, \hspace{0.03cm} \dot{j}} (K^{\ms{a}}_{\mc{B}_{(1,2)}}(p))^{\hspace{0.05cm} i \hspace{0.04cm} \dot{i} }_{ \hspace{0.03cm} j \dot{j}} 
\hspace{0.2cm} \bb{E}_{ij} \ot \bb{E}_{\dot{i}\dot{j}},  \\
&& \hspace{-0.65cm} 
\bb{A}^{\ms{a}}_{\mc{B}_{(1,2)}}(p) = \bb{A}^{\ms{a}}_{\mc{B}_{1}}(p) \otimes \dot{\bb{A}}^{\ms{a}}_{\mc{B}_{2}}(p) = \sum_{i, \hspace{0.03cm} j, \hspace{0.03cm} \dot{i}, \hspace{0.03cm} \dot{j}} (A^{\ms{a}}_{\mc{B}_{(1,2)}}(p))^{\hspace{0.05cm} i \hspace{0.04cm} \dot{i} }_{ \hspace{0.03cm} j \dot{j}} 
\hspace{0.2cm} \bb{E}_{ij} \ot \bb{E}_{\dot{i}\dot{j}},
\end{eqnarray}
with 
\[ (K^{\ms{a}}_{\mc{B}_{(1,2)}}(p) )^{\hspace{0.05cm} i \hspace{0.04cm} \dot{i}}_{ \hspace{0.03cm} j \dot{j}}  
\, = \,  (K^{\ms{a}}_{\mc{B}_{1}}(p))^{\hspace{0.05cm} i }_{ \hspace{0.03cm}j} \hspace{0.15cm} 
(K^{\ms{a}}_{\mc{B}_{2}}(p))^{ \hspace{0.05cm} \dot{i} }_{ \hspace{0.03cm} \dot{j}}, \hspace{0.85cm}
(A^{\ms{a}}_{\mc{B}_{(1,2)}}(p) )^{\hspace{0.05cm} i \hspace{0.04cm} \dot{i}}_{ \hspace{0.03cm} j \dot{j}}  
\, = \,  (A^{\ms{a}}_{\mc{B}_{1}}(p))^{\hspace{0.05cm} i }_{ \hspace{0.03cm}j} \hspace{0.15cm} 
(A^{\ms{a}}_{\mc{B}_{2}}(p))^{ \hspace{0.05cm} \dot{i} }_{ \hspace{0.03cm} \dot{j}}.  \]
We notice that $\bb{K}_{\mc{B}_{1}}$(p) and $\bb{K}_{\mc{B}_{2}}(p)$ must both be chiral or must both be achiral for a non-zero reflection matrix $\bb{K}_{\mc{B}_{(1,2)}}(p)$ built in this way. 

\paragraph{Constraints from the central elements.}

Again, boundary subalgebras associated with singlet boundaries may not contain the central elements $\mf{P}$ and $\mf{P}^{\dag}$.  The inclusion of $\mf{H}_{\sL}$ or $\mf{H}_{\sR}$ in the boundary algebra $\mc{B}$ once more implies a chiral $K$-matrix, although achiral $K$-matrices are allowed if only $\mf{H}=\mf{H}_{\sL}+\mf{H}_{\sR}$ is contained in $\mc{B}$. 

\subsubsection{Boundary subalgebras and $K$-matrices}

\paragraph{Left, right and mixed half-supersymmetric boundary algebras.}

The left, right and mixed half-supersymmetric boundary superalgebras, $\mc{B}_{(\sL,\sL)}$, $\mc{B}_{(\sR,\sR)}$, $\mc{B}_{(\sL,\sR)}$ and $\mc{B}_{(\sR,\sL)}$, are defined to be coideal subalgebras of $\mc{A}$ generated as follows:
\begin{eqnarray}
&& \nonumber \hspace{-0.75cm} \mc{B}_{(\sL,\sL)} = \big\lan \mf{Q}_{\sL 1}, \, \mf{S}_{\sL 1}, \, \mf{Q}_{\sL 2}, \, \mf{S}_{\sL 2}, \, \mf{H}_{\ms{\sL }}, \, \mf{H}_{\ms{\sR}} \big\ran, \hspace{0.9cm}
\mc{B}_{(\sR,\sR)} = \big\lan \mf{Q}_{\sR 1}, \, \mf{S}_{\sR 1}, \, \mf{Q}_{\sR 2}, \, \mf{S}_{\sR 2}, \, \mf{H}_{\ms{\sL }}, \, \mf{H}_{\ms{\sR}} \big\ran, \\
&& \hspace{-0.75cm} \mc{B}_{(\sL,\sR)} = \big\lan \mf{Q}_{\sL 1}, \, \mf{S}_{\sL 1}, \, \mf{Q}_{\sR 2}, \, \mf{S}_{\sR 2}, \, \mf{H}_{\ms{\sL }}, \, \mf{H}_{\ms{\sR}} \big\ran, 
\hspace{0.85cm}
\mc{B}_{(\sR,\sL)} = \big\lan \mf{Q}_{\sR 1}, \, \mf{S}_{\sR 1}, \, \mf{Q}_{\sL 2}, \, \mf{S}_{\sL 2}, \, \mf{H}_{\ms{\sL }}, \, \mf{H}_{\ms{\sR}} \big\ran . 
\end{eqnarray}
That $\mc{B}_{(\ms{a},\ms{b})}$ is an integrable boundary algebra follows from the fact that $(\mf{g},\mf{h}_{(\ms{a},\ms{b})})$ forms a symmetric pair of Lie algebras, if we define
\begin{eqnarray} 
&& \nonumber \hspace{-0.65cm} \mf{g} \equiv L(\mc{A})=[\mf{psu}(1|1)^{2}\oplus \mf{u}(1)]^{2}_{\ms{c}}, \\
&& \nonumber \hspace{-0.65cm} \mf{h}_{(\sL,\sL)} \equiv L(\mc{B}_{(\sL,\sL)})=[\mf{psu}(1|1)^{2}\oplus \mf{u}(1)]_{\sL} \op \mf{u}(1)_\sR, \hspace{0.5cm}
\mf{h}_{(\sR,\sR)} \equiv L(\mc{B}_{(\sR,\sR)})=\mf{u}(1)_{\sL} \op [\mf{psu}(1|1)^{2}\oplus \mf{u}(1)]_{\sR}, \hspace{0.35cm} \\
&& \hspace{-0.65cm} \mf{h}_{(\sL,\sR)} \equiv  L(\mc{B}_{(\sL,\sR)})=\mf{su}(1|1)_{\sL} \op \mf{su}(1|1)_{\sR}, \hspace{1.73cm}
\mf{h}_{(\sR,\sL)} \hspace{0.05cm}\equiv  L(\mc{B}_{(\sR,\sL)}) =\mf{su}(1|1)_{\sL} \op \mf{su}(1|1)_{\sR}.
\end{eqnarray}
Here $\mf{g} = \mf{h}_{(\ms{a},\ms{b})} \oplus \mf{m}_{(\ms{a},\ms{b})}$, with $\mf{m}_{(\sL,\sL)}$, $\mf{m}_{(\sR,\sR)}$, $\mf{m}_{(\sL,\sR)}$ and $\mf{m}_{(\sR,\sL)}$ generated by 
$\{ \mf{Q}_{\sR 1}, \mf{S}_{\sR 1}, \mf{Q}_{\sR 2}, \mf{S}_{\sR 2}, \mf{P}, \mf{P}^{\dag}\}$,
$\{ \mf{Q}_{\sL 1}, \mf{S}_{\sL 1}, \mf{Q}_{\sL 2}, \mf{S}_{\sL 2}, \mf{P}, \mf{P}^{\dag}\}$,
$\{ \mf{Q}_{\sR 1}, \mf{S}_{\sR 1}, \mf{Q}_{\sL 2}, \mf{S}_{\sL 2}, \mf{P}, \mf{P}^{\dag}\}$ and
$\{ \mf{Q}_{\sL 1}, \mf{S}_{\sL 1}, \mf{Q}_{\sR 2}, \mf{S}_{\sR 2}, \mf{P}, \mf{P}^{\dag}\}$.
The boundary Lie superalgebras $\mf{h}_{(\sL,\sL)}$ and $\mf{h}_{(\sR,\sR)}$ are associated with open superstrings on $AdS_{3} \times S^{3} \times T^{4}$ ending on $Y=0$ and $\bar{Y}=0$ D1- or D5-brane maximal giant gravitons, respectively.

The $K$-matrix solutions of the relevant boundary intertwining equations are given by
\begin{equation} 
\bb{K}_{\mc{B}_{(\ms{a},\ms{b})}}(p) = \bb{K}_{\mc{B}_{(\ms{a},\ms{b})}}^{\sL}(p) \oplus \bb{K}_{\mc{B}_{(\ms{a},\ms{b})}}^{\sR}(p), \hspace{0.5cm} \text{with} \hspace{0.35cm} 
\bb{K}^{\ms{c}}_{\mc{B}_{(\ms{a},\ms{b})}}(p) = \bb{K}^{\ms{c}}_{\mc{B}_{\ms{a}}}(p) \otimes \dot{\bb{K}}^{\ms{c}}_{\mc{B}_{\ms{b}}}(p), 
\end{equation}
for $\ms{a},\ms{b}, \ms{c} \in \{\SL,\SR\}$, in terms of solutions in Section \ref{part3-sec2-2-1}. This satisfies the reflection equation (\ref{RE-partial}).

\paragraph{Non-supersymmetric chiral boundary algebra.}

A boundary coideal subalgebra which contains no supercharges, but leads to boundary scattering processes which preserve chirality, is
\begin{equation}
\mc{B}_{(\sN\sC,\sN\sC)} = \big\lan \mf{H}_\sL , \mf{H}_\sR \big\ran.
\end{equation}
The boundary Lie algebra $L(\mc{B}_{(\sN\sC,\sN\sC)})=\mf{u}(1)_{\sL}\oplus\mf{u}(1)_{\sR}$ is that of an open superstring on $AdS_{3} \times S^{3} \times T^{4}$ ending on the $\bar{Z}=0$ D1- or D5-brane maximal giant graviton. 
A $K$-matrix solution of the boundary intertwining equation for $\mc{B}_{(\sN\sC,\sN\sC)}$ and the reflection equation is  
\begin{equation} \label{K-NC-NC}
\bb{K}_{\mc{B}_{(\sN\sC,\sN\sC)}}(p) = \bb{K}_{\mc{B}_{(\sN\sC,\sN\sC)}}^{\sL}(p) \oplus \bb{K}_{\mc{B}_{(\sN\sC,\sN\sC)}}^{\sR}(p), \hspace{0.65cm} \text{with} \hspace{0.5cm} 
\bb{K}^{\ms{a}}_{\mc{B}_{(\sN\sC,\sN\sC)}}(p) = \bb{K}^{\ms{a}}_{\mc{B}_{\sN\sC}}(p) \otimes  \dot{\bb{K}}^{\ms{a}}_{\mc{B}_{\sN\sC}}(p).
\end{equation} 
Let us now show that there is a {\it hidden symmetry}. A diagonally supersymmetric boundary algebra is defined by
\begin{equation}
\mc{B}_{(\sD,\sD)} = \big\lan \, \mf{q}_{+1}, \, \mf{q}_{- 1},  \, \mf{q}_{+2}, \, \mf{q}_{- 2}, \, \mf{d} , \, \tilde{\mf{d}} \, \big\ran,  \label{BD}
\end{equation}
which is a coideal subalgebra of $\mc{A}$.  The $K$-matrix (\ref{K-NC-NC}) intertwines representations of these hidden boundary symmetries $\mc{B}_{(\sD,\sD)}$.  Here
\begin{eqnarray}
\nonumber && \hspace{-0.75cm} \mf{q}_{+ j} = \mf{P}^{\dag} \mf{Q}_{\sL j} + i c \, \mf{P} \, \mf{S}_{\sR j}, \hspace{1.0cm} 
\mf{d} = \big( \mf{H}_{\sL} - c^2 \mf{H}_\sR + i c \, (\mf{P}+\mf{P}^\dag) \big) \, \mf{K}, \\
&& \hspace{-0.75cm} \mf{q}_{- j} = \mf{P} \, \mf{S}_{\sL j} + i c \, \mf{P}^{\dag} \mf{Q}_{\sR j}, \hspace{1.0cm} 
\tilde{\mf{d}} = \big( \mf{H}_{\sL} - c^2 \mf{H}_\sR - i c \, (\mf{P}+\mf{P}^\dag) \big) \, \mf{K}, 
\end{eqnarray}
with $\mf{K}=\mf{P}\,\mf{P}^\dag$. The space $\mc{M}_{(\sD,\sD)}$ is generated by 
$\{  \mf{s}_{+ 1}, \, \mf{s}_{- 1}, \, \mf{s}_{+ 2}, \, \mf{s}_{- 2}, \, \mf{n}, \, \tilde{\mf{n}} \}$, where we define
\begin{eqnarray}
\nonumber && \hspace{-0.75cm} \mf{s}_{+ j} = \mf{P}^{\dag} \mf{Q}_{\sL j} - i c \, \mf{P}\,\mf{S}_{\sR j}, \hspace{1.0cm} 
\mf{n} = \big(\mf{H}_{\sL} + c^2 \mf{H}_{\sR} + i c(\mf{P}-\mf{P}^\dag)\big)\,\mf{K}, \\
&& \hspace{-0.75cm} \mf{s}_{- j} = \mf{P} \, \mf{S}_{\sL j} - i c \, \mf{P}^{\dag} \mf{Q}_{\sR j}, \hspace{1.0cm} 
\tilde{\mf{n}} = \big(\mf{H}_{\sL} + c^2 \mf{H}_{\sR} - i c(\mf{P}-\mf{P}^\dag)\big)\,\mf{K}.
\end{eqnarray} 
The generators of $\mc{B}_{(\sD,\sD)}$ and $\mc{M}_{(\sD,\sD)}$ satisfy 
\spl{
\{ \mf{q}_{+i}, \mf{q}_{-j} \} &= \mf{d} \, \delta_{ij},  \hspace{0.85cm}
\{ \mf{q}_{+i}, \mf{s}_{-j} \} = \mf{n} \, \delta_{ij},  \hspace{0.85cm}
\{ \mf{q}_{+i}, \mf{s}_{+j} \} = 0, \\
\{ \mf{s}_{+i}, \mf{s}_{-j} \} &= \tilde{\mf{d}} \, \delta_{ij},  \hspace{0.85cm}
\{ \mf{q}_{-i}, \mf{s}_{+j} \} = \tilde{\mf{n}} \, \delta_{ij},  \hspace{0.85cm}
\{ \mf{q}_{-i}, \mf{s}_{-j} \} = 0,
}
with $\mf{d}$, $\tilde{\mf{d}}$ and $\mf{n}$, $\tilde{\mf{n}}$ central elements. These relations are identical to those in \eqref{extended-algebra-psu}. The associated Lie superalgebra 
$L(\mc{B}_{(\sD,\sD)}\oplus \mc{M}_{(\sD,\sD)})$ is isomorphic to $L(\mc{A})$, while here
$L(\mc{B}_{(\sD,\sD)})$ is generated by 
$\{\mf{q}_{+1},\,\mf{q}_{-1},\,\mf{q}_{+2},\,\mf{q}_{-2},\,\mf{d}, \, \tilde{\mf{d}} \}$. 
Now $(\mf{g},\mf{h}_{(\sD,\sD)})$ defines a symmetric pair, with
\begin{eqnarray} 
&& \hspace{-0.65cm} \nonumber \mf{g} \equiv L(\mc{B}_{(\sD,\sD)}\oplus \mc{M}_{(\sD,\sD)}) \cong L(\mc{A})=[\mf{psu}(1|1)^{2}\oplus \mf{u}(1)]^{2}_{\ms{c}}, \\ 
&& \hspace{-0.65cm} \mf{h}_{(\sD,\sD)} \equiv L(\mc{B}_{(\sD,\sD)})=[\mf{psu}(1|1)^{2}\oplus \mf{u}(1)]_{\sD}\op \mf{u}(1)_{\sD}. 
\end{eqnarray}

\paragraph{Left and right quarter-supersymmetric boundary algebras.}

The following left and right quarter-supersymmetric boundary superalgebras can also be defined
\begin{eqnarray}
&& \nonumber \hspace{-0.65cm} \mc{B}_{(\sL,\sN\sC)} = \big\lan \mf{Q}_{\sL 1}, \, \mf{S}_{\sL 1}, \, \mf{H}_{\ms{\sL }}, \, \mf{H}_{\ms{\sR}} \big\ran,  \hspace{0.925cm}
\mc{B}_{(\sN\sC,\sL)} = \big\lan \mf{Q}_{\sL 2}, \, \mf{S}_{\sL 2}, \, \mf{H}_{\ms{\sL }}, \, \mf{H}_{\ms{\sR}} \big\ran, \\
&& \hspace{-0.65cm} \mc{B}_{(\sR, \sN\sC)} = \big\lan \mf{Q}_{\sR 1}, \, \mf{S}_{\sR 1}, \, \mf{H}_{\ms{\sL }}, \, \mf{H}_{\ms{\sR}} \big\ran,  \hspace{0.85cm}
\mc{B}_{(\sN\sC,\sR)} = \big\lan \mf{Q}_{\sR 2}, \, \mf{S}_{\sR 2}, \, \mf{H}_{\ms{\sL }}, \, \mf{H}_{\ms{\sR}} \big\ran.
\end{eqnarray}
Here the associated boundary Lie superalgebras are $L(\mc{B}_{(\sL,\sN\sC)})=L(\mc{B}_{(\sN\sC,\sL)})=\mf{su}(1|1)_{\sL} \oplus \mf{u}(1)_{\sR}$ and $L(\mc{B}_{(\sR,\sN\sC)})=L(\mc{B}_{(\sN\sC,\sR)})=\mf{u}(1)_{\sL} \oplus \mf{su}(1|1)_{\sR}$.
The $K$-matrices which satisfy the boundary intertwining equations of $\mc{B}_{(\ms{a},\sN\sC)}$ and $\mc{B}_{(\sN\sC, \ms{a})}$, respectively, and the reflection equation are
\begin{eqnarray} 
&& \nonumber \hspace{-0.65cm} 
\bb{K}_{\mc{B}_{(\ms{a},\sN\sC)}}(p) = \bb{K}_{\mc{B}_{(\ms{a},\sN\sC)}}^{\sL}(p) \oplus \bb{K}_{\mc{B}_{(\ms{a},\sN\sC)}}^{\sR}(p), \hspace{0.65cm} \text{with} \hspace{0.5cm} 
\bb{K}^{\ms{b}}_{\mc{B}_{(\ms{a},\sN\sC)}}(p) = \bb{K}^{\ms{b}}_{\mc{B}_{\ms{a}}}(p) \otimes \dot{\bb{K}}^{\ms{b}}_{\mc{B}_{\sN\sC}}(p), \\
&& \hspace{-0.65cm} 
\bb{K}_{\mc{B}_{(\sN\sC,\ms{a})}}(p) = \bb{K}_{\mc{B}_{(\sN\sC,\ms{a})}}^{\sL}(p) \oplus \bb{K}_{\mc{B}_{(\sN\sC,\ms{a})}}^{\sR}(p), \hspace{0.65cm} \text{with} \hspace{0.5cm} 
\bb{K}^{\ms{b}}_{\mc{B}_{(\sN\sC,\ms{a})}}(p) = \bb{K}^{\ms{b}}_{\mc{B}_{\sN\sC}}(p) \otimes \dot{\bb{K}}^{\ms{b}}_{\mc{B}_{\ms{a}}}(p), \hspace{0.5cm}
\end{eqnarray} 
for $\ms{a},\ms{b} \in \{\SL,\SR\}$.

There are again {\it hidden symmetries} in the Hopf superalgebra.  We define coideal subalgebras of $\mc{A}$ which take the form 
\begin{eqnarray}
&& \nonumber \hspace{-0.65cm} \mc{B}_{(\sL,\sD)} = \big\lan \mf{Q}_{\sL 1}, \mf{S}_{\sL 1}, \, \mf{q}_{+2}, \, \mf{q}_{-2}, \, \mf{d}, \, \tilde{\mf{d}}, \, \mf{H}_{\sL}, \, \mf{H}_{\sR}  \big\ran,
\hspace{0.5cm}
\mc{B}_{(\sD,\sL)} = \big\lan \mf{q}_{+1}, \, \mf{q}_{-1}, \, \mf{Q}_{\sL 2}, \mf{S}_{\sL 2}, \, \mf{d}, \, \tilde{\mf{d}}, \, \mf{H}_{\sL}, \, \mf{H}_{\sR}  \big\ran,  \\
&& \hspace{-0.65cm} \mc{B}_{(\sR,\sD)} = \big\lan \mf{Q}_{\sR 1}, \mf{S}_{\sR 1}, \mf{q}_{+2}, \, \mf{q}_{-2}, \, \mf{d}, \, \tilde{\mf{d}}, \, \mf{H}_{\sL}, \, \mf{H}_{\sR}  \big\ran,
\hspace{0.5cm}
\mc{B}_{(\sD,\sR)} = \big\lan \mf{q}_{+1}, \, \mf{q}_{-1}, \, \mf{Q}_{\sR 2}, \mf{S}_{\sR 2}, \mf{d}, \, \tilde{\mf{d}}, \, \mf{H}_{\sL}, \, \mf{H}_{\sR}  \big\ran. \hspace{1.0cm}   
\end{eqnarray}
We  define $\mc{M}_{(\sL,\sD)}$, $\mc{M}_{(\sD,\sL)}$, $\mc{M}_{(\sR,\sD)}$ and $\mc{M}_{(\sD,\sR)}$ to be spaces generated by 
$\{ \mf{Q}_{\sR 1}, \, \mf{S}_{\sR 1}, \, \mf{s}_{+ 2}, \, \mf{s}_{- 2}, \, \mf{n}, \, \tilde{\mf{n}} \}$,
$\{ \mf{s}_{+ 1}, \, \mf{s}_{- 1}, \, \mf{Q}_{\sR 2}, \, \mf{S}_{\sR 2}, \,  \mf{n}, \, \tilde{\mf{n}} \}$,
$\{ \mf{Q}_{\sL 1}, \, \mf{S}_{\sL 1}, \, \mf{s}_{+ 2}, \, \mf{s}_{- 2}, \, \mf{n}, \, \tilde{\mf{n}} \}$ and 
$\{ \mf{s}_{+ 1}, \, \mf{s}_{- 1}, \, \mf{Q}_{\sL 2}, \, \mf{S}_{\sL 2}, \,  \mf{n}, \, \tilde{\mf{n}} \}$.
Here we can construct symmetric pairs $(\mf{g},\mf{h}_{(\ms{a},\sN\sC)})$  and  $(\mf{g},\mf{h}_{(\sN\sC,\ms{a})})$ of Lie algebras:
\begin{eqnarray} 
&& \hspace{-0.75cm} \nonumber \mf{g} \equiv L(\mc{B}_{(\ms{a},\sD)} \oplus \mc{M}_{(\ms{a},\sD)}) =  L(\mc{B}_{(\sD,\ms{a})} \oplus \mc{M}_{(\sD,\ms{a})}) \cong L(\mc{A}) 
=[\mf{psu}(1|1)^{2}\oplus \mf{u}(1)]^{2}_{\ms{c}}, \\
&& \hspace{-0.75cm} \mf{h}_{(\ms{a},\sD)} \equiv L(\mc{B}_{(\ms{a},\sD)})=\mf{su}(1|1)_{\ms{a}} \op \mf{su}(1|1)_{\sD}, \hspace{0.75cm} \mf{h}_{(\sD,\ms{a})} \equiv L(\mc{B}_{(\sD,\ms{a})})=\mf{su}(1|1)_{\sD} \op \mf{su}(1|1)_{\ms{a}}. \hspace{0.5cm}
\end{eqnarray}

\paragraph{Non-supersymmetric achiral boundary algebra.}

A boundary algebra which contains no supercharges and breaks chiral symmetry is
\begin{equation}
\mc{B}_{(\sN\sA,\sN\sA)} = \big\lan \mf{H} \big\ran.
\end{equation}
The boundary Lie algebra is $L(\mc{B}_{(\sN\sA,\sN\sA)})=\mf{u}(1)_{+}$. A $K$-matrix solution of the boundary intertwining equation of $\mc{B}_{(\sN\sA,\sN\sA)}$ and reflection equation is  
\begin{equation}
\bb{K}_{(\sN\sA,\sN\sA)}(p) \, = \, 
\begin{pmatrix} 0 &  \bb{A}_{(\sN\sA,\sN\sA)}^{\sR}(p) \\ 
\bb{A}_{(\sN\sA,\sN\sA)}^{\sL}(p) &  0
\end{pmatrix},
 \hspace{0.65cm} \text{with} \hspace{0.5cm} 
\bb{A}^{\ms{a}}_{(\sN\sA,\sN\sA)}(p) = \bb{A}^{\ms{a}}_{\sN\sA}(p) \otimes \dot{\bb{A}}^{\ms{a}}_{\sN\sA}(p).
\end{equation} 
It is not clear if there is a hidden symmetry in this case.

\subsection{Vector boundaries} \label{part4-sec2-3} 

The totally supersymmetric boundary superalgebra $\mc{B}_{(\sT,\sT)}$ is the coideal subalgebra of $\mc{A}$ generated as
\begin{equation}
\mc{B}_{(\sT,\sT)} = \big\lan \mf{Q}_{\sL 1},\mf{S}_{\sL 1}, \mf{Q}_{\sL 2},\mf{S}_{\sL 2}, \mf{Q}_{\sR 1},\mf{S}_{\sR 1}, \mf{Q}_{\sR 2},\mf{S}_{\sR 2},\mf{H}_\sR, \mf{H}_\sL, \mf{P}, \mf{P}^\dag \big\ran,
\end{equation}
with $\mf{U},\mf{U}^{-1}\notin\mc{B}_{(\sT,\sT)}$. 

The vector boundary state transforms in the left- or right-moving representation, $\pi_{\sB}$ or $\bar{\pi}_{\sB}$, of the boundary Lie superalgebra $L(\mc{B}_{(\sT,\sT)}) = [\mf{psu}(1|1)^{2}\oplus \mf{u}(1)]^{2}_{\ms{c}}$. Table \ref{table-D1D5} shows that this is the boundary superalgebra of an open superstring on $AdS_{3}\times S^{3}\times T^{4}$ attached to the $Z=0$ D1- or D5-brane maximal giant graviton.

\paragraph{Boundary representations $\pi_\sB$ and $\bar{\pi}_\sB$.}

We define the vector spaces associated with left and right-moving boundary vector states in the same way as for the magnons in the bulk:
\[
\mc{V}_{\sB}={\rm span}_{\C}\{ \ket{\varphi^{+\dot{+}}_{\sB}}, \ket{\varphi^{+\dot{-}}_{\sB}}, \ket{\varphi^{-\dot{+}}_{\sB}}, \ket{\varphi^{-\dot{-}}_{\sB}} \},  \hspace{0.75cm} 
\bar{\mc{V}}_{\sB}={\rm span}_{\C}\{  \ket{\bar{\varphi}^{+\dot{+}}_{\sB}}, \ket{\bar{\varphi}^{+\dot{-}}_{\sB}}, \ket{\bar{\varphi}^{-\dot{+}}_{\sB}}, \ket{\bar{\varphi}^{-\dot{-}}_{\sB}} \},
\]  
both isomorphic to $\C^{1|1} \ot \dot{\C}^{1|1}$. We set $\mc{W}_{\sB} = \mc{V}_{\sB} \oplus \bar{\mc{V}}_{\sB}$. The left and right boundary representations $\pi_\sB : \mc{B}_{(\sT,\sT)} \to \End(\C^{1|1} \ot \dot{\C}^{1|1})$ and $\bar\pi_\sB : \mc{B}_{(\sT,\sT)} \to \End(\C^{1|1} \ot \dot{\C}^{1|1})$ 
are given by \eqref{Rep:L-T4} and \eqref{Rep:R-T4}, with the subindex $p$ replaced by $\SB$.  We choose the parametrization (\ref{abcd:boundary}) for $a_{\sB}$, $b_{\sB}$, $c_{\sB}$ and $d_{\sB}$, with $m_{\sB}$ the boundary mass parameter.

\paragraph{Boundary intertwining equations and $K$-matrix.}

The $K$-matrix $K(p, \SB) \in \End (\mc{W}_p \ot \mc{W}_{\sB})$ decomposes into a direct sum of $K$-matrices in four sectors $K^{\ms{ab}}_{\mc{B}_{(\sT,\sT)}}(p,\SB)$, which correspond to the partial $K$-matrices
$\bb{K}^{\ms{ab}}_{\mc{B}_{(\sT,\sT)}}(p,\SB) \in \End((\C^{1|1} \otimes \dot{\C}^{1|1})\otimes (\C^{1|1} \otimes \dot{\C}^{1|1}))$.
Here the complete $K$-matrix 
\begin{equation}
\bb{K}_{\mc{B}_{(\sT,\sT)}}(p,\SB) = \bb{K}^{\sL \sL}_{\mc{B}_{(\sT,\sT)}}(p,\SB) \oplus \bb{K}^{\sL \sR}_{\mc{B}_{(\sT,\sT)}}(p,\SB) \oplus \bb{K}^{\sR \sL}_{\mc{B}_{(\sT,\sT)}}(p,\SB) \oplus \bb{K}^{\sR \sR}_{\mc{B}_{(\sT,\sT)}}(p,\SB)
\end{equation}
satisfies the boundary intertwining equations (\ref{intw:b2}) for all $\mf{b}\in [\mf{psu}(1|1)^{2}\oplus \mf{u}(1)]^{2}_{\ms{c}}$, with $\pi_{p}$ and $\bar{\pi}_{p}$ the representations (\ref{Rep:L-T4}) and (\ref{Rep:R-T4}), and the reflected representations $\pi_{-p}$ and $\bar{\pi}_{-p}$ and boundary representations $\pi_{\sB}$ and $\bar{\pi}_{\sB}$ similarly defined.  The complete $K$-matrix is required to be unitary.

This $K$-matrix of a $\mf{psu}(1,1|2)^{2}$ spin-chain with a vector boundary can be built from two copies of the $K$-matrix of a $\mf{d}(2,1;\alpha)^{2}$ spin-chain with a similar boundary given in Section \ref{sec:4.3} -- which have partial $K$-matrices of the form
\[
\bb{K}^{\hspace{0.025cm}\ms{ab}}_{\mc{B}_{\sT}}(p,\SB) 
= \sum_{i, \hspace{0.03cm} j, \hspace{0.03cm} k, \hspace{0.03cm} l} (K^{\hspace{0.025cm}\ms{ab}}_{\mc{B}_{\sT}}(p,\SB))^{\hspace{0.05cm} i \hspace{0.05cm} k}_{\hspace{0.05cm} j \hspace{0.05cm} l} 
\hspace{0.2cm} \bb{E}_{ij}\ot \bb{E}_{kl}, \hspace{0.85cm}
\dot{\bb{K}}^{\hspace{0.025cm} \ms{ab}}_{\mc{B}_{\sT}}(p,\SB) 
= \sum_{\dot{i}, \hspace{0.03cm} \dot{j}, \hspace{0.03cm} \dot{k}, \hspace{0.03cm} \dot{l}} (K^{\hspace{0.025cm} \ms{ab}}_{\mc{B}_{\sT}}(p,\SB))^{\hspace{0.05cm} \dot{i} \hspace{0.05cm} \dot{k}}_{\hspace{0.05cm} \dot{j} \hspace{0.05cm} \dot{l}} 
\hspace{0.2cm} \bb{E}_{\dot{i} \dot{j}}\ot \bb{E}_{\dot{k} \dot{l}}.
\]
\vspace{-0.6cm}

The $K$-matrix $\bb{K}_{\mc{B}_{(\sT,\sT)}}(p,\SB)$ then has partial $K$-matrices
\begin{equation} 
\bb{K}^{\hspace{0.025cm}\ms{ab}}_{\mc{B}_{(\sT,\sT)}}(p,\SB) \, = \sum_{^{i,\dot{i},j,\dot{j},}_{k ,\dot{k},\ell,\dot{\ell}}} (K^{\hspace{0.025cm}\ms{ab}}_{\mc{B}_{(\sT,\sT)}}(p,\SB) )^{i \hspace{0.03cm} \dot{i}, \hspace{0.05cm} k \dot{k}}_{j \dot{j}, \hspace{0.05cm} \ell \hspace{0.01cm} \dot{\ell}} 
\hspace{0.2cm} (\bb{E}_{ij}\ot \bb{E}_{\dot{i}\dot{j}}) \otimes  (\bb{E}_{k\ell}\ot \bb{E}_{\dot{k} \dot{\ell}}),  
\end{equation}
\vspace{-0.6cm}

where
\begin{equation} (K^{\ms{ab}}_{\mc{B}_{(\sT,\sT)}}(p,\SB) )^{ i \hspace{0.04cm} \dot{i}, \hspace{0.05cm} k  \dot{k}}_{ j \dot{j}, \hspace{0.05cm} \ell \hspace{0.01cm} \dot{\ell}}  
\, = \, (-1)^{(k+\ell)(\dot{i}+\dot{j})} \hspace{0.15cm} (K^{\ms{ab}}_{\mc{B}_{\sT}}(p,\SB) )^{i \hspace{0.03cm} k}_{j \hspace{0.01cm} \ell} \hspace{0.15cm} 
(K^{\ms{ab}}_{\mc{B}_{\sT}}(p,\SB) )^{\dot{i} \hspace{0.04cm} \dot{k}}_{\dot{j} \hspace{0.01cm} \dot{\ell}},
\end{equation}
and satisfies the reflection equation (\ref{RE-complete}).


\part{Discussion}  \label{part5}

We have derived integrable boundary $S$-matrices which describe magnon scattering off vector and singlet boundaries for 
 $\mf{d}(2,1;\al)^{2}$ and $\mf{psu}(1,1|2)^{2}$ open spin-chains in AdS$_{3}$/CFT$_{2}$. These massive magnon excitations have $\mf{su}(1|1)^{2}_{\ms{c}}$ and $[\mf{psu}(1|1)^{2} \op \mf{u}(1)]^{2}_{\ms{c}}$ bulk symmetries, which are the level-0 Lie superalgebras of bulk Hopf superalgebras. The matrix parts of these boundary $S$-matrices are reflection $K$-matrices which are solutions of the boundary Yang-Baxter equation and the boundary intertwining equations associated with a coideal subalgebra $\mc{B}$ of the bulk superalgebra $\mc{A}$. 

\pagebreak

In the case of the $\mf{d}(2,1;\al)^{2}$ open spin-chain, we find {\it chiral integrable reflections} associated with
\vspace{-0.35cm}
\begin{itemize}
\item[$\ast$]{a totally supersymmetric boundary algebra, $\mc{B}_{\sT}$;}
\vspace{-0.2cm}
\item[$\ast$]{left and right half-supersymmetric boundary algebras, $\mc{B}_{\sL}$ and $\mc{B}_{\sR}$;} 
\vspace{-0.2cm}
\item[$\ast$]{a non-supersymmetric boundary algebra, $\mc{B}_{\sN\sC}$.}
\end{itemize} 
\vspace{-0.3cm}
We also derive an {\it achiral integrable reflection} corresponding to a non-supersymmetric boundary algebra $\mc{B}_{\sN\sA}$ generated by the magnon Hamiltonian only. These all match to D1- and D5-brane maximal giant graviton boundaries in 
$AdS_3 \times S^3 \times {S^{\prime}}^{3} \times S^1$.  We uncover a {\it hidden symmetry} which enhances the non-supersymmetric chiral boundary $\mc{B}_{\sN\sC}$ to a diagonally supersymmetric coideal subalgebra $\mc{B}_{\sD}$ of $\mc{A}$.  This hidden symmetry $\mc{B}_{\sD}$ has no known analogue in AdS$_{5}$/CFT$_{4}$.

In the case of the $\mf{psu}(1,1|2)^{2}$ open spin-chain, the integrable bulk $S$-matrix was found in \cite{BOSST13} to be essentially two copies of the bulk $S$-matrix of the $\mf{d}(2,1;\al)^{2}$ spin-chain \cite{BSS13}.  The same is true for the integrable boundary $S$-matrices.
We can put together two $K$-matrices for the $\mf{d}(2,1;\al)^{2}$ spin-chain, which are associated with boundary coideal subalgebras $\mc{B}_{1}$ and $\mc{B}_{2}$, to form a $K$-matrix for a $\mf{psu}(1,1|2)^{2}$ spin-chain corresponding to a boundary algebra denoted $\mc{B}_{(1,2)}$. In this way, we derived {\it chiral integrable reflections} associated with
\vspace{-0.35cm}
\begin{itemize}
\item[$\ast$]{a totally supersymmetric boundary algebra, $\mc{B}_{(\sT,\sT)}$;}
\vspace{-0.2cm}
\item[$\ast$]{left, right and mixed half-supersymmetric boundary algebras, $\mc{B}_{(\sL,\sL)}$, $\mc{B}_{(\sR,\sR)}$, $\mc{B}_{(\sL,\sR)}$ and $\mc{B}_{(\sR,\sL)}$;} 
\vspace{-0.2cm}
\item[$\ast$]{left and right quarter-supersymmetric boundary algebras, $\mc{B}_{(\sL,\sN\sC)}$, $\mc{B}_{(\sN\sC,\sL)}$,$\mc{B}_{(\sR,\sN\sC)}$ and $\mc{B}_{(\sN\sC,\sR)}$;} 
\vspace{-0.2cm}
\item[$\ast$]{a non-supersymmetric boundary algebra, $\mc{B}_{(\sN\sC,\sN\sC)}$.}
\end{itemize} 
\vspace{-0.3cm}
Now only $\mc{B}_{(\sT,\sT)}$,  $\mc{B}_{(\sL,\sL)}$, $\mc{B}_{(\sR,\sR)}$ and $\mc{B}_{(\sN\sC,\sN\sC)}$ have obvious interpretations as D1- and D5-brane maximal giant gravitons in $AdS_{3}\times S^{3}\times T^{4}$.  There are {\it hidden symmetries} enhancing the quarter supersymmetric and non-supersymmetric chiral boundary algebras to $\mc{B}_{(\sL,\sD)}$, $\mc{B}_{(\sD,\sL)}$,$\mc{B}_{(\sR,\sD)}$, $\mc{B}_{(\sD,\sR)}$ and
$\mc{B}_{(\sD,\sD)}$.  There is also an {\it achiral integrable reflection} with a non-supersymmetric boundary algebra $\mc{B}_{(\sN\sA,\sN\sA)}$, which now has no clear D-brane interpretation.

It is well known that $\mf{su}(1|1)^{2}_{\ms{c}}$ $R$-matrices can be identified with certain subsectors of the $\mf{su}(2|2)_{\ms{c}}$ $R$-matrix \cite{Beisert:2008, Beisert:2006, ALT-univ-blocks}. We find that our $K$-matrices for the totally supersymmetric boundary can be identified with the corresponding subsectors of the $K$-matrix associated with the $Z=0$ giant graviton \cite{HM:2007} and the right factor of the $K$-matrix associated with the D7-brane \cite{CY:2008}. The $K$-matrices for the half-supersymmetric boundaries can be identified with the corresponding subsectors of the $K$-matrix associated with the $Y=0$ giant graviton \cite{HM:2007} and its dual \cite{MR:2010}. The $K$-matrix for the non-supersymmetric chiral boundary is essentially two copies of the reflection matrix of \cite{Murgan:2008}, with its free parameter $a$ identified with our $c$ and $-\tfrac{1}{c}$ in these copies. For certain values of the parameter $c$, this $K$-matrix can also be identified with the corresponding subsectors of the left factor of the $K$-matrix associated with the D7-brane \cite{CY:2008}. The remaining $K$-matrices have no such analogues.

This work takes the first steps in an exploration of boundary integrability in AdS$_3$/CFT$_2$. There are many important questions which may now be addressed.  We expect to present the boundary crossing symmetry relations and an analysis of boundary bound-states in future research. The underlying boundary Yangian symmetries and the boundary Bethe equations in AdS$_3$/CFT$_2$ still remain to be studied. Recently, the authors of \cite{BSSS:2015} were able to incorporate massless magnons into the AdS$_3$/CFT$_2$ bulk scattering picture. It would be interesting to extend our analysis to include the scattering of massless excitations off integrable boundaries.  The Wilson loop computations of \cite{Drukker:2012,CMS:2012} relied upon results \cite{CRY:2011} from AdS$_5$/CFT$_4$ boundary scattering and our work may prove useful should similar computations be undertaken in AdS$_3$/CFT$_2$ dualities.  Finally, it would be interesting to study integrable boundaries for open superstrings on $AdS_{3}$ supergravity backgrounds with mixed NS-NS and R-R flux \cite{BSSS:2015,Cagnazzo:2012se,Hoare:2013pma,Hoare:2013ida,Delduc:2013qra,Arutynov:2014ota,Arutyunov:2013ega,Babichenko:2014yaa,Lloyd:2014bsa}.

\vspace{-0.25cm}
\section*{Acknowledgements}
\vspace{-0.25cm}

The authors would like to thank Rafael Nepomechie and Bogdan Stefa\'{n}ski for useful discussions.
V.R. thanks the EPSRC for a Postdoctoral Fellowship under the Grant Project No. EP/K031805/1, ``New Algebraic Structures Inspired by Gauge/Gravity Dualities''.
A.T. thanks the EPSRC for funding under the First Grant Project No. EP/K014412/1, ``Exotic Quantum Groups, Lie Superalgebras and Integrable Systems''.
AT also thanks the STFC for support under the Consolidated Grant Project No. ST/L000490/1, ``Fundamental Implications of Fields, Strings and Gravity''. A.P. and A.T. acknowledge useful conversations with the participants of the ESF and STFC supported workshop ``Permutations and Gauge-String Duality'' (under the HoloGrav network activity Grant No. 5124, and STFC Grant No. 4070083442) at Queen Mary College, University of London in July 2014. No new data was created during this study.

\appendix


\part{Appendices}


\section{Spinor Conventions}  \label{app:A}

Here we use the following 10D gamma matrices\footnote{We make use of the conventions of \cite{Babichenko:2010} with a rearrangement of the Pauli matrices.}
\begin{eqnarray}
\nonumber && \hspace{-0.65cm} 
\Gamma^{\mu} = \sigma^{1} \, \otimes \, \sigma^{2} \, \otimes \, \gamma^{\mu} \, \otimes \, \mathbb{I} \, \otimes \, \mathbb{I},   
\hspace{1.0cm}  \Gamma^{n} = \sigma^{1} \, \otimes \, \sigma^{1} \, \otimes \, \mathbb{I} \, \otimes \, \gamma^{n} \, \otimes \, \mathbb{I},   \\
&& \hspace{-0.65cm}
\Gamma^{\dot{n}} = \sigma^{1} \, \otimes \, \sigma^{3} \, \otimes \, \mathbb{I} \, \otimes \, \mathbb{I} \, \otimes \, \gamma^{\dot{n}},  \hspace{1.0cm}
 \Gamma^{9} \hspace{0.04cm} = \sigma^{2} \, \otimes \, \mathbb{I} \, \otimes \, \mathbb{I} \, \otimes \, \mathbb{I} \, \otimes \, \mathbb{I}, 
\end{eqnarray}
where we now choose $\gamma^{\mu} = (i \sigma^{3}, \sigma^{2}, \sigma^{1})$, $\gamma^{n} = (\sigma^{1}, \sigma^{3}, \sigma^{2})$ and
$\gamma^{\dot{n}} = (\sigma^{1}, \sigma^{3}, \sigma^{2})$.   Hence
\begin{equation}
 \Gamma^{012} =  \sigma^{1} \, \otimes \, \sigma^{2} \, \otimes \, \mathbb{I} \, \otimes \, \mathbb{I} \, \otimes \, \mathbb{I},  
 \hspace{0.35cm}
 \Gamma^{345} = - \hspace{0.05cm} i \hspace{0.15cm} \sigma^{1} \, \otimes \, \sigma^{1} \, \otimes \, \mathbb{I} \, \otimes \, \mathbb{I} \, \otimes \, \mathbb{I}, 
 \hspace{0.35cm}
 \Gamma^{678} = - \hspace{0.05cm} i \hspace{0.15cm} \sigma^{1} \, \otimes \, \sigma^{3} \, \otimes \, \mathbb{I} \, \otimes \, \mathbb{I} \, \otimes \, \mathbb{I}.
\end{equation}
The Weyl condition $\Gamma \varepsilon = \varepsilon$ is written in terms of the chirality matrix as
\begin{eqnarray}
 && \Gamma \equiv \Gamma_{0123456789}
 = \sigma^{3} \, \otimes \, \mathbb{I} \, \otimes \, \mathbb{I} \, \otimes \, \mathbb{I} \, \otimes \, \mathbb{I}.
\end{eqnarray}
The Majorana condition on left- and right-moving spinors is $(B\varepsilon^{\sL})^{*} = \varepsilon^{\sL}$ and 
$(B\varepsilon^{\sR})^{*} = -\varepsilon^{\sR}$, with
\begin{equation}
B \equiv \Gamma^{2} \, \Gamma^{5} \, \Gamma^{8} \, \Gamma^{9} =  - \, \sigma^{3} \, \otimes \, \mathbb{I} \, \otimes \, \sigma^{1} \, \otimes \, \sigma^{2} \, \otimes \, \sigma^{2},
\end{equation}
which satisfies $B \, \Gamma^{M} \, B^{-1} =  (\Gamma^{M} )^{\ast}$, with $B^{-1} = B$.  The charge conjugation matrix is defined to be
\begin{equation}
C \equiv B \, \Gamma^{0}  =  i \hspace{0.15cm} \sigma_{2} \otimes \sigma_{2} \otimes \sigma_{2} \otimes \sigma_{2} \otimes \sigma_{2},  
\end{equation}
satisfying $C \, \Gamma^{M} \, C^{-1} =  - \left(\Gamma_{M}\right)^{\ast}$, with $C^{-1} = - \, C$.
We can compute the bilinears
\begin{equation}
 \Gamma_{12} 
= - \hspace{0.05cm} i \hspace{0.15cm} \mathbb{I} \, \otimes \, \mathbb{I} \, \otimes \, \sigma^{3} \, \otimes \, \mathbb{I} \,  \otimes \, \mathbb{I}, 
\hspace{0.5cm} 
\Gamma_{35} 
= \, i \hspace{0.15cm} \mathbb{I} \, \otimes \, \mathbb{I} \, \otimes \, \mathbb{I} \, \otimes \, \sigma^{3} \,  \otimes \, \mathbb{I}, \hspace{0.5cm}
\Gamma_{68} 
= \, i \hspace{0.15cm} \mathbb{I} \, \otimes \, \mathbb{I} \, \otimes \, \mathbb{I} \, \otimes \, \mathbb{I} \,  \otimes \, \sigma^{3}. \hspace{0.1cm} \label{bilinears}
\end{equation}
We define kappa symmetry projection operators for the $AdS_{3} \times S^{3} \times S^{3} \times S^{1}$ background as
\begin{eqnarray}
\nonumber K^{\pm}(\alpha) & \equiv & \frac{1}{2} \left[ 1 \, \pm \, \left( \sqrt{\alpha} \hspace{0.15cm} \Gamma^{012} \, \Gamma^{345} + \sqrt{1-\alpha} \hspace{0.15cm} \Gamma^{012} \, \Gamma^{678} \right) \right] \\
& = & \frac{1}{2} \hspace{0.15cm} \mathbb{I} \, \otimes \, \left[ \mathbb{I} \, \mp \, \left( \sqrt{\alpha} \hspace{0.15cm} \sigma^{3} - \sqrt{1-\alpha} \hspace{0.15cm} \sigma^{1} \right) \right] \, \otimes \, \mathbb{I} \, \otimes \, \mathbb{I} \, \otimes \, \mathbb{I},
\end{eqnarray}
which are dependent on the parameter $\alpha$, which controls the relative size of the 3-spheres and appears also in the superconformal algebra 
$\mf{d}(2,1;\alpha)_{\sL} \oplus \mf{d}(2,1;\alpha)_{\sR} \oplus \mf{u}(1)$.  In the limit as $\alpha \to 1$, we obtain the kappa symmetry projectors for the 
the $AdS_{3} \times S^{3} \times T^{4}$ background
\begin{equation}  
K^{\pm} \equiv \frac{1}{2} \left( 1 \, \pm \, \Gamma^{012} \, \Gamma^{345} \right)
= \frac{1}{2} \hspace{0.15cm} \mathbb{I} \, \otimes \, \left( \mathbb{I} \, \mp \, \sigma^{3} \right) \, \otimes \, \mathbb{I} \, \otimes \, \mathbb{I} \, \otimes \, \mathbb{I},
\end{equation}
with the superconformal algebra now $\mf{psu}(1,1|2)_{\sL} \oplus \mf{psu}(1,1|2)_{\sR} \oplus \mf{u}(1)^{4}$.


\section{Representations of $\mf{d}(2,1;\al)$ and $\mf{psu}(1,1|2)$} \label{app:B}

\subsection{$\mf{d}(2,1;\al)$ superalgebra and BPS representations} 

Let us briefly review the representation theory of the exceptional Lie superalgebra $\mf{d}(2,1;\al)$ based on \cite{Babichenko:2010,BSS13,VanDerJeugt:1985hq}.
The bosonic subalgebra of $\mf{d}(2,1;\al)$ is $\mf{su}(1,1)\op\mf{su}(2)\op \mf{su}(2)^{\prime}$.  The bosonic generators are denoted
\[
\mf{J}_{\mu} \in \mf{su}(1,1), \hspace{0.85cm} \mf{L}_{m} \in \mf{su}(2), \hspace{0.85cm}  \mf{R}_{\dot{m}} \in \mf{su}(2)^{\prime},
\]
with $\mu \in \{0,\pm\}$, $m \in \{5,\pm\}$, $\dot{m} \in \{8,\pm\}$, and the fermionic generators $\mf{Q}_{b\beta\dot{\beta}}$, with $\pm$ indices.
The full $\mf{d}(2,1;\al)$ superalgebra is given by
\begin{eqnarray}
\nonumber && \hspace{-0.65cm} [\mf{J}_{0}, \, \mf{J}_{\pm}]=\pm \, \mf{J}_{\pm},  \hspace{1.15cm} 
[\mf{J}_{+}, \, \mf{J}_{-}] = 2 \hspace{0.05cm} \mf{J}_{0},  \hspace{0.985cm}
[\mf{J}_{0}, \, \mf{Q}_{\pm \hspace{0.01cm} \beta \hspace{0.01cm} \dot{\beta} }] = \pm \, \tfrac{1}{2} \hspace{0.05cm} \mf{Q}_{\pm \hspace{0.01cm} \beta \hspace{0.01cm} \dot{\beta}}, \hspace{0.85cm} 
[\mf{J}_{\pm}, \, \mf{Q}_{\mp\beta\dot{\beta}}] = \mf{Q}_{\pm\beta\dot{\beta}}, \\
\nonumber && \hspace{-0.65cm} [\mf{L}_{5}, \, \mf{L}_{\pm}]=\pm \, \mf{L}_{\pm},  \hspace{1.02cm} 
[\mf{L}_{+}, \, \mf{L}_{-}] = 2 \hspace{0.05cm} \mf{L}_{5}, \hspace{0.85cm}
[\mf{L}_{5}, \, \mf{Q}_{b \hspace{0.03cm} \pm \hspace{0.01cm} \dot{\beta}}]=\pm \, \tfrac{1}{2} \hspace{0.05cm} \mf{Q}_{b \hspace{0.03cm} \pm \hspace{0.01cm} \dot{\beta}},  \hspace{0.885cm} 
[\mf{L}_{\pm}, \, \mf{Q}_{b \hspace{0.03cm} \mp \hspace{0.01cm} \dot{\beta}}] = \mf{Q}_{b \hspace{0.03cm} \pm \hspace{0.01cm} \beta}, \\
\nonumber && \hspace{-0.65cm}  [\mf{R}_{8}, \, \mf{R}_{\pm}]=\pm \, \mf{R}_{\pm}, \hspace{0.84cm} [\mf{R}_{+}, \, \mf{R}_{-}] = \mf{R}_{8}, \hspace{0.895cm}
[\mf{R}_{8}, \, \mf{Q}_{b \hspace{0.03cm} \beta \hspace{0.01cm} \pm}]=\pm \, \tfrac{1}{2} \hspace{0.05cm} \mf{Q}_{b \hspace{0.03cm} \beta \hspace{0.01cm} \pm},  \hspace{0.85cm} 
[\mf{R}_{\pm}, \, \mf{Q}_{b \hspace{0.03cm} \beta \pm}] = \mf{Q}_{b \hspace{0.03cm} \beta \pm}, \\
\nonumber && \\
&& \hspace{-0.65cm} \{ \mf{Q}_{\pm + +}, \, \mf{Q}_{\pm - -} \} = \pm \hspace{0.1cm} \mf{J}_{\pm}, \hspace{4.32cm}
\{ \mf{Q}_{\pm + -}, \, \mf{Q}_{\pm - +} \} = \mp \hspace{0.1cm} \mf{J}_{\pm},   \\
\nonumber && \hspace{-0.65cm} \{ \mf{Q}_{+ \pm +}, \, \mf{Q}_{- \pm -} \} = \mp \, \alpha \, \mf{L}_{\pm}, \hspace{4.0cm}
\{ \mf{Q}_{+ \pm -},  \, \mf{Q}_{- \pm +} \} = \pm \, \alpha \hspace{0.1cm} \mf{L}_{\pm},   \\
\nonumber && \hspace{-0.65cm} \{ \mf{Q}_{+ + \pm}, \, \mf{Q}_{- - \pm} \} = \mp \, (1-\alpha) \hspace{0.1cm} \mf{R}_{\pm}, \hspace{2.96cm}
\{ \mf{Q}_{+ - \pm},  \, \mf{Q}_{- + \pm} \} = \pm \, (1-\alpha) \hspace{0.1cm} \mf{R}_{\pm},   \\
\nonumber && \hspace{-0.65cm} \{ \mf{Q}_{+ \pm \pm}, \, \mf{Q}_{-\mp\mp} \} = - \hspace{0.1cm} \mf{J}_{0} \pm \alpha \, \mf{L}_{5} \pm (1-\alpha) \, \mf{R}_{8}, \hspace{1.0cm}
\{ \mf{Q}_{+ \pm \mp}, \, \mf{Q}_{-\mp\pm} \} = \mf{J}_{0} \mp \alpha \, \mf{L}_{5} \pm (1-\alpha) \, \mf{R}_{8}.
\end{eqnarray}
The bosonic and fermionic generators $\{\mf{J}_{+}, \mf{L}_{+}, \mf{R}_{+}\}$ and $\{\mf{Q}_{+++}, \mf{Q}_{++-}, \mf{Q}_{+-+}, \mf{Q}_{+--}\}$ are raising operators.  The Cartan subalgebra is generated by $\{ \mf{H}, \mf{L}_{5}, \mf{R}_{8} \}$, with 
$\mf{H}=-\mf{J}_{0}-\alpha \mf{L}_{5}-(1-\alpha) \, \mf{R}_{8}$.

\paragraph{Half-BPS representation $(-\tfrac{\alpha}{2},\tfrac{1}{2},0)$.}
The module $M^{\alpha}(-\tfrac{\alpha}{2},\tfrac{1}{2},0)$ on which this representation of $\mf{d}(2,1;\alpha)$ acts is spanned by bosons $\ket{\phi_{\pm}^{(n)}}$, which transform in the $\bf{2}$ of $\mf{su}(2)$, and fermions $\ket{\psi_{\pm}^{(n)}}$, which transform in the $\bf{2}$ of $\mf{su}(2)^{\prime}$, both transforming non-trivially also under $\mf{su}(1,1)$.
The non-trivial action of $\mf{d}(2,1;\al)$ on this module is given by
\begin{eqnarray}
\nonumber && \hspace{-0.65cm} \hspace{0.51cm} \mf{L}_{5} \hspace{0.02cm} \ket{\phi_{\pm}^{(n)}} \hspace{0.02cm} = \pm \hspace{0.05cm} \tfrac{1}{2} \hspace{0.05cm} \ket{\phi_{\pm}^{(n)}}, \hspace{4.1cm} \mf{L}_\pm \ket{\phi_{\mp}^{(n)}} \hspace{0.01cm} = \ket{\phi_{\pm}^{(n)}},  \\
\nonumber &&  \hspace{-0.65cm} \hspace{0.52cm} \mf{R}_{8} \ket{\psi_{\pm}^{(n)}} = \pm \hspace{0.05cm} \tfrac{1}{2} \hspace{0.05cm} \ket{\psi_{\pm}^{(n)}}, \hspace{3.95cm}
 \mf{R}_\pm \ket{\psi_{\mp}^{(n)}} = \ket{\psi_\pm^{(n)}},  \\
\nonumber && \hspace{-0.65cm} \hspace{0.59cm} \mf{J}_0 \ket{\phi_{\beta}^{(n)}} \hspace{0.02cm} = - \hspace{0.05cm} ( \tfrac{\alpha}{2} + n ) \, \ket{\phi_{\beta}^{(n)}}, \hspace{3.15cm} 
 \mf{J}_0 \hspace{0.03cm} \ket{\psi_{\dot\beta}^{(n)}} = - \hspace{0.05cm} ( \tfrac{\alpha}{2} + \tfrac{1}{2} + n ) \hspace{0.05cm} \ket{\psi_{\dot\beta}^{(n)}},  \\
\nonumber && \hspace{-0.65cm} \hspace{0.5cm} \mf{J}_\pm \ket{\phi_{\beta}^{(n)}} \hspace{0.02cm} = \pm\sqrt{n(n -\tfrac{1}{2} \mp \tfrac{1}{2} + \alpha)} \hspace{0.1cm} \ket{\phi_{\beta}^{(n\mp1)}}, \hspace{0.75cm}
  \mf{J}_\pm \ket{\psi_{\dot\beta}^{(n)}} = \pm\sqrt{n(n +\tfrac{1}{2} \mp \tfrac{1}{2} + \alpha)} \, \ket{\psi_{\dot\beta}^{(n\mp1)}}, \\
\nonumber && \hspace{-0.65cm} \mf{Q}_{-\pm\dot\beta} \ket{\phi_{\mp}^{(n)}} \hspace{0.02cm} = \pm \sqrt{n+\alpha} \hspace{0.1cm} \ket{\psi_{\dot\beta}^{(n)}}, \hspace{2.57cm} \mf{Q}_{+\pm\dot\beta} \ket{\phi_{\mp}^{(n)}} \hspace{0.01cm} = \pm \sqrt{n} \hspace{0.1cm} \ket{\psi_{\dot\beta}^{(n-1)}},  \\
&& \hspace{-0.65cm} \mf{Q}_{-\beta\pm} \ket{\psi_{\mp}^{(n)}} = \mp \sqrt{n+1}\ket{\phi_{\beta}^{(n+1)}}, \hspace{2.38cm}
   \mf{Q}_{+\beta\pm} \ket{\psi_{\mp}^{(n)}} = \mp \sqrt{n+\alpha} \hspace{0.1cm} \ket{\phi_{\beta}^{(n)}}.
\end{eqnarray}
The highest weight state $\ket{\phi_{+}^{(0)}}$ is annihilated by all the raising operators, as well as by $\mf{R}_{-}$ and by two of the four fermionic lowering operators,
$\mf{Q}_{-++}$ and $\mf{Q}_{-+-}$.  This representation is therefore half-BPS with the shortening condition
\begin{equation} 
\{ \mf{Q}_{+ - \mp}, \, \mf{Q}_{- + \pm} \} \, \ket{\phi_{+}^{(0)}} =  (\mp \, \mf{J}_{0} \mp \alpha \, \mf{L}_{5} + (1-\alpha) \, \mf{R}_{8}) \, \ket{\phi_{+}^{(0)}} = 0, 
\end{equation}
which implies $\mf{H} \ket{\phi_{+}^{(0)}}  = 0$. The lowering operators act non-trivially on this highest weight state as
\begin{equation}
\mf{L}_{-} \ket{\phi_{+}^{(0)}} = \ket{\phi_{-}^{(0)}}, \hspace{0.35cm} \mf{J}_{-} \ket{\phi_{+}^{(0)}} = - \ket{\phi_{+}^{(1)}}, \hspace{0.35cm}
\mf{Q}_{--+} \ket{\phi_{+}^{(0)}} = - \sqrt{\alpha} \ket{\psi_{+}^{(0)}}, \hspace{0.35cm} \mf{Q}_{---} \ket{\phi_{+}^{(0)}} = - \sqrt{\alpha} \ket{\psi_{-}^{(0)}}.
\end{equation}

\paragraph{Half-BPS representation $(-\tfrac{1-\alpha}{2},0,\tfrac{1}{2})$.}
The module $M^{\prime \, (1-\alpha)}(-\tfrac{1-\alpha}{2},\tfrac{1}{2},0)$ on which this representation of $\mf{d}(2,1;\alpha)$ acts is spanned by bosons $\ket{\phi_{\pm}^{\prime \, (n)}}$, which transform in the $\bf{2}$ of $\mf{su}(2)^{\prime}$, and fermions $\ket{\psi_{\pm}^{\prime \,(n)}}$, which transform in the $\bf{2}$ of $\mf{su}(2)$, both transforming non-trivially under $\mf{su}(1,1)$.
The non-trivial action of $\mf{d}(2,1;\al)$ on this module is given by
\begin{eqnarray}
\nonumber && \hspace{-0.65cm} \hspace{0.51cm} \mf{R}_{8} \hspace{0.02cm} \ket{\phi_{\pm}^{\prime \,(n)}} \hspace{0.02cm} = \pm \hspace{0.05cm} \tfrac{1}{2} \hspace{0.05cm} \ket{\phi_{\pm}^{\prime \,(n)}}, \hspace{3.96cm} \mf{R}_\pm \ket{\phi_{\mp}^{\prime \,(n)}} \hspace{0.01cm} = \ket{\phi_{\pm}^{\prime \,(n)}},  \\
\nonumber &&  \hspace{-0.65cm} \hspace{0.48cm} \mf{L}_{5} \ket{\psi_{\pm}^{\prime \,(n)}} = \pm \hspace{0.05cm} \tfrac{1}{2} \hspace{0.05cm} \ket{\psi_{\pm}^{\prime \,(n)}}, \hspace{4.1cm} \mf{L}_\pm \ket{\psi_{\mp}^{\prime \,(n)}} = \ket{\psi_\pm^{\prime \, (n)}},  \\
\nonumber && \hspace{-0.65cm} \hspace{0.53cm} \mf{J}_0 \ket{\phi_{\dot{\beta}}^{\prime \,(n)}} \hspace{0.02cm} = - \hspace{0.05cm} ( \tfrac{1-\alpha}{2} + n ) \, \ket{\phi_{\dot{\beta}}^{\prime \,(n)}}, \hspace{2.81cm} 
 \mf{J}_0 \hspace{0.03cm} \ket{\psi_{\beta}^{\prime \,(n)}} = - \hspace{0.05cm} ( \tfrac{1-\alpha}{2} + \tfrac{1}{2} + n ) \hspace{0.1cm} \ket{\psi_{\beta}^{\prime \,(n)}},  \\
\nonumber && \hspace{-0.65cm} \hspace{0.5cm} \mf{J}_\pm \ket{\phi_{\beta}^{\prime \,(n)}} \hspace{0.02cm} = \pm\sqrt{n(n +\tfrac{1}{2} \mp \tfrac{1}{2} - \alpha)} \hspace{0.1cm} \ket{\phi_{\beta}^{\prime \,(n\mp1)}}, \hspace{0.75cm}
  \mf{J}_\pm \ket{\psi_{\dot\beta}^{\prime \,(n)}} = \pm\sqrt{n(n +\tfrac{3}{2} \mp \tfrac{1}{2} - \alpha)} \hspace{0.1cm} \ket{\psi_{\dot\beta}^{\prime \,(n\mp1)}}, \\
\nonumber && \hspace{-0.65cm} \mf{Q}_{-\beta\pm} \ket{\phi_{\mp}^{\prime \,(n)}} \hspace{0.02cm} = \pm \sqrt{n+1-\alpha} \hspace{0.1cm} \ket{\psi_{\beta}^{\prime \,(n)}}, \hspace{1.89cm} \mf{Q}_{+\beta\pm} \ket{\phi_{\mp}^{\prime \,(n)}} \hspace{0.01cm} = \pm \sqrt{n} \hspace{0.1cm} \ket{\psi_{\beta}^{\prime \,(n-1)}},  \\
&& \hspace{-0.65cm} \mf{Q}_{-\pm\dot{\beta}} \ket{\psi_{\mp}^{\prime \,(n)}} = \mp \sqrt{n+1} \hspace{0.1cm} \ket{\phi_{\dot{\beta}}^{\prime \,(n+1)}}, \hspace{2.25cm} \mf{Q}_{+\pm\dot{\beta}} \ket{\psi_{\mp}^{\prime \,(n)}} = \mp \sqrt{n+1-\alpha} \hspace{0.1cm} \ket{\phi_{\dot{\beta}}^{\prime \,(n)}}.
\end{eqnarray}
The highest weight state $\ket{\phi_{+}^{\prime \,(0)}}$ is annihilated by all the raising operators, as well as by $\mf{L}_{-}$ and by two of the four fermionic lowering operators, $\mf{Q}_{-++}$ and $\mf{Q}_{--+}$.  
 This representation is therefore half-BPS with the shortening condition
\begin{equation} 
\{ \mf{Q}_{+ \mp -}, \, \mf{Q}_{- \pm +} \} \, \ket{\phi_{+}^{\prime \,(0)}} =  (\mp \, \mf{J}_{0} - \alpha \, \mf{L}_{5} \mp (1-\alpha) \, \mf{R}_{8}) \, \ket{\phi_{+}^{\prime \,(0)}} = 0, 
\end{equation}
which implies $\mf{H} \ket{\phi_{+}^{(0)}}  = 0$.
The lowering operators act non-trivially on this highest weight state as
\begin{eqnarray}
&& \hspace{-0.75cm} \nonumber 
\mf{L}_{-} \ket{\phi_{+}^{\prime \,(0)}} = \ket{\phi_{-}^{\prime \,(0)}}, \hspace{3.2cm} 
\mf{J}_{-} \ket{\phi_{+}^{\prime \,(0)}} = - \ket{\phi_{+}^{\prime \,(1)}}, \\
&& \hspace{-0.75cm}
\mf{Q}_{-+-} \ket{\phi_{+}^{\prime \,(0)}} = - \sqrt{1-\alpha} \, \ket{\psi_{+}^{\prime \,(0)}}, \hspace{1.0cm} \mf{Q}_{---} \ket{\phi_{+}^{\prime \,(0)}} = - \sqrt{1-\alpha} \, \ket{\psi_{-}^{\prime \,(0)}}.
\end{eqnarray}

\paragraph{Quarter-BPS representation $(-\tfrac{\alpha}{2},\tfrac{1}{2},0) \otimes (-\tfrac{1-\alpha}{2},0,\tfrac{1}{2})$.}

The tensor product representation of the two half-BPS representations consists of vectors in the module $M=M^{\alpha} \otimes M^{\prime \,(1-\alpha)}$. The highest weight state is $\ket{\phi_{+}^{(0)} \hspace{0.05cm} \phi_{+}^{\prime \,(0)}}$  which is annihilated by all the raising operators and one of the four fermionic lowering operators, $\mf{Q}_{-++}$.  This representation is quarter-BPS with the shortening condition
\begin{equation} 
\{ \mf{Q}_{+ - -}, \, \mf{Q}_{- + +} \} \, \ket{\phi_{+}^{(0)} \hspace{0.05cm} \phi_{+}^{\prime \,(0)}} = - \, ( \mf{J}_{0} + \alpha \, \mf{L}_{5} + (1-\alpha) \, \mf{R}_{8}) \, \ket{\phi_{+}^{(0)} \hspace{0.05cm} \phi_{+}^{\prime \,(0)}} 
= - \, \mf{H} \, \ket{\phi_{+}^{(0)} \hspace{0.05cm} \phi_{+}^{\prime \,(0)}} = 0, \hspace{0.1cm}
\end{equation}
and the other elements of the Cartan subalgebra act on the highest weight state as
\begin{equation}
\mf{L}_{5} \ket{\phi_{+}^{(0)} \hspace{0.05cm} \phi_{+}^{\prime \,(0)}} = \tfrac{1}{2} \hspace{0.05cm} \ket{\phi_{+}^{(0)} \hspace{0.05cm} \phi_{+}^{\prime \,(0)}}, \hspace{1.0cm} 
\mf{R}_{8} \ket{\phi_{+}^{(0)} \hspace{0.05cm} \phi_{+}^{\prime \,(0)}} = \tfrac{1}{2} \hspace{0.05cm} \ket{\phi_{+}^{(0)} \hspace{0.05cm} \phi_{+}^{\prime \,(0)}}. \end{equation}
The non-trivial action of the bosonic lowering operators on this highest weight state is
\begin{eqnarray} 
&& \hspace{-0.75cm} \nonumber \mf{J}_{-} \ket{\phi_{+}^{(0)} \hspace{0.05cm} \phi_{+}^{\prime \,(0)}} 
= - \ket{\phi_{+}^{(1)} \hspace{0.05cm} \phi_{+}^{\prime \,(0)}} - \ket{\phi_{+}^{(0)} \hspace{0.05cm} \phi_{+}^{\prime \,(1)}}, \\
&& \hspace{-0.75cm} \mf{L}_{-} \ket{\phi_{+}^{(0)} \hspace{0.05cm} \phi_{+}^{\prime \,(0)}} = \ket{\phi_{-}^{(0)} \hspace{0.05cm} \phi_{+}^{\prime \,(0)}}, \hspace{0.85cm}
\mf{R}_{-} \ket{\phi_{+}^{(0)} \hspace{0.05cm} \phi_{+}^{\prime \,(0)}} = \ket{\phi_{+}^{(0)} \hspace{0.05cm} \phi_{-}^{\prime \,(0)}},
\end{eqnarray}
and the non-trivial action of the fermionic lowering operators is
\begin{eqnarray} 
\nonumber && \hspace{-0.75cm} \mf{Q}_{--+} \ket{\phi_{+}^{(0)} \hspace{0.05cm} \phi_{+}^{\prime \,(0)}} = - \sqrt{\alpha} \hspace{0.15cm} 
\ket{\psi_{+}^{(0)} \hspace{0.05cm} \phi_{+}^{\prime \,(0)}}, \\
\nonumber && \hspace{-0.75cm} \mf{Q}_{-+-} \ket{\phi_{+}^{(0)} \hspace{0.05cm} \phi_{+}^{\prime \,(0)}} = - \sqrt{1-\alpha} \hspace{0.15cm} 
\ket{\phi_{+}^{(0)} \hspace{0.05cm} \psi_{+}^{\prime \,(0)}}, \\
&& \hspace{-0.75cm} \mf{Q}_{---}  \ket{\phi_{+}^{(0)} \hspace{0.05cm} \phi_{+}^{\prime \,(0)}} 
= -\sqrt{\alpha} \, \ket{\psi_{-}^{(0)} \hspace{0.05cm} \phi_{+}^{\prime \,(0)}} - \sqrt{1-\alpha} \hspace{0.15cm} 
\ket{\phi_{+}^{(0)} \hspace{0.05cm} \psi_{-}^{\prime \,(0)}}.
\end{eqnarray}

\subsection{$\mf{psu}(1,1|2)$ superalgebra and BPS representations} 

Here we review, based on \cite{BOSST13}, the representation theory of the Lie superalgebra $\mf{psu}(1,1|2)$, which has
the bosonic subalgebra of $\mf{d}(2,1;\al)$ is $\mf{su}(1,1)\op\mf{su}(2)$.  The bosonic generators are
\[
\mf{J}_{\mu} \in \mf{su}(1,1), \hspace{1.0cm} \mf{L}_{m} \in \mf{su}(2),
\]
with $\mu \in \{0,\pm\}$, $m \in \{5,\pm\}$, and the fermionic generators $\mf{Q}_{b\beta\dot{\beta}}$, with $\pm$ indices.
The full $\mf{psu}(1,1|2)$ superalgebra is  
\begin{eqnarray}
\nonumber && \hspace{-0.75cm} [\mf{J}_{0}, \, \mf{J}_{\pm}]=\pm \, \mf{J}_{\pm},  \hspace{0.99cm} 
[\mf{J}_{+}, \, \mf{J}_{-}] = 2 \hspace{0.05cm} \mf{J}_{0},  \hspace{0.98cm}
[\mf{J}_{0}, \, \mf{Q}_{\pm \hspace{0.01cm} \beta \hspace{0.01cm} \dot{\beta} }] = \pm \, \tfrac{1}{2} \hspace{0.05cm} \mf{Q}_{\pm \hspace{0.01cm} \beta \hspace{0.01cm} \dot{\beta}}, \hspace{0.85cm} 
[\mf{J}_{\pm}, \, \mf{Q}_{\mp\beta\dot{\beta}}] = \mf{Q}_{\pm\beta\dot{\beta}}, \\
\nonumber && \hspace{-0.75cm} [\mf{L}_{5}, \, \mf{L}_{\pm}]=\pm \, \mf{L}_{\pm},  \hspace{0.85cm} 
[\mf{L}_{+}, \, \mf{L}_{-}] = 2 \hspace{0.05cm} \mf{L}_{5}, \hspace{0.85cm}
[\mf{L}_{5}, \, \mf{Q}_{b \hspace{0.03cm} \pm \hspace{0.01cm} \dot{\beta}}]=\pm \, \tfrac{1}{2} \hspace{0.05cm} \mf{Q}_{b \hspace{0.03cm} \pm \hspace{0.01cm} \dot{\beta}},  \hspace{0.88cm} [\mf{L}_{\pm}, \, \mf{Q}_{b \hspace{0.03cm} \mp \hspace{0.01cm} \dot{\beta}}] = \mf{Q}_{b \hspace{0.03cm} \pm \hspace{0.01cm} \beta}, \\
\nonumber && \\
\nonumber && \hspace{-0.75cm} \{ \mf{Q}_{\pm + +}, \, \mf{Q}_{\pm - -} \} = \pm \hspace{0.1cm} \mf{J}_{\pm}, \hspace{3.24cm}
\{ \mf{Q}_{\pm + -}, \, \mf{Q}_{\pm - +} \} = \mp \hspace{0.1cm} \mf{J}_{\pm},   \\
\nonumber && \hspace{-0.75cm} \{ \mf{Q}_{+ \pm +}, \, \mf{Q}_{- \pm -} \} = \mp \, \mf{L}_{\pm}, \hspace{3.22cm}
\{ \mf{Q}_{+ \pm -},  \, \mf{Q}_{- \pm +} \} = \pm \hspace{0.1cm} \mf{L}_{\pm},   \\
&& \hspace{-0.75cm} \{ \mf{Q}_{+ \pm \pm}, \, \mf{Q}_{-\mp\mp} \} = - \hspace{0.1cm} \mf{J}_{0} \pm \mf{L}_{5}, \hspace{2.42cm}
\{ \mf{Q}_{+ \pm \mp}, \, \mf{Q}_{-\mp\pm} \} = \mf{J}_{0} \mp \mf{L}_{5}.
\end{eqnarray}
The bosonic generators $\{ \mf{J}_{+}, \mf{L}_{+}\}$ and fermionic generators $\{ \mf{Q}_{+++}, \mf{Q}_{++-}, \mf{Q}_{+-+}, \mf{Q}_{+--} \}$ are the raising operators.  The Cartan subalgebra is generated by $\{ \mf{H}, \mf{L}_{5} \}$, with $\mf{H}=-(\mf{J}_{0} + \mf{L}_{5})$.
There is also a $\mf{u}(1)$ automorphism $\mf{R}_{8}$ which satisfies
\begin{equation} 
[ \mf{R}_{8}, \mf{Q}_{b\beta\dot{\beta}} ] = \pm \, \tfrac{1}{2} \hspace{0.05cm} \mf{Q}_{b\beta\dot{\beta}}. 
\end{equation}
This $\mf{psu}(1,1|2)$ superalgebra can be seen as the $\alpha \to 1$ limit of the $\mf{d}(2,1;\al)$ superalgebra.

\paragraph{Half-BPS representation $(-\tfrac{1}{2},\tfrac{1}{2})$.}
This representation of $\mf{psu}(1,1|2)$ acts on the module $M(-\tfrac{1}{2},\tfrac{1}{2})$ which is spanned by bosons $\ket{\phi_{\pm}^{(n)}}$, transforming in the $\bf{2}$ of $\mf{su}(2)$, and fermions $\ket{\psi_{\pm}^{(n)}}$ which are singlets. Both transform non-trivially under $\mf{su}(1,1)$.
The superalgebra $\mf{psu}(1,1|2)$  acts as
\begin{eqnarray}
\nonumber && \hspace{-0.65cm} \hspace{0.51cm} \mf{L}_{5} \hspace{0.02cm} \ket{\phi_{\pm}^{(n)}} \hspace{0.02cm} = \pm \hspace{0.05cm} \tfrac{1}{2} \hspace{0.05cm} \ket{\phi_{\pm}^{(n)}}, \hspace{3.24cm} \mf{L}_\pm \ket{\phi_{\mp}^{(n)}} \hspace{0.01cm} = \ket{\phi_{\pm}^{(n)}},  \\
\nonumber && \hspace{-0.65cm} \hspace{0.59cm} \mf{J}_0 \ket{\phi_{\beta}^{(n)}} \hspace{0.02cm} = - \hspace{0.05cm} ( \tfrac{1}{2} + n ) \, \ket{\phi_{\beta}^{(n)}}, \hspace{2.3cm} 
 \mf{J}_0 \hspace{0.03cm} \ket{\psi_{\dot\beta}^{(n)}} = - \hspace{0.05cm} ( 1 + n ) \hspace{0.05cm} \ket{\psi_{\dot\beta}^{(n)}},  \\
\nonumber && \hspace{-0.65cm} \hspace{0.5cm} \mf{J}_\pm \ket{\phi_{\beta}^{(n)}} \hspace{0.02cm} = \pm \hspace{0.05cm} (n + \tfrac{1}{2} \mp \tfrac{1}{2}) \, \ket{\phi_{\beta}^{(n\mp1)}}, \hspace{1.2cm}
  \mf{J}_\pm \ket{\psi_{\dot\beta}^{(n)}} = \pm\sqrt{(n + \tfrac{1}{2} \mp \tfrac{1}{2})(n + \tfrac{3}{2} \mp \tfrac{1}{2})} \, \ket{\psi_{\dot\beta}^{(n\mp1)}}, \\
\nonumber && \hspace{-0.65cm} \mf{Q}_{-\pm\dot\beta} \ket{\phi_{\mp}^{(n)}} \hspace{0.02cm} = \pm \sqrt{n+1} \hspace{0.1cm} \ket{\psi_{\dot\beta}^{(n)}}, \hspace{1.76cm} \mf{Q}_{+\pm\dot\beta} \ket{\phi_{\mp}^{(n)}} \hspace{0.01cm} = \pm \sqrt{n} \hspace{0.1cm} \ket{\psi_{\dot\beta}^{(n-1)}},  \\
 && \hspace{-0.65cm} \mf{Q}_{-\beta\pm} \ket{\psi_{\mp}^{(n)}} = \mp \sqrt{n+1}\ket{\phi_{\beta}^{(n+1)}}, \hspace{1.505cm}
   \mf{Q}_{+\beta\pm} \ket{\psi_{\mp}^{(n)}} = \mp \sqrt{n+1} \hspace{0.1cm} \ket{\phi_{\beta}^{(n)}}.
\end{eqnarray}
The highest weight state $\ket{\phi_{+}^{(0)}}$ is annihilated by the raising operators, as well as by two of the four fermionic lowering operators,
$\mf{Q}_{-++}$ and $\mf{Q}_{-+-}$.  This representation is half-BPS with the shortening condition  
\begin{equation} 
\{ \mf{Q}_{+ - \mp}, \, \mf{Q}_{- + \pm} \} \, \ket{\phi_{+}^{(0)}} = \mp \hspace{0.05cm} (\mf{J}_{0} + \mf{L}_{5}) \, \ket{\phi_{+}^{(0)}} 
= \mp \hspace{0.05cm} \mf{H} \, \ket{\phi_{+}^{(0)}} = 0.
\end{equation}
The lowering operators act non-trivially on this highest weight state as
\begin{equation} \hspace{-0.18cm}
\mf{L}_{-} \ket{\phi_{+}^{(0)}} = \ket{\phi_{-}^{(0)}}, \hspace{0.3cm} \mf{J}_{-} \ket{\phi_{+}^{(0)}} = - \ket{\phi_{+}^{(1)}}, \hspace{0.3cm}
\mf{Q}_{--+} \ket{\phi_{+}^{(0)}} = - \ket{\psi_{+}^{(0)}}, \hspace{0.3cm} \mf{Q}_{---} \ket{\phi_{+}^{(0)}} = - \ket{\psi_{-}^{(0)}}.
\end{equation}


\section{Bosonic Symmetries}  \label{app:C} 

\subsection{$SO(2,2)$ isometry group}

\paragraph{$\mf{so}(2,2)$ splitting.}
Let us specify the Lie algebra splitting $\mf{so}(2,2) = \mf{su}(1,1)_{\sL} \oplus \mf{su}(1,1)_{\sR}$.  Here $SO(2,2)$ group transformations
act on $x^{\mu} = (x^1, x^2, x^3, x^4) \in AdS_{3} \subset \mathbb{R}^{2+2}$.  If we combine the vector components into a quaternion
\begin{equation}
\underline{x} \equiv x^{\mu} \hspace{0.025cm} \tau_{\mu}, \hspace{0.75cm} \text{with} \hspace{0.5cm} 
\tau_{\mu} = (i \hspace{0.05cm} \mathbb{I}, \, i \hspace{0.025cm}\sigma_3, \, \sigma_1, \, \sigma_2) = (i \hspace{0.05cm} \mathbb{I}, \, t_{k}),
\end{equation}
then $SO(2,2)$ transformations can be realised as a $SU(1,1)_{\sL} \times SU(1,1)_{\sR}$ transformation
\[
\underline{x} \longrightarrow U_\sL \,\, \underline{x} \,\, U_\sR^{-1} \, \approx \, \underline{x} \, + \, \delta \underline{x}, \hspace{0.5cm} \text{with} \hspace{0.35cm}
U_{\sL} = e^{\alpha^{k} \,  t_{k}} \in SU(2)_{\sL} \hspace{0.35cm} \text{and} \hspace{0.35cm}
U_{\sR} = e^{- \,\bar{\alpha}^{k} \, t_{k}} \in SU(2)_{\sR},
\]
where thus
\begin{equation}  
\delta\underline{x} \, = \alpha^{k} \, t_{k} \, \underline{x} - \bar{\alpha}^{k} \, \underline{x} \, t_{k} 
\, = \, \delta x^{\mu} \,\tau_{\mu}. 
\end{equation}
The double-covering nature of this relationship corresponds to the fact that $(\pm \, U_\sL, \pm \, U_\sR)$ generate the same $SO(2,2)$ rotation.
The rotation angles and boost parameters are
\begin{eqnarray}
\nonumber && \hspace{-0.75cm}  \theta_{12} = -\theta_{21} = \alpha^{3} - \bar{\alpha}^{3}, \hspace{1.59cm}
\beta_{14} = \beta_{14} = \alpha^{2} - \bar{\alpha}^{2}, \hspace{1.59cm}
  \beta_{13} = \beta_{13} = \alpha^{1} - \bar{\alpha}^{1}, \\
&& \hspace{-0.75cm} \theta_{34} = -\theta_{43} = -(\alpha^{3} + \bar{\alpha}^{3}),  \hspace{1.0cm}
 \beta_{24} = \beta_{42} = -(\alpha^{2} + \bar{\alpha}^{2}), \hspace{1.0cm}   
\beta_{23} = \beta_{32} = \alpha^{1} + \bar{\alpha}^{1}. \hspace{0.75cm}
\end{eqnarray}
Here $e^{i \,\alpha^{3}  \sigma_{3}}$ and $e^{i \,\bar{\alpha}^{3} \sigma_{3}}$ are in the Cartan subgroups $U(1)_{\sL}\subset SU(1,1)_{\sL}$ and $U(1)_{\sR}\subset SU(1,1)_{\sR}$.  
The $\mf{u}(1)\oplus \mf{u}(1)$ generators of rotations by $\theta_{12}$ and $\theta_{34}$ can be written as $\mf{J}_{\sL \, 0} - \mf{J}_{\sR \, 0}$ and  $-(\mf{J}_{\sL \, 0} + \mf{J}_{\sR \, 0})$  in terms of the left and right generators, $\mf{J}_{\ms{a} \, 0}$, of the Cartan subalgebra $\mf{u}(1)_{\sL} \oplus \mf{u}(1)_{\sR}$.

\subsection{$SO(4)$ isometry group}

\paragraph{$\mf{so}(4)$ splitting.}
Let us specify the Lie algebra splitting $\mf{so}(4) \sim \mf{su}(2)_{\sL} \oplus \mf{su}(2)_{\sR}$.   A $SO(4)$ rotation acts on $x^{\tK} = (x^1, x^2, x^3, x^4) \in S^3 \subset \mathbb{R}^4$. If we combine the vector components into a quaternion
\begin{equation}
\underline{x} \equiv x^{\tK} \hspace{0.025cm} \tau_{\tK}, \hspace{0.75cm} \text{with} \hspace{0.5cm} 
\tau_{\tK} = (\mathbb{I}, \, i\sigma_3, \, i\sigma_1, \, i\sigma_2), 
\end{equation}
then any rotation in $SO(4)$ can be realised (in two ways) as a $SU(2)_{\sL} \times SU(2)_{\sR}$ transformation
\[
\underline{x} \longrightarrow U_\sL \,\, \underline{x} \,\, U_\sR^{-1} \, \approx \, \underline{x} \, + \, \delta \underline{x}, \hspace{0.5cm} \text{with} \hspace{0.35cm}
U_{\sL} = e^{i \,\alpha^{k}  \sigma_{k}} \in SU(2)_{\sL} \hspace{0.35cm} \text{and} \hspace{0.35cm}
U_{\sR} = e^{i \,\bar{\alpha}^{k} \sigma_{k}} \in SU(2)_{\sR},
\]
where
\begin{equation} 
\delta\underline{x} \, = i \, ( \alpha^{k} \, \sigma_{k} \, \underline{x} -  \bar{\alpha}^{k} \, \underline{x} \, \sigma_{k} )
\, = \, \delta x^{\tK} \,\tau_{\tK}. 
\end{equation}
The rotation angles are given by
\begin{eqnarray}
\nonumber && \hspace{-0.75cm}  \theta_{12} = -\theta_{21} = \alpha^{3} - \bar{\alpha}^{3}, \hspace{1.39cm}  
\theta_{14} = -\theta_{41} = \alpha^{2} - \bar{\alpha}^{2}, \hspace{1.39cm}
\theta_{13} = -\theta_{42} = \alpha^{1} - \bar{\alpha}^{1},  \\
&& \hspace{-0.75cm} \theta_{34} = -\theta_{43} = -(\alpha^{3} + \bar{\alpha}^{3}), \hspace{0.8cm}
 \theta_{23} = -\theta_{32} = - (\alpha^{2} + \bar{\alpha}^{2}), \hspace{0.8cm}
\theta_{24} = -\theta_{42} = \alpha^{1} + \bar{\alpha}^{1}. \hspace{0.75cm}
\end{eqnarray}
Here $e^{i \,\alpha^{3}  \sigma_{3}}$ and $e^{i \,\bar{\alpha}^{3} \sigma_{3}}$ are elements of the Cartan subgroups $U(1)_{\sL} \subset SU(2)_{\sL}$ and $U(1)_{\sR} \subset SU(2)_{\sR}$. 
The $\mf{u}(1)\oplus \mf{u}(1)$ generators of rotations by $\theta_{12}$ and $\theta_{34}$ can be written as $\mf{L}_{\sL \, 5} - \mf{L}_{\sL \, 5}$ and  $-(\mf{L}_{\sL \, 5} + \mf{L}_{\sR \, 5})$ in terms of the left and right generators, $\mf{L}_{\ms{a} \, 5}$, of the Cartan subalgebra $\mf{u}(1)_{\sL} \oplus \mf{u}(1)_{\sR}$.

\paragraph{$\mf{so}(4)^{\prime}$ splitting.}

There is a similar splitting $\mf{so}(4)^{\prime} \sim \mf{su}(2)_{\sL}^{\prime} \oplus \mf{su}(2)_{\sR}^{\prime}$ with the $\mf{u}(1)\oplus \mf{u}(1)$ generators, $\mf{R}_{\sL \, 5} - \mf{R}_{\sL \, 5}$ and  $-(\mf{R}_{\sL \, 5} + \mf{R}_{\sR \, 5})$, of rotations by $\theta_{12}^{\prime}$ and $\theta_{34}^{\prime}$ written in terms of left and right generators, $\mf{R}_{\ms{a} \, 8}$, of the Cartan subalgebra $\mf{u}(1)_{\sL}^{\prime} \oplus \mf{u}(1)_{\sR}^{\prime}$.

\paragraph{$\mf{d}(2,1;\alpha)^{2}$ spin-chain fields.}

This $\mf{d}(2,1;\alpha)^{2}$ spin-chain described in Chapter \ref{part3-sec1} contains left- and right-moving fields which transform under $\mf{so}(4) \oplus \mf{so}(4)^{\prime} \sim (\mf{su}(2)_{\sL} \oplus \mf{su}(2)_{\sR}) \oplus (\mf{su}(2)^{\prime}_{\sL} \oplus \mf{su}(2)^{\prime}_{\sR})$.

Let us define $Z = x_{1} + i x_{2}$ and $Y = x_{3} + i x_{4}$, which transform under $SO(4)$.  We notice that
\begin{equation} 
\underline{x} = \begin{pmatrix} x_{1} + ix_{2} & i \, (x_{3} - ix_{4}) \\ i \, (x_{3} + ix_{4}) & x_{1} - ix_{2} \end{pmatrix} 
= \begin{pmatrix} Z & i \hspace{0.025cm} \bar{Y} \\ i \hspace{0.025cm} Y & \bar{Z} \end{pmatrix}. 
\end{equation}
We can write a similar relation for $Z^{\prime} = x^{\prime}_{1} + i x^{\prime}_{2}$ and $Y^{\prime} = x^{\prime}_{3} + i x^{\prime}_{4}$ transforming under $SO(4)^{\prime}$.

Now $\phi_{\sL \, \beta}^{(0)} = (\phi_{\sL \, +}^{(0)}, \phi_{\sL \, -}^{(0)})$ and 
$\phi_{\sR \, \dot{\beta}}^{(0)} = (\phi_{\sR \, +}^{(0)}, \phi_{\sR \, -}^{(0)})$ transform non-trivially under $SU(2)_{\sL}$ and $SU(2)_{\sR}$, respectively.
Thus $Z$ and $Y$ (and their complex conjugates $\bar{Z}$ and $\bar{Y}$) transform under the rotational symmetry $SO(4) \sim SU(2)_{\sL} \times SU(2)_{\sR}$ as
\[ Z \sim \begin{pmatrix} \phi^{(0)}_{\sL \, +} \\ \phi^{(0)}_{\sR \, +} \end{pmatrix},   \hspace{0.75cm} 
\bar{Z} \sim \begin{pmatrix} \phi^{(0)}_{\sL \, -} \\ \phi^{(0)}_{\sR \, -} \end{pmatrix},  \hspace{0.75cm}  
Y \sim \begin{pmatrix} \phi^{(0)}_{\sL \, -} \\ \phi^{(0)}_{\sR \, +} \end{pmatrix},  \hspace{0.75cm} 
\bar{Y} \sim \begin{pmatrix} \phi^{(0)}_{\sL \, +} \\ \phi^{(0)}_{\sR \, -} \end{pmatrix} \]
in the notation of the fields in our double-row spin-chain. Similarly, for the primed fields associated with the $SO(4)'$ rotational symmetry group, 
\[ Z' \sim \begin{pmatrix} \phi^{\prime \, (0)}_{\sL \, +} \\ \phi^{\prime \, (0)}_{\sR \, +} \end{pmatrix}, \hspace{0.75cm} 
\bar{Z}' \sim \begin{pmatrix} \phi^{\prime \, (0)}_{\sL \, -} \\ \phi^{\prime \, (0)}_{\sR \, -} \end{pmatrix}, \hspace{0.75cm} 
Y' \sim \begin{pmatrix} \phi^{\prime \, (0)}_{\sL \, -} \\ \phi^{\prime \, (0)}_{\sR \, +} \end{pmatrix}, \hspace{0.75cm}  
\bar{Y}' \sim \begin{pmatrix} \phi^{\prime \, (0)}_{\sL \, +} \\ \phi^{\prime \, (0)}_{\sR \, -} \end{pmatrix}. \]
The vacuum of the $\mf{d}(2,1;\alpha)^2$ spin-chain therefore transforms as
\[ \mathcal{Z} = {\small\begin{pmatrix} Z_{\sL} \\ Z_{\sR} \end{pmatrix}} = \begin{pmatrix} (\phi^{(0)}_{\sL \, +},\phi^{\prime \, (0)}_{\sL \, +}) \\ (\phi^{(0)}_{\sR \, +},\phi^{\prime \, (0)}_{\sR \, +}) \end{pmatrix} \sim  ZZ', \]
while the fundamental bosonic excitations transform as
\begin{eqnarray} 
&& \hspace{-0.75cm} \nonumber \phi = {\small\begin{pmatrix} \phi_{\sL} \\ Z_{\sR} \end{pmatrix}} = \begin{pmatrix} (\phi^{(0)}_{\sL \, -},\phi^{\prime \, (0)}_{\sL \, +}), \\ (\phi^{(0)}_{\sR \, +},\phi^{\prime \, (0)}_{\sR \, +}) \end{pmatrix} \sim YZ',  \hspace{1.0cm}
\phi^{\prime} = {\small\begin{pmatrix} \phi^{\prime}_{\sL} \\ Z_{\sR} \end{pmatrix}} = \begin{pmatrix} (\phi^{(0)}_{\sL \, +},\phi^{\prime \, (0)}_{\sL \, -}) \\ (\phi^{(0)}_{\sR \, +},\phi^{\prime \,(0)}_{\sR \, +}) \end{pmatrix} 
 \sim ZY', \\
&& \hspace{-0.75cm} \nonumber \bar{\phi} = {\small\begin{pmatrix} Z_{\sL} \\ \phi_{\sR} \end{pmatrix}} = \begin{pmatrix} (\phi^{(0)}_{\sL \, -},\phi^{\prime \, (0)}_{\sL \, +}) \\ (\phi^{(0)}_{\sR \, +},\phi^{\prime \, (0)}_{\sR \, +}) \end{pmatrix} \sim \bar{Y} Z', \hspace{1.1cm}
\bar{\phi}^{\prime} = {\small\begin{pmatrix} Z_{\sL} \\ \phi^{\prime}_{\sR} \end{pmatrix}} = \begin{pmatrix} (\phi^{(0)}_{\sL \, +},\phi^{\prime \, (0)}_{\sL \, -}) \\ (\phi^{(0)}_{\sR \, +},\phi^{\prime \, (0)}_{\sR \, +}) \end{pmatrix}  \sim Z \bar{Y'}.
\end{eqnarray}
Here a composite state of the $\phi$ and $\phi^{\prime}$ excitations would transform as $Y Y^{\prime}$.

\paragraph{$\mf{psu}(1,1|2)^{2}$ spin-chain fields.}

The $\mf{psu}(1,1|2)^{2}$ spin-chain described in Chapter \ref{part4-sec1} contains left and right-moving fields which transform under $\mf{so}(4) \sim \mf{su}(2)_{\sL} \oplus \mf{su}(2)_{\sR}$.

Defining again $Z = x_{1} + i x_{2}$ and $Y = x_{3} + i x_{4}$, we find that the vacuum of the $\mf{psu}(1,1|2)^2$ spin-chain transforms as
\[ \mathcal{Z} = {\small\begin{pmatrix} Z_{\sL} \\ Z_{\sR} \end{pmatrix}} = \begin{pmatrix} \phi^{(0)}_{\sL \, +}  \\ \phi^{(0)}_{\sR \, +} \end{pmatrix} \sim  Z, \]
while, for the first bosonic excitations,
\[ 
\varphi^{+\dot{+}} = {\small\begin{pmatrix} \varphi^{+\dot{+}}_{\sL} \\ Z_{\sR} \end{pmatrix}} = \begin{pmatrix} \phi^{(0)}_{\sL \, -} \\ \phi^{(0)}_{\sR \, +} \end{pmatrix} \sim Y,  \hspace{1.0cm}
\bar{\varphi}^{+\dot{+}} = {\small\begin{pmatrix} Z_{\sL} \\ \varphi^{+\dot{+}}_{\sR}  \end{pmatrix}} = \begin{pmatrix} \phi^{(0)}_{\sL \, +} \\ \phi^{(0)}_{\sR \, -} \end{pmatrix} \sim \bar{Y}.
\]





\begin{thebibliography}{99} 

\newcommand{\nlin}[2]{{\tt [\href{http://xxx.lanl.gov/abs/nlin/#2}{\tt nlin.#1/#2}]}}
\newcommand{\hepth}[1]{{\tt [\href{http://xxx.lanl.gov/abs/hep-th/#1}{\tt hep-th/#1}]}}
\newcommand{\arx}[1]{{\tt [\href{http://arxiv.org/abs/#1}{\tt arXiv:#1}]}}

\bibitem{Frolov:2009}
G.~Arutyunov, S.~Frolov, 
{\it Foundations of the $AdS_{5} \times S^{5}$ superstring: Part I}, J. Phys. {\bf A 42} (2009) 254003 
\arx{0901.4937}.

\bibitem{Review}
N.~Beisert, C.~Ahn, L.~F.~Alday, Z.~Bajnok et al.,
{\it Review of AdS/CFT Integrability: An Overview},
Lett. Math. Phys. {\bf 99} (2012) 3
\hepth{1012.3982}. 

\bibitem{Maldacena:1997}
J.~M.~Maldacena,
{\it The large N limit of superconformal field theories and supergravity},
Adv. Theor. Math. Phys. {\bf 2} (1998) 231
\hepth{9711200}.

\bibitem{Drinfeld}
V.~G.~Drinfeld, 
{\it Quantum groups}, 
Proceedings of the International Congress of Mathematicians, Berkley, 1986 (AMS, Providence, 1987) 798.

\bibitem{Mackay04}
N.~Mackay,
{\it Introduction to Yangian symmetry in integrable field theory},
Int. J. Mod. Phys. {\bf A 20} (2005) 7189
\hepth{0409183}.

\bibitem{Beisert:2008}
N.~Beisert,
{\it The $SU(2|2)$ dynamic S-matrix},
Adv. Theor. Math. Phys. {\bf 12} (2008) 948
\hepth{0511082}.

\bibitem{HM:2006}
D.~M.~Hofman, J.~M.~Maldacena,
{\it Giant magnons},
J. Phys. {\bf A 39} (2006) 13095
\hepth{0604135}.

\bibitem{HM:2007}
D.~M.~Hofman, J.~M.~Maldacena, 
{\it Reflecting magnons}, 
J. High Energy Phys. {\bf 11} (2007) 050 
\arx{0708.2272}.

\bibitem{CY:2008}
D.~H.~Correa, C.~A.~S.~Young, 
{\it Reflecting magnons from D7 and D5 branes}, 
J.~Phys. {\bf A 41} (2008) 455401 
\arx{0808.0452}.

\bibitem{Murgan:2008}
R.~Murgan, R.~I.~Nepomechie, 
{\it Open-chain transfer matrices for AdS/CFT},
J. High Energy Phys. {\bf 09} (2008) 085 
\arx{0808.2629}. 

\bibitem{CRY:2011}
D.~H.~Correa, V.~Regelskis, C.~A.~S.~Young, 
{\it Integrable achiral D5-brane reflections and asymptotic Bethe equations}, 
J.\ Phys.\ {\bf A 44} (2011) 325403 
\arx{1105.3707}.

\bibitem{Sk:1988}
E.~Sklyanin,
{\it Boundary conditions for integrable quantum systems},
J. Phys. {\bf A 21} (1988) 2375.
 	
\bibitem{Babichenko:2010}
A.~Babichenko, B.~Stefa\'{n}ski, Jr., K.~Zarembo, 
{\it Integrability and the $AdS_{3}/CFT_{2}$ correspondence}, 
J. High Energy Phys. {\bf 03} (2010) 058
\arx{0912.1723}.

\bibitem{Strominger:2004} 
S.~Gukov, E.~Martinec, G.~W.~Moore, A.~Strominger,	
{\it The Search for a holographic dual to $AdS_{3} \times S^{3} \times S^{3} \times S^{1}$},
Adv. Theor. Math. Phys. {\bf 9} (2005) 435  
\hepth{0403090}.

\bibitem{Pakman:2009}
A.~Pakman, L.~Rastelli, S.~S.~Razamat,
{\it A spin-chain for the symmetric product CFT$_{2}$},
J. High Energy Phys. {\bf 05} (2010) 099
\arx{0912.0959}.

\bibitem{SSS-fieldtheory:2014}
O.~O.~Sax, A.~Sfondrini, B.~Stefa\'{n}ski, Jr.,
{\it Integrability and the conformal field theory of the Higgs branch}
\arx{1411.3676}. 

\bibitem{Tong:2014}
D.~Tong,
{\it The holographic dual of $AdS_{3} \times S^{3} \times S^{3} \times S^{1}$},
J. High Energy Phys. {\bf 04} (2014) 193
\arx{1402.5135}.

\bibitem{Cowdall:1998}
P.~M.~Cowdall, P.~K.~Townsend,
{\it Gauged supergravity vacua from intersecting branes},
Phys. Lett. {\bf B 429} (1998) 281; {\bf B 434} (1998) 458(E)
\hepth{9801165}.

\bibitem{Gauntlett:1998}
J.~P.~Gauntlett, R.~C.~Myers, P.~K.~Townsend,
{\it Supersymmetry of rotating branes},
Phys. Rev. {\bf D 59} (1998) 025001
\hepth{9809065}.

\bibitem{BSS13}
R.~Borsato, O.~O.~Sax, A.~Sfondrini, 
{\it A dynamic $su(1|1)^2$ S-matrix for AdS$_{3}$/CFT$_{2}$}, 
J. High Energy Phys. {\bf 04} (2013) 113 
\arx{1211.5119}.

\bibitem{BOSST13}
R.~Borsato, O.~O.~Sax, A.~Sfondrini, B.~Stefa\'{n}ski, Jr., A.~Torrielli,
{\it The all-loop integrable spin-chain for strings on $AdS_3\times S^3\times T^4$: The massive sector}, 
J. High Energy Phys. {\bf 08} (2013) 043
\arx{1303.5995}.

\bibitem{Sfondrini:2014}
A.~Sfondrini, 
{\it Towards integrability for AdS$_{3}$/CFT$_{2}$}, 
J. Phys. {\bf A 48} (2015) 023001 
\arx{1406.2971}. 

\bibitem{David:2008yk}
J.~R.~David, B.~Sahoo,
{\it Giant magnons in the D1-D5 system},
J. High Energy Phys. {\bf 07} (2008) 033
\arx{0804.3267}.

\bibitem{David:2010yg}
J.~R.~David, B.~Sahoo,
{\it S-matrix for magnons in the D1-D5 system},
J. High Energy Phys. {\bf 10} (2010) 112
\arx{1005.0501}.

\bibitem{AB:2013}
C. Ahn, D. Bombardelli,
{\it Exact S-matrices for AdS$_{3}$/CFT$_{2}$},
Int. J. Mod. Phys. {\bf A 28} (2013) 1350168
\arx{1211.4512}.
 
\bibitem{Rughoonauth:2012qd}
N.~Rughoonauth, P.~Sundin, L.~Wulff,
{\it Near BMN dynamics of the $AdS_3 \times S^3 \times S^3 \times S^1$ superstring},
J. High Energy Phys. {\bf 07} (2012) 159
\arx{1204.4742}.

\bibitem{Sundin:2012gc}
P.~Sundin, L.~Wulff,
{\it Classical integrability and quantum aspects of the $AdS_3 \times S^3 \times S^3 \times S^1$ superstring},
J. High Energy Phys. {\bf 10} (2012) 109
\arx{1207.5531}.
  
\bibitem{Abbott:2012dd}
M.~C.~Abbott,
{\it Comment on strings in $AdS_3 \times S^3 \times S^3 \times S^1$ at one loop},
J. High Energy Phys. {\bf 02} (2013) 102
\arx{1211.5587}.

\bibitem{Beccaria:2012kb}
M.~Beccaria, F.~Levkovich-Maslyuk, G.~Macorini, A.~A.~Tseytlin,
{\it Quantum corrections to spinning superstrings in $AdS_3 \times S^3 \times M^4$: determining the dressing phase},
J. High Energy Phys. {\bf 04} (2013) 006
\arx{1211.6090}.
  
\bibitem{Beccaria:2012pm}
M.~Beccaria, G.~Macorini,
{\it Quantum corrections to short folded superstring in $AdS_3 \times S^3 \times M^4$},
J. High Energy Phys. {\bf 03} (2013) 040
\arx{1212.5672}.

\bibitem{Sundin:2013ypa}
P.~Sundin, L.~Wulff,
{\it Worldsheet scattering in AdS$_{3}$/CFT$_{2}$},
J. High Energy Phys. {\bf 07} (2013) 007
\arx{1302.5349}.
 
\bibitem{Bianchi:2013nra}
L.~Bianchi, V.~Forini, B.~Hoare,
{\it Two-dimensional S-matrices from unitarity cuts},
J. High Energy Phys. {\bf 07} (2013) 088
\arx{1304.1798}.

\bibitem{Engelund:2013fja}
O.~T.~Engelund, R.~W.~McKeown, R.~Roiban,
{\it Generalized unitarity and the worldsheet $S$-matrix in $AdS_n \times S^n \times M^{10-2n}$},
J. High Energy Phys. {\bf 08} (2013) 023
\arx{1304.4281}.

\bibitem{Abbott:2013ixa}
M.~C.~Abbott,
{\it The $AdS_3 \times S^3 \times S^3 \times S^1$ Hern\'{a}ndez-L\'{o}pez phases: A semiclassical derivation},
J.\  Phys. {\bf A 46} (2013) 445401
\arx{1306.5106}.

\bibitem{Sundin:2013uca}
P.~Sundin, L.~Wulff,
{\it The low energy limit of the $AdS_3 \times S^3 \times M^4$ spinning string},
J. High Energy Phys. {\bf 10} (2013) 111
\arx{1306.6918}.

\bibitem{Bianchi:2014rfa}
L.~Bianchi, B.~Hoare,
{\it $AdS_3 \times S^3 \times M^4$ string S-matrices from unitarity cuts},
J. High Energy Phys. {\bf 08} (2014) 097
\arx{1405.7947}.
  
\bibitem{Wulff:2015mwa}
L.~Wulff,
{\it On integrability of strings on symmetric spaces}
\arx{1505.03525}.

\bibitem{Borsato:2013hoa}
R.~Borsato, O.~O.~Sax, A.~Sfondrini, B.~Stefa\'{n}ski, Jr., A.~Torrielli,
{\it  Dressing phases of AdS$_{3}$/CFT$_{2}$},
Phys.\ Rev.\ {\bf D 88} (2013) 066004
\arx{1306.2512}.

\bibitem{Abbott:2014pia}
M.~C.~Abbott, I.~Aniceto,
{\it  An improved AFS phase for AdS$_3$ string integrability},
Phys.\ Lett.\ {\bf B 743} (2015) 61
\arx{1412.6863}.
  
\bibitem{SST13}
O.~O.~Sax, B.~Stefa\'{n}ski, Jr., A.~Torrielli,
{\it On the massless modes of the AdS$_{3}$/CFT$_{2}$ integrable systems}, J. High Energy Phys. {\bf 03} (2013) 109
\arx{1211.1952}.

\bibitem{Lloyd:2013wza}
T.~Lloyd, B.~Stefa\'{n}ski, Jr.,
{\it $AdS_3/CFT_2$, finite-gap equations and massless modes},
J. High Energy Phys. {\bf 04} (2014) 179
\arx{1312.3268}.

\bibitem{Borsato:2014}
R.~Borsato, O.~O.~Sax, A.~Sfondrini, B.~Stefa\'{n}ski, Jr.,
{\it All-loop worldsheet S matrix for $AdS_3 \times S^3 \times T^4$}, 
Phys. Rev. Lett. {\bf 113} (2014) 131601
\arx{1403.4543}.

\bibitem{BSS14}
R.~Borsato, O.~O.~Sax, A.~Sfondrini, B.~Stefa\'{n}ski, Jr.,
{\it The complete $AdS_3 \times S^3 \times T^4$ worldsheet S-matrix},
J. High Energy Phys. {\bf 10} (2014) 66
\arx{1406.0453}.

\bibitem{Abbott:2014rca}
M.~C.~Abbott, I.~Aniceto,
{\it Macroscopic (and microscopic) massless modes},
Nucl.\ Phys. {\bf B 894} (2015) 75
\arx{1412.6380}.
  
\bibitem{GH06}
C.~Gomez, R.~Hernandez,
{\it The magnon kinematics of the AdS/CFT correspondence}, 
J. High Energy Phys. {\bf 11} (2006) 021 
\hepth{0608029}.
  
\bibitem{PST06}
J.~Plefka, F.~Spill, A.~Torrielli,
{\it On the Hopf algebra structure of the AdS/CFT S-matrix}, 
Phys. Rev. {\bf D 74} (2006) 066008 
\hepth{0608038}.

\bibitem{Be:2006}
N.~Beisert,
{\it The S-matrix of AdS/CFT and Yangian symmetry}, Proceedings of Science, Solvay (2006) 002
\arx{0704.0400}.

\bibitem{ST:2009}
F.~Spill, A.~Torrielli,	
{\it On Drinfeld's second realization of the AdS/CFT su(2$|$2) Yangian},
J. Geom. Phys. {\bf 59} (2009) 489
\arx{0803.3194}.

\bibitem{BL:2014}
N.~Beisert, M.~de Leeuw,
{\it The RTT-realization for the deformed gl(2$|$2) Yangian},
J. Phys. {\bf A 47} (2014) 305201 
\arx{1401.7691}.

\bibitem{PTW14}
A.~Pittelli, A.~Torrielli, M.~Wolf,
{\it Secret symmetries of type IIB superstring theory on $AdS_{3} \times S^{3} \times M^{4}$},
J. Phys. {\bf A 47} (2014)  455402
\arx{1406.2840}.

\bibitem{Re}
V. Regelskis,
{\it Yangian of $AdS_3/CFT_2$ and its deformation} 
\arx{1503.03799}.

\bibitem{ALT:2009}
G.~Arutyunov, M.~de Leeuw, A.~Torrielli,
{\it The bound state S-matrix for $AdS_5 \times S_5$ superstring}
Nucl. Phys. {\bf B 819} (2009) 319350 
\arx{0902.0183}.

\bibitem{Sax:2011}
O.~O.~Sax, B.~Stefa\'{n}ski, Jr.,
{\it Integrability, spin-chains and the AdS$_{3}$/CFT$_{2}$ correspondence}, 
J. High Energy Phys. {\bf 08} (2011) 029
\arx{1106.2558}.

\bibitem{Borsatz:2012:all-loop} 	
R.~Borsato, O.~O.~Sax, A.~Sfondrini,
{\it All-loop Bethe ansatz equations for AdS$_{3}$/CFT$_{2}$}, 
J. High Energy Phys. {\bf 04} (2013) 116
\arx{1212.0505}.

\bibitem{DS:2009}
A.~Doikou, K.~Sfetsos,
\textit{Contracted and expanded integrable structures},
J. Phys. {\bf A 42} (2009) 475204
\arx{0904.3437}. 

\bibitem{AN:2010}	
C.~Ahn, R.~Nepomechie,
{\it Yangian symmetry and bound states in AdS/CFT boundary scattering},    
J. High Energy Phys. {\bf 05} (2010) 05 
\arx{1003.3361}.
   
\bibitem{MR:2010}
N.~MacKay, V.~Regelskis, 
{\it Yangian symmetry of the Y=0 maximal giant graviton}, 
J. High Energy Phys. {\bf 12} (2010) 076
\arx{1010.3761}.

\bibitem{MR:2012}
N.~MacKay, V.~Regelskis, 
\textit{Reflection algebra, Yangian symmetry and bound-states in AdS/CFT}, 
J. High Energy Phys. {\bf 01} (2012) 134 
\arx{1101.6062}.

\bibitem{Pa:2011}
L.~Palla, 
{\it Yangian symmetry of boundary scattering in AdS/CFT and the explicit form of bound-state reflection matrices},
J. High Energy Phys. {\bf 04} (2011) 022 
\arx{1102.0122}.

\bibitem{MR:2011}
N.~MacKay, V.~Regelskis,
\textit{Achiral boundaries and the twisted Yangian of the D5-brane},
J. High Energy Phys. {\bf 08} (2011) 019 
\arx{1105.4128}.
 
\bibitem{Ne:2009}
R.~Nepomechie,
{\it Bethe ansatz equations for open spin chains from giant gravitons},
J. High Energy Phys. {\bf 05} (2009) 100 
\arx{0903.1646}.
	
\bibitem{Ga:2009}
W.~Galleas,
{\it The Bethe ansatz equations for reflecting magnons},
Nucl. Phys. {\bf B 820} (2009) 664681 
\arx{0902.1681}.
	
\bibitem{CY:2010:Bethe}
D.~H.~Correa, C.~A.~S.~Young,
{\it Asymptotic Bethe equations for open boundaries in planar AdS/CFT},
J. Phys. {\bf A 43} (2010) 145401 
\arx{0912.0627}.

\bibitem{Nepomechie-et-al:2015}
X.~Zhang, J.~Cao, S.~Cui, R.~I.~Nepomechie, W-L.~Yang, K.~Shi, Y.~Wang,
{\it Bethe ansatz for an AdS/CFT open spin chain with non-diagonal boundaries},
J. High Energy Phys. {\bf 10} (2015) 133
\arx{1507.08866}.
 			
\bibitem{Raju-et-al:2008}
G.~Mandal, S.~Raju, M.~Smedback, 
{\it Supersymmetric giant graviton solutions in AdS$_{3}$},
Phys. Rev. {\bf D 77} (2008) 046011 
\arx{0709.1168}.

\bibitem{Janssen:2005}
B.~Janssen, Y.~Lozano, D.~Rodriguez-Gomez,
{\it Giant gravitons in $AdS_{3} \times S^{3} \times T^{4}$ as fuzzy cylinders},
Nucl. Phys. {\bf B 711} (2005) 392
\hepth{0406148}.

\bibitem{Prinsloo:2014}
A.~Prinsloo,
{\it D1 and D5-brane giant gravitons on $AdS_{3} \times S^{3} \times S^{3} \times S^{1}$}, 
J. High Energy Phys. {\bf 12} (2014) 094
\arx{1406.6134}.

\bibitem{GZ93}
S.~Ghoshal, A.~Zamolodchikov,
{\it Boundary S-matrix and boundary state in two-dimensional integrable quantum field theory},
Int. J. Mod. Phys. {\bf A 09} (1994) 3841;  {\bf A 09} (1994) 4353(E)
\hepth{9306002}.

\bibitem{ABR07}
C.~Ahn, D.~Bak, S-J.~Rey,
{\it Reflecting magnon bound states},
J. High Energy Phys. {\bf 04} (2008) 50
\arx{0712.4144}.

\bibitem{Beisert:2006}
N.~Beisert,
{\it An $SU(1|1)$-invariant $S$-matrix with dynamic representations},
Bulg. J. Phys. {\bf 33S1} (2006) 371
\hepth{0511013}.

\bibitem{ALT-univ-blocks}
G.~Arutyunov, M.~de Leeuw, A.~Torrielli,
{\it Universal blocks of the AdS/CFT scattering matrix},
J. High Energy Phys. {\bf 05} (2009) 086
\arx{0903.1833}.

\bibitem{BSSS:2015}
R.~Borsato, O.O.~Sax, A.~Sfondrini, B.~Stefa\'{n}ski, Jr.,
{\it The $AdS_{3}\times S_{3} \times S_{3} \times S_{1}$ worldsheet S-matrix},
J. Phys. {\bf A 48} (2015) 415401
\arx{1506.00218}.

\bibitem{Drukker:2012}
N.~Drukker,
{\it Integrable Wilson loops},
J. High Energy Phys. {\bf 10} (2013) 135
\arx{1203.1617}. 

\bibitem{CMS:2012}
D.~Correa, J.~Maldacena, A.~Sever,
{\it The quark anti-quark potential and the cusp anomalous dimension from a TBA equation},
J. High Energy Phys. {\bf 08} (2012) 134
\arx{1203.1913}.

\bibitem{Cagnazzo:2012se}
A.~Cagnazzo, K.~Zarembo,
{\it B-field in AdS$_3$/CFT$_2$ correspondence and integrability},
J. High Energy Phys. {\bf 11} (2012) 133
\arx{1209.4049}.

\bibitem{Hoare:2013pma}
B.~Hoare, A.~A.~Tseytlin,
{\it On string theory on $AdS_3 \times S^3 \times T^4$ with mixed 3-form flux: Tree-level S-matrix},
Nucl.\ Phys. {\bf B 873} (2013) 682
\arx{1303.1037}.

\bibitem{Hoare:2013ida}
B.~Hoare, A.~A.~Tseytlin,
{\it Massive S-matrix of $AdS_3 \times S^3 \times T^4$ superstring theory with mixed 3-form flux},
Nucl.\ Phys. {\bf B 873} (2013) 395
\arx{1304.4099}. 

\bibitem{Delduc:2013qra}
F.~Delduc, M.~Magro, B.~Vicedo,
{\it An integrable deformation of the $AdS_5 \times S^5$ superstring action},
Phys.\ Rev.\ Lett.\  {\bf 112} (2014)  051601
\arx{1309.5850}.

\bibitem{Arutynov:2014ota}
G.~Arutyunov, M.~de Leeuw, S.~J.~van Tongeren,
{\it The exact spectrum and mirror duality of the $(AdS_{5} \times S^5)_{\eta}$ superstring},
Theor.\ Math.\ Phys.\  {\bf 182} (2015)  23
\arx{1403.6104}.

\bibitem{Arutyunov:2013ega}
G.~Arutyunov, R.~Borsato, S.~Frolov,
{\it S-matrix for strings on $\eta$-deformed $AdS_{5} \times S^5$},
J. High Energy Phys. {\bf 04} (2014) 002
\arx{1312.3542}.

\bibitem{Babichenko:2014yaa}
A.~Babichenko, A.~Dekel, O.~O.~Sax,
{\it Finite-gap equations for strings on $AdS_{3} \times S^{3} \times T^{4}$ with mixed 3-form flux},
J. High Energy Phys. {\bf 11} (2014) 122
\arx{1405.6087}.

\bibitem{Lloyd:2014bsa}
T.~Lloyd, O.~O.~Sax, A.~Sfondrini, B.~Stefa\'{n}ski, Jr.,
{\it The complete worldsheet S matrix of superstrings on $AdS_3 \times S^3 \times T^4$ with mixed three-form flux},
Nucl.\ Phys. {\bf B 891} (2015) 570
\arx{1410.0866}.
 	
\bibitem{VanDerJeugt:1985hq}
J.~Van der Jeugt, 
{\it Irreducible representations of the exceptional Lie superalgebras D(2,1,a)}, 
J.\ Math.\ Phys.\  {\bf 26} (1985) 913.
 	
\end{thebibliography}
\end{document}